\documentclass[12pt]{book}
\usepackage{graphicx}
\usepackage{epsfig}
\usepackage{bm}
\usepackage{fancyhdr}
\usepackage{longtable}

\newcommand{\text}{\rm}
\def\>{{\rangle}}
\def\<{{\langle}} 
\def\ba{\bm{a}}
\def\bk{{\bm{k}}}
\def\bq{{\bm{q}}}
\def\bp{{\bm{p}}}
\def\bn{{\bm{n}}}
\def\bN{{\bm{N}}}

\def\muhat{\hat{\bm{\mu}}}

\def\nnn{{NNN}}
\def\bmm{{\bm{m}}}
\def\br{{\bm{r}}}

\def\Tr{{\rm Tr}\,}
\def\tr{{\text tr}\,}
\def\Det{{\text Det}\,}
\def\Eq#1{Eq.~(\ref{#1})}
\def\e{{\text e}}
\def\o{{\text o}}

\renewcommand\Im{{\text Im}\,}
\renewcommand\Re{{\text Re}\,}

\renewcommand{\chaptermark}[1]{         
\markboth{\chaptername\ \thechapter.\ #1}{}} %
\renewcommand{\sectionmark}[1]{         
\markright{\thesection.\ #1}}         %
\makeatletter
\def\cleardoublepage{\clearpage\if@twoside \ifodd\c@page\else%
\hbox{}%
\thispagestyle{empty}
\newpage%
\if@twocolumn\hbox{}\newpage\fi\fi\fi}
\makeatother

\begin{document}

\bibliographystyle{unsrt}

\pagestyle{empty}

\begin{center}

\vskip 1cm

{\Huge  Dense Baryonic Matter in Strong}
\vskip 0.5cm
{\Huge Coupling Lattice Gauge Theory}

\vskip 2cm

{\Large Thesis submitted towards the degree of}
\vskip 0.25 cm
{\Large ``Doctor of Philosophy''
\vskip 1cm
by}
\vskip 1cm

{\Large \bf Barak Bringoltz}

\vskip 1.5cm

{\Large This work was carried out under the supervision of
\\ Professor Benjamin Svetitsky
\vskip 0.5cm
July, 2004}

\end{center}
\clearpage

\newpage
\quad
\newpage

\quad
\vskip 3.5cm 
{\large To my parents Yehudit and Naum}

\newpage
\quad
\newpage

{
\topmargin -1.7cm

\begin{center} 
{\large \bf  Abstract} 
\end{center}

\vskip 0.35cm

We investigate the strong coupling limit of quantum chromodynamics on a lattice (lattice QCD) for systems with high baryon density. 
Our starting point is the Hamiltonian of lattice QCD for quarks with $N_c$ colors and $N_f$ flavors at strong coupling. We regard this Hamiltonian as an effective Hamiltonian that describes QCD at large distances, and expand it in orders of the inverse coupling. In leading order one finds a Hamiltonian that looks like an antiferromagnet. It has interactions between nearest-neighbor and next-nearest-neighbor sites, and is invariant under
 $U(N_f)\times U(N_f)$. 
Physically this Hamiltonian describes meson dynamics with a fixed background of baryon density. Our goal is to extract the ground state and low lying excitations of this Hamiltonian for systems with non-zero baryon density.

We first write the partition function of the Hamiltonian with boson coherent states and take the limit of large $N_c$ and $N_f$ with the ratio $r=N_c/N_f$ kept fixed. In this limit we make a mean field ansatz to extract the ground state. For zero density we extend a condensed matter physics result to three dimensions, and show that above a critical value of $r$
 the ground state breaks global
 symmetry. Departing from zero density
we show that for certain cases, where the baryon density is periodic, 
the critical value of $r$ increases with increasing density.
In cases where the baryon positions are random, 
 we average over them (in a quenched calculation that is valid only for low density) and show that the mean field ansatz is misleading near the critical $r$. 
Nevertheless, following the condensed matter analog of this quenched calculation, we expect that in the presence of baryons, low lying excitations will have non-zero widths and reduced velocities.

Writing the partition function with generalized spin coherent states leads to a nonlinear sigma model, which we investigate in the rest of the thesis. We first analyze the classical counterpart of a toy sigma model with mean field theory, 
and find 
that the critical temperature, under which the symmetry of the model is spontaneously broken, decreases with increasing density.

The quantum sigma model is treated in the limit of large $N_c$ with $N_f$ fixed. 
We find that the 
ground state is classically degenerate. For zero density
the degeneracy is discrete. In the case of 
non-zero uniform density the ground state is continuously and locally degenerate
, like classical ground states of frustrated magnetic systems. 

When we consider quantum corrections of order $1/N_c$ we find the phenomenon of ``order from disorder,'' also present in frustrated magnetic systems. 
For zero density we extend an old result to the next-nearest-neighbor theory, and show that fluctuations remove the discrete degeneracy
to choose a ground state that spontaneously breaks $U(N_f)\times U(N_f)$ to $U(N_f)_V$.
For non-zero density 
we see that fluctuations remove the local degeneracy and choose a state that spontaneously breaks 
$U(N_f)\times U(N_f)$ as well as discrete lattice rotations in a way that depends on $N_f$ and the baryon density. 


Finally we find a rich variety of low lying excitations that includes type I and type II Goldstone bosons. The latter emerge only for non-zero density and only when quantum fluctuations are taken into account. Their energies are of order $1/N_c$, and are quadratic in momentum. Bosons of either type can develop anisotropic dispersion relations.


}

\newpage
\quad
\newpage

\newpage
\quad
\newpage

\setcounter{page}{1}
\renewcommand{\thepage}{\roman{page}}

\pagestyle{fancy}

\tableofcontents
\listoffigures
\listoftables


\chapter{Introduction}
\label{chap:Intro}
\setcounter{page}{1}
\renewcommand{\thepage}{\arabic{page}}

The theoretical framework that describes the strong interactions of quarks and gluons is the theory of quantum chromodynamics (QCD). It is of fundamental interest to study what QCD predicts for nuclear matter at extreme conditions. This includes systems at very high temperatures ($\sim10^{12}$~K) accessible in ultra-relativistic collisions of heavy ions, and systems with high nuclear densities ($\sim 10^{15}$-$10^{16} \, \text{g} \, \text{cm}^{-3}$) that can be found in the cores of neutron stars. In this work we are concerned with the latter, nuclear systems of high density.

The study of quantum chromodynamics at high density is almost as old as the theory itself, and goes hand in hand with the discovery of asymptotic freedom \cite{Gross:1973id,Politzer:1973fx}. 
The first to approach the problem were Collins and Perry \cite{Collins}. They argued that at high baryon densities of the order of 
ten to a hundred times nuclear density, the hadrons overlap and lose their individuality. At these densities asymptotic freedom weakens the color interactions, and in the absence of long range forces the system can be considered as a weakly interacting ``quark soup'', with quarks arranged in a Fermi sea.

Next, Barrois \cite{Barrois_paper,Barrois_thesis} suggested that the Fermi quark sea is unstable, and shows some phenomenon analogous to superconductivity. He suggested that the instability is caused by gluon exchange between quarks in the antisymmetric $\bar{\underbar{3}}$ channel, which is attractive. Nevertheless Barrois argued in \cite{Barrois_paper} that the ground state does not break $SU(3)$ gauge symmetry, and is rather characterized by condensates of six quark operators, which are color invariants. 

The next step is the work of Bailin and Love \cite{Bailin}, that studied relativistic fermionic systems in similar lines to the study of fermions in condensed matter systems that exhibit non-perturbative phenomena. Using similar arguments to those in \cite{Collins} they justify that quarks in neutron stars weakly interact and that they are the relevant degrees of freedom, rather than the neutrons themselves. They describe the superfluid, superconductor, and color superconductor ground states of this dense quark matter, generated by one-gluon exchange between quarks in the $\bar{\underbar{3}}$ channel. In contrast to Barrois they treat condensates of quark-quark cooper pairs, rather than higher condensates. These condensates are not gauge invariant and spontaneously break $SU(3)_{\text{color}}$. This is the phenomenon of color superconductivity (CSC).

Although these are fascinating theoretical predictions, one finds that the the gap and critical superconducting temperature $T_c$ are in the MeV range. This is disappointing, since it means that the color superconductive part of the QCD phase diagram is very narrow. For that reason the study of color superconductivity was at a halt for over a decade. 
The
stimulus for the revival of the idea of CSC was the observation \cite{Alford,Rapp}
that the instanton-induced quark--quark interaction can be much
stronger than that induced by simple one-gluon exchange, and can
thus give a transition temperature on the order of 100~MeV.
Subsequent work \cite{Son} showed that the perturbative
color-magnetic interaction also gives rise to a strong pairing
interaction.
These and other dynamical considerations \cite{SchaferWilczek}
underlie a picture of the ground state of high-density QCD in
which the $SU(3)$ gauge symmetry is spontaneously broken by a
BCS-like condensate. The details of the breaking, which include
both the Higgs (or Meissner) effect and the rearrangement of
global symmetries and Goldstone bosons, depend on quark masses,
chemical potentials, and temperature. Prominent in the list of
possibilities are those of color-flavor locking in three-flavor
QCD \cite{Alford1} and crystalline superconductivity---with broken
translation invariance---when there are two flavors with different
densities \cite{Alford2}. For a review see \cite{Krishna}.

As noted, CSC at high density is so far a prediction of
weak-coupling analysis. One expects the coupling to become weak
only at high densities, and in fact it turns out that reliable
calculations demand extremely high densities \cite{Shuster}. The
use of weak-coupling methods to make predictions for moderate
densities is thus not an application of QCD, but of a model based
on it.  It is imperative to confirm these predictions by
non-perturbative methods.  Standard lattice Monte Carlo methods,
unfortunately, fall afoul of well-known technical problems when
the chemical potential is made non-zero, although we do note
remarkable progress made recently in the small-$\mu$ regime
\cite{Fodor,Owe,Ejiri}. For a review of these methods see for example \cite{Stephanov}.

In this work we study high-density quark matter
based on lattice QCD in the strong-coupling limit, which we regard as an effective theory that describes QCD at large distances. We work in the Hamiltonian formalism, which is
more amenable than the Euclidean formalism to strong-coupling
perturbation theory and to qualitative study of the ensuing
effective theory \cite{BSK,SDQW,Smit,BS}.  The fermion kinetic
Hamiltonian is a perturbation that mixes the zero-flux states that
are the ground-state sector of the electric term in the gauge
Hamiltonian. In second order, it moves color-singlet fermion pairs
around the lattice; the effective Hamiltonian for these pairs is a
generalized antiferromagnet, with spin operators constructed of
fermion bilinears.

We depart from studies of the vacuum by allowing a background baryon
density, which is perforce static in second order in perturbation
theory.  Our aim is to discover the ground state of
the theory with this background.  In third order (when $N_c=3$)
the baryons become dynamical; we display the effective Hamiltonian but make
no attempt to treat it.

The symmetry group of the effective antiferromagnet is the same as
the global symmetry group of the original gauge theory.  This
depends on the formulation chosen for the lattice fermions.
Following \cite{SDQW}, we begin with naive, nearest-neighbor fermions, which suffer from species doubling \cite{NN} and possess
a global $U(4N_f)$ symmetry group that contains the ordinary
chiral symmetries [as well as the axial $U(1)$]  as subgroups.  We
subsequently break the too-large symmetry group with
next-nearest-neighbor (NNN) couplings along the axes in the
fermion hopping Hamiltonian.  A glance at the menu of fermion
formulations reveals the reasons for our choice.  Wilson fermions
\cite{Wilson} have no chiral symmetry and make comparison of
results to continuum CSC difficult if not impossible.  Staggered
fermions \cite{Susskind} likewise possess only a reduced axial
symmetry while suffering a reduced doubling problem.  The overlap
action \cite{overlap} is non-local in time and hence possesses no
Hamiltonian; attempts \cite{Creutz} to construct an overlap
Hamiltonian directly have not borne fruit.  Finally, domain-wall
fermions \cite{Kaplan,Shamir} have been shown \cite{BS} to lose
chiral symmetry and regain doubling when the coupling is strong.

While the NNN theory still exhibits
doubling in the free fermion spectrum, we are not interested in
the perturbative fermion propagator but in the spectrum of the
confining theory. We take it as a positive sign that the unbroken
symmetry is now $U(N_f) \times U(N_f)$. \footnote{The breaking of the
naive fermions' symmetry by longer-range terms is a feature
\cite{SDQW} of SLAC fermions \cite{DWY} and also occurs if naive
fermions are placed on a {\em bcc\/} lattice \cite{Myint}.} This symmetry is what we
want for the continuum theory, except for the axial $U(1)$.  The
latter can still be broken by hand \cite{U1A}.

Our emphasis on the global symmetries is a consequence of the fact
that the gauge field is not present in the ground-state sector and
does not reappear in strong-coupling perturbation theory.  In
other words, confinement is a {\em kinematic\/} feature of the
theory, leaving no possibility of seeing the Higgs-Meissner effect
directly.  This is but an instance of confinement-Higgs duality,
typical of gauge theories with matter fields in the fundamental
representation \cite{duality}.  Our aim is thus to identify the
pattern of spontaneous breaking of global symmetries.  For various
values of $N_c$ and $N_f$, this can be compared to weak-coupling
results \cite{Schafer}.

Other groups have recently studied the strong-coupling limit of QCD in the Hamiltonian and Euclidean formalisms 
at non-zero chemical potential and baryon density \cite{Azcoiti,Yaouanc,Gregory,Umino,Karsch,Nishida,Fang}.
We differ from most approaches in eschewing mean field theory in favor
of the exact transformation to a different representations, which is amenable
to semiclassical treatment. Also we base our program on
NNN fermions; we also work at fixed baryon density. 

The outline of this dissertation is as follows. In Chapter~\ref{chap:Heff} we first give a brief introduction to Hamiltonian lattice QCD, and discuss our approach that regards the strong coupling limit of the Hamiltonian as an effective model for QCD at large distances. Next, in Section~\ref{sec:expansion}, we present the derivation of the effective Hamiltonian of lattice QCD in strong-coupling perturbation theory \cite{SDQW,Smit}. 
The second-order Hamiltonian [$O(1/g^2)$] is an antiferromagnet with $U(4N_f)$ spins; the global symmetry group is $U(4N_f)$ for the nearest-neighbor theory, broken to $U(N_f)\times U(N_f)$ by NNN terms. The baryon number at each site determines the representation of $U(4N_f)$ carried by the spin at that site. In
second order, baryon number is static; it becomes mobile in the next order, where (for $N_c=3$) the new term in the effective Hamiltonian is a baryon hopping term.

\begin{figure}[htb]
\begin{center}
        \epsfig{width=5cm,file=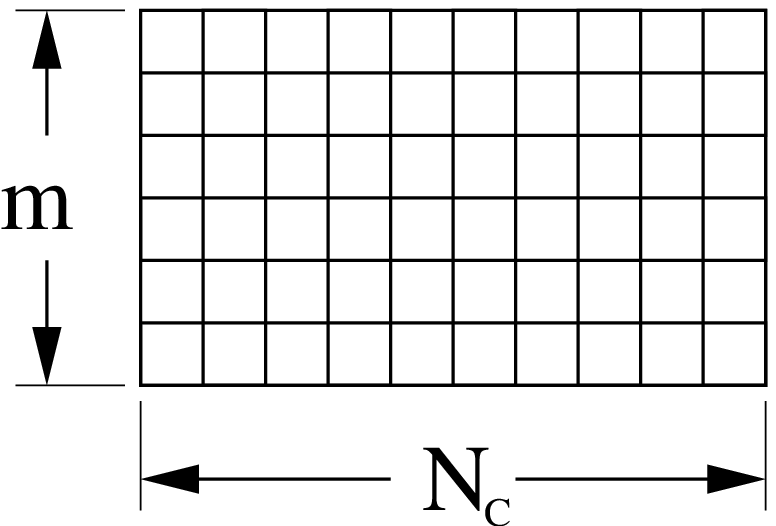} \caption[The representation of $U(4N_f)$]{The
        representation of $U(N)$ carried by the spin in the
        effective antiferromagnet [See Chapter~\ref{chap:Heff}]. $m$ is related to the baryon
        number $B$ at the site according to $m=B+2N_f$, with
        $|B|\leq 2N_f$.}\label{fig:young1}
\end{center}
\end{figure}

In the remainder of this thesis, we work only to $O(1/g^2)$, where
the baryons are fixed in position. Motivated by the similarity of
our Hamiltonian to the Heisenberg antiferromagnet, we apply
condensed matter methods developed for that problem. Indeed,
condensed matter physicists have generalized the $SU(2)$,
spin-$1/2$ Heisenberg model to $SU(N)$ in many representations
\cite{Haldane,Affleck,MA,AA,RS1,RS2,Arovas,Salam,Auerbach_book}, which
corresponds to adding flavor and color degrees of freedom to the
electrons.\footnote{We refer the reader to the paper by Read and
Sachdev \cite{RS1} for a survey, including a phase diagram in the
$(N,N_c)$ plane.} These are exactly the generalizations needed for
our effective Hamiltonian. With $N_c$ colors and $N$
(single-component) flavors, a site of the lattice can be
constrained to contain a color-singlet combination of $mN_c$
particles. The flavor indices of the spin then make up a
representation of $SU(N)$ whose Young diagram has $N_c$ columns
and $m$ rows (see Fig.~\ref{fig:young1}). We set
\begin{equation}
N=4N_f
\end{equation}
and the correspondence is complete (until we include NNN terms
in the Hamiltonian).

In Chapter~\ref{chap:MF_schwinger} we analyze the effective Hamiltonian by following Arovas and Auerbach \cite{AA} and write it with boson operators that obey appropriate constraints to restrict quantum states to the Hilbert space represented by Fig.~\ref{fig:young1}. After introducing the boson operators and their corresponding path integral in Section~\ref{sec:schwinger}, we apply the Hubbard-Stratonovich transformation in Section~\ref{sec:HS} and decouple the quartic boson interaction to a quadratic interaction. The result of integrating the bosons is a path integral tractable in the limit of large $N_c$ and $N_f$ with a fixed ratio $r=N_c/N_f$. We take this limit in Sections~\ref{sec:MF_zero}--\ref{sec:MF_finite} and study a mean field ansatz for the ground state of a set of baryon number configurations. In Section~\ref{sec:MF_zero} we extend the zero density results of \cite{AA} from dimension $d=2$ to $d=3$. As for $d=2$, one finds a Bose-Einstein condensate (BEC) above some critical value of $r$. This BEC corresponds to a N\'eel ground state which represents a spontaneous breaking of chiral symmetry, as expected. For QCD, with $N_c=3$, the system is safely in the {\em ordered\/} phase for any reasonable number of flavors. In Section~\ref{sec:MF_finite} we depart zero density by inserting non-zero baryon number on some of the sites. More precisely we fill some of the sites with the maximum allowed baryon number per site. The $U(N)$ representation on these sites then has $m=N$ rows, and is a singlet. Analyzing again a mean field ansatz we discuss how these singlets affect the free energy and critical value of $r$. In this chapter we mainly restrict to the disordered side of the phase transition. 

In Chapter~\ref{chap:NLSM}, and in the rest of the thesis, we 
treat the system deep in its ordered phase. Also the non-zero baryon number configuration we investigate are different than those of Chapter~\ref{chap:MF_schwinger}, and we mainly concentrate on uniform baryon configuration, with the same baryon number $B$ on all sites. In Section~\ref{sec:coherent}--\ref{sec:action_NLSM} we follow Read and Sachdev \cite{RS1} and employ spin coherent states
\cite{Klauder} to write the partition function as a path integral for a nonlinear sigma model. $N_f$ and $m$
determine the target space of the sigma model to be the
symmetric space $U(N)/[U(m)\times U(N-m)]$; the number of colors
$N_c$ becomes an overall coefficient of the action.
As for the quantum
Hamiltonian, the nearest-neighbor theory is symmetric under $U(N)$
while the NNN terms break the symmetry to $U(N_f)\times U(N_f)$
(while leaving the manifold unchanged).

In Section~\ref{sec:classical} we make a pause from the real sigma model, and treat its classical counterpart, with a variety of baryon number configurations. This analysis presumably describes the high temperature region of the system, where quantum fluctuations are not very important. It also serves as a tool to see how the symmetry realization changes as we add baryons to the lattice. Although the classical sigma model can be treated numerically (its action is real), we postpone this to future research and use mean field theory instead for a toy model that has $N=3$. In that case calculations can be done analytically, and we extract the {\em classical} phase diagram in temperature-density plane.

Next we return to discuss the real quantum sigma model. The factor of $N_c$ that multiplies the action invites a large-$N_c$ analysis which we make in Section~\ref{sec:Nc_infty}. We start by studying this limit for zero baryon number in Section~\ref{sec:zeroB}. This leads us to an exercise proposed and solved by Smit \cite{Smit}, in generalizing the zero baryon number sector to allow baryon number $\pm B$ on alternating sites; this means specifying conjugate representations of $U(N)$ on alternating sites, with respectively $m$ and $N-m$ rows. We find the ground state of the nearest-neighbor theory to be discretely degenerate as seen in \cite{Smit}. We extend this result to the NNN theory and show how the degeneracy is removed by the NNN terms.

We turn to non-zero baryon density in Section~\ref{sec:nonzeroB}. As mentioned, we study homogeneous states, in which all sites carry the same representation of $U(N)$, with $m>N/2$. 
We begin by studying the two-site problem, and we learn that its ground state is continuously degenerate. This degeneracy is additional to the global degeneracy of rotating the sigma fields on the two sites together. Replicating this ground state to the infinite lattice makes the lattice vacuum locally degenerate.
This makes our system similar to some frustrated models of magnetic systems \cite{Aharony,Ord_disorder,double_Xchange,Kagome,Sachdev,Henley}. In contrast to the zero density case this local degeneracy is not removed by the NNN terms.

In order to remove this degeneracy we consider quantum fluctuations in Chapter~\ref{chap:Qfluctuation}. We first treat the case of zero baryon number in Section~\ref{sec:Rem_B0}, where we see that $1/N_c$ corrections to the free energy removes the discrete degeneracy of the nearest-neighbor theory\footnote{This result was obtained by Smit using a Holstein-Primakoff transformation on the quantum Hamiltonian.}.
We find that fluctuations choose the same ground state chosen by the NNN terms.

We treat fluctuations for the non-zero density system in Section~\ref{sec:Rem_Bn0}. In Section~\ref{sec:O1} we first show that the local degeneracy is expressed by having zero modes in the $O(1)$ dispersion relations of excitations around the classical ground state. These have zero energy for all momenta. In Section~\ref{sec:Self_E}, we show that these zero modes get non-zero self energy of order $1/N_c$, thus removing the classical local degeneracy.
This is the phenomenon of ``order from disorder'' also present in the frustrated spin systems mentioned above. 
We also find that the energies produced are quadratic in momentum, which makes the zero modes type II Goldstone bosons.

The analysis in Chapters~\ref{chap:MF_schwinger}--\ref{chap:Qfluctuation} is mainly concerned with the nearest-neighbor theory that is symmetric under the too large group of $U(4N_f)$. In Chapter~\ref{chap:NNN} we take into account next-nearest-neighbor interactions, and treat them as a perturbation on the nearest-neighbor ground state. In Section~\ref{sec:NNN_action} we introduce the action of the sigma model which is now invariant only under $U(N_f)\times U(N_f)$. In Section~\ref{sec:gs}, we show that some of the global degeneracy of the nearest-neighbor ground state is removed, and discuss how the vacuum expectation values of the sigma fields break chiral symmetry and discrete lattice rotations. 
In Section~\ref{sec:nnn_excite} we calculate the effects of these next-nearest-neighbor interactions on the spectrum of the nearest-neighbor theory. 

In Chapter~\ref{chap:summary} we summarize the thesis and point to directions of future research.

All the work presented is this dissertation, except for Section~\ref{sec:MF_finite}, and some of the technical appendices, has already been published. The analysis of Chapter~\ref{chap:Heff} has appeared already in earlier work \cite{BSK,SDQW,Smit,BS,DWY} and in \cite{Boston,paper1}. The results of Section~\ref{sec:MF_zero} (without the analysis itself) and Chapter~\ref{chap:NLSM} were published in \cite{Boston} and \cite{paper1}. Sections~\ref{sec:Rem_B0} and~\ref{sec:Rem_Bn0} were also published in \cite{paper1}, whereas the remainder of Chapter~\ref{chap:Qfluctuation} was published in \cite{odo}. Finally the treatment of NNN interactions of Chapter~\ref{chap:NNN} was published in \cite{NNN} and \cite{NNNexcite}. 

\clearpage
\chapter{The effective Hamiltonian}
\label{chap:Heff}

In this chapter we derive the effective Hamiltonian we use to investigate QCD at high densities. We begin by giving a short review on the Hamiltonian approach to lattice QCD in Section~\ref{sec:HLQCD}, where we stress important points relevant to this work, and fix some of the notations. Aiming to work at strong coupling, and aware of the fact that the continuum limit is at weak coupling, we take the point of view that the investigated Hamiltonian is an effective one that describes QCD at large distances. We elaborate on this formal point in Section~\ref{sec:idiology}. We then proceed to derive the effective Hamiltonian by strong coupling expansion in Section~\ref{sec:expansion}. We conclude by emphasizing the similarity of the Hamiltonian to the $t$--$J$ model Hamiltonian of condensed matter physics.

\section{Hamiltonian approach to lattice QCD}
\label{sec:HLQCD}

In this section we introduce the Hamiltonian formalism of lattice QCD.  This formalism is more amenable to strong-coupling
perturbation theory and to qualitative study of the ensuing
effective theory \cite{BSK,SDQW,Smit,BS,DWY} than the Euclidean formalism.

The Hamiltonian formalism was first introduced by Kogut and Susskind in 1975 \cite{Kogut75}, shortly after Wilson's Euclidean formulation \cite{Wilson74}. This formalism defines the theory of strong interactions on a three dimensional space lattice with lattice spacing $a$ and continuous time $t$. A lattice site is denoted by a three dimensional vector $\bn=\left(n_x,n_y,n_z\right)$ taking integer values, and a lattice link is denoted by $(\bn\muhat)$, where $\mu=\pm x,\pm y,\pm z$.


The quantum fields that describe quarks are the fermion fields $\psi^{af\alpha}_{\bn}$ that live on the sites $\bn$ of the lattice. They have color indices $a=1,\dots,N_c$, Dirac indices $\alpha=1,\dots,4$, and flavor indices $f=1,\dots,N_f$. It is more convenient to group the Dirac and flavor indices to a single index $\alpha$ that takes values from $1$ to $4N_f$. The Fermi fields obey the following anti-commutation relations
\begin{equation}
\left\{ \psi^{a\alpha}_\bn,\psi^{\dag b \beta}_\bmm \right\} = \delta_{\bn\bmm}\delta_{ab}\delta_{\alpha\beta}. \label{anticomm}
\end{equation}

For the gauge fields, that describe the gluons, one chooses to work in the temporal gauge that fixes $A_0=0$, and removes one degree of freedom (and its conjugate momentum). As a result, one is left only with the link operators $\left( U_{\bn \muhat} \right)_{ab}$, and $\left( U^\dag_{\bn \muhat} \right)_{ab}=\left(U_{\bn+\muhat,-\muhat}\right)_{ab}$. These represent gauge fields on the link $(\bn\muhat)$, whose two edges are colored with colors $a$ and $b$. There are $6$ links emanating from each site $\bn$. On each link reside the gauge fields $\left(U_{\bn\muhat}\right)_{ab}$ and their conjugate momenta $E^i_{\bn\muhat}$, $i=1,\dots,N_c^2-1$. The following are the commutation relations of this set of operators,
\begin{eqnarray}
\left[ E^i_{\bn \muhat} , \left( U_{\bn\muhat} \right)_{ab} \right] &=& \left( \lambda^i U_{\bn\muhat} \right)_{ab}, \label{comm_plus1} \\
\left[ E^i_{\bn \muhat} , E^j_{\bn \muhat} \right] &=& i f^{ijk} E^k_{\bn \muhat}, \label{comm_plus2}
\end{eqnarray}
for the set of the links $(\bn\muhat)$, and 
\begin{eqnarray}
\left[ E^i_{\bn, -\muhat} , \left(U_{\bn,-\muhat} \right)_{ab} \right] &=& -\left( U_{\bn,-\muhat} \lambda^i\right)_{ab} \label{comm_minus1} \\
\left[ E^i_{\bn,- \muhat} , E^j_{\bn,- \muhat} \right] &=& -i f^{ijk} E^k_{\bn,- \muhat}, \label{comm_minus2}
\end{eqnarray}
for the set $(\bn,-\muhat)$.
Here $\lambda^i$ are matrices that represent the generators of $SU(N_c)$ in the fundamental representation, and $f^{ijk}$ are the structure constants of $SU(N_c)$.
For $N_c=2$, Eqs.~(\ref{comm_plus1})--(\ref{comm_minus2}) are exactly the commutations relations that appear in the classical problem of the rotating top, where $E^i_{\bn \muhat}$, and $E^i_{\bn,-\muhat}$ generate rotations of space fixed and body fixed coordinate systems. One uses this analogy to write 
\begin{equation}
E^i_{\bn+\muhat,- \muhat} = - \left(U^{\text{Adj.}}_{\bn\muhat}\right)_{ij} E^j_{\bn,-\muhat}, \label{Eminus}
\end{equation}
where $U^{\text{Adj.}}_{\bn\muhat}$ is the rotation $U_{\bn\muhat}$ in the adjoint representation of $SU(N_c)$. For a more detailed discussion see \cite{Kogut75}.


We now discuss the lattice Hilbert space. A state $|\Omega \rangle$ is the direct product on all lattice sites,
\begin{equation}
|\Omega \rangle = | \Omega \rangle_G \otimes | \Omega \rangle_{\Psi}.
\end{equation}
Here the first factor is the projection of the state $|\Omega\>$ to the gauge field sector, while the second factor describes the fermionic sector. 

Any state $|\Omega\>_{\Psi}$ is the following direct product
\begin{equation}
|\Omega\>_{\Psi} = \prod_{\otimes\bn} \left( |\Omega\>_{\Psi} \right)_\bn.
\end{equation}
We now concentrate on the Hilbert space of each site. The ``lowest'' state is the no-quantum ``drained'' state $|\textrm{Dr}\rangle_\bn$, defined as
\begin{equation}
\psi^{a\alpha}_\bn|\textrm{Dr}\rangle_\bn = 0. \label{drained}
\end{equation}
Applications of various $\psi^{\dag a\alpha}_\bn$ create the corresponding quarks on that site, 
\begin{equation}
|a\alpha \rangle_{\bn\Psi} = \psi^{\dag a\alpha}_\bn |\textrm{Dr}\rangle_\bn.
\end{equation}
For the free theory, $\alpha=1,\dots,2N_f$ correspond to creation of positive energy excitations, i.e. quarks, whereas $\alpha=2N_f,\dots,4N_f$ corresponds to creation of negative energy excitations, i.e. annihilation of anti-quarks. To use the usual quark--anti-quark language we write for each site, color, and flavor,
\begin{equation}
\psi=\left( \begin{array}{c} b_{\uparrow} \\  b_{\downarrow} \\  d^{\dag}_{\downarrow} \\  d^{\dag}_{\uparrow} \end{array} \right).
\end{equation}
$b^{\dag}$ create a quark and $d^{\dag}$ an anti-quark. From \Eq{drained} we see that $b$ and $d^{\dag}$ annihilate the drained state, which means that this state is empty of quarks, and filled with anti-quarks. The operator of local baryon number is
\begin{equation}
B_\bn = \frac1{N_c}\sum_{a=1}^{N_c}\sum_{s=\uparrow,\downarrow}\sum_{f=1}^{N_f} \left[ b^{\dag af}_{s} b^{af}_{s} - d^{\dag af}_{s} d^{af}_{s} \right]_\bn =\frac1{N_c} \psi^{\dag}_\bn \psi_\bn - 2N_f. \label{B}
\end{equation}
According to \Eq{B}, the baryon number of the drained state is $-2N_f$, corresponding to filling the site with anti-quarks. The vacuum $|0\rangle$ is the state with no quarks and no anti-quarks. This state is the filled Dirac sea on a single site and obeys,
\begin{equation}
b|0\rangle=d|0\rangle=0.
\end{equation}
The baryon number of this state is $B=0$. Because of Pauli exclusion principle we cannot put too many fermions on a single site. The maximum number of local baryon number will be $2N_f$, and is found only in the state $|\textrm{filled} \rangle$
\begin{equation}
b^{\dag}|\textrm{filled}\rangle=d|\textrm{filled}\rangle=0.
\end{equation}

Below we will see that gauge invariance puts more restrictions on the single site Hilbert space in order that it be color neutral.


Moving to the gauge Hilbert space, we also write it as a direct product of the form
\begin{equation}
|\Omega\>_G = \prod_{\otimes\bn\muhat} \left( |\Omega\>_G \right)_{\bn\muhat}.
\end{equation}

One denotes the state with no electric field $E$ by $|0\>_G$,
\begin{equation}
 E^i| 0 \>_G = 0, \qquad \forall i.
\end{equation}
Any application of the link operators $\left( U^{\dag}_{\bn\muhat} \right)_{ab}$ on $|0\>_G$, creates states which correspond to flux lines on the link $(\bn\muhat)$. The state $|0 \>_G$ is the only state with no flux at all. Using the fact that the electric field operators generate a $SU(N_c)$ algebra, one can distinguish between the different quantum states created by the link operators as follows. Define the quadratic Casimir operator
\begin{equation}
\vec{E}^2_{\bn\muhat}\equiv\sum_{i=1}^{N_c^2-1}E^{i2}_{\bn\muhat}. \label{E2}
\end{equation}
It is clear that the fluxless state $| 0 \>_G$ is an eigenstate of this Casimir, with zero eigenvalue.
Next the commutations of the Casimir with the link operators are verified from Eqs.~(\ref{comm_plus1})--(\ref{comm_minus2}) to be
\begin{equation}
\left[ \vec{E}^2_{\bn\muhat},\left( U^{\dag}_{\bn\muhat} \right)_{ab} \right] = C_F\left( U^{\dag}_{\bn\muhat} \right)_{ab},
\end{equation}
where $C_F$ is the Casimir operator in the fundamental representation, and is 
\begin{equation}
C_F=\frac{N_c^2-1}{2N_c}
\end{equation}
for $SU(N_c)$.

This means that the state $\left[ \left( U^{\dag}_{\bn\muhat} \right)_{ab}| 0 \>_G \right]$ is also an eigenstate of~(\ref{E2}), with eigenvalue equal to $C_F$. One can now classify the states in $|\Omega \>_G$ according to their $\vec{E}^2$ eigenvalue. The result is a Hilbert space with a ladder-like structure. The lowest state is $|0\>_G$, with zero flux, and is a singlet of $SU(N_c)$. Repeated applications of the gauge field operators $\left( U^{\dag}_{\bn\muhat} \right)_{ab}$ create states with higher and higher values of flux and the operators $\vec{E}^2_{\bn \muhat}$ measures the flux on the link $(\bn\muhat)$. Indeed we shall see shortly that it is proportional to the (kinetic) energy of the gauge fields.


To complete the picture we now discuss gauge invariance. First recall that the starting point of this formalism was to choose the timelike gauge. This leaves only time-independent gauge transformations as a symmetry. The fermion operators belong to the fundamental representation of the gauge symmetry and transform as
\begin{equation}
\psi^a_\bn \rightarrow \left( V_\bn \right)^{ab} \psi^b_\bn, \label{gauge_F1}
\end{equation}
with $V\in SU(N_c)$ given in general by
\begin{equation}
V=\exp \left[ i\sum_{i=1}^{N_c^2-1} \theta^i_\bn \lambda^i\right].
\end{equation} 
Note that here, and in the following discussion, we suppress the Dirac-flavor indices, irrelevant to gauge invariance. Using the anticommutation relations~(\ref{anticomm}), one can show that the quantum operator ${\cal V}_F$ that realizes \Eq{gauge_F1} in Hilbert space as
\begin{equation}
\psi^a_\bn \rightarrow {\cal V}_F \psi^a_\bn {\cal V}^\dag_F,
\end{equation}
is 
\begin{equation}
{\cal V}_F = \exp \left[ i\sum_\bn \sum_{i=1}^{N_c^2-1} \theta^i_\bn \left( \psi^{\dag a}_\bn \lambda^i_{ab} \psi^b_\bn \right) \right]. \label{gauge_F2}
\end{equation}

The gauge fields transform according to 
\begin{equation}
\left( U_{\bn \muhat} \right)_{ab} \rightarrow \left( V_\bn U_{\bn\muhat}  V^\dag_{\bn+\muhat} \right)_{ab}.
\end{equation}

Using the commutation relations~(\ref{comm_plus1})--(\ref{comm_minus2}), one shows that the quantum operator ${\cal V}_G$ that generates these rotations is given by
\begin{equation}
{\cal V}_G = \exp \left[ i\sum_\bn \sum_{i=1}^{N_c^2-1} \theta^i_\bn  \left( \sum_{\pm \muhat} E^i_{\bn \muhat} \right)\right]. \label{gauge_G}
\end{equation}

Finally putting \Eq{gauge_F2}, and \Eq{gauge_G} together and using \Eq{Eminus}, we see that operator that induces gauge transformations is 
\begin{equation}
{\cal V}=\exp \left[ i\sum_\bn \sum_{i=1}^{N_c^2-1} \theta^i_\bn  \left( \rho^i_G + \rho^i_F \right)_\bn\right],
\end{equation}
where $\rho^i_{F\bn}$, and $\rho^i_{G\bn}$ are the color charge densities of the fermions, and gauge  fields. These two quantities are given by
\begin{eqnarray}
\rho^i_{F\bn}&=&\psi^{\dag a}_\bn \lambda^i_{ab} \psi^b_\bn, \\
\rho^i_{G,\bn}&=&\sum_{\mu} \left( E^i_{\bn,\muhat}-\left( U^{\text{Adj.}}_{\bn-\muhat,\muhat}\right)_{ij} E^j_{\bn-\muhat,\muhat}\right). \label{rhoG} 
\end{eqnarray}
Note that \Eq{rhoG} can be written as the covariant divergence of the electric field,
\begin{equation}
\rho^i_{G\bn}=\vec{D}\cdot \vec{E}^i_\bn \equiv  \vec{\nabla}_L\cdot \vec{E}^i_\bn + \sum_{\mu} \left( 1-U^{\text{Adj.}}_{\bn-\muhat,\muhat}\right)_{ij} E^j_{\bn-\muhat,\muhat}, 
\end{equation}
where the lattice divergence is given by 
\begin{equation}
\vec{\nabla}_L\cdot \vec{E}^i_\bn \equiv \sum_\mu \left( E^i_{\bn\mu}-E^i_{\bn-\mu,\mu} \right).
\end{equation}

Since the lattice Hamiltonian is gauge invariant, we know that the generators of the gauge transformations commute with the Hamiltonian
\begin{equation}
\left[ H , \rho^i_{G\bn} + \rho^i_{F\bn} \right]=0, \hskip 1cm \forall i \quad \forall \bn,
\end{equation}
This means that we can choose to work with a basis that block diagonalizes $\rho^i_G+\rho^i_F$. This breaks the Hilbert space to separate sectors classified by their $\rho^i_G+\rho^i_F$ representations on the lattice. Each set of operators $\rho^i_{{\text ext},\bn}$ describes a different physical case, with a different {\em external} distribution of color charge (that can correspond, for example, to infinitely heavy quarks etc.). To describe the physics of quarks with finite masses, and zero external gauge fields, we work with the choice $\rho^i_{\text{ext}}=0$.
Working in this subspace means that {\em all} physical states must be color singlets, since all gauge transformations are trivial.


Finally we describe the lattice Hamiltonian $H$ of $SU(N_c)$ gauge theory with $N_f$ flavors of fermions. $H$ is given by
\begin{equation}
H=H_E+H_G+H_F. \label{eq:H_initial}
\end{equation}
Here $H_E$ is the electric term, a sum over links $(\bn\mu)$ of
the $SU(N_c)$ Casimir operator
\begin{equation}
H_E=\frac12g^2\sum_{\bn\mu}\vec{E}^2_{\bn\mu}.
\end{equation}
Next is the magnetic term, a sum over plaquettes,
\begin{equation}
H_G=\frac1{2g^2}\sum_p\left(N_c-\Tr U_p\right).
\end{equation}
Finally we have the fermion Hamiltonian,
\begin{equation}
H_F=-i\sum_{\bn\mu}\psi^{\dag af}_\bn \alpha_\mu
\sum_{j>0}D(j)\left(\prod_{k=0}^{j-1}U_{\bn+k\muhat,\mu}\right)
_{ab} \psi^{bf}_{\bn+j\muhat}+h.c. \label{eq:H_F}
\end{equation}
In \Eq{eq:H_F} we suppress Dirac and color indices, and denote explicitly the flavor index $f$. The function $D(j)$ is a kernel that
defines the lattice fermion derivative. It should be sufficiently local, namely at least falling exponentially with distance. For example it can yield a naive, nearest-neighbor action if $D(j)=\frac12\delta_{j,1}$; a next-nearest-neighbor kernel with $D(j\ge3)=0$; a long-range SLAC derivative \cite{DWY} with $D(j)=-(-1)^j/j$ will however not be good enough, since it falls only algebraically. 

The choice of the fermion kernel is important, since it fixes the global symmetries of the Dirac Hamiltonian~(\ref{eq:H_F}), and in turn the global symmetries of the effective Hamiltonian that we derive in Section~\ref{sec:expansion}. Superficially, the Hamiltonian has only the chiral symmetry of $U(N_f)_R\times U(N_f)_L$ regardless of the kernel used. This is however not true. For a naive kernel, one can spin diagonalize the fermions, and show that the Dirac $\alpha_\mu$ matrices disappear. This means that the Dirac $\alpha=1,\dots,4$ indices become just a different kind of flavor, and the symmetry is then $U(4N_f)$. Applying the same procedure to the next-nearest-neighbor kernel, one sees that the Dirac structure does not disappear, and the symmetry is indeed only chiral symmetry $U(N_f)\times U(N_f)$. We postpone further discussion in symmetries to the level of the effective Hamiltonian calculated in Section~\ref{sec:expansion}.

\section{The strong coupling limit}
\label{sec:idiology}

Formulating a quantum field theory on a lattice is in fact a way of regulating the theory by providing it with a cutoff in coordinate space. As in any other regularization scheme, the cutoff must be removed at the end in favor of observable quantities. On the lattice, the removal of the cutoff is basically the procedure of taking the continuum limit of the theory, i.e., taking the lattice spacing $a$ to zero. In order to extract continuum physics, one calculates a physical observable as a function of the bare parameters of the theory (e.g. $a$ and the bare coupling $g_0$), and demands that the $a\rightarrow 0$ continuum limit of the observable will be its measured value. This procedure tells us how the bare coupling $g_0$ behaves as a function of the cutoff $a$. In the case of asymptotically free theories (and in particular in the case of QCD), the bare coupling $g_0(a)$ approaches zero as $a\rightarrow 0$. This means that ideally we should examine the lattice theory at small coupling and small lattice spacings. The regime where these both are small enough is determined by the scaling behavior of physical observables one calculates. This approach has become standard, and is realized by doing Monte Carlo simulations of the lattice theory. These, performed close to the continuum limit, give us information on non-perturbative aspects of QCD.

As mentioned in Chapter~\ref{chap:Intro}, these Monte Carlo approaches fail when the theory is defined at finite chemical potential. In order to examine the dense system, one is forced to resort to other methods of calculation. 
We turn to the strong coupling regime of QCD, which is far from the continuum limit. As a result one cannot trust numerical figures that come out of strong-coupling calculations (especially since we use only the lowest order and do not perform any extrapolation to small coupling). Despite this, examining the strong-coupling regime gives insight on qualitative features of the dense nuclear system. The physical picture we have in mind is that if one would perform a block-spin transformation of the lattice theory near its continuum limit, and integrate out high momentum degrees of freedom, then one will be left with a lattice theory for large distances (of the order of hadronic scale distances of $1$ fm.) The resulting effective theory will be strongly-coupled, and have the symmetries of QCD. In particular it will have an $SU(3)$ gauge group (or in general an $SU(N_c)$, where $N_c$ is the number of colors). It will also have the global symmetry
\begin{equation}
SU(N_f)_L \times SU(N_f)_R \times U(1)_B.
\end{equation}

Its dynamical physical features should include confinement and spontaneous chiral symmetry breaking (at zero density). All these important features are present in lattice QCD with strong bare coupling, and a suitable definition of the fermion kernel, which we take to be our starting point. In this work we show that the Hamiltonian formulation of this theory is tractable even with non-zero baryon density.

The only property that the theory explored here does not have is the anomaly of the axial $U(1)$ symmetry. In QCD, this symmetry is classically conserved, but broken in the presence of gauge fields. This is expressed in the fact that the corresponding singlet axial current has a non-zero divergence. This fact is  most easily seen in perturbation theory in continuum QCD \cite{anomaly}, and is verified experimentally by the large mass of the $\eta'$ meson, which should have been as small as the pion masses in the absence of this anomaly. 
The same calculation can be done with weak-coupling perturbative methods of lattice QCD \cite{KarstenSmit}. There, the lattice doubling of the fermions plays a crucial role and leads to conservation of the singlet axial current; the doubled fermions fall into two groups with opposite chiral charges, and the contribution of each group to the anomaly is canceled. In order to obtain the anomaly in a lattice theory, one must break chiral symmetry  explicitly (for example by a Wilson term) to remove the doubling. The continuum limit is then taken carefully to restore all chiral symmetries except the axial $U(1)$. 

We do not use Wilson fermions, and as mentioned also do not work in the continuum limit. Our fermion kernel leads to full species doubling in the weak-coupled continuum limit, but this fact by itself is not a real problem, since we work at strong coupling, where we do not have quark excitations. Our Hamiltonian will {\it not} lead to doubling of mesonic excitations, as long as it has the correct symmetry. Nevertheless, we can not avoid the absence of the anomaly; $U(1)_A$ is not anomalous, and is conserved at the quantum level as well. At this point we stress the fact that we regard our theory as an effective theory for QCD at large distances. In the process of the block-spin transformation that begins from a theory with no axial singlet symmetry, a $U(1)_A$ violating term will have to be generated. Therefore in order to have physical features related to this term, we will have to add it by hand, without knowledge of its exact strength. Note that in addition, we know neither the exact value of the coupling, nor the exact way the fermion kernel behave (we do assume that it falls fast enough).

In practice we can add the $U(1)_A$ violating term in any stage of the calculation. At the level of the effective lattice Hamiltonian introduced below, it will be a 't~Hooft vertex. 
For two flavors it can also be expressed in terms of the color singlet objects of Section~\ref{sec:AFM}, and for any number of flavors it is easiest to add this term at the level of the non-linear sigma model derived in Chapter~\ref{chap:NLSM} \cite{U1A}. This term will have some simple effects on the low energy spectrum we derive, and we discuss them in Chapter~\ref{chap:NNN}.

\section{Strong coupling perturbation theory}
\label{sec:expansion}

Our starting point is therefore the Hamiltonian~(\ref{eq:H_initial}) with large bare coupling. For $g\gg1$ the ground state of $H$ is determined by $H_E$ alone
to be any state with zero gauge flux, whatever its fermion
content,
\begin{equation}
|0\rangle=\left[\prod_{\bn\mu}|E_{\bn\mu}^2=0\rangle\right]_G
|\Psi\rangle.
\end{equation}
These states have energy $\epsilon_0=0$ and are degenerate with
respect to all the fermionic degrees of freedom. We consider
perturbation theory in $V=H_G+H_F$. Both $H_G$ and $H_F$ are sums
of operators that are strictly bounded, independent of $g$ except
for the explicit coefficient in $H_G$. We can dismiss first-order
perturbations by noting that $H_G$ and $H_F$ are multilinear in
link operators $U$ and $U^\dag$, which are raising/lowering
operators for the electric field; thus there are no non-zero
matrix elements within the zero-field sector.

We proceed to higher orders, and seek an effective Hamiltonian
that acts in the zero-field sector \cite{Kato}. Define $P_0$ to be
the projector onto the subspace of all the $E=0$ states. Then
perturbation theory in $V$ gives an effective Hamiltonian,
\begin{equation}
H_{\text{eff}}=P_0VQDVP_0+P_0VQDVQDVP_0+\cdots.
\end{equation}
Here $Q=1-P_0$ projects onto the subspace orthogonal to the $E=0$
states; the operator $D\equiv (\epsilon_0-H_E)^{-1}$ supplies
energy denominators, so that
\begin{equation}
QD=\sum_{E\neq 0\, \rm{states}}|\lambda\rangle\frac1{\epsilon_0-\epsilon_\lambda}\langle
\lambda|.
\end{equation}

The intermediate states $|\lambda\rangle$ contain flux
excitations. In second and third order the patterns of flux can
only be strings of length $j$ in the fundamental representation of
the color group. Thus the energy denominators are
\begin{equation}
\epsilon_0-\epsilon_\lambda=-\frac12g^2C_F|j|.
\end{equation}

The perturbations $H_G$ and $H_F$ are explicitly of $O(1/g^2)$ and
$O(1)$, respectively; each energy denominator gives a factor of
$1/g^2$. Thus to $O(1/g^4)$ we can forget about $H_G$. Our result
to this order is
\begin{equation}
H_{\textrm{eff}}=P_0H_FDH_FP_0+P_0H_FDH_FDH_FP_0. \label{Heff}
\end{equation}
Since $H_F$ has no non-zero matrix elements within the $E=0$
sector, we have dispensed with $Q$ in \Eq{Heff}. The first term in
\Eq{Heff} arises for any value of $N_c$ and is $O(1/g^2)$; the
case $N_c=2$ must be treated carefully, but all cases $N_c>2$ are
generic. The second term is special to $N_c=3$ and is $O(1/g^4)$.

\subsection{Second order: the antiferromagnet}
\label{sec:AFM}

We calculate explicitly the first term in $H_{\text{eff}}$. Each
term in $H_F$ creates a string of flux of length $j$, which must
be destroyed by the conjugate term. Thus
\begin{equation}
H_{\text{eff}}^{(2)}=2\sum_{j>0} [-K(j)]\sum_{\bn\mu}
\left(\psi_\bn^{\dag a f}\alpha_\mu\psi_{\bn+j\muhat}^{b
f}\right) \langle0|\left(\prod U\right)_{a b} \left(\prod
U^\dag\right)_{cd}|0\rangle
\left(\psi_{\bn+j\muhat}^{\dag c g}\alpha_\mu\psi_\bn^{d
g}\right), \label{H2eff1}
\end{equation}
where we define
\begin{equation}
K(j)= \frac{\left[D(j)\right]^2} {\frac12g^2C_F|j|}>0.
\end{equation}
The matrix element of the gauge fields yields
$\frac1{N_c}\delta_{ad}\delta_{bc}$,
independent of $j$.

As they appear in \Eq{H2eff1}, each $\psi^\dag$ is next to a
$\psi$ on a different site. This invites a Fierz transformation on
the product of fermion fields, which we write generally as
\begin{equation}
\left(\psi^{\dag}_i\alpha_\mu\psi_j\right)\left(\psi^{\dag}_k\alpha_\mu\psi_l\right)
= \delta_{jk}\psi^{\dag}_i\psi_l -\frac14\sum_{A}s_{A}^{\mu}
\left(\psi^{\dag}_i\Gamma^A\psi_l\right)
\left(\psi^{\dag}_k\Gamma^A\psi_j\right). \label{befFierz}
\end{equation}
Here $i,j,k,l$ are combined site, flavor, and color indices, and
we have assumed that $k$ and $l$ are always different while $j$
and $k$ might be equal [as in \Eq{H2eff1}]. The matrices
$\Gamma^A$ are the 16 Dirac matrices, normalized to
$(\Gamma^A)^2=\bm1$, and we have defined
\begin{equation}
s_{A}^{\mu}=\frac14\Tr\Gamma^A\alpha_\mu\Gamma^A\alpha_\mu=\pm1.
\end{equation}
This sign factor is $\pm1$ according to whether $\Gamma^A$
commutes or anticommutes with $\alpha_\mu$; it will be a constant
companion in our calculations. As they appear in
$H_{\text{eff}}^{(2)}$, the indices $i,l$ are the same site and
color but different flavors, and likewise $j,k$. Leaving the
flavor indices explicit, we obtain
\begin{equation}
H_{\text{eff}}^{(2)}= \frac1{4N_c}\sum_{\bn\mu\atop j\not=0} K(j)
s_A^\mu \left(\psi^{\dag f}\Gamma^A\psi^g\right)_\bn
\left(\psi^{\dag g}\Gamma^A\psi^f\right)_{\bn+j\muhat}
-d\sum_{j\not=0}K(j) \sum_\bn\left(\psi^{\dag f}\psi^f\right)_\bn. \label{eq:H}
\end{equation}
Each fermion bilinear in parentheses is a color singlet located at
a given site. The second term contains the baryon
density\footnote{ This baryon number is positive semidefinite, and
is zero for the drained state. The conventional baryon
number $B_\bn$ is zero in the half-filled state, and thus
$B_\bn'=B_\bn+2N_f$.} $B'_\bn=N_c^{-1}\psi^{\dag}_\bn\psi_\bn$,
and the sum $\sum_\bn B'_\bn$ is the total baryon number $B'$. A final important note is that one can easily verify that the effective Hamiltonian $H^{(2)}_{\text eff}$ commutes with the local baryon number $B_\bn$ (and $B'_\bn$). This is a result of the fact that each of the local operators in \Eq{eq:H} does not change the fermion occupation of a site.

We now combine the Dirac indices with the flavor indices and write
\begin{equation}
\left(\psi^{\dag f}\Gamma^A\psi^g\right)_\bn \left(\psi^{\dag
g}\Gamma^A\psi^f\right)_{\bn'} =8\left(\psi^\dag
M^\eta\psi\right)_\bn\left(\psi^\dag M^\eta\psi\right)_{\bn'}.
\end{equation}
We have defined new matrices $M^\eta$ as direct products of the
$4\times4$ Dirac matrices and the $U(N_f)$ flavor generators,
\begin{equation}
M^\eta=\Gamma^A\otimes\lambda^a, \label{dirprod}
\end{equation}
and we have normalized them conventionally according to
\begin{equation}
\Tr M^\eta M^{\eta'}=\frac12\delta^{\eta\eta'}.
\end{equation}
The $M^\eta$ generate a $U(N)$ algebra, with $N\equiv4N_f$.

An alternating flip
\begin{equation}
\psi_\bn\to \left[\prod_\mu (\alpha_\mu)^{n_\mu}\right]\psi_\bn
\end{equation}
(spin diagonalization \cite{SDQW}) removes the $\alpha_\mu$
matrices from the odd-$j$ terms in $H_F$, and hence removes the
sign factors $s_A^\mu$ from the odd-$j$ terms in
$H_{\text{eff}}^{(2)}$. We have finally
\begin{equation}
H_{\text{eff}}^{(2)}= \frac2{N_c}\sum_{\bn\mu j} K(j)
\left(s_\eta^\mu\right)_{\text{even}\, j \atop\text{only}}
\left(\psi^{\dag}M^\eta\psi\right)_\bn
\left(\psi^{\dag}M^\eta\psi\right)_{\bn+j\muhat}
-\left(dN_c\sum_jK(j)\right)B'. \label{H2eff2}
\end{equation}
The odd-$j$ terms are of the form $\bm{M}\cdot\bm{M}$ which can be
written in any basis for the $U(4N_f)$ algebra. The even-$j$
terms, however, contain $s_\eta^\mu$ which is defined only in the
original basis (\ref{dirprod}).


We now turn to describe the zero-field sector Hilbert space in which the Hamiltonian~(\ref{H2eff2}) works. In this sector in which we work, Gauss' Law constrains
the fermion state at each site to be a color singlet. The drained
state $|\text{Dr}\rangle$, with $\psi^{a f \alpha}_\bn|\text{Dr}\rangle=0$ for all $(a f \alpha)$, is the unique
state with $B'=0$. The other color singlet states may be generated
by repeated application of the baryon creation operator,
\begin{equation}
b^\dag_{\alpha \beta \cdots}= \epsilon_{ab\cdots}\psi^{\dag a \alpha}_\bn \psi^{\dag b \beta}_\bn\cdots,
\end{equation}
with $N_c$ operators $\psi$. (Here and henceforth, the indices
$\alpha,\beta,\ldots$ combine the flavor and Dirac indices.)

\begin{figure}[htb]
\begin{center}
        \epsfig{width=2cm,file=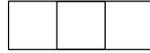}
        \caption[$U(4N_f)$ representation of a baryon operator with $N_c=3$]{Young diagram of the representation of $U(4N_f)$
carried by the baryon operator}
        \label{fig:Young3}
\end{center}
\end{figure}

At each site $\bn$ the operators
\begin{equation}
Q^\eta_\bn=\psi^\dag_\bn M^\eta \psi_\bn = \psi^{\dag a \alpha}_\bn
M^\eta_{\alpha \beta}\psi_\bn^{a \beta} \label{eq:Q}
\end{equation}
generate a $U(N)$ algebra, with $N=4N_f$. The drained state is
obviously a singlet under this algebra. The creation operator
$b^\dag_{\alpha \beta \cdots}$ is symmetric to any exchange of its $N_c$ Dirac-flavor indices. As a result it is in the symmetric representation of $U(N)$
with one row and $N_c$ columns (see Fig.~\ref{fig:Young3}).

\begin{figure}[htb]
\begin{center}
        \epsfig{width=3.5cm,file=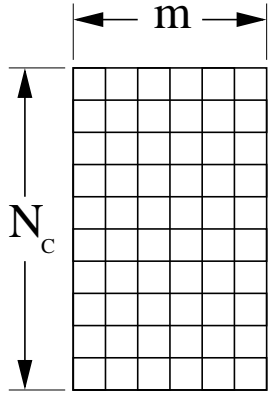} 
\caption[$SU(N_c)$ representation carried of a single site state]{The representation of $SU(N_c)$ carried by the color indices of the singlet site state, that has $mN_c$ fermions. The baryon number is $B=m-2N_f$.}\label{fig:Young0}
\end{center}
\end{figure}

Repeated application of $b^\dag_{\alpha \beta \cdots}$ to the drained state
gives the state
\begin{equation}
b^\dag b^\dag \cdots|\text{dr}\rangle. \label{singlet}
\end{equation}
The form~(\ref{singlet}) places the single site state within an irreducible representation \linebreak of $SU(4N_f)$. 
Let us elaborate on this point. The single site state~(\ref{singlet}) is generated by an operator that has two sets of $n$ indices; $SU(N_c)$ color indices, and $U(4N_f)$ Dirac-flavor indices. The number $n$ is the number of fermions occupying the site. Gauge invariance restricts the color indices to be in an $SU(N_c)$ singlet, which means that 
\begin{equation}
n=m\cdot N_c,
\end{equation}
where $m$ is an integer. In fact, $m$ is the number of $b^\dag$ applications in \Eq{singlet}. More precisely, gauge invariance restricts the color indices to belong to the singlet representation $\underbar{s}$ given in Fig~\ref{fig:Young0}. 

\begin{figure}[htb]
\begin{center}
        \epsfig{width=2.cm,file=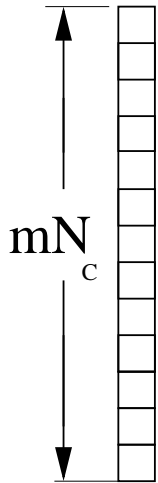} 
\caption[$S_n$ representation of a single site state]{The representation of $S_n$ carried by the complete set of indices of the singlet site state.}\label{fig:Young_S}
\end{center}
\end{figure}

Next, we recall that states in an irreducible representation of any special unitary group $SU(k)$, are also an irreducible representation of the permutation group. In our case, this means that the $n$ color indices belong also to the $\underbar{s}$ irreducible representation of $S_n$. Our aim is to determine the irreducible representation $\underbar{r}$ of the Dirac-flavor indices, with regard to the $SU(4N_f)$ group. Since this is also the $S_n$ irreducible representation of the $n$ Dirac-flavor indices, we find that the state~(\ref{singlet}) can belong to any possible part of the {\it reducible} representation $\underbar{r} \otimes \underbar{s}$ of $S_n$. Finally, we consider the fermionic nature of the state, and realize that the state belongs {\it only} to the completely antisymmetric representation $\underbar{A}$ of $S_n$ included in $\bar{r} \otimes \bar{s}$ and given in Fig.~(\ref{fig:Young_S}). 

\begin{figure}[htb]
\begin{center}
        \epsfig{width=5cm,file=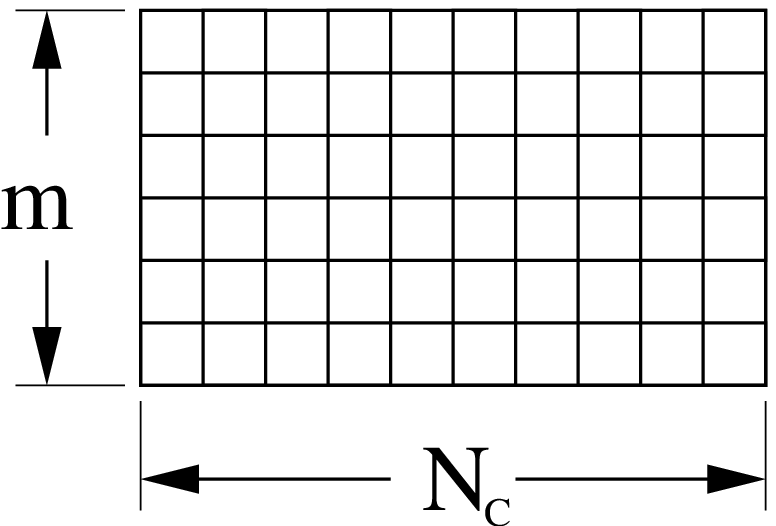} \caption[$U(4N_f)$ representation of a single site state]{The representation of $U(4N_f)$ carried by the spin in the effective antiferromagnet.  $m$ is related to the baryon number at the site according to $m=B+2N_f$, with $|B|\leq 2N_f$.}\label{fig:Young}
\end{center}
\end{figure}

For given $\underbar{s}$ and $\underbar{r}$, $\underbar{A}$ occurs in $\underbar{s}\otimes \underbar{r}$ if and only if $\underbar{r}$ is the $S_n$ conjugate representation of $\underbar{s}$ (which means flipping the Young tableau's rows and columns). Moreover the representation $\underbar{A}$ occurs only once in the decomposition of $\underbar{s}\otimes \text{conjugate}(\underbar{s})$. The conclusion is that the $S_n$ (and therefore the $SU(4N_f)$) representation of the Dirac-flavor indices is given in Fig.~\ref{fig:Young}.

We conclude by emphasizing the importance of the result of this section. Since the second-order effective Hamiltonian $H_{\text{eff}}^{(2)}$
preserves $B'_\bn$, the baryon number on each site, then any
distribution of $B'_\bn$ defines a sector within which
$H_{\text{eff}}^{(2)}$ is to be diagonalized. In other words,
baryons constitute a fixed background in which to study
``mesonic'' dynamics. The baryon number at each site fixes the
representation of $U(N)$ at that site, which is the space of
states in which the charges $Q^\eta_\bn$ act.


We now describe the global symmetries of the Hamiltonian. The $j=1$ terms in \Eq{H2eff2} are of the form $Q^\eta_\bn
Q^\eta_{\bn+\muhat}$, and they commute with the generators
\begin{equation}
Q^\eta=\sum_\bn Q^\eta_\bn
\end{equation}
of a global $U(N)$ symmetry group. This symmetry is in fact
familiar from the lattice Hamiltonian of naive, nearest-neighbor
fermions: Spin diagonalization of $N_f$ naive Dirac fermions
transforms the Hamiltonian into that of $4N_f$ staggered fermions.
In the weak coupling limit, there are in fact $8N_f$ fermion
flavors---the doubling problem. This doubling is partially
reflected in the accidental $U(4N_f)$ symmetry, which is intact in
the $g\to\infty$ limit and is respected by the effective
Hamiltonian. Retaining terms in the fermion Hamiltonian (and thus
in $H^{(2)}_{\text{eff}}$) that involve odd separations $j$ does
not break this symmetry.

The Nielsen-Ninomiya theorem \cite{NN} guarantees that any fermion
Hamiltonian of finite range will possess the full doubling
problem. This is a statement, however, about weak coupling only,
since the dispersion relation of free fermions is irrelevant if
the coupling is strong and the fermions are confined. It is
interesting that the accidental $U(4N_f)$ symmetry nonetheless
survives into strong coupling as a vestige of doubling.

The terms in \Eq{H2eff2} with {\em even\/} $j$, on the other hand,
break the $U(N)$ symmetry, as do even-$j$ terms in the original
fermion Hamiltonian. It is easy to see via spin diagonalization,
which leaves the even-$j$ terms unchanged, that the only
generators left unbroken are the $Q^\eta$ corresponding to
\begin{equation}
M^\eta=\bm{1}\otimes\lambda^a\quad\text{and}\quad
\gamma_5\otimes\lambda^a,
\end{equation}
which form the $U(N_f)_L\times U(N_f)_R$ chiral algebra. This of
course makes no difference to the Nielsen-Ninomiya theorem, which
will enforce 8-fold doubling in the perturbative propagator even
without the $U(4N_f)$ symmetry. If we are interested in the
realization of the global symmetries of the continuum theory,
though, we can study this lattice theory which has the same
symmetry. The simplest theory one may study is thus one containing
nearest-neighbor and next-nearest-neighbor terms. We shall proceed to discard
terms with longer range; we shall begin with the nearest-neighbor
theory, with its accidental doubling symmetry, and later break
this symmetry to $U(N_f)_L\times U(N_f)_R$ with the next-nearest-neighbor terms.

Two essential differences will always remain between this lattice
theory and the continuum theory. One is the presence of the axial
$U(1)$ symmetry on the lattice. This symmetry is exact, broken by
no anomaly, and may make the drawing of conclusions for the
continuum theory less than straightforward unless it is broken by
hand. The other difference is the fact that the effective
Hamiltonian for baryons (see below) is also a short-ranged hopping
Hamiltonian. If the baryons were almost free, we would say that
they are surely doubled like the original quarks. The fact that
the simplicity of the hopping terms is only apparent, and that the
baryons are still coupled strongly to mesonic excitations, offers
the possibility that doubling may not return.

\subsection{Third order: the baryon kinetic term}
\label{sec:3rd}

Finally we show here the form of the third-order contribution to $H_{\text{eff}}$, which only exists in the
case of $N_c=3$. This term is calculated via
\begin{equation}
H_{\text{eff}}^{(3)}=P_0V_FDV_FDV_FP_0.
\end{equation}
For a single link, we have
\begin{equation}
\langle0|U_{ab}U_{a'b'}U_{\dot a\dot b}|0\rangle=
\frac16
\epsilon_{aa'\dot a}\epsilon_{bb'\dot b},
\end{equation}
and the same can be proved for a chain of links,
\begin{equation}
\langle0| \left(\prod U\right)_{ab} \left(\prod
U\right)_{a'b'} \left(\prod U\right)_{\dot a\dot b}
|0\rangle= \frac16
\epsilon_{aa'\dot a}\epsilon_{bb'\dot b}.
\end{equation}
Thus [$f,g,h,\ldots$ are here flavor indices],
\begin{equation}
H_{\text{eff}}^{(3)}=-i\sum_{j>0}\tilde K(j) \sum_{\bn\mu}
\left(\psi^{\dag a f}_\bn \alpha_\mu
\psi^{b f}_{\bn+j\muhat}\right) \left(\psi^{\dag a' g}_\bn
\alpha_\mu \psi^{b' g}_{\bn+j\muhat}\right) \left(\psi^{\dag \dot a h}_\bn \alpha_\mu \psi^{\dot b h}_{\bn+j\muhat}\right)
\epsilon_{aa'\dot a}\epsilon_{bb'\dot b}
+h.c.
\end{equation}
The kernel is
\begin{equation}
\tilde K(j)=\frac{(D(j))^3}{6\left(\frac12g^2C_F|j|\right)^2}.
\end{equation}
Again, spin diagonalization simplifies the odd-$j$ terms, but not
the even-$j$ terms. The result is
\begin{equation}
H_{\text{eff}}^{(3)}=H_{\text{odd}}^{(3)}+H_{\text{even}}^{(3)},
\end{equation}
with
\begin{equation}
H_{\text{odd}}^{(3)}= -i\sum_{j>0\atop j \rm{odd}} \tilde K(j)
\sum_{\bn\mu}b_\bn^{\dag I}b_{\bn+j\muhat}^I
\eta_\mu(\bn)+h.c.,
\end{equation}
where $\eta_\mu(\bn)$ is the usual staggered-fermion sign factor,
and
\begin{equation}
H_{\text{even}}^{(3)}= -i\sum_{j>0\atop j \rm{even}} \tilde
K(j) \sum_{\bn\mu}b_\bn^{\dag I}\left[
\alpha_\mu\otimes\alpha_\mu\otimes\alpha_\mu\right]_{II'}
b_{\bn+j\muhat}^{I'}\zeta_\mu(\bn)+h.c.,
\end{equation}
where $\zeta_\mu(\bn)=(-1)^{\sum_{\nu\not=\mu}n_\nu}$. The baryon
operators are
\begin{equation}
b^I=\epsilon_{abc}\psi^{a \alpha}\psi^{b \beta}\psi^{c \gamma},
\end{equation}
where we have written $I$ to represent the compound index
$\{\alpha\beta\gamma\}$, taking values in the symmetric three-index
representation of $U(N)$ in Fig.~\ref{fig:Young3}. More precisely, 
\begin{center}
        \[ b^I = \left\{ \begin{array}{lr}
        \psi^{1}_\alpha \psi^{2}_\alpha \psi^{3}_\alpha        &       I=(\alpha,\alpha,\alpha)       \\
        \frac{1}{\sqrt{3}} (\psi^{1}_{( \alpha} \psi^{2}_\alpha \psi^{3}_{\beta )})   &
I=(\alpha,\alpha,\beta)       \\
        \frac{1}{\sqrt{3!}} (\psi^{1}_{( \alpha} \psi^{2}_\beta \psi^{3}_{\gamma )})   &   I=(\alpha,\beta,\gamma)
\end{array} \right. \]
\end{center}
$\alpha$, $\beta$ and $\gamma$ are indices of the fundamental representation of $U(4N_f)$, and the numbers are color indices. Note that these baryonic operators do not obey the canonical anticommutation relations of fundamental fermions \cite{Pittel:aa,Meyer:mu,Hadjimichef:1996cv,Sambataro:1995ma,Hadjimichef:1998rx,Girardeau:kw}, but rather obey 
\begin{equation}
\left\{b^I_\bn,b^{\dag
I'}_{\bn'}\right\}=\Delta_{II'}\delta_{\bn\bn'}\neq\delta_{II'}\delta_{\bn\bn'} .
\end{equation}
The operator $\Delta_{II'}$ has a complex index structure and is given in Appendix~\ref{app:anticomm}.

The odd-$j$ part of $H^{(3)}_{\text{eff}}$,
like that of $H^{(2)}_{\text{eff}}$, is symmetric under the $U(N)$
doubling symmetry. The even-$j$ part breaks $U(N)$ to
$U(N_f)_L\times U(N_f)_R$.
$H^{(3)}_{\text{eff}}$ is a baryon hopping term. As mentioned in
the introduction, however, its simplicity is deceptive. 

The separation of $H^{(3)}_{\text{eff}}$ into a canonical kinetic
energy and an interaction term is a challenge for the future.

The baryon operators responsible for the hopping are composite
operators that do not obey canonical anti-commutation relations.
If this were not the case, then the effective Hamiltonian in third
order would strongly resemble that of the $t$--$J$ model of condensed matter physics
\cite{Auerbach_book},
\begin{equation}
H_{t-J}=-t\sum_{\langle ij\rangle\atop
s}c^{\dag}_{is}c_{js} +J\sum_{\langle ij\rangle}\left(\bm
S_i\cdot\bm S_j-\frac{n_in_j}4\right)+\mathcal J'.
\end{equation}
Here $c_{js}$ is an annihilation operator for an electron at site
$j$ with spin $s$, and the number operators $n_i=c^\dag_ic_i$ and
spin operators $\bm S_i=\frac12c^\dag_i\bm\sigma c_i$ are
constructed from it. The added term $\mathcal J'$ is a more
complicated hopping and interaction term.  The $t$--$J$ model 
describes a doped antiferromagnet; it arises as the strong-binding
limit of Hubbard model, a popular model for itinerant magnetism
and possibly for high-$T_c$ superconductivity. The model is not
particularly tractable and, absent new theoretical developments,
does not offer much hope for progress in our finite-density
problem. It is nonetheless worth pondering the fact that a model
connected, however tentatively, with superconductivity appears in
a study of high-density nuclear matter.

A first step will be to replace these baryon operators with simpler operators that mimic only some of their features. Reintroducing all the properties of the real baryons can be done using methods of Hilbert space mapping (see for example \cite{Meyer:mu}). These methods have been introduced in the context of atomic and nuclear physics to treat quantum fields that are composite operators and therefore do not obey canonical commutation relations. One adds to the physical, original Hilbert space an auxiliary Hilbert space that is occupied by canonical operators, that replace the composite (non-canonical) operators. The price for this replacement is mainly very complicated interactions between the composite operators (baryons in our case), that has up to six-dimension operators. One can simplify these interactions by solving a non-Hermitian effective Hamiltonian, whose spectrum can provide us with insight on the real spectrum only if its projection on the physical spectrum is clustered at low energies.

\section{Summary}
\label{sec:Heff_summary}

To conclude we gather the results of the previous sections and present the effective Hamiltonian derived from the strong coupling limit of lattice QCD. We restrict to the second order contribution to the effective Hamiltonian. At this order the {\em local} baryon number operator commutes with the Hamiltonian. As a result, the baryon number on each of the sites is fixed, and the Hamiltonian acting on the degenerate subspace of fluxless states takes a block diagonal form, in which each block corresponds to a certain distribution of baryon number on the lattice. This also means that the last term in \Eq{eq:H} is just a constant and can be dropped. One must now diagonalize $H_{\text{eff}}^{(2)}$ within each of the blocks, i.e., for a certain distribution of baryon number, and therefore a certain distribution of representations of the $Q_\bn$ operators. 

The effective Hamiltonian has the structure of an antiferromagnet with interactions that connect sites over $j$ lattice links, with $j=1,2,3,\dots$. It has the following form (we drop a constant proportional to $B'$).
\begin{equation}
\label{eq:Hqm} H=\sum_{\bn \mu ,\eta \atop j=1,2,3,\dots} \left( s_\mu^{\eta} \right)^{j+1} \, J_j \, Q_{\bn}^{\eta}
Q_{\bn+j\muhat}^{\eta} .
\end{equation}
Here $Q_{\bn}^{\eta}$ are $U(N)$ charges at site $\bn$, with
$\eta=1,\ldots,N^2$. This Hamiltonian moves mesonic excitations
around the lattice, leaving the baryon density fixed. The states
at $\bn$ comprise a $U(N)$ representation whose Young tableau has
$N_c$ columns. The number of rows $m$ depends on the baryon number
$B$ at $\bn$ according to
\begin{equation}
m=B_\bn+2N_f.
\end{equation}
The sign factors $s^{\eta}_\mu$ are given by
\begin{equation}
\label{sign_factor} s^{\eta}_\mu=2\, \Tr M^{\eta} \alpha_\mu M^{\eta}
\alpha_\mu.
\end{equation}
Here $M^{\eta}=\Gamma^A\otimes \lambda^i$, where $\Gamma^A$ are the $16$ Dirac matrices, and $\lambda^i$ are the $N_f^2$ flavor generators. This makes $M^\eta ;\,\eta=1,\dots,16N_f^2$, the generators of $U(N)$ in the fundamental
representation. We fix their normalization by,
$\Tr M^{\eta} M^{\eta'} = \frac12 \delta_{\eta \eta'}$. The
matrices $\alpha_\mu$ are the $4\times4$ Dirac matrices times the
unit matrix in flavor space.

 In terms of the fermion kernel $D(j)$ the effective couplings $J_j$ are
\begin{equation}
J_j=\frac{4|D(j)|^2}{g^2C_FN_c\cdot j}=\frac{8|D(j)|^2}{g^2(N^2_c-1)\cdot j}>0. \label{eq:J_j}
\end{equation}

Recalling that we regard the strong coupling Hamiltonian as an effective Hamiltonian that describes QCD at large distances, then a physical fermion kernel must fall with $j$ at least exponentially. In view of the symmetry structure of the even and odd terms in $H_{\text{eff}}$ it is sufficient to take into account only nearest-neighbor and next-nearest-neighbor terms, and fix $D(j>2)=0$. The real values of the exchange coupling should be determined by the block-spin transformation discussed in Section~\ref{sec:idiology}.

\clearpage
\chapter{Mean field theory with Schwinger bosons}
\label{chap:MF_schwinger}

In this chapter we follow Arovas and Auerbach \cite{AA} and express the nearest-neighbor Hamiltonian derived in Chapter~\ref{chap:Heff} in terms of Schwinger bosons\footnote{This paper is concerned with $SU(N)$ antiferromagnets, in order to understand the $SU(2)$ antiferromagnet. For more work in that context see \cite{Affleck,MA,RS1,RS2,Arovas,Salam,Auerbach_book}.}.
We write the partition function as a path integral whose action is quartic in the bosons. Using a Hubbard-Stratonovich (HS) transformation, the interaction is decoupled, and the action becomes a quadratic form. After integrating the bosons we get an effective action for the HS fields. This action is proportional to the number $N=4N_f$ of boson flavors, and can be approached in the large $N$ limit. (To be precise the number $N_c$ of colors is also large in this limit, and $N_c/N$ is kept fixed.) We restrict to uniform and time-independent configurations of the HS fields, and replace the action by the mean field (MF) action $S_{\text{MF}}$. The latter depends on the MF values of the HS fields, and its minimization yields the ground state, which we analyze.

We first extend the analysis of \cite{AA} to three dimensions and show that at zero density and for large enough values of $N_c/N$ the ground state spontaneously breaks the global $U(N)$ symmetry. The ordered phase is the N\'eel phase of the antiferromagnet and in terms of chiral symmetry respects only vector rotations.\footnote{This is a subtle point, since the initial global symmetry $U(N)$ is much larger than chiral symmetry. We discuss this point in chapters~\ref{chap:NLSM} and \ref{chap:NNN}.}

Next we examine how the MF approach changes when we insert a non-zero concentration $c$ of baryons into the lattice. The baryons cannot move, and are represented by sites filled completely with baryon number. On such a site there is only one state, which is an $SU(N)$ singlet. This means that the generalized ``spin'' operators defined in \Eq{eq:Q_schwinger} obey $Q^{\eta}=0$, and that there are missing links in the nearest-neighbor antiferromagnetic Hamiltonian~(\ref{eq:Hqm}). Therefore the system becomes similar to an antiferromagnet with magnetic vacancies. We calculate the MF action to this system by two methods. For concentrations $c=\frac18$ and $\frac14$ we exactly minimize the MF action and see that the N\'eel phase shrinks as $c$ increases. For other values of $c\ll 1$ we disorder the baryon background and use a quenched approximation \cite{Brenig}.

\section{Schwinger bosons}
\label{sec:schwinger}

Our starting point is the effective nearest-neighbor Hamiltonian derived in Chapter~\ref{chap:Heff},
\begin{equation}
H=\frac{\tilde J_1}{N} \sum_{\bn}\sum_{\mu>0} \vec{Q}_\bn \cdot \vec{Q}_{\bn+\muhat}, \label{eq:H_AFM}
\end{equation}
with $\tilde J_1/N=J_1$ from Chapter~\ref{chap:Heff}. For brevity we drop the tilde sign in the rest of the chapter. The operators $Q^{\eta}$ generate the $U(N)$ algebra 
\begin{equation}
\left[ Q^\eta_\bn , Q^\rho_\bmm \right] = if^{\eta \rho \sigma} Q^\sigma_\bn \delta_{\bn \bmm}, \label{eq:commutation}
\end{equation}
and are defined as
\begin{equation}
Q^\eta_\bn=\sum_{\alpha \beta,a}\psi^{\dag a \alpha}_{\bn} M^\eta_{\alpha \beta} \psi^{a \beta}_{\bn}. \label{eq:Q_schwinger}
\end{equation}
Here $\alpha$ and $\beta$ are Dirac-flavor indices that range from $1$ to $N$, while $a$ is a color index that ranges from $1$ to $N_c$. The matrices $M^\eta$ are the generators of $U(N)$ in the fundamental representation. 
The Hilbert space on site $\bn$, occupied by $B$ baryons, is an irreducible representation of $U(N)$ that corresponds to a rectangular Young tableau of the form of Fig.~\ref{fig:Young}. It has $N_c$ columns, and its number of rows $m$ obeys $m=B+N/2$.
Following \cite{AA,RS1} we first restrict to configurations with zero average baryon number. In particular we examine configurations with baryon number $B$ on the even sites and $-B$ on the odd sites. This means we put conjugate $U(N)$ representations on adjacent sites, represented by Young tableaux with $m$ rows on the even sites, and with $(N-m)$ on the odd sites. The case of exactly zero baryon number on each site is the case of $m=N/2$. 

We now show how to construct the Hilbert spaces using bosons. On an even site $\bn$, we define a boson operator $b_{\alpha p,\bn}$. $\alpha$ is again a Dirac-flavor index that ranges from $1$ to $N$, but the index $p$ is now an auxiliary ``color'' index that ranges from $1$ to $m$. The bosons belong to the fundamental representations of both $SU(N)$ and $SU(m)$, and obey
\begin{equation}
\left[ b_{\alpha p\bn} , b^\dag_{\beta q\bmm} \right] = \delta_{\alpha\beta} \delta_{pq} \delta_{\bn \bmm}. \label{eq:comm_bosons}
\end{equation}
The Hilbert space on this site is constructed by $mN_c$  applications of the operators $b^{\dag}_\bn$ on the no-quantum state $|0\>$ 
\begin{equation}
|\alpha\beta \cdots\, ; pq\cdots \> \sim b^\dag_{\alpha p,\bn} b^\dag_{\beta q,\bn} \cdots |0\>. \label{eq:boson_state}
\end{equation}
We denote the $SU(N)$ representation of the Dirac-flavor indices $(\alpha,\beta,\dots)$ as ${\underbar r}$, and the $SU(m)$ representation of the color indices $(p,q,\dots)$ as $\underbar{r}'$. This means that the state~(\ref{eq:boson_state}) belongs to a representation $\underbar{r}'' \in \underbar{r}\times \underbar{r}'$ of the permutation group $S_{mN_c}$.
In particular, since the state $|\alpha\beta\cdots\,;pq\cdots\>$ is created with bosons, $\underbar{r}''$ is the completely symmetric representation of $S_{mN_c}$. This means that $\underbar{r}'$ must obey $\underbar{r}=\underbar{r}'$. 
As a result we find that  in order to constraint the $SU(N)$ indices of \Eq{eq:boson_state} to belong to Fig.~\ref{fig:Young}, it suffices to restrict its $SU(m)$ indices to a representation with the same tableau. This is easy. Because Fig.~\ref{fig:Young} has $m$ rows, it is a singlet of $SU(m)$, so any state $|\Omega \>$ must obey
\begin{equation}
\sum_{\alpha=1}^{N} b^{\dag}_{\alpha p,\bn} b_{\alpha q,\bn} |\Omega\>= N_c \delta_{pq}|\Omega\>. \label{eq:constraint}
\end{equation}

On the odd sites we choose the bosons to be in the conjugate-to-fundamental representations of $SU(N)$ and $SU(m)$, and consider them as hole operators. The commutation relations~(\ref{eq:comm_bosons}) are the same, and so is the restriction~(\ref{eq:constraint}), which is now interpreted as counting the number of holes created by applying the hole operators $b^{\dag}$ to the filled singlet state with $N$ rows in its Young tableaux.

In term of the bosons, the operators $Q^\eta_\bn$ are given by 
\begin{equation}
Q^\eta_\bn=\left\{\begin{array}{ll}
                        b^{\dag}_\bn \cdot M^\eta \cdot b_\bn    &       \qquad \bn \in \text{even},        \\
                        b^{\dag}_\bn \cdot \left(-M^{\eta}\right)^* \cdot b_\bn &       \qquad \bn \in \text{odd}.
\end{array} \right. \label{eq:Q_eo}
\end{equation}
Here we have suppressed all indices.
The difference between the even and odd definitions of $Q^\eta_\bn$ above is inserted in order that on both sublattices the operators $Q^\eta_\bn$ belong to the adjoint representation of $SU(N)$; the fact that $b$ transforms in the fundamental representation on the even sites and in the conjugate-to-fundamental representation on the odd sites is compensated by the fact that while $M^\eta$ generates the fundamental, $-M^{\eta *}$ generates its conjugate.
Using these definitions, the Hamiltonian becomes the following normal ordered quartic form,
\begin{eqnarray}
H&=&-\frac{J_1}{2N} \sum_{\bn \mu} \Tr \left[ {\cal A}^\dag_{\bn \muhat}  {\cal A}_{\bn \muhat} \right], \label{eq:H_A} \\
\left( {\cal A}_{\bn \muhat} \right)_{pq}&\equiv &\sum_{\alpha=1}^{N} b_{\alpha p,\bn} b_{\alpha q,\bn+\muhat}.
\end{eqnarray}
Here we have used the hermiticity of $M^\eta$ and the relation
\begin{equation}
\sum_\eta M^\eta_{\alpha \beta} M^{\eta}_{\alpha' \beta'} = \frac12 \delta_{\alpha \beta'}\delta_{\beta \alpha'}.
\end{equation}


We now introduce bosonic coherent states (see for example \cite{Auerbach_book}). This enables us to write the partition function as a path integral.
 
For each bosonic operator $b$ (we suppress all site, Dirac-flavor and color indices), one defines the bosonic coherent state $|z\>$ as
\begin{equation}
|z\>= e^{z b^\dag } |0 \>. \label{eq:coherent_z}
\end{equation}
Here $|0\>$ is the state annihilated by $b$, and $z$ is a complex number. This state obeys
\begin{eqnarray}
b|z\> &=& z |z \>, \label{eq:eigen} \\
\< z_1 | z_2 \> &=& e^{z^*_1 z_2}, \label{eq:overlap}  \\
{\bm 1} &=& \int d^2 z \, e^{-|z|^2 } |z \> \< z |, \label{eq:unity} \\
\Tr \left( \hat{O} \right) &=& \int d^2 z \, e^{-|z|^2 }\< z | \hat{O} | z \>. \label{eq:trace}
\end{eqnarray}
Here $\hat{O}$ is an operator that depends on $b$ and $b^\dag$. We use \Eq{eq:trace} to write the partition function 
\begin{equation}
Z=\Tr' e^{-\beta H}, \label{eq:Z1}
\end{equation}
where $\beta=1/T$, as  
\begin{equation}
Z=\int' d^2 z_0 \, \< z_0 | e^{-\beta H}|z_0\> \, e^{-|z_0|^2}. \label{eq:Z1_5}
\end{equation}
Here the trace in \Eq{eq:Z1} and integral of \Eq{eq:Z1_5} are constrained to contain bosonic states that obey \Eq{eq:constraint} only.

Next, we regard the inverse temperature $\beta$ as Euclidean time and slice it $N_\tau$ times,
\begin{equation}
\<z_0 | e^{-\beta H} | z_0 \> = \< z_0 | e^{-\epsilon H}  e^{-\epsilon H}  e^{-\epsilon H} \cdots  e^{-\epsilon H} | z_0 \>. \label{eq:slicing}
\end{equation}
Here $\epsilon=\beta/N_\tau$ and is small enough to write
\begin{equation}
e^{-\epsilon H} = \left( 1-\epsilon H \right). \label{eq:expand_exp}
\end{equation}
Inserting the unity~(\ref{eq:unity}) in between each of the $1-\epsilon H$ factors in \Eq{eq:slicing} we have
\begin{equation}
Z=\prod_{\tau=0}^{N_\tau-1} \int d^2 z_\tau e^{-z^*_\tau z_\tau} \< z_\tau | 1-\epsilon H |z_{\tau+1} \>,
\end{equation}
where $z_{N_\tau}=z_0$.
Each of the unities inserted is denoted by a different Euclidean time index $\tau=0,\dots,N_\tau-1$, hence the notations. Next we evaluate the matrix elements of \Eq{eq:expand_exp} to $O(\epsilon)$ using Eqs.~(\ref{eq:eigen})--(\ref{eq:overlap}). The result is  
\begin{equation}
\< z_\tau | 1-\epsilon H |z_{\tau+1} \> = e^{z^*_\tau z_{\tau+1}-\epsilon H(z^*_\tau,z_{\tau+1})}. \label{eq:matrix_el}
\end{equation}
Here $H(z^*_\tau,z_{\tau+1})$ is given by replacing all $b^\dag$ with $z^*$ and all $b$ with $z$ in the quantum Hamiltonian. This procedure gives in the following path integral.
\begin{eqnarray}
Z&=&\int_{z_0=z_{N_\tau}} Dz Dz^* \, e^{-S} \prod_{\bn p q}\delta\left( \sum_\alpha z^*_{\alpha p,\bn} z_{\alpha q,\bn} - N_c \delta_{pq} \-\right), \label{eq:Z2} \\
S&=&\epsilon \sum_{\tau=1}^{N_\tau-1} - z^*_{\tau} \partial_\tau z_{\tau} + H(z^*_\tau,z_\tau), \label{eq:S} \\
H&=&-\frac{J_1}{2N} \sum_{\bn \mu } \Tr \left[ {\cal A}^\dag_{\bn \mu}(\tau) {\cal A}_{\bn \mu}(\tau) \right], \\
\left({\cal A}_{\bn \mu} (\tau) \right)_{pq} &=& \sum_{\alpha=1}^{N} z_{\alpha p,\bn} z_{\alpha q,\bn+\muhat}.
\end{eqnarray}  
Here one assumes smooth Euclidean time configurations of the fields, and we have suppressed site, Dirac-flavor and color indices in the first term of \Eq{eq:S}. A note is needed regarding the definition of the temporal derivative of \Eq{eq:S}. To be unambiguous we write 
\begin{equation}
\epsilon \, \partial_\tau f(\tau) \equiv f_{\tau+1} - f_{\tau}. \label{eq:partial}
\end{equation}
In appendix~\ref{app:discretization} we show that ignoring this slicing procedure, and taking the continuum limit of the derivative, leads to slightly different results for energies. The difference is usually ignored, but in our context it remains important. 

Finally let us note the $\delta$ function in \Eq{eq:Z2}. This constrains the boson fields and keeps the Hilbert space explored by the path integration within the quantum constraint~(\ref{eq:constraint}). 

\section{Hubbard-Stratonovich transformation}
\label{sec:HS}

We proceed to implement a Hubbard-Stratonovich (HS) transformation to decouple the quartic interaction \Eq{eq:S} into quadratic forms. We use the identity
\begin{equation}
\int DQ DQ^\dag \, \exp \left( -\Tr \int_0^{\beta} d\tau \left( Q_{\bn \mu}-(J_1/2N){\cal A}_{\bn \mu} \right)^\dag \frac{2N}{J_1} \left( Q_{\bn \mu} - (J_1/2N){\cal A}_{\bn \mu}\right) \right)=1, \label{eq:HS}
\end{equation}
where we suppress all auxiliary color indices, and symbolically write the temporal sums as an integration over the variable $\tau$. The HS fields $Q^{pq}_{\bn \mu}$ are complex matrices in the color indices $p$ and $q$, and live on links of the lattice. \Eq{eq:HS} leads to 
\begin{eqnarray}
\exp \left( \frac{J_1}{2N} \Tr \int d\tau \, {\cal A}^\dag_{\bn \mu}(\tau) {\cal A}_{\bn \mu}(\tau) \right) &=& \int DQ DQ^\dag \exp \left[ \Tr \left( -\frac{2N}{J_1} Q^\dag_{\bn \mu} Q_{\bn \mu} \right. \right. \nonumber \\
&& \left. \left. +Q^\dag_{\bn \mu} {\cal A}_{\bn \mu} + {\cal A}^\dag_{\bn \mu} Q_{\bn \mu} \right) \right]. \label{eq:HS1}
\end{eqnarray}
We also write the constraint in \Eq{eq:Z2} as 
\begin{equation}
\int D\lambda \exp \left[ i \sum_{pq} \lambda^{pq}_\bn(
\tau) \left( \sum_\alpha z^*_{\alpha p,\bn} z_{\alpha q,\bn} - N_c \delta_{pq} \right) \right], \label{eq:constraint_Lam}
\end{equation}
where $\lambda^{pq}_\bn(\tau)$ is a real valued matrix field. Using \Eq{eq:HS1} and \Eq{eq:constraint_Lam} the partition function becomes 
\begin{eqnarray}
        Z &=& \int Dz Dz^* DQ DQ^\dag D\lambda \, e^{-S},  \label{eq:Z_HS} \\
        S &=& \int d\tau \, \Tr \left\{ \sum_{\bn \mu} \frac{2N}{J_1} Q^\dag_{\bn \mu} Q_{\bn \mu} + i \sum_\bn  N_c \lambda_\bn -\sum_\bn z^\dag_\bn \left[\partial_{\tau} + i \lambda_\bn \right] z_\bn  \right. \nonumber \\
&& \hskip 1.7cm \left. - \sum_{\bn \mu} \left[ Q^\dag_{\bn \mu} {\cal A}_{\bn \mu} + {\cal A}^\dag_{\bn \mu} Q_{\bn \mu} \right] \right \}.     \label{eq:S_HS}
\end{eqnarray}

The action is now quadratic in the bosons, and they can be integrated. This gives the following effective action for the $Q$ and $\lambda$.
\begin{equation}
S=N\int d\tau \, \Tr \left\{ \sum_{\bn \mu} \frac{2}{J_1} Q^\dag_{\bn \mu} Q_{\bn \mu} + i \sum_\bn  \kappa \lambda_\bn + \log G^{-1} \right \}. \label{eq:Seff}
\end{equation}
Here $G^{-1}$ is the propagator of a {\em single} boson in the presence of $Q$ and $\lambda$, while the general propagator is proportional to unity in the Dirac-flavor indices. Therefore the boson integration gives an overall factor of $N$, and $\log G^{-1}$ is of $O(1)$. Also note that we have introduced $\kappa \equiv N_c/N$. \footnote{$\kappa$ is equal to $r/4$, where $r$ is the ratio $N_c/N_f$ mentioned in Chapter~\ref{chap:Intro}, and in the abstract and summary of the thesis.} This will be useful in the following sections when we take the large $N$ limit and keep $\kappa$ fixed.

Let us make a final remark on the HS transformation. This exact transformation is based on \Eq{eq:HS}, which is always true for $J_1>0$. As a result we see that we can perform a HS transformation only if the interaction term in \Eq{eq:S} can be written in the form~(\ref{eq:H_A}). The latter was received with the aid of \Eq{eq:Q_eo}, which is true only for the special baryon number configuration of conjugate representations on adjacent sites (that has zero total baryon number). There are two additional simple cases in which one can write the magnetic interaction in the form of \Eq{eq:H_A} \cite{AA}. The first is the ferromagnet with identical representations on all sites, and the second is the antiferromagnet, written with {\it fermions} and zero baryon number on all the sites. The HS transformation for the fermions was examined in the large $N$ limit by many authors (for example \cite{RS1}), and seems to lead to a disordered phase. This limit is of no interest to us, since we expect that at zero density, our Hamiltonian describes an ordered phase with spontaneously broken chiral symmetry.

Antiferromagnetic interactions for other baryon distributions can also be written in the form of \Eq{eq:H_A}. The key point is that for each link one must have a representation of the generators analogous to \Eq{eq:Q_eo}. This however makes calculations more difficult. For example let us take uniform baryon density, with the same baryon number $B$ on all sites. This distribution is realized by having the same $U(N)$ representations on all sites, and therefore the same number of rows $m$ in the Young tableaux of all sites. We now build the Hilbert space on even and odd sites differently, making the boson operators on the even and odd sites belong to the fundamental and conjugate-to-fundemental representation of $SU(m)$ and $SU(N-m)$ respectively. This choice of bosons indeed leads to a relation similar to \Eq{eq:Q_eo}. 
This means that while the bosons on the even sites have an auxiliary color index $p$ that takes values in $[1,m]$, the auxiliary color index $q$ on the odd sites will take values in $[1,N-m]$. The bosons on the two sublattices are of a different kind and calculations become more difficult. For example, the HS field will now be rectangular matrices with $m$ rows and $N-m$ columns.


\newpage
\section{Mean field theory at zero density}
\label{sec:MF_zero}

Taking the large $N$ limit with $\kappa$ fixed allows a stationary approximation for the path integral. We therefore proceed to work with the following ansatz,
\begin{eqnarray}
Q^{pq}_{\bn \mu}(\tau)&=&Q \, \delta^{pq} , \label{eq:ansatz1}\\
\lambda^{pq}_\bn(\tau)&=& i \lambda\,  \delta^{pq}, \label{eq:ansatz2}
\end{eqnarray}
where we take $Q$ and $\lambda$ to be real. Using this ansatz we find that the MF action $S_{\text{MF}}$ defined by 
\begin{equation}
S_{\text{MF}}=(Nm\beta N_s)^{-1}\log Z_{\text{MF}},
\end{equation}
becomes
\begin{equation}
S_{\text{MF}}(Q,\lambda) = \frac{2dQ^2}{J_1} -\kappa \lambda + \frac{1}{\beta N_s} \Tr \log G^{-1}(Q,\lambda) .   \label{eq:S_MF}
\end{equation}

To evaluate the last term, we perform the integration over the bosons with the constraint and HS fields replaced by their MF values. (Here we have a single flavor and ``color'' field.)
\begin{equation}
\det G(Q,\lambda) = \int Dz Dz^\dag \exp \left\{ \int d\tau \sum_{\bn} \left[ z^*_\bn \left(\partial_{\tau} - \lambda \right) z_\bn  + Q \sum_{\mu=1}^d \left( z_\bn z_{\bn+\muhat} + z^*_\bn z^*_{\bn+\muhat} \right) \right] \right\}. \label{eq:det}
\end{equation}
We redefine the bosons on the even and odd sites as $z^{e,o}_\bN$, where $\bN$ is an index denoting sites on the even sublattice, which is an fcc. We now have two degrees of freedom on each site of the fcc. The first is the original boson $z^e$ on  $\bN$, and we use the convention that the second is the boson residing on the odd site $\bN+\hat{z}/2$, one link above the even site occupied by $z^e_\bN$.\footnote{We use the convention that the lattice spacing of the fcc lattice is $1$.}
We now apply the Fourier transform,
\begin{equation}
z^a_\bN(\tau) = \sqrt{\frac2{N_s \beta}} \sum_{\bk,n} z_{\bk,n} e^{i\left( \bk \cdot \bN - \omega_n \tau \right) } \times \left\{ \begin{array}{lc} 1 & a=e \\
        e^{-ik_z/2} & a=o \end{array} \right. , \label{eq:FT}
\end{equation}
where $\omega_n=2\pi n /T$ are Matsubara frequencies, and $\bk$ belongs to the first Brillouin zone of an fcc (i.e. to the unit cell of a bcc lattice.)  In momentum space the inverse propagator is
\begin{equation}
G^{-1}_\bk(\omega_n) = \left( \begin{array}{cc}
        \lambda + i\omega_n     &       -2dQ\gamma_\bk  \\
        -2dQ\gamma_\bk  &       \lambda - i\omega_n     \end{array}     \right),        \label{eq:invG}
\end{equation}
with the $2\times 2$ internal space defined by
\begin{equation}
f_{\bk,n} \equiv \left( \begin{array}{c} z^{a}_{\bk,n} \\  \\ z^{*o}_{-\bk,-n} \end{array} \right), \label{eq:f_nk}
\end{equation}
and $\gamma_\bk = \frac{1}{d}\sum_{\mu=1}^d \cos(k_{\mu}/2). $
The spectrum of this mean field theory is given by the poles of the propagator, 
\begin{equation}
\omega^2_n + \lambda^2-4d^2Q^2\gamma^2_\bk=0. \label{eq:dispersion}
\end{equation}
Near $\bk=0$, the energy described in \Eq{eq:dispersion} is $E^2_\bk = \Delta^2+c^2\bk^2$, with the squared mass $\Delta^2=\lambda^2-4d^2Q^2$, and the velocity $c^2=dQ^2$. This spectrum becomes massless for $\lambda^2=4d^2Q^2$. Evaluating the determinant of $G$ we get 
\begin{equation}
S_{\text{MF}} = \frac{2dQ^2}{J_1} -\lambda \left(\kappa + \frac{1}{2}\right) + \frac{2}{N_s\beta} \sum_\bk \log 2\sinh \left( \frac{\beta \omega_\bk}{2} \right) . \label{eq:S_MF1}
\end{equation}
Here $\omega_\bk=\sqrt{\lambda^2-4d^2Q^2\gamma^2_\bk}$, and $\lambda$ is assumed to be positive. 
Note that the transition from \Eq{eq:S_MF} and \Eq{eq:invG} to \Eq{eq:S_MF1} included performing a Matsubara sum. We stress that a naive summation will {\it not} lead to the $+\frac12$ in the second term of \Eq{eq:S_MF1}. This factor is received only by performing the determinant of the inverse propagator exactly, with the correct discretization of Euclidean time given in \Eq{eq:matrix_el}. We give a detailed explanation in Appendix~\ref{app:discretization}, where we derive the following rule of thumb for evaluations of a two-sublattice determinants of the form~(\ref{eq:det}): Evaluate the determinant ``naively'' by replacing the partial derivative $\partial_\tau$ with its simple continuum form $-i\omega_n$, and multiply the result by a factor of $e^{\beta \lambda/2}$ for each degree of freedom. For example, in the case considered, the number of degrees of freedom is $NmN_s$, hence the factor $\frac12$ in \Eq{eq:S_MF1}. 

We proceed to write the mean field equations for the action~(\ref{eq:S_MF1}),\footnote{Note that in principle one should also check whether the point $(Q_{\text{MF}},\lambda_{\text{MF}})$ is a minimum. If it is not, however, one can take some of the variables (in this case $Q$ or $\lambda$) to be imaginary, and make all curvatures positive. This is actually the reason for the ansatz~(\ref{eq:ansatz2})}
\begin{equation}
\frac{\partial S_{\text{MF}}}{\partial \lambda} = \frac{\partial S_{\text{MF}}}{\partial Q} = 0,
\end{equation}
that become
\begin{eqnarray}
\frac{Q}{2dJ_1}&=&\frac2{N_s} \sum_{\bk} \frac{Q\gamma^2_\bk}{\omega_\bk} \left( n(\omega_\bk) + \frac12 \right), \label{eq:MF1} \\
\kappa+\frac12 &=& \frac2{N_s} \sum_\bk \frac{\lambda}{\omega_\bk} \left( n(\omega_\bk)+\frac12 \right) . \label{eq:MF2}
\end{eqnarray}
Here $n(\omega)$ is the Bose function $(e^{\beta \omega}-1)^{-1}$. From this point on we restrict to $T=0$. 
The first solution (solution A) to Eqs.~(\ref{eq:MF1})--(\ref{eq:MF2}) is
\begin{eqnarray}
Q_A&=&0, \\
\lambda_A&=&\frac2{\beta} \coth ^{-1} (2\kappa+1).
\end{eqnarray}
This gives $Q_A=\lambda_A=0$ at $T=0$, and a value of $S_{\text{MF}}=-\beta^{-1}\log \frac{(\kappa+1)^{\kappa+1}}{{\kappa}^{\kappa}}\rightarrow 0^-$. 

\begin{figure}[htb]
\centerline{
\epsfig{width=12cm,file=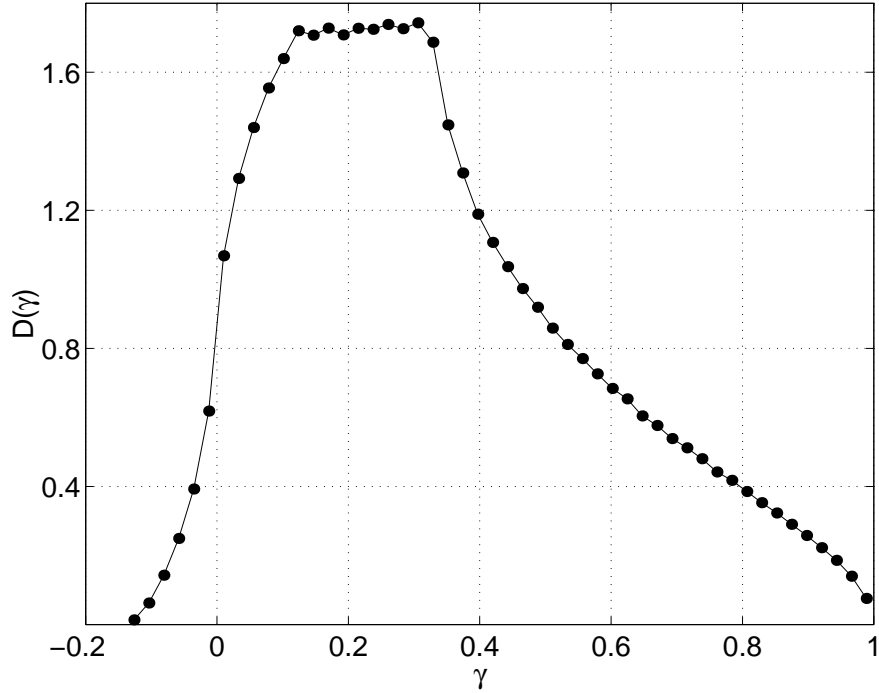}  
}
\caption{The integration measure of $\gamma$ for $d=3$.}
\label{fig:Dgamma}
\end{figure}

The second solution has $Q\neq0$ and \Eq{eq:MF2} becomes
\begin{eqnarray}
\kappa &=& f(\eta) + \frac{n(\Delta)}{N_s/2} \frac{\lambda}{\Delta}, \label{eq:MF2_sep} \\
f(\eta)&=&\int d\gamma \, D(\gamma) \, \frac12\left[ \frac1{\sqrt{1-\eta^2\gamma^2}} - 1 \right], \label{eq:f_eta}
\end{eqnarray}
with $\eta\equiv \lambda/(2dQ)$. Here the momentum integration was replaced by an integration over the function $\gamma_\bk$, provided with a integration measure $D(\gamma)$ given in Fig.~\ref{fig:Dgamma}, normalized to $\int d\gamma D(\gamma) = 1 $. 

\begin{figure}[htb]
\begin{center}
        \epsfig{width=12cm,file=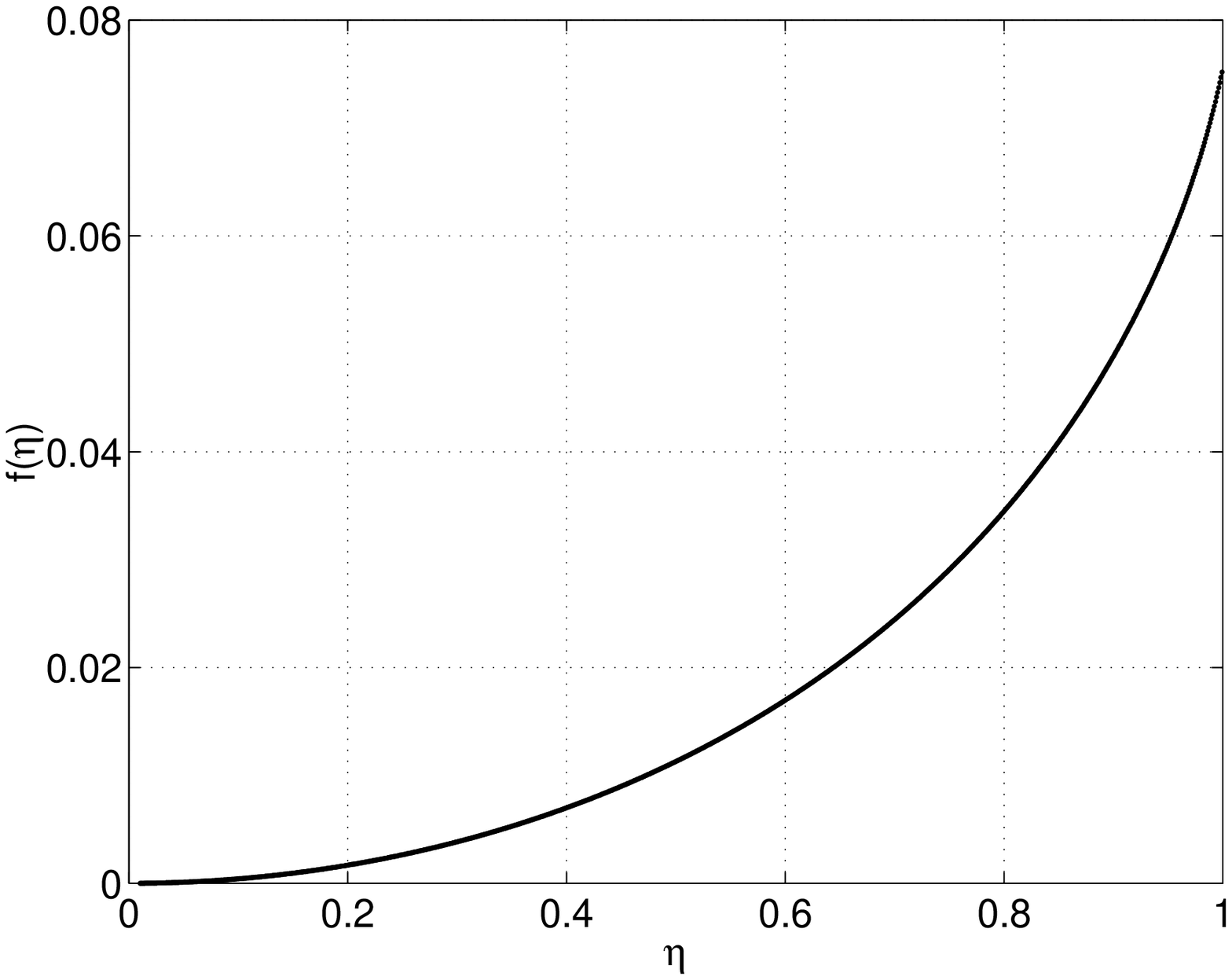}
        \caption[$f(\eta)$ -- right hand side of MF equation at zero baryon number.]{The right hand side of \Eq{eq:MF2_sep} excluding the condensation term for $d=3$.}
        \label{fig:f_eta}
\end{center}
\end{figure}

Note that we have separated the $\bk=0$ ($\gamma=1$) mode from the others, to allow a Bose-Einstein condensate (BEC). The second term in \Eq{eq:MF2_sep} will be non-zero at the thermodynamic limit only if $n(\Delta)/\sqrt{1-\eta^2}\sim O(N_s)$, which in turn means that $\eta\rightarrow 1$. Note that since $\gamma^2 \le 1$, the range of $\eta$ is $[0,1]$. The function $f(\eta)$ is given in Fig.~\ref{fig:f_eta}. 
It is monotonically increasing, and at $\eta=1$ takes the following values:
\begin{equation}
f(1)=\left[ \begin{array}{lc} \infty & d=1, \\
        0.19 & d=2, \\
        0.0787 & d=3. \end{array} \right. \label{eq:f1}
\end{equation}
We divide the solution of \Eq{eq:MF2_sep} into two cases.

\vskip 0.5cm
{\bf \large \underline{Disordered phase (solution B)}}

\vskip 0.5cm 

Here one assumes that $\Delta\neq0$ at the thermodynamic limit and drops the second term in the right hand side of \Eq{eq:MF2_sep}. The solution then reduces to
 \begin{eqnarray}
\kappa&=&f(\eta_B), \label{eq:MF_eta_B} \\
\lambda_B/dJ_1&=& \int d\gamma \, D(\gamma) \, \frac{\gamma^2}{\sqrt{1-\eta^2\gamma^2}}, \label{eq:lam2} \\
Q_B&=&\eta_B\lambda_B/2d.
\end{eqnarray}
This solution is guaranteed as long as $\kappa \le f(1)$. Recalling \Eq{eq:f1}, this solution exists always for $d=1$, but only for $\kappa \le \kappa_c$, with $\kappa_c\simeq 0.19$ for $d=2$, and $\kappa_c\simeq0.0787$ for $d=3$. The MF action of this solution is $-4dQ^2_B/J_1<0$. 

\vskip 0.5cm
{\bf \large \underline{Ordered phase (solution C)}}

\vskip 0.5cm 


For $\kappa>\kappa_c$, \Eq{eq:MF2_sep} cannot be solved without the extra condensation term. A consistent solution will be 
\begin{eqnarray}
\lambda_C&=&\lambda_B(1)+dJ_1(\kappa-\kappa_c),\\
Q_C&=&\lambda_C/2d, \\
\eta_C&=&\sqrt{1-\frac{T}{N_s\lambda_C(\kappa-\kappa_c)}} \simeq 1.
\end{eqnarray}
Here $\lambda_B(1)$ means the value of \Eq{eq:lam2} at $\eta=1$. The action of this solution is $-4dQ^2_C/J_1$, so solutions $B$ and $C$ have lower action than solution $A$, and are continuous at $\kappa=\kappa_c$. Nevertheless, the derivative of the mass $\Delta$ at $\kappa=\kappa_c$ is not continuous.
Recalling that $\kappa \equiv N_c/4N_f$, one finds that for $d=3$, and at $T=0$ there is a N\'eel phase for $\frac{N_c}{N_f} > 0.31$. This means that for $N_c=3$, and $N_f \leq 9.6$ the $U(4N_f)$ is spontaneously broken (for calculations of the staggered magnetization that show that the BEC indeed corresponds to a N\'eel phase \cite{AA,Auerbach_book}). 

\begin{figure}[htb]
\begin{center}
\epsfig{width=13cm,file=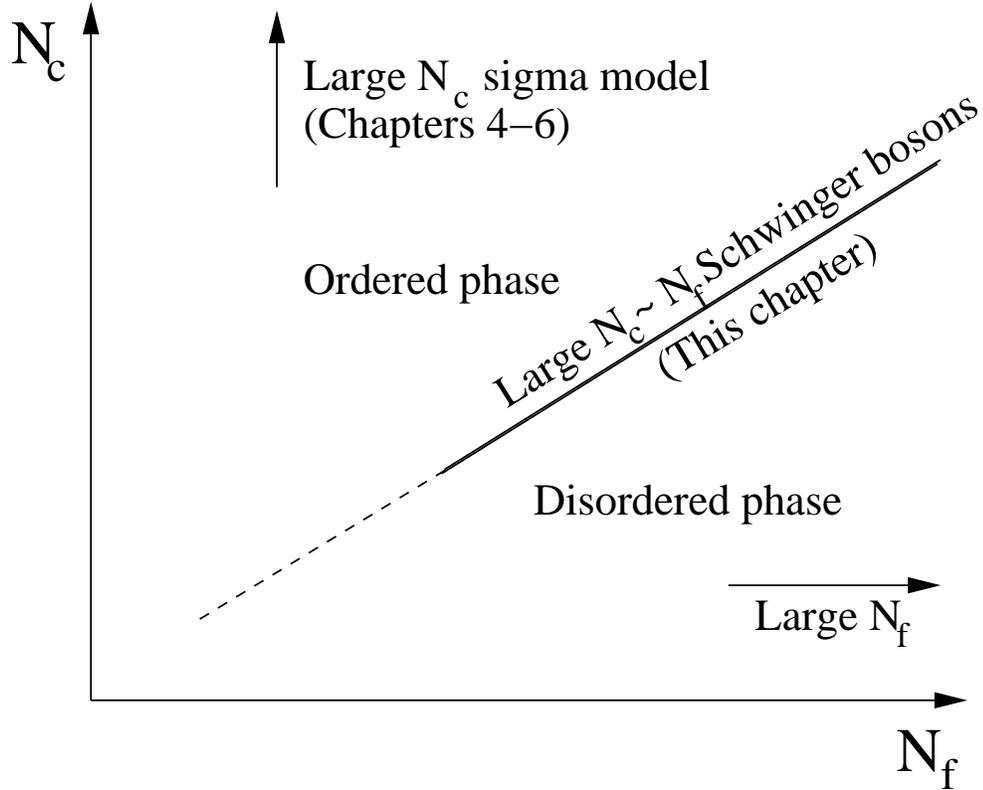}
\caption[Phase diagram of Schwinger bosons MF theory at $T=0$ and zero density.]{Phase diagram for the nearest-neighbor antiferromagnet at $T=0$, and zero density in $N_c$--$N_f$ plane, for MF theory with Schwinger bosons.}


\label{fig:MF_PT_schwinger}
\end{center}
\end{figure}

We summarize the zero density problem in the schematic phase diagram in plane $N_c$--$N_f$ in Fig.~\ref{fig:MF_PT_schwinger}, borrowed from \cite{AA}.
The large-$N_f$ phase is disordered, and will not concern us.
The location of the phase boundary is a line of constant slope, which turns out to be the line $N_c/N_f=0.31$; this means that QCD is safely in the
{\em ordered\/} phase for any reasonable number of flavors.


\section{Mean field theory at non-zero density}
\label{sec:MF_finite}
In this section we investigate how a non-zero density of baryons influences the MF equations and changes the boundary of the N\'eel phase in the $N_c$--$N_f$ plane. The MF approach discussed above can be used to investigate the following configuration of non-zero baryon number. On $M$ of the $N_s$ lattice sites we replace the $U(N)$ representations with the singlet of $SU(N)$, that has $B=2N_f$. This means that the $Q^\eta_\bn$ operators in these sites are zero (except $Q^0$ which corresponds to baryon number). The density $\rho$ of these configurations is $\rho=mc$, where 
\begin{equation}
c=\frac{M}{N_s},
\end{equation}
is the concentration of the singlet representations. 
In the presence of the ``impurities'', the effective Hamiltonian~(\ref{eq:H_AFM}) suffers from a set of missing links and becomes
\begin{equation}
H=\frac{J_1}{2N} \sum_{\bn,\bn+\muhat \not \in\{ \br\} } \vec{Q}_{\bn} \vec{Q}_{\bn+\muhat}. \label{eq:H_Imp}
\end{equation}
$\{\br\}$ denotes the locations of the $M$ impurities. We again make the MF ansatz (\ref{eq:ansatz1})--(\ref{eq:ansatz2}) and turn to calculate the MF action and MF equations. First note that $cN_s$ sites do not participate in the Hamiltonian. Also if there are no neighboring impurities, then there are $2c\, d N_s$ missing links. This means that the first two terms in \Eq{eq:S_MF} are now multiplied by $(1-c)$ and $(1-2c)$.
\begin{eqnarray}
Q^2 &\rightarrow & (1-c)\, Q^2 , \nonumber \\
\kappa \lambda &\rightarrow & (1-2c)\, \kappa \lambda. \label{eq:change1}
\end{eqnarray}
Also, the evaluation of the functional determinant is different. We have analyzed this problem using two techniques. The first is an analysis of certain configurations of impurities, in which the system does not lose translation invariance. We solve only the simplest cases of $c=\frac18, \, \frac14$ in section~\ref{sec:c1_84}. In section~\ref{sec:quenched} we make a quenched average of the impurities' positions which restores translation invariance. In this method we find that the corrections to the MF action for low values of $c$ can be written as a power series in $c$, and we the present results of the first order. 

\subsection{Periodic density with baryon concentration $c=\frac18,\,\frac14$}
\label{sec:c1_84}

In general a random distribution of baryon impurities destroys translation invariance found at zero baryon number, and makes the methods used in the previous section inapplicable. In particular, a Fourier transform does not diagonalize the propagator of the bosons, making the evaluation of the determinant very difficult.
There are however densities that can be realized with a translation invariant configuration of impurities. In three dimensions the simplest configurations have $c=\frac18$, and $c=\frac14$. In this section we calculate the MF action of these configurations, and proceed to solve the corresponding MF equations.

\subsubsection{\underline{Concentration $c=\frac18$}}
\begin{figure}
\vspace{0.5cm}
\begin{center}
\epsfig{width=1.5in,file=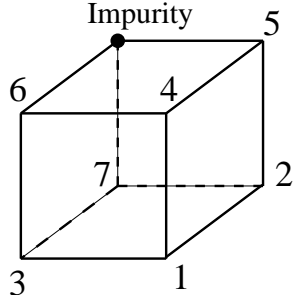}
\caption[A unit cell with $c=\frac18$ concentration of ``magnetic'' vacancies]{A unit cell for impurity concentration of $c=\frac18$.}
\label{fig:c18_3d}
\end{center}
\end{figure}

\begin{figure}
\vspace{0.5cm}
\begin{center}
\epsfig{width=12cm,file=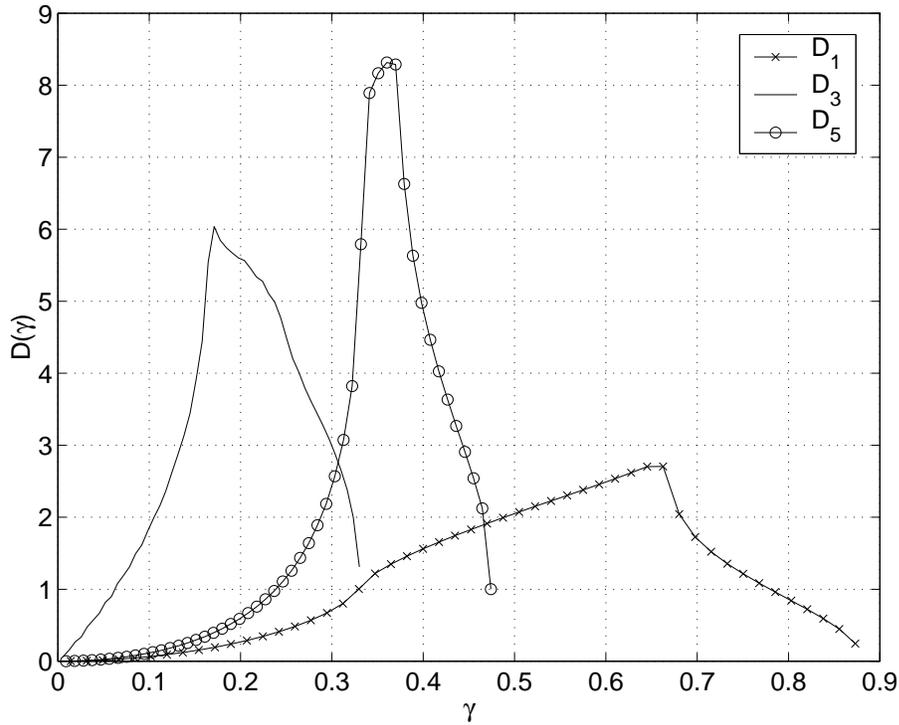}
\caption[The integration measures of $\gamma_{1,3,5}$ for $c=\frac18$.]{The integration measures for $c=\frac18$.}
\label{fig:D135_c18}
\end{center}
\end{figure}

We divide the lattice into unit cells of the form shown in Fig.~\ref{fig:c18_3d}. 
Each includes seven ordinary sites and one impurity. We place the latter on the upper-left-back corner of the cell, and denote the bosons on the other sites by $z_{\bn ,\rho}$. Here $\bn$ is the location of the cell in a simple cubic lattice of spacing 2. The index $\rho=1,\dots,7$, is an internal index denoting the sites in each cell, corresponding to the numbering in Fig.~\ref{fig:c18_3d}. Using these notations, the MF action is 
\begin{eqnarray}
S_{\text{MF}}&=&\frac{9Q^2}{2J_1}-\frac78\lambda \kappa + \frac1{N_s\beta}\Tr \log G^{-1}, \label{eq:MF_action_c18}\\
\det G &=& \int Dz Dz^* \exp \left[ -\int d\tau \sum_{\bn \rho}  z^*_{\bn \rho} \left(-\partial_{\tau} + \lambda \right) z_{\bn \rho} -  S'(Q) \right] , \label{eq:S_bosons_c18} \\
S'(Q)&=&-Q\int d\tau \sum_\bn \left[ z_{\bn,2}(z_{\bn,1}+z_{\bn-\hat{x},1}) + z_{\bn,3}(z_{\bn,1}+z_{\bn-\hat{y},1}) \right. \nonumber \\
&&\hskip 2cm +z_{\bn,4}(z_{\bn,1}+z_{\bn-\hat{z},1})+z_{\bn,5}(z_{\bn,2}+z_{\bn-\hat{z},2}) \nonumber \\
&&\hskip 2cm +z_{\bn,6}(z_{\bn,3}+z_{\bn-\hat{z},3})+z_{\bn,7}(z_{\bn,2}+z_{\bn-\hat{y},2}) \nonumber \\
&&\hskip 2cm +z_{\bn,7}(z_{\bn,3}+z_{\bn-\hat{x},3})+z_{\bn,5}(z_{\bn,4}+z_{\bn-\hat{x},4}) \nonumber \\
&&\left. \hskip 2cm +z_{\bn,6}(z_{\bn,4}+z_{\bn-\hat{y},4}) \right] + c.c.
\end{eqnarray} 
Next we introduce the Fourier transform 
\begin{equation}
z_{\bn,\rho}=\sqrt{\frac{8}{N_s}}\sum_{\bk,n} z_{\bk n,\rho}e^{i(\bn\cdot \bk-\omega_n\tau)},
\end{equation}
where $\bk$ belongs to the first Brillouin zone of the simple cubic lattice. In momentum space $S'(Q)$ becomes
\begin{equation}
S'(Q)=-\frac12 \left( 6 Q \, z^T_\bk \, \Gamma_\bk \, z_{-\bk} + c.c. \right), \label{eq:S_prime}
\end{equation}
where $z$ and $\Gamma$ are 
\begin{eqnarray}
z_\bk \equiv \left( \begin{array}{c} z_{\bk,1} \\ \vdots \\ z_{\bk,7} \end{array} \right), \qquad
\Gamma_\bk = \left( \begin{array}{ccccccc}
 0 & X & Y & Z & 0 & 0 & 0 \\
 X^* & 0 & 0 & 0 & Z & 0 & Y \\
 Y^* & 0 & 0 & 0 & 0 & Z & X \\
 Z^* & 0 & 0 & 0 & X & Y & 0 \\
 0 & Z^* & 0 & X^* & 0 & 0 & 0 \\
 0 & 0 & Z^* & Y^* & 0 & 0 & 0 \\
 0 & Y^* & X^* & 0 & 0 & 0 & 0 \end{array} \right), \label{eq:Gamma}
\end{eqnarray}
and $X_i = \frac16(1+e^{-ik_i})$. We find that the eigenvalues $\gamma_\rho$ of $\Gamma_\bk$ obey 
\begin{eqnarray}
0&=&\gamma\left[ \gamma^6 +a_2 \gamma^4 + a_1 \gamma^2 +a_0 \right], \label{eq:polynom_c18} \\
a_2 &=&-3 \sum_i \rho^2_i, \\
a_1 &=&3 \sum_i \rho^4_i + 4\sum_{i>j} \rho^2_i\rho^2_j -2 \, {\text Re}\left( (XY^*)^2 +\rho^2_3\, (X^2+Y^2) \right), \\
a_0 &=& \sum_{i\neq j} \rho^4_i \rho^2_j - \sum_i \rho^6_i - 4 \rho^2_x \rho^2_y \rho^2_z + 2\, {\text Re} \left[ (XY^*)^2 (\rho^2_x+\rho^2_y-\rho^2_z) \right. \nonumber \\
&& \qquad \qquad \qquad \left. + (X^2 - Y^2)(\rho^2_z \rho^2_x - \rho^2_z \rho^2_y ) + (X^2+Y^2) \rho^4_z \right],
\end{eqnarray}
with $\rho^2_i=|X_i|^2$. 
We have verified that the solutions to \Eq{eq:polynom_c18} are real and obey $\gamma_{0}=0$, $\gamma_{2}=-\gamma_{3}$, $\gamma_{4}=-\gamma_{5}$, and $\gamma_{6}=-\gamma_{7}$, and are all even in the momentum $\bk$. Since the first term in the action of \Eq{eq:S_bosons_c18} is unity in the $7$-dimensional internal space, we write
\begin{eqnarray}
\det G &=& \int Dz Dz^* \exp \left\{ - \sum_{k_z>0\atop \rho} \int d\tau \left( \begin{array}{cc} z^*_{\bk \rho} & z_{-\bk,\rho} \end{array} \right) \left( \begin{array}{cc} \partial_{\tau} + \lambda & -6Q\gamma_{\bk \rho} \\ -6Q \gamma_{\bk \rho} & -\partial_{\tau} + \lambda \end{array} \right) \left( \begin{array}{c} z_{\bk \rho} \\ z^*_{-\bk \rho} \end{array} \right) \right\}. \nonumber \\ \label{eq:detG18}
\end{eqnarray}
Here the factor $1/2$ of \Eq{eq:S_prime} is replaced by summing over only half of the Brillouin zone (the remainder of the Brillouin zone gives the same contribution due to the coupling of $z_{\bk}$ and $z_{-\bk}$).
As \Eq{eq:detG18} stands, all the fields are independent and can be integrated over. This gives
\begin{equation}
S_{\text{MF}}=\frac{9Q^2}{2J_1}-\frac78 \lambda \left( \kappa + \frac12 \right) + \frac18 \sum_{\rho} \frac1{\beta} \int_\rho d\gamma D_\rho(\gamma) \log 2\sinh \left( \frac{\beta \omega(\gamma)}{2} \right), \label{eq:S_MF_c_18}
\end{equation}
where $\omega(\gamma) \equiv \sqrt{\lambda^2-36Q^2\gamma^2}$, and $D_\rho(\gamma)$ is the integration measure of $\gamma_{\bk,\rho}$ given in Fig.~\ref{fig:D135_c18}, normalized to $\int d\gamma D_\rho(\gamma)=1$. The factor $\frac18$ accounts for the $N_s/8$ sites found in the simple cubic lattice. Also note that the third term, $-\frac78 \lambda \frac12$, is added according to the rule of thumb prescription of Appendix~\ref{app:discretization}.

\begin{figure}[htb]
\vspace{0.5cm}
\centerline{\epsfig{width=12cm,file=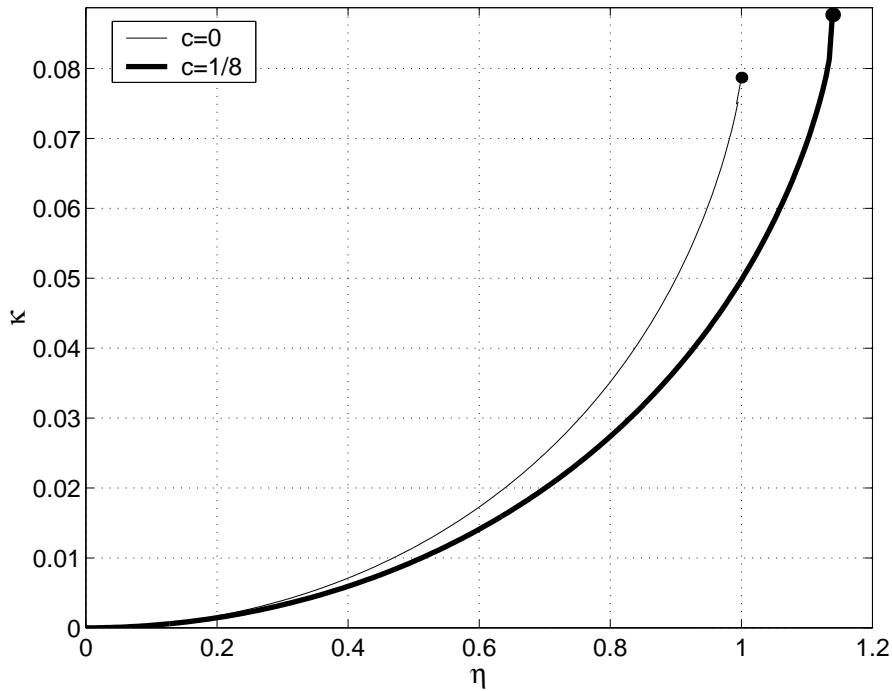}}
\caption[Right hand side of the MF equation for $c=\frac18$.]{Right hand side of the MF equations \Eq{eq:MF_eta_B} and \Eq{eq:MF_eq_18}, with $c=0$ and $c=\frac18$. The solution $\eta(\kappa)$ is given by intersection of the curves with horizontal lines. Note that the curves have an end at $\eta=1$ and $\eta=1.14$, marked by the full circles there.}
\label{fig:compare_18_0}
\end{figure}

We proceed to concentrate on the MF equation 
\begin{equation}
\frac{\partial S_{\text{MF}}}{\partial \lambda}=0,
\end{equation}
which, excluding BEC, becomes 
\begin{equation}
\frac78 (\kappa+\frac12) = \frac1{8} \left(n(\lambda) +\frac12 \right) + \frac14 \sum_{\rho=1,3,5} \int d\gamma D_\rho(\gamma) \frac{1}{\sqrt{1-\eta^2 \gamma^2}} \left( n(\omega) +\frac12 \right).
\end{equation}
At $T=0$ this simplifies to
\begin{equation}
\kappa =  \frac1{7} \sum_{\rho=1,3,5} \int d\gamma D_\rho(\gamma) \left[\frac{1}{\sqrt{1-\eta^2 \gamma^2}} - 1\right]. \label{eq:MF_eq_18}
\end{equation}
In Fig.~\ref{fig:compare_18_0} we show the right hand side of the MF equations (\ref{eq:MF_eta_B}) and (\ref{eq:MF_eq_18}) for comparison. First note that for $c=\frac18$, $\eta$ takes values up to $1.14$. This is because the maximum value $\gamma$ can have is $0.877$, and not $1$ as for $c=0$. Next we find that the right hand side has a maximum of $0.0881$ at $\eta= 1.14$. This is the point where BEC occurs. The physical result is therefore that the N\'eel phase at $c=\frac18$ shrinks, and symmetry restoration occurs above $\kappa_c=0.0881$. For QCD with $N_c=3$ this means that the symmetry is restored already at $N_f\simeq 8.5$ instead at $N_f\simeq 9.6$ at $c=0$.

\subsubsection{\underline{Concentration $c=\frac14$}}

\begin{figure}
\vspace{0.5cm}
\begin{center}
\epsfig{width=12cm,file=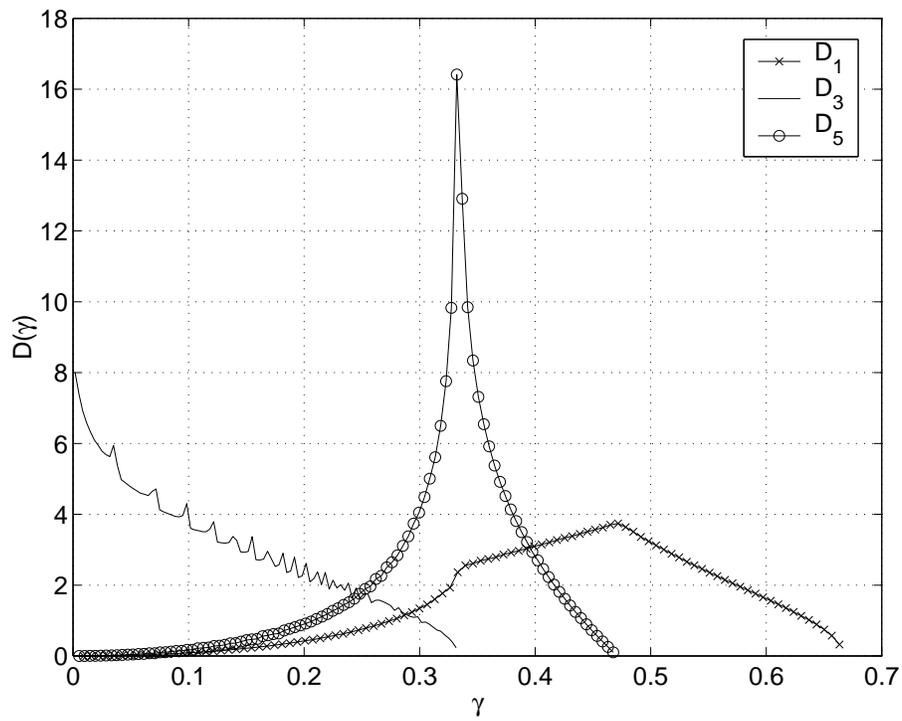}
\caption[The integration measures of $\gamma_{1,3,5}$ for $c=\frac14$]{The integration measures $D_\rho(\gamma)$. The oscillations in $D_3$ are an artifact of the resolution of the momentum mesh.}
\label{fig:D135_c14}
\end{center}
\end{figure}

In this case we put an impurity on site no.~1 in Fig.~\ref{fig:c18_3d} as well. This means that $c=\frac14$, and \Eq{eq:MF_action_c18} is changed to
\begin{equation}
S_{\text{MF}}=\frac{3Q^2}{J_1}-\frac34\lambda \kappa + \frac1{N_s\beta}\Tr \log G^{-1}, \label{eq:MF_action_c14}
\end{equation}
In evaluating the last term, we again use \Eq{eq:S_bosons_c18}, but with the boson field $z_{\bn,1}$ omitted. This means that the matrix $\Gamma_\bk$ in \Eq{eq:Gamma} has lost its first row and column and is now $6$-dimensional. We find that its eigenvalues obey
\begin{equation}
\gamma^6 +a_2 \gamma^4 + a_1 \gamma^2 +a_0 =0, \label{polynom_c14}
\end{equation}
with
\begin{equation}
a_2 =-2 \sum_i \rho^2_i, \qquad a_1 = \left(\sum_i \rho^2_i \right)^2, \qquad a_0 = - 4 \rho^2_x \rho^2_y \rho^2_z.
\end{equation}
Again we find that the solutions $\gamma_\rho$ are real and obey $\gamma_1=-\gamma_2$, $\gamma_3=-\gamma_4$, and $\gamma_5=-\gamma_6$. At $T=0$, the MF equation becomes 
\begin{equation}
\kappa =  \frac1{6} \sum_{\rho=1,3,5} \int d\gamma D_\rho(\gamma) \left[\frac{1}{\sqrt{1-\eta^2 \gamma^2}} - 1\right], \label{eq:MF_eq_14}
\end{equation}
with the measures given in Fig.~\ref{fig:D135_c14}.

\begin{figure}[htb]
\vspace{0.5cm}
\centerline{\epsfig{width=12cm,file=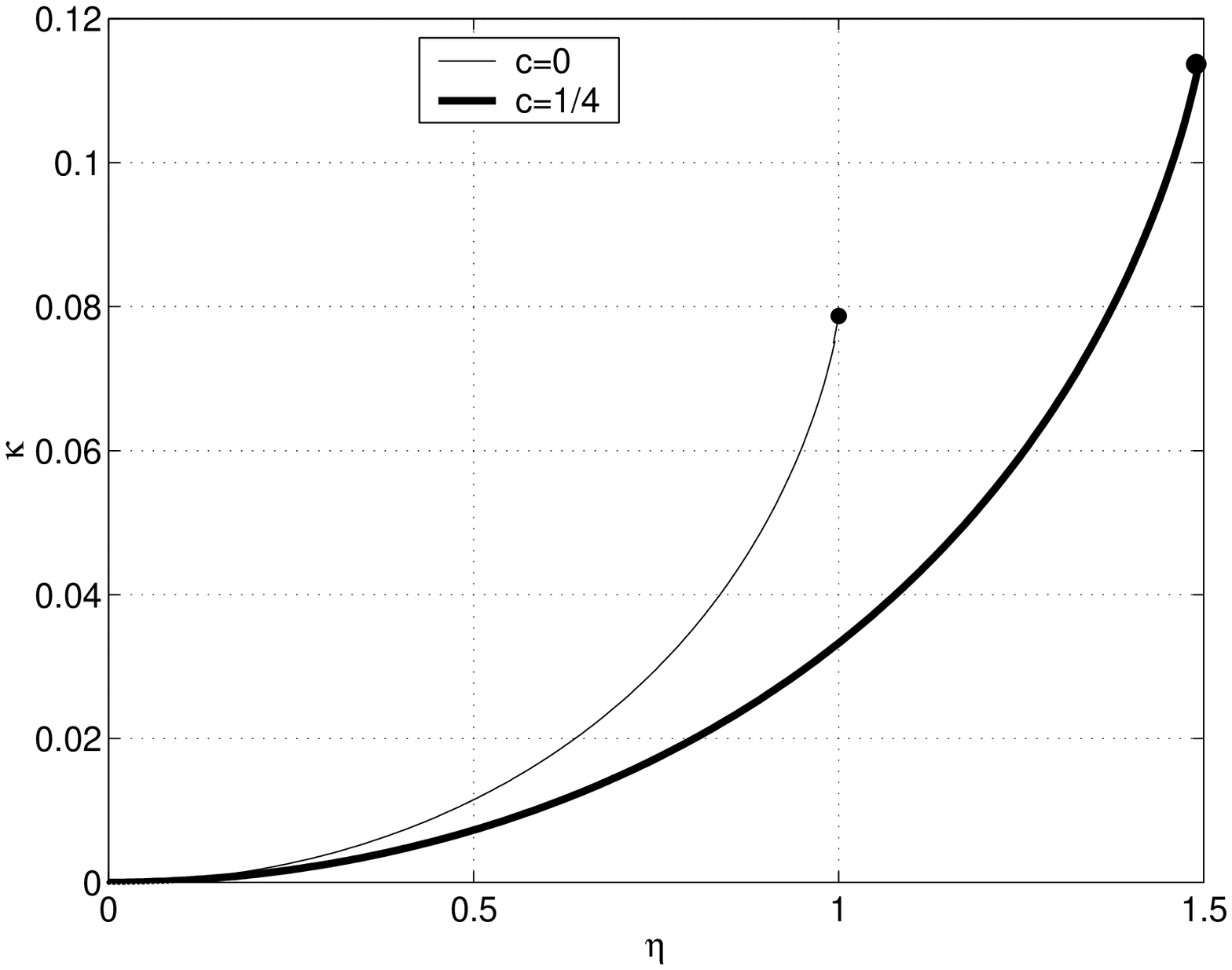}}
\caption[Right hand side of the MF equation for $c=\frac14$.]{The right hand side of the MF equations \Eq{eq:MF_eta_B} and \Eq{eq:MF_eq_14}, with $c=0$ and $c=\frac14$. Note that the curves have an end at $\eta=1$ and $\eta=1.5$, marked by the full circles there.}
\label{fig:compare_14_0}
\end{figure}

In Fig.~\ref{fig:compare_14_0} we show the right hand side of the MF equations~(\ref{eq:MF_eta_B}) and~(\ref{eq:MF_eq_14}) for comparison. In this case, because the maximum value of $\gamma$ is $0.66$, $\eta$ takes values up to $1.5$. We find that the right hand side at the BEC point $\eta=1.5$ is $0.117$. The physical result is therefore that the N\'eel phase at $c=\frac14$ shrinks even more than in $c=\frac18$, and symmetry restoration occurs above $\kappa_c=0.117$, or for QCD with $N_c=3$ already above $N_f\simeq 6.4$.

\subsection{Random baryon positions -- the quenched approximation}
\label{sec:quenched}

In this section we solve the MF equation of a quenched random distribution of baryon impurities. We adopt the formalism of \cite{Brenig} developed to treat the affect of magnetic impurities on an $SU(2)$ magnet. Here the analog to the impurities are sites with $SU(4N_f)$ singlet representations. We denote the number of of singlet representations on the even and odd sites as $M_e$ and $M_o$, and present calculations to first order in the concentration $M_e/N_s$, and $M_o/N_s$. Again the tricky part is to calculate the functional determinant~(\ref{eq:det}), because now the boson propagator is no longer diagonal in momentum space.

We begin by modifying \Eq{eq:det} as follows,
\begin{equation}
\det G(Q,\lambda) \left[ \int \left( Dz Dz^* \right)_{\bn \in \{ \br \}} e^{-S'} \right]= \int Dz Dz^\dag \exp \left\{-S_0-S_{\text{Imp}} \right\}, \label{eq:det1}
\end{equation}
where $S_0$ is the `pure' action of \Eq{eq:det}
\begin{equation}
S_0=\int d\tau \sum_{\bn} \left[ z^*_\bn \left(-\partial_{\tau} + \lambda\right) z_\bn  - Q \sum_{\mu>0} \left( z_\bn z_{\bn+\muhat} + z^*_\bn z^*_{\bn+\muhat} \right) \right].
\end{equation}
The auxiliary action $S'$ is 
\begin{equation}
S'=\int d\tau \sum_{\bn\in \{ \br \}} z^*_\bn \left(-\partial_{\tau} + \lambda+\lambda' \right) z_\bn , 
\end{equation}
where $\lambda'$ is real and obeys $\lambda+\lambda'>0$ for stability. Also we choose the action $S'$ to be defined in the continuum temporal limit, i.e. with the ``naive'' definition of $\partial_\tau$ (in Matsubara space it is just $-i\omega_n$). This means that the expression in the square brackets in the left hand side of \Eq{eq:det1} is given by
\begin{equation}
\log Z'=-c\, N_s \log 2 \sinh \frac{\beta(\lambda+\lambda')}{2}. \label{eq:Z'}
\end{equation}
The $\lambda'$ term was inserted in order that $S_{\text{Imp}}$ will have the following form, 
\begin{equation}
S_{\text{Imp}}=\int d\tau \sum_{\bn \in \{ \br \}}  \left[ \lambda' z^*_\bn z_\bn + Q \sum_{\mu=\pm1}^{\pm d} \left( z_\bn z_{\bn+\muhat} + z^*_\bn z^*_{\bn+\muhat} \right) \right] \equiv \int d\tau \sum_{\bn, \bmm}  z^*_\bn V_{\bn \bmm} z_\bmm . \label{eq:S_imp}
\end{equation}

We now write the MF action for the Hamiltonian~(\ref{eq:H_Imp}) as
\begin{eqnarray}
S_{\text{MF}}&=&(1-2c) \frac{2dQ^2}{J_1} -c\, \lambda(\kappa +\frac12 ) - c \log \left( 1-e^{-\beta(\lambda+\lambda')} \right) \\ 
&& \hskip 0.5cm + \frac1{N_s\beta} \Tr \log G^{-1} + \Delta S,  \label{eq:S_MF2}\\
\Delta S &=&  \frac1{N_s\beta} \Tr \log \left( {\bm 1} + G V\right). \label{eq:delS}
\end{eqnarray}
Here $G^{-1}(Q,\lambda)$ is the same inverse propagator that appears in the zero density problem in section \ref{sec:MF_zero}, and we proceed to evaluate $\Delta S$. Using the Fourier transform~(\ref{eq:FT}) we write $V$ in momentum space as 
\begin{eqnarray}
S_{\text{Imp}}&=&\sum_{\omega_n} \sum_{\bk,\bk'} f^\dag_{\bk,n} V_{\bk,\bk'} f_{\bk',n}, \\
V_{\bk,\bk'}&=& 2dQ \, \frac2{N_s} \left[ \sum_{\bmm\in \{ \br_e \}}  e^{i(\bk'-\bk)\bmm} \left( \begin{array}{cc} b^2 & \gamma_{\bk'} \\ \gamma_\bk & 0 \end{array} \right) + \sum_{\bmm\in \{ \br_o \}}  e^{i(\bk'-\bk)\bmm} \left( \begin{array}{cc} 0 & \gamma_{\bk} \\ \gamma_\bk' & b^2 \end{array} \right) \right]. \nonumber \\
\end{eqnarray}
Here $f_{\bk,n}$ is given in \Eq{eq:f_nk}, $\bmm$ are the positions of the impurity sites in terms of the original simple cubic lattice, $\{\br_e \}$ and $\{\br_o \}$ denote the impurities on even and odd sublattices, and $b^2\equiv \lambda'/2dQ$. Fortunately, the vertex $V_{\bk,\bk'}$ is reducible to the form \footnote{This reducibility is achieved only for $b\neq0$, and is in fact the reason we introduced $\lambda'$.}
\begin{eqnarray}
V_{\bk,\bk'}&=&\frac{1}{N_s}\sum_{\bmm \in \{ \br \}_e}  {\cal I}^e_\bk {\cal J}^e_{\bk'} e^{i(\bk'-\bk)\bmm} + \frac{1}{N_s} \sum_{\bmm \in \{ \br \}_o}  {\cal I}^o_\bk {\cal J}^o_{\bk'} e^{i(\bk'-\bk)\bmm} , \label{eq:V_reduced}\\
{\cal I}^e_\bk&=&\sqrt{4dQ}\left( \begin{array}{cc} b & 0 \\ \gamma_\bk/b &  -\gamma_\bk/b \end{array} \right), \hskip 0.5cm {\cal J}^e_{\bk'} = \sqrt{4dQ}\left( \begin{array}{cc} b & \gamma_{\bk'}/b \\ 0 & \gamma_\bk'/b \end{array} \right), \label{eq:V_reduced_e} \\
{\cal I}^o_\bk&=&\sqrt{4dQ}\left( \begin{array}{cc} -\gamma_\bk/b & \gamma_\bk/b \\ 0 &  b \end{array} \right), \hskip 0.5cm {\cal J}^o_{\bk'} = \sqrt{4dQ}\left( \begin{array}{cc} \gamma_{\bk'}/ b & 0 \\ \gamma_\bk'/b & b \end{array} \right).  \label{eq:V_reduced_o}
\end{eqnarray}

With the above definitions the evaluation of $\Delta S$ becomes tractable, and takes the form of a power series in the impurity concentration. We proceed to write 
\begin{equation}
N_s \beta \Delta S = \sum_{p=1}^{\infty} \frac{(-1)^{p+1}}{p} \, \Tr (GV  )^p, \label{eq:delS1}
\end{equation}
where we implicitly assume that the sum~(\ref{eq:delS1}) is convergent. The trace is over the momentum and internal even-odd indices. Using Eqs.~(\ref{eq:V_reduced_e})--(\ref{eq:V_reduced_o}) and \Eq{eq:invG} we find that \Eq{eq:delS1} contains contributions of the form
\begin{equation}
\Tr \sum_{\bk_1,\dots,\bk_p \atop \bmm_1,\dots,\bmm_p \in \{ \br \}} {\cal A}^1_{\bk_1} {\cal A}^2_{\bk_1} \cdots {\cal A}^p_{\bk_p} e^{i\left[ \left(\bk_1-\bk_2\right)\bmm_1 + \dots + \left(\bk_p-\bk_1\right)\bmm_p \right]}. \label{eq:loop_p}
\end{equation}
Here ${\cal A}^i_\bk$ can be any one of the four matrices 
\begin{eqnarray}
{\cal A}^{ee}&=&\frac{1}{N_s}{\cal J}^e_\bk G_\bk {\cal I}^e_\bk, \qquad {\cal A}^{oo} =\frac{1}{N_s}{\cal J}^o_\bk G_\bk {\cal I}^o_\bk, \\
{\cal A}^{eo}&=&\frac{1}{N_s}{\cal J}^e_\bk G_\bk {\cal I}^o_\bk, \qquad {\cal A}^{oe} = \frac{1}{N_s}{\cal J}^e_\bk G_\bk {\cal I}^e_\bk. 
\end{eqnarray}
A diagrammatic method is useful to treat the evaluation of \Eq{eq:loop_p}, and we define the following Feynman rules.
\begin{enumerate}
\item   The propagator $G$ given by \Eq{eq:invG} is represented by Fig.~\ref{fig:G_imp}. It is diagonal in momentum and Matsubara frequency. $a,b=e,o$ are the even-odd indices.
\begin{figure}[hb]
\begin{center}
\includegraphics[clip]{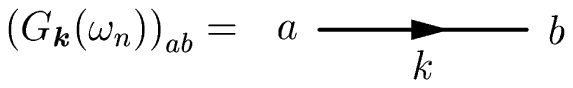}
        \caption{The boson propagator.}
        \label{fig:G_imp}
\end{center}
\end{figure}
\item   The interaction are given in \Eq{eq:V_reduced} is a sum of $M_e+M_o$ vertices, one for each of the impurities. The vertex that corresponds to the impurity on site $\br$ is given in Fig.~\ref{fig:V_imp}.
\begin{figure}[htb]
\begin{center}
\epsfig{width=13cm,file=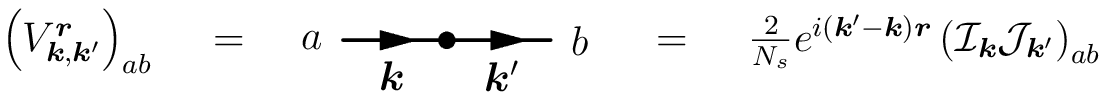}
        \caption[The impurity vertex]{The vertex of the impurity that resides on the site $\br$. Here ${\cal I,J}$ are defined by either \Eq{eq:V_reduced_e} or \Eq{eq:V_reduced_o}.}
        \label{fig:V_imp}
\end{center}
\end{figure}

\item   The expression~(\ref{eq:loop_p}) is represented by the diagram in Fig.~\ref{fig:loop_imp}, which describes the scattering of the bosons off three impurities.
\begin{figure}[htb]
\begin{center}
\epsfig{width=7.25cm,file=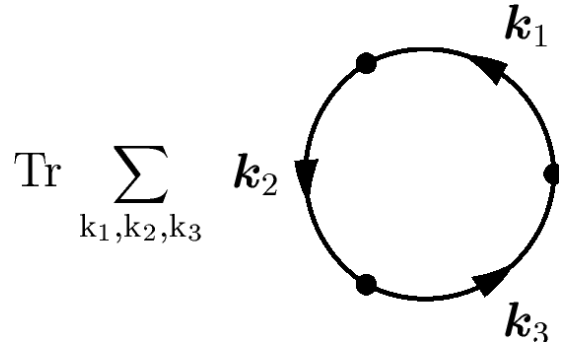}
        \caption[Scattering of bosons off three impurities]{The loop diagram that corresponds to the expression in \Eq{eq:loop_p} with $p=3$.}
        \label{fig:loop_imp}
\end{center}
\end{figure}
Note that this diagram stands for all possible ways to scatter three times off $M_e+M_o$ available impurities. 
\item   We adopt the notations in \cite{Brenig}, and write the diagram in Fig.~\ref{fig:loop_imp} as the sum in Fig.~\ref{fig:loop_imp1}.
\begin{figure}[htb]
\epsfig{width=17cm,file=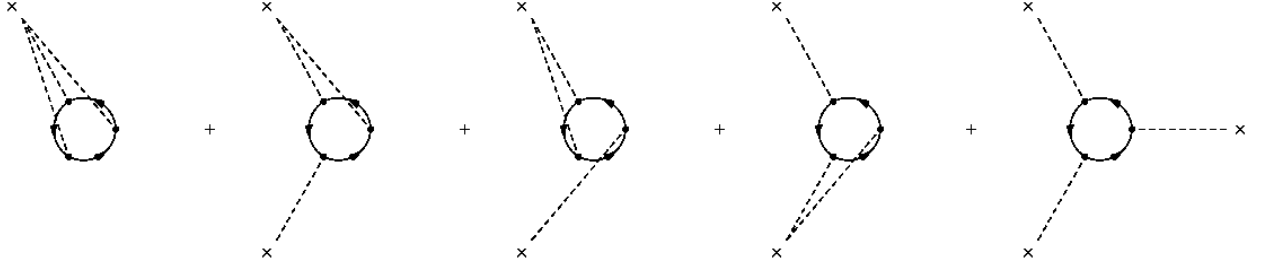}
        \caption[The loop diagram that corresponds to the expression~(\ref{eq:loop_p}).]{The loop diagram that corresponds to the expression~(\ref{eq:loop_p}) with $p=3$. The leftmost diagram describes a boson scatters three times off the same impurity, while the rightmost diagram describes a boson that scatters three times off three different impurities.}
        \label{fig:loop_imp1}
\end{figure}
A diagram with $q$ external points (each denoted by an ${\bm \times}$) corresponds to a process in which the bosons scatter off $q$ {\it different} impurities. The lines connecting the external points with the vertices denote which of the impurities was hit in a given vertex. For example in the leftmost diagram, the boson collides three times off the same impurity, while in the rightmost diagram, it collides three times off three different impurities. Finally note that each of the diagrams in Fig.~\ref{fig:loop_imp1} stands for all possibilities to choose the impurities that the bosons scatter off.
\end{enumerate}

The next step is to use the quenched approximation, and average \Eq{eq:loop_p} over the position of each impurity. This assumes that the impurities cannot move on the time scale that the bosons move, and is exact in the order we work to within the strong coupling expansion. We define the averaging procedure over the locations of the set $\{ \br \}$ as
\begin{equation}
\<{\cal O}\> \equiv \left( \frac{2}{N_s} \right)^{M_e+M_o} \, \sum_{\{\br_{e,o}\}=1}^{N_s/2} {\cal O}((\{\br_e\};\{\br_o\}).
\end{equation}

This relies on the assumption that the concentration of impurities is sufficiently small that there are no links between any two impurities. This assumption is invalid at high $c$. 

\begin{figure}[htb]
\begin{center}
        \epsfig{width=12cm,file=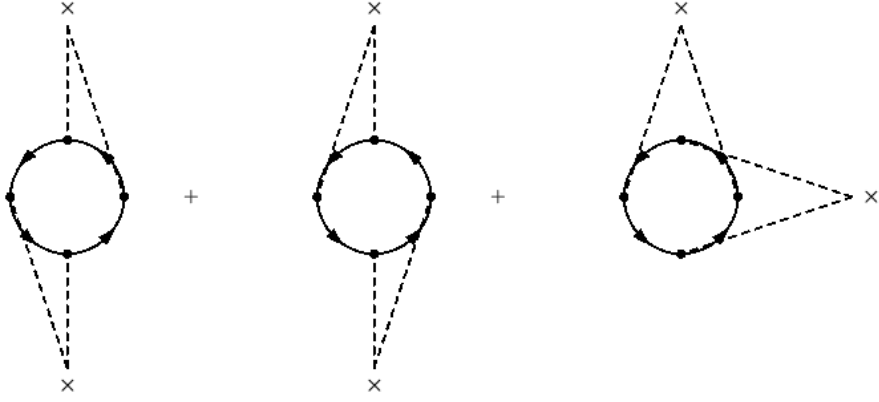}
        \caption{The loop diagrams that corresponds to $p=4$, $q=q_e=2$, and $s=1,2,3$.}
        \label{fig:example_s}
\end{center}
\end{figure}

Applying the averaging procedure to each of the diagrams in Fig.~\ref{fig:loop_imp1} results in the following power series for expression~(\ref{eq:loop_p}),
\begin{equation}
N_s \sum_{q=1}^{\text{Min}[p,M]} \sum_{q_e=0}^q c_e^{q_e} c_o^{q-q_e} {\cal B}^{(q,q_e)}_p.
\end{equation}
Here $q_e$ and $q-q_e$ are, respectively, the number of external points that correspond to an impurity residing on an even or odd site, and $c_{e,o}$ are defined as 
\begin{equation}
c_a=M_a/N_s, \qquad a=e,o.
\end{equation}
We proceed to prove this point. 
First we denote a diagram with $p$ vertices and a total of $q$ external points, out of which $q_e$ represent impurities on even sites, by ${\cal D}^{(s)}_{p,q,q_e}$. $s$ is an additional index that denotes the different diagrams that have the same values of $p$, $q$, and $q_e$. For example for $p=4$, and $q=q_e=2$, $s$ takes the values $1,2$, and $3$, and corresponds to the three diagrams in Fig.~\ref{fig:example_s}.

Looking at the structure of the exponentials in \Eq{eq:loop_p} we see that when we evaluate ${\cal D}^{(s)}_{p,q,q_e}$, where $\bmm_i$ take $q$ different values, the exponential turn into
\begin{equation}
e^{i(\bk'_1\br_1+\bk'_2\br_2+\dots+\bk'_{q-1}\br_{q-1})}. \label{eq:exp}
\end{equation}
Here $\{\bk'_i\}$ are linear combinations of the momenta that flow in the diagram. 
The averaging procedure is in fact applied only to the exponentials~(\ref{eq:exp}), and results in $q-1$ momenta conservation rules,
\begin{equation}
\< e^{i(\bk'_1\br_1+\bk'_2\br_2+\dots+\bk'_{q-1}\br_{q-1})} \> = \delta_{\bk'_1}\delta_{\bk'_2} \cdots \delta_{\bk'_{q-1}} . \label{eq:av_exp}
\end{equation}
Substituting \Eq{eq:av_exp} into \Eq{eq:loop_p}, and taking into account of the number of possibilities to choose which impurities are represented by the external points in $D^{(s)}_{p,q,q_e}$ 
we have
\begin{equation}
\< {\cal D}^{(s)}_{p,q,q_e} \>=\, \left( \begin{array}{c} M_e \\ q_e \end{array} \right) \left( \begin{array}{c} M_o \\ q-q_e \end{array} \right) \Tr \sum_{\bk_1,\dots,\bk_p } {\cal A}^1_{\bk_1}\cdots {\cal A}^p_{\bk_p} \delta_{\bk'_1}\delta_{\bk'_2} \cdots \delta_{\bk'_{q-1}}. \label{eq:av_D}
\end{equation}
The next step is to assume that the terms in \Eq{eq:delS1} with $p>M_{e,o}$ are negligible. ($M_{e,o}$ are of the order of $N_s$) Recalling that ${\cal A}^i_\bk \propto N_s^{-1}$, we find the following continuum limit of \Eq{eq:av_D} for $p,q_e,q-q_e\ll M_{e,o}$.
\begin{eqnarray}
\< {\cal D}^{(s)}_{p,q,q_e} \> &=& \frac{M^{q_e}_e}{q_e!} \frac{M^{q-q_e}_o}{(q-q_e)!} N_s^{-p} \left( N_s \int d\bk \right)^{p-q+1} F^{(s)}_{p,q,q_e}(\{ \bk \}) \nonumber \\
&=& N_s \frac{(c_e)^{q_e}(c_o)^{q_o}}{q_e!(q-q_e)!} \left( \int d\bk \right)^{p-q+1} F^{(s)}_{p,q,q_e}(\{ \bk \}), \label{eq:av_D1}
\end{eqnarray}
where the $(p-q+1)$ momentum integrals are over the Brillouin zone and are finite. We can now identify that 
\begin{equation}
{\cal B}^{(q,q_e)}_p = \frac{(c_e)^{q_e}(c_o)^{q-q_e}}{q_e!(q-q_e)!}\sum_s \left( \int d\bk \right)^{p-q+1} F^{(s)}_{p,q,q_e}(\{ \bk \}),
\end{equation}
for $p\ll M_{e,o}$.
\quad Q.E.D

\subsubsection{\underline{MF action to $O(c)$}}

We now evaluate $\Delta S$ to first order in the concentration $c$. We therefore take into account diagrams with $q=1$ only. In that case $q_e=0,1$, and $s=1$ (there is only one possible diagram for each value of $q_e$). In this case the diagrams that contribute to the $p$-th term in \Eq{eq:delS1} correspond to $p$ scatterings off the same impurity, which resides on either sublattice. The structure of \Eq{eq:delS1} is now simple. For $q_e=1$ all ${\cal A}^i={\cal A}^{ee}$, and for $q_e=0$ all ${\cal A}^i={\cal A}^{oo}$. The result is
\begin{eqnarray}
N_s \beta \Delta S &\simeq& N_s \, \sum_{\omega_n}  \Tr \sum_{p\ll M_{e,o}} \frac{(-1)^{p+1}}{p} \left[ c_e {\cal B}_p^{(1,1)} + c_o {\cal B}_p^{(1,0)} \right] \nonumber \\
&=& N_s \, \sum_{\omega_n} \sum_{a=e,o} c_a \Tr \sum_{p\ll M_{e,o}} \frac{(-1)^{p+1}}{p}  \left[ \int \left(\frac{d\bk}{2\pi} \right)^d {\cal J}^a_\bk G_\bk(\omega_n) {\cal I}^a_\bk   \right]^p \nonumber \\
&\simeq& N_s \, \sum_n \sum_a c_a \Tr \log \left[ {\bm 1} + \int \left(\frac{d\bk}{2\pi} \right)^d {\cal J}^a_\bk G_\bk(\omega_n) {\cal I}^a_\bk \right]. \label{eq:delS2}
\end{eqnarray}
We have verified that the contribution of the $\Tr$ is independent of $a$, so $\Delta S$ becomes
\begin{eqnarray}
\beta \Delta S(\lambda,\eta;b) &=& c \, \sum_{\epsilon_n} \, \log \left\{ \det \left[ {\bm 1} + \int d\gamma \, D(\gamma) \, \frac{\eta}{1+\epsilon^2_n-\eta^2\gamma^2}\times \right. \right. \nonumber \\ 
&& \left. \left. \left(\begin{array}{cc} b^2(1-i\epsilon) +2\eta\gamma^2 + \gamma^2(1+i\epsilon_n)/b^2 & -\gamma^2(\eta+(1+i\epsilon_n)/b^2) \\ \gamma^2(\eta + (1+i\epsilon_n)/b^2) & -(1+i\epsilon_n)\gamma^2/b^2\end{array} \right)\right] \right\}. \nonumber \\ \label{eq:delS3}
\end{eqnarray} 
Here the overall concentration $c=c_e+c_o=M/N_s$, $\epsilon_n\equiv \omega_n/\lambda$, and the momentum integration was again replaced by an integration over the variable $\gamma$ with the measure shown in Fig.~\ref{fig:Dgamma}.
We have calculated $\Delta S(\lambda,\eta;b)$ and confirmed that for $|\eta|\le 1$ it is real. Substituting \Eq{eq:delS3} into \Eq{eq:delS} we have 
\begin{eqnarray}
S_{\text{MF}} &=& \frac{2d(1-2c)}{J_1}Q^2 -  (1-c)\lambda\kappa \nonumber \\ 
&& + \frac1{\beta} \int d\gamma D(\gamma) \log 2\sinh \frac{\beta\omega(\gamma)}{2} - \frac12 \lambda (1-c) \nonumber \\ 
&& + \, \frac{c}{\beta}\left(- \log 2\sinh \frac{\beta \lambda(1+\eta b^2)}{2} + \Delta S(\lambda,\eta;b) \right). \label{eq:S_final}
\end{eqnarray}
The first row of \Eq{eq:S_final} is the result of~(\ref{eq:change1}). The second row is $\Tr \log G^{-1}$ for the pure system. (Note the term proportional to $(1-c)$ is exactly the rule of thumb prescription from Appendix~\ref{app:discretization}) The contributions to the third row are the continuum evaluation of $\Delta S$ and $-\log Z'$, evaluated in the continuum limit of Euclidean time. 
We have verified that the contributions in the third row do not depend on the value of the unphysical constant $b$. 
 
\subsubsection{\underline{MF equations to $O(c)$}}

We concentrate here on the $T=0$ case. There we have
\begin{eqnarray}
\frac1{N_s N m \beta} S_{\text{MF}} &=& \frac{2d(1-2c)}{J_1}Q^2 -  \lambda\left(\kappa(1-c)+\frac12\right) + \lambda {\cal F}(\eta), \\
{\cal F}(\eta)&\equiv&\frac12 \int d\gamma D(\gamma)\, \sqrt{1-\eta^2\gamma^2} + c \left( \lim_{\beta \lambda \rightarrow \infty} \left[ \frac{1}{\beta\lambda}\Delta S (\lambda,\eta;b) \right] -\frac12 \eta b^2 \right). \nonumber \\
\label{eq:S_T0}
\end{eqnarray}
Where we have evaluated the $\beta\lambda \rightarrow \infty$ limit of the second term in ${\cal F}(\eta)$ as an integral over $\epsilon$. 
In order to investigate the boundary of the Ne\'el phase with non-zero value of $c$, we concentrate on the equation 
\begin{equation}
\frac{\partial S_{\text{MF}}}{\partial \lambda}=0, \label{eq:MF_lam}
\end{equation}
and assume no condensates. This becomes
\begin{equation}
\kappa = \frac1{1-c} \left[ f(\eta) + c \, {\cal G}_{\text{loop}} \right]. \label{eq:MF_eq_c}
\end{equation}
Here $f(\eta)$ is the right hand side of the MF equation for $c=0$, \Eq{eq:MF2_sep}, as plotted in Fig.~\ref{fig:f_eta}. ${\cal G}_{\text{loop}}$ includes all other loop contributions.

\begin{figure}[htb]
\begin{center}
\epsfig{width=15cm,file=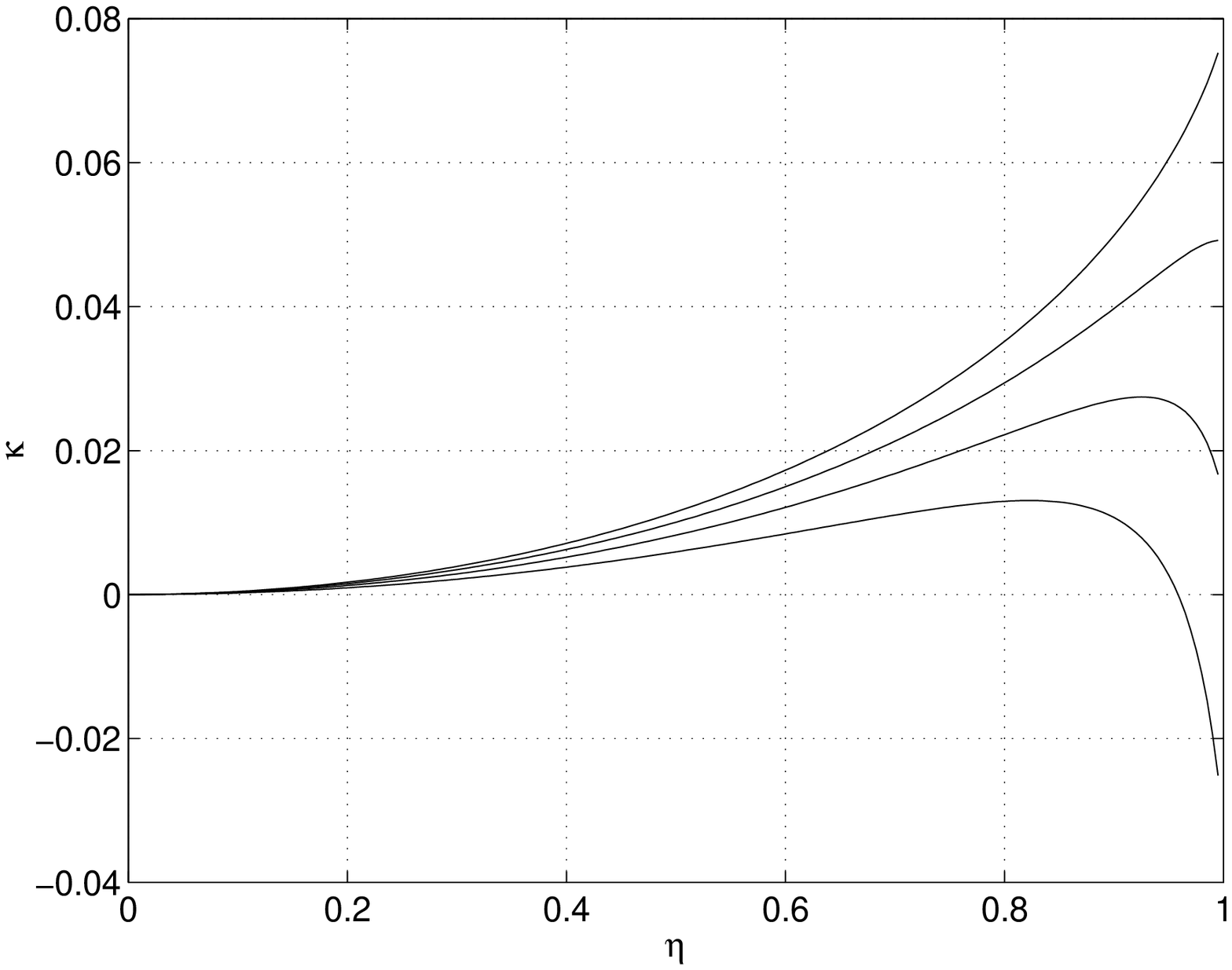}
\caption[Right hand side of the MF equation with $c\neq0$ in the quenched approximation.]{The right hand side of \Eq{eq:MF_eq_c} for $c=0$ (highest curve), $0.1$, $0.2$, and $0.3$ (lowest curve). }
\label{fig:rhs_c}
\end{center}
\end{figure}

We plot the right hand side of \Eq{eq:MF_eq_c} in Fig.~\ref{fig:rhs_c} and see that as the concentration $c$ increases, the maximum value of $\kappa$ {\it decreases}. 
This means that the N\'eel phase is expanding, which appears to contradict the results of Section~\ref{sec:c1_84}. 

\begin{figure}[htb]
\begin{center}
\epsfig{width=15cm,file=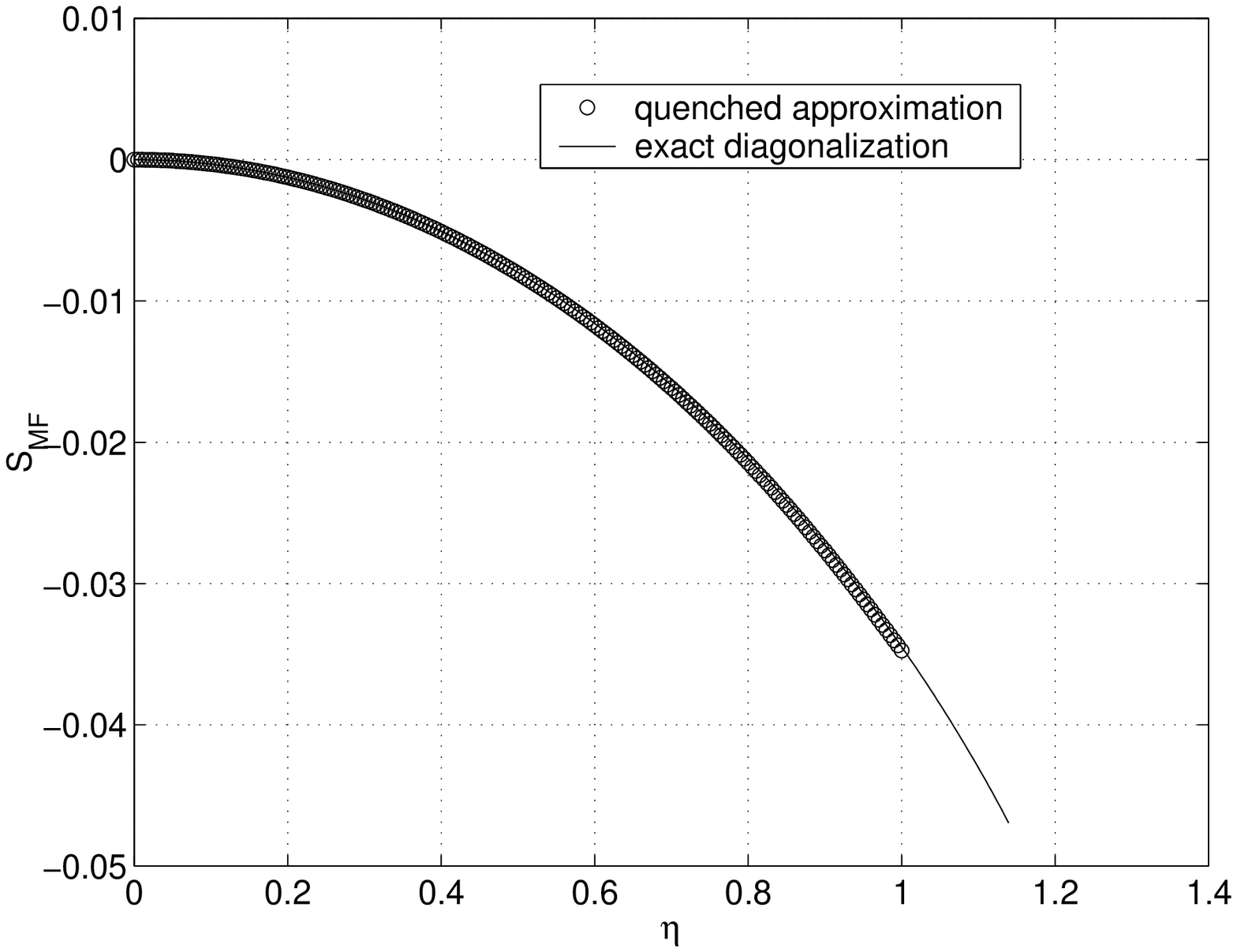}
\caption[MF action for $c=\frac18$; quenched approximation vs. exact diagonalization]{The MF effective action for $c=\frac18$ in the quenched approximation, and the exact diagonalization.}
\label{fig:S_quenched_exact18}
\end{center}
\end{figure}

\begin{figure}[htb]
\begin{center}
\epsfig{width=15cm,file=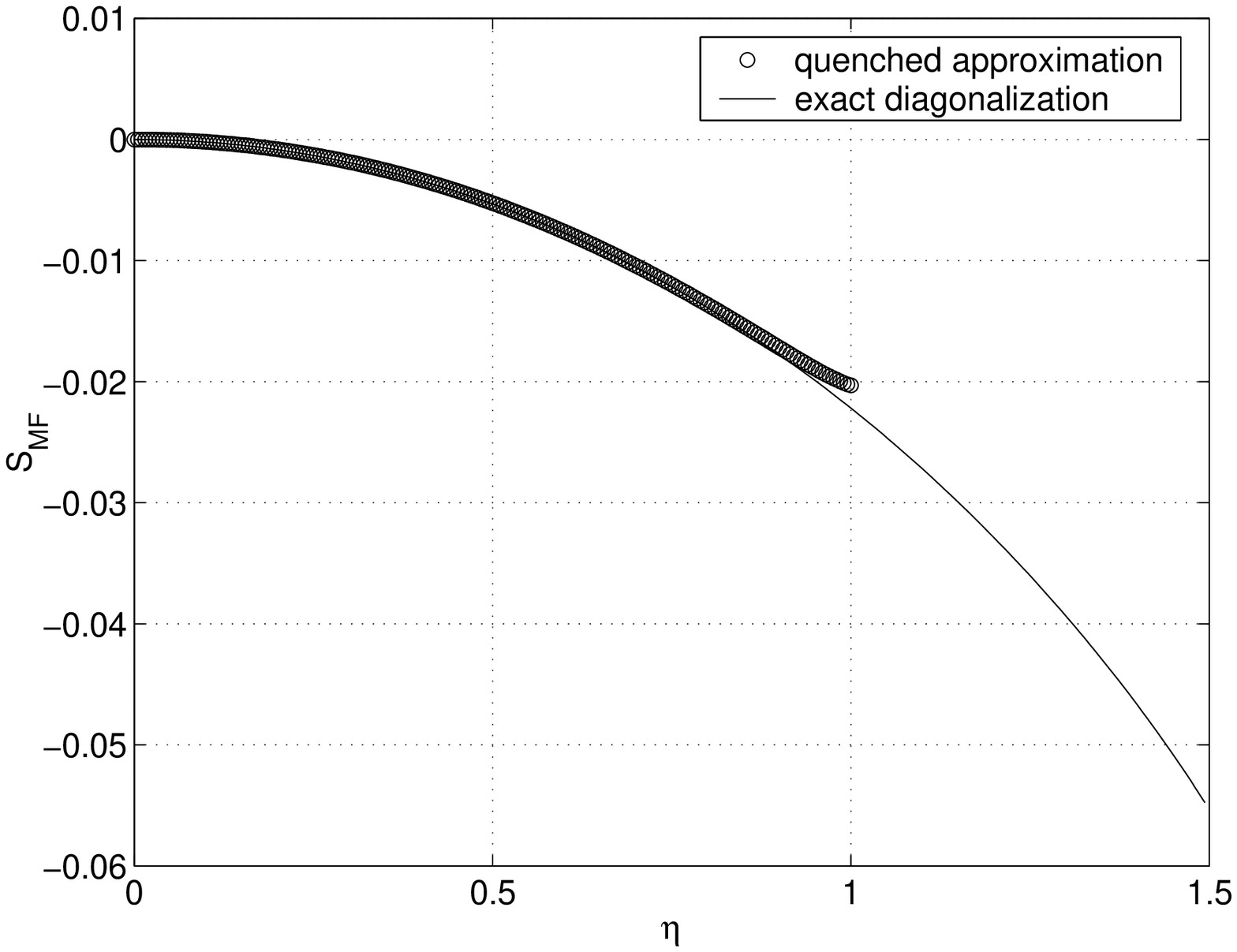}
\caption[MF action for $c=\frac14$; quenched approximation vs. exact diagonalization]{The MF effective action for $c=\frac14$ in the quenched approximation, and the exact diagonalization.}
\label{fig:S_quenched_exact14}
\end{center}
\end{figure}

\begin{figure}[htb]
\begin{center}
\epsfig{width=15cm,file=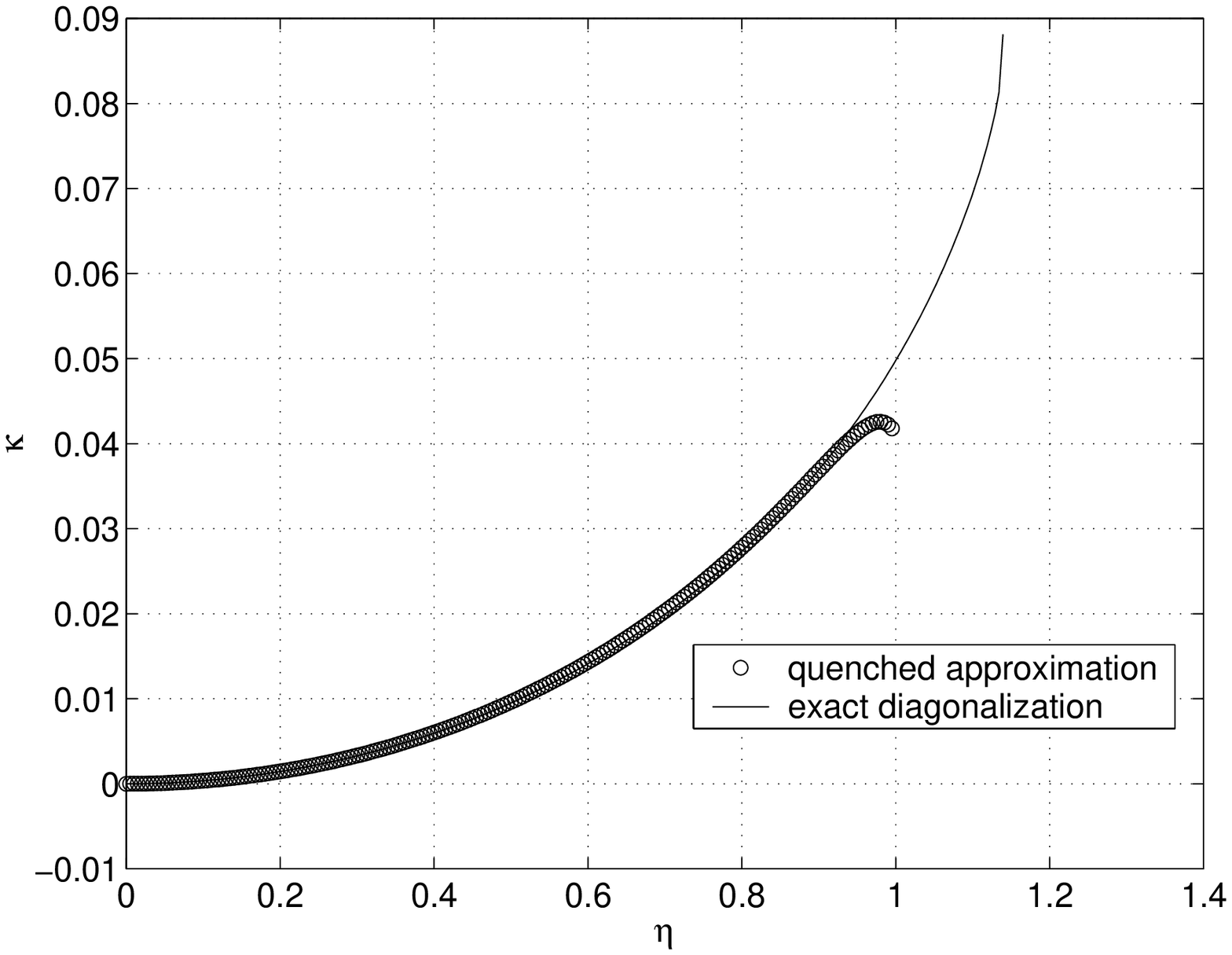}
\caption[Right hand side of MF equation for $c=\frac18$; quenched approximation vs. exact diagonalization]{The right hand side of the MF equation for $c=\frac18$ in the quenched approximation, and the exact diagonalization.}
\label{fig:rhs_quenched_exact18}
\end{center}
\end{figure}

\begin{figure}[htb]
\begin{center}
\epsfig{width=15cm,file=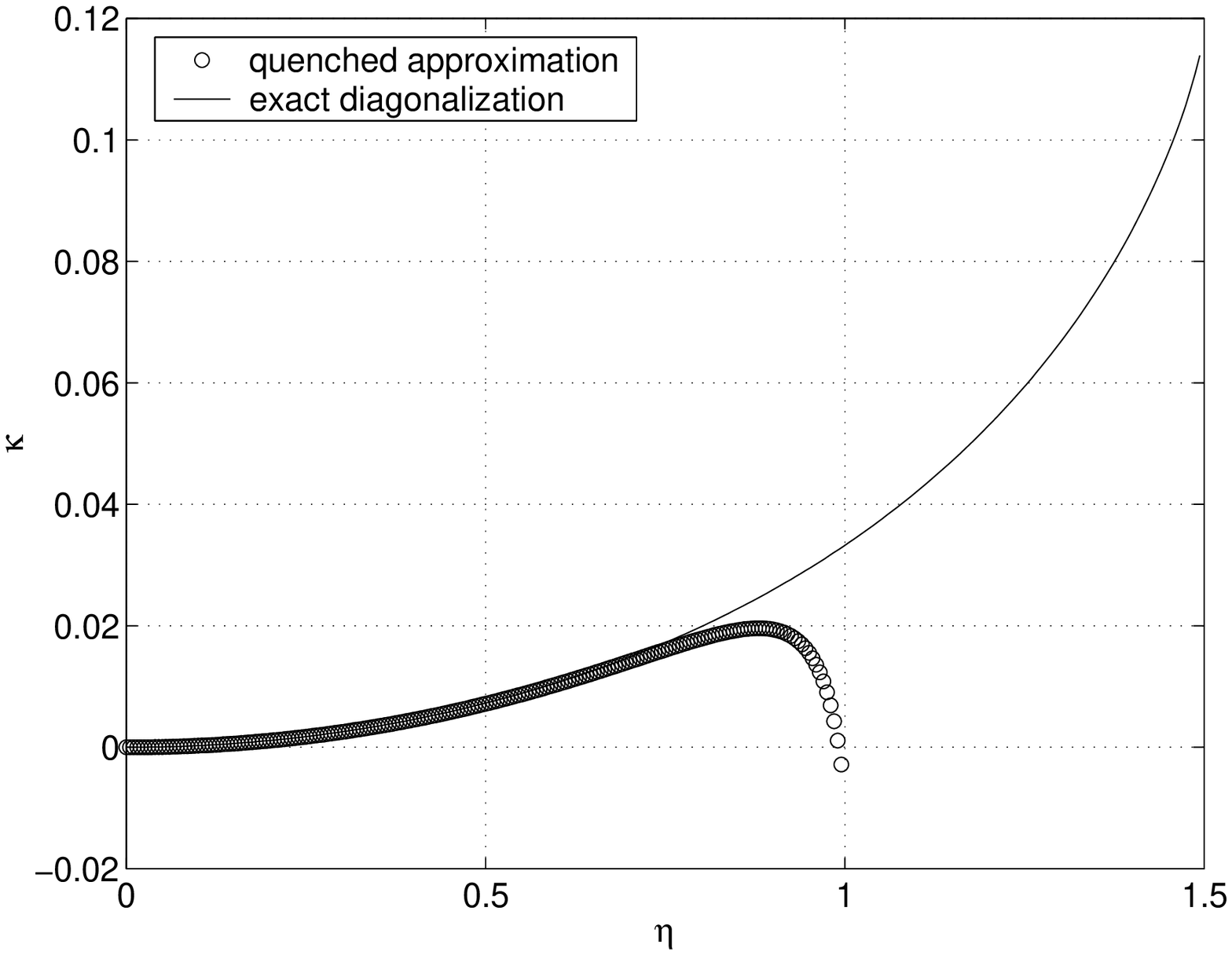}
\caption[Right hand side of MF equation for $c=\frac14$; quenched approximation vs. exact diagonalization]{The right hand side of the MF equation for $c=\frac14$ in the quenched approximation, and the exact diagonalization.}
\label{fig:rhs_quenched_exact14}
\end{center}
\end{figure}

In order to resolve this problem we compare the results of the quenched calculation for $c=\frac18$ and $c=\frac14$ with the exact diagonalization in Section~\ref{sec:c1_84}. We present the results of the two calculations in Figs.~\ref{fig:S_quenched_exact18}--\ref{fig:rhs_quenched_exact14}. The results are very similar. We see that the errors of the quenched results are larger for $c=\frac14$ than for $c=\frac18$. This is presumably because of the restriction to first order in $c$. Also we note that the quenched perturbative approximation is inapplicable for $\eta>1$, since the unperturbed ground state does not admit such values of $\eta$. Nevertheless, the result of Section~\ref{sec:c1_84} lets $\eta$ be larger than $1$, and in fact it is in this extra interval that the right hand side of the MF equation continues to grow, and the N\'eel phase shrinks. For both $c=\frac18$ and $c=\frac14$, there is a value of $\eta_{\text{max}}>1$ at which BEC occurs. This regime of MF parameters is unfortunately not accessible in the quenched approach.


To conclude we see that the MF quenched approximation in first order in $c$ does not reproduce the MF theory near the phase transition. The deviation between the quenched approximation and the exact diagonalization grows as one approaches the BEC. This approach is not applicable to extract information about the phase transition, simply because it is a perturbative approach whose starting point is the $c=0$ system. One can use this approach to extract quantitative properties of the disordered phase, and to see how these change when baryons are included. For example, the propagators of bosons will then obtain a self-energy contribution also calculable via the diagrammatic tools described above. It is expected that this will show that the baryon impurities slow down the bosons, and even make them decay. This is in analogy to what happens in condensed matter magnetic systems with lattice vacancies \cite{Brenig,Lee,Kotov}. 

Finally we note that although this MF analysis was devoted to the regime $\kappa < \kappa_c$ and  explored the disordered phase (while emphasizing the location of the phase transition), it is also possible to investigate the ordered phase. This can be done in the large $N_c$ limit by using a Holstein-Primakoff representation of the $Q^{\eta}$ operators (which is presented for $m=1$ in Appendix~\ref{app:Hamiltonian_odo}). The quenched approach can be also applied in this limit, and in analogy to \cite{Brenig} we expect that the Goldstone bosons in this phase (the pions and kaons in our system) will have a smaller velocity, and will obtain finite decay widths.
\clearpage
\chapter{Nonlinear sigma model and the classical ground state}
\label{chap:NLSM}

In this chapter we use a coherent state basis to write the partition function of the Hamiltonian~(\ref{eq:Hqm}) as a path integral. The result is a nonlinear sigma model.
We first analyze the properties of the classical counterpart of a toy sigma model that has $N=3$. We use mean field theory to find its ground state as a function of temperature for a set of baryon number configurations. Next we approach the full quantum sigma model. The action of the model has an overall factor of $N_c$, and is tractable in the large $N_c$ limit. We take this limit and examine the classical ground state for a set of uniform baryon number configurations.

\section{Coherent states and sigma fields}
\label{sec:coherent}
We employ a generalization of spin coherent states \cite{Klauder}
to derive a path integral for the spin model of
$H^{(2)}_{\text{eff}}$. We recall that a given site carries
generators $Q^\eta_\bn$ of $U(N)$ in a representation with $N_c$
columns and $m$ rows, with $B_\bn=m-2N_f$.

First we choose a basis for the Lie algebra of $U(N)$. This
consists of the generators $S^\alpha_\beta$, with $\alpha,\beta=1,\ldots,N$, whose
matrix elements in the fundamental representation are
\begin{equation}
(S^\alpha_\beta)_{\gamma \delta}= \delta_{\alpha \gamma}\delta_{\beta \delta}.
\end{equation}
The corresponding charges are
\begin{eqnarray}
Q^\alpha_\beta&=&\sum_a \psi^{\dag}_a S^\alpha_\beta
\psi_a-\frac12N_c\delta^\alpha_\beta \nonumber\\
&=&\sum_\alpha \psi^{\dag}_{a \alpha}\psi_{a \beta}-\frac12N_c\delta^\alpha_\beta, \label{Shat}
\end{eqnarray}
where we have subtracted a constant for convenience. The Cartan
subalgebra consists of the operators
\begin{equation}
H_\alpha=Q^\alpha_\alpha. \label{Cartan}
\end{equation}

We build the coherent states from the state of highest weight $|\Psi_0\>$
\begin{equation}
|\Psi_0\> = \prod_{\alpha=1}^m \left( \psi^\dag_{\alpha,1} \psi^\dag_{\alpha,2} \cdots \psi^\dag_{\alpha,N_c} \right) |\text{Dr}\>, \label{eq:highest}
\end{equation}
where the numbers are the color indices. This state is an eigenstate of the Cartan generators,
\begin{equation}
H_\alpha|\Psi_0\rangle=\left\{
\begin{array}{cl}
(N_c/2)|\Psi_0\rangle&{\rm for}\ \alpha=1,\ldots,m,\\[2pt]
-(N_c/2)|\Psi_0\rangle&{\rm for}\ \alpha=m+1,\ldots,N.
\end{array}\right.
\label{CartanH}
\end{equation}
The expectation values of the generators in this state obey
\begin{equation}
\langle \Psi_0|Q^\alpha_\beta|\Psi_0\rangle=\frac12N_c\Lambda_{\alpha \beta},
\label{SQ1}
\end{equation}
with
\begin{equation}
\Lambda=\left(
\begin{array}{cc}
\bm1_m&0\\
0&-\bm1_{N-m}
\end{array}\right).
\end{equation}
In order to obtain the other states in the rep, we perform the following $U(N)$ rotations on $|\Psi_0\rangle$,
\begin{equation}
|a\rangle=\exp\left(\sum_{\lambda=1}^m \sum_{\mu=m+1}^N
(a_\mu^\lambda Q^\mu_\lambda -a_\mu^{*\lambda}Q_\mu^\lambda)
\right)|\Psi_0\rangle. \label{cohstate}
\end{equation}
The only generators $Q^\mu_\lambda$ that appear in \Eq{cohstate}
are those that lower an $H_\alpha$ that starts from $N_c/2$ in
\Eq{CartanH} while raising another $H_\alpha$ that starts from
$-N_c/2$. Any other generator would annihilate $|\Psi_0\rangle$
and thus give no effect in the exponential.

The coherent states are normalized,
\begin{equation}
\langle a|a\rangle=1,
\end{equation}
and over-complete,
\begin{equation}
\int d\mu(a)\,|a\rangle\langle a|={\bm 1}. \label{complete}
\end{equation}
In \Eq{complete} the integral is over the coset space
$U(N)/[U(m)\times U(N-m)]$ (see below). Matrix elements of the
generators are given by
\begin{equation}
\langle a|Q^\alpha_\beta|a\rangle=\frac12N_c\sigma_{\alpha \beta}, \label{SQ}
\end{equation}
where the matrix $\sigma_{\alpha \beta}$ is given by a unitary rotation from
$\Lambda$,
\begin{equation}
\sigma=U(a)\Lambda U(a)^{\dag}. \label{QULU}
\end{equation}
The matrix $U(a)$ is built out of the $m\times(N-m)$ matrix
$a_\mu^\lambda$,
\begin{equation}
U=e^A = \exp\left[\left(
\begin{array}{cc}
0&a\\
-a^{\dag}&0
\end{array}\right)\right].
\label{Ua}
\end{equation}
$\sigma$ is both Hermitian and unitary.

The manifold of matrices $\sigma$ is the coset space
$U(N)/[U(m)\times U(N-m)]$, a sub-manifold of $U(N)$. This is
because for any matrix $U(a)$, one can generate an orbit $U(a)V$
by multiplying with a matrix
\begin{equation}
V=\left(
\begin{array}{cc}
X&0\\0&Y
\end{array}\right),
\end{equation}
where $X\in U(m)$ and $Y\in U(N-m)$. All matrices in the orbit
will give the same matrix $\sigma$ when inserted into \Eq{QULU},
and thus in integrating over the configuration space of $\sigma$
one must choose only a single representative of each orbit. This
set of representatives, the coset space $U(N)/[U(m)\times
U(N-m)]$, is the quotient space of the non-invariant subgroup
$U(m)\times U(N-m)$. Basically the origin of this manifold is the fact that the quantum state $|\Psi_0 \>$ is invariant (up to a phase) to $U(N)$ rotations that belong to the subgroup $U(m)\times U(N-m)$. This can be understood by the form of the Young tableau in Fig.~\ref{fig:Young}. This representation of $SU(N)$ is a singlet of a certain $SU(m)\times SU(N-m)$ subgroup, more precisely, of the rotations that do not involve indices from the $[m+1,N]$ range and that leave the $[1,m]$ indices of the state invariant. 

The measure over the coset space must be invariant under unitary
rotations,
\begin{equation}
|a\rangle\to R(V)|a\rangle,
\end{equation}
where $R(V)$ represents the rotation $V$ in Hilbert space.
Equation~(\ref{SQ}) shows that this is a rotation
\begin{equation}
\sigma\to V\sigma V^{\dag}
\end{equation}
and by \Eq{QULU}, this means that a measure in $U$ must be
invariant under $U\to VU$. This fixes the measure uniquely to be
the Haar measure in $U(N)$, and thus one can integrate over the
coset space by integrating with respect to $U$ over $U(N)$ and
using \Eq{QULU}.

A representation whose Young diagram has $N-m$ rows is the
conjugate to the representation with $m$ rows. Its coherent state
space can be constructed to look the same, with only a sign
difference. To do this we start with the {\em lowest\/}-weight
state, which satisfies [cf.~\Eq{CartanH}]
\begin{equation}
H_\alpha|\Psi_0\rangle=\left\{
\begin{array}{cl}
-(N_c/2)|\Psi_0\rangle&{\rm for}\ \alpha=1,\ldots,m\\[2pt]
(N_c/2)|\Psi_0\rangle&{\rm for}\ \alpha=m+1,\ldots,N.
\end{array}\right.
\end{equation}
This introduces a minus sign into \Eq{SQ1}. The subsequent steps
are identical, with only the replacement of \Eq{SQ} by
\begin{equation}
\langle a|Q^\alpha_\beta|a\rangle=-\frac12N_c\sigma_{\alpha \beta}. \label{SQminus}
\end{equation}
Here, too, $\sigma$ is given in terms of $\Lambda$ and $U$ by
\Eq{QULU}.

Using the Taylor expansion of the exponentials in \Eq{Ua} and noting that the exponential argument $A$ anticommutes with $\Lambda$ we have \footnote{This generalizes a parametrization found in
\cite{SW}.}
\begin{equation}
\sigma(a)=\left( \begin{array}{cc} \cos\left(2\sqrt{a
a^\dag}\right) & -a\frac{\displaystyle\sin\left(2\sqrt{a^\dag
a}\right)}
      {\displaystyle\sqrt{a^\dag a}}       \\[2pt]
-\frac{\displaystyle\sin\left(2\sqrt{a^\dag a}\right)}
            {\displaystyle\sqrt{a a^\dag}}\,a^\dag &
-\cos\left(2\sqrt{a^\dag a}\right)
\end{array}\right).
\label{eq:sigma_a}
\end{equation}
Here $a^{\dag} a$ is a square matrix of dimension $N-m$ and $aa^{\dag}$ is a square matrix of dimension $m$. Using the notation
\begin{equation}
\phi=a\frac{\sin \left( \sqrt{a^\dag a}\right) }{\sqrt{a^\dag a}},
\end{equation}
we have
\begin{eqnarray}
U(\phi)&=&\left( \begin{array}{cc} \sqrt{1-\phi\phi^\dag} & -\phi \\ \phi^\dag & \sqrt{1-\phi^\dag \phi} \end{array}\right), \label{eq:Uphi} \\
\sigma(\phi)&=&\left( \begin{array}{cc} 1-2\phi\phi^\dag & -2\phi \sqrt{1-\phi^\dag \phi} \\ -2\sqrt{1-\phi^\dag \phi}\phi^\dag & -1+2\phi^\dag\phi \end{array} \right). \label{eq:sig_phi}
\end{eqnarray}
The representation given in Eqs.~(\ref{eq:Uphi})--(\ref{eq:sig_phi}) is useful to treat fluctuations of order $1/N_c$, as will be seen in next chapter. Also for $m=N-1$ it coincides with the form given in \cite{Salam}, which in turn can be used to infer a generalized Holstein-Primakoff realization of the $U(N)$ generators (the $\phi$ operators then become the Holstein-Primakoff bosons). We note in passing that another useful representation given in \cite{colorflavortransf} is related to Eqs.~(\ref{eq:Uphi})--(\ref{eq:sig_phi}) $\phi$ by identifying $Z^\dag=a \tan \sqrt{a^\dag a}/\sqrt{a^\dag a}=\phi/\sqrt{1-\phi^\dag \phi}$. 

\section{Partition function and action}
\label{sec:action_NLSM}
The partition function $Z=\Tr e^{-\beta H}$ with $\beta=1/T$, can be written as a
path integral by inserting the completeness relation
(\ref{complete}) at every slice of imaginary time. This gives
\begin{equation}
Z=\int D\sigma \,\exp -S, \label{partitionfunction}
\end{equation}
where the the action is
\begin{equation}
S=\int_0^\beta d\tau\left[\frac{1-\langle
a(\tau)|a(\tau+d\tau)\rangle}{d\tau} +H(\sigma(\tau))\right].
\end{equation}
The Hamiltonian $H(\sigma)$ is a transcription of  the quantum
Hamiltonian to the classical $\sigma$ matrices. Starting with the
quantum operator $Q^\eta_{\bn}$, we have
\begin{eqnarray}
Q^\eta_{\bn}&=&\psi^{\dag}_{\bn}M^\eta\psi_{\bn}
=M^\eta_{\alpha \beta}\psi^{\dag}_{\bn}S^\alpha_\beta\psi_{\bn}\nonumber\\
&=&M^\eta_{\alpha \beta}Q^\alpha_\beta(\bn)+\frac12N_c\Tr M^\eta.
\end{eqnarray}
Expressed in these variables, the quantum Hamiltonian
is\footnote{The $B'$ term from \Eq{H2eff2} indeed disappears. We
have dropped an additive constant that is independent of $B'$.}
\begin{equation}
H_{\text{eff}}^{(2)}= \sum_{\bn\mu \atop j\not=0}J_j Q^\eta_{\bn}
Q^\eta_{\bn+j\muhat} \left(s^\mu_\eta\right)^{j+1}.
\label{Hquantum}
\end{equation}
where $J_j=(2/N_c)K(j)$. We transcribe this according to \Eq{SQ}
to obtain the classical Hamiltonian,
\begin{equation}
H(\sigma)=\left(\frac{N_c}2\right)^2\sum_{\bn\mu \atop j\not=0}J_j
\sigma^\eta_{\bn}\sigma^\eta_{\bn+j\muhat}\left(s^\mu_\eta\right)^{j+1},
\label{Hsigma}
\end{equation}
where
\begin{equation}
\sigma^\eta_{\bn}=\Tr M^{\eta T} \sigma_\bn.
\end{equation}
Recall that each $\sigma_\bn$ is an $N\times N$ matrix ranging
over the coset space appropriate to site $\bn$. We find it useful to write the interaction as
\begin{equation}
H(\sigma)=\frac12 \left( \frac{N_c}{2} \right)^2 \sum_{\bn,\mu} \left[ \sum_{j\in \text{odd}} J_j \, \Tr \sigma_\bn \sigma_{\bn+j\muhat} +  \sum_{j\in \text{even}} J_j\,  \Tr \sigma_\bn \alpha_\mu \sigma_{\bn+j\muhat} \alpha_\mu \right], \label{eq:H_sigma}
\end{equation}
where the three Dirac matrices $\alpha_\mu$ have replaced the sign factors $s^\eta_\mu$. 

The time-derivative term in $S$ is a Berry phase \cite{RS1}. It
can be expressed as follows
\begin{equation}
S_B=-\frac{N_c}{4} \int_0^{\beta} d\tau \int_0^1 du \, \Tr \left[ \, \sigma(\tau,u) \, \partial_u \sigma(\tau,u) \, \partial_\tau \sigma(\tau,u) \, \right]. \label{eq:S_B1}
\end{equation}
Here $\sigma(\tau,u)=U(\tau,u) \Lambda \, U^\dag (\tau,u)$ with 
\begin{equation}
U(\tau,u)=e^{u A(\tau)}, \label{eq:Uutau}
\end{equation}
and obeys
\begin{eqnarray}
\sigma(\tau,0)&=&\sigma(\tau',0), \qquad \forall \tau, \tau', \label{eq:restrict1} \\
\sigma(\tau,1)&=&\sigma(\tau), \hskip 1.3cm \forall \tau, \label{eq:restrict2} \\
\sigma(0,u)&=&\sigma(\beta,u). \hskip 0.95cm \forall u, \label{eq:restrict3}
\end{eqnarray}
This means that the $\tau$--$u$ plane is a disk with unit radius. $\tau$ plays the role of an angular variable, while $u$ is the radial coordinate. Using Stokes theorem \Eq{eq:S_B1} is brought to the form
\begin{equation}
S_B(U)=-\frac{N_c}{2} \int_0^{\beta} d\tau \Tr \left( \Lambda U^\dag(\tau) \, \partial_\tau U(\tau) \right). \label{eq:S_B2}
\end{equation}
It is interesting that \Eq{eq:S_B2} gives the same result even if $U$ is a general $U(N)$ matrix, not restricted to the form of \Eq{Ua}. To see that we write $U'(\tau,u)=U(\tau,u)V(\tau,u)$. Here $U'\in U(N)$, $U$ is given by \Eq{eq:Uutau}, and $V\in U(m)\times U(N-m)$,  
\begin{equation}
V=\left( \begin{array}{cc} \exp \left( u A^{(1)}_0(\tau) \right) & 0 \\ 0 & \exp \left( u A^{(2)}_0(\tau) \right) \end{array} \right).
\end{equation}
$U'$ and therefore $V$ are continuous functions on the disk. Substituting $U'$ into \Eq{eq:S_B2} one finds that
\begin{equation}
\Delta S_B = S_B(U')-S_B(U) = \frac{N_c}{2} \int \partial_\tau \left( \Tr \left( A^{(1)}_0\right) - \Tr \left( A^{(2)}_0 \right) \right), \label{eq:delS_B1}
\end{equation}
This means that $\Delta S_B $ is a an integer multiple of $i\pi N_c$. A non-zero value of $\Delta S_B$ is a result of winding the $U(1)$ factors of $V$ on the disk boundary $u=1$. Such winding is however forbidden, since it means that $V$ must have a discontinuity inside the disk. In fact the continuity of $V$ means that $\det (V)$ must also be continuous, so $\forall u\in [0,1]$ one has
\begin{equation}
0=\int d\tau \, \partial_\tau \, \det V(\tau,u) = \int d\tau \, \partial_\tau \left( \log \det \exp \left( uA^{(1)}_0\right) + \log \det \exp \left( uA^{(2)}_0 \right) \right). \label{eq:delS_B2}
\end{equation}
Substituting \Eq{eq:delS_B2} with $u=1$ into \Eq{eq:delS_B1} we find that $\Delta S_B$ is an integer multiple of $i 2\pi N_c $. Since $N_c$ is an integer, $\exp \Delta S_B=1$, and $\exp S_B(U)=\exp S_B(U')$.\footnote{Nevertheless we will concentrate on $U$ that has the form given in \Eq{Ua}.} 

Finally the lattice action is 
\begin{equation}
S=\frac{N_c}2 \int_0^\beta d\tau\left[-\sum_{\bn}\Tr\Lambda_\bn
U^{\dag}_{\bn}
\partial_\tau U_{\bn}
+\frac2{N_c} H(\sigma(\tau))\right]. \label{eq:action_sigma}
\end{equation}
$\Lambda_\bn$ will vary from site to site if $m$ does. If one
takes the route of \Eq{SQminus} for a representation with $N-m$
rows, then the kinetic term for that site acquires a minus sign
(see below).

The number of colors has largely dropped out of the problem, since
$\sigma$ is just an $N\times N$ unitary matrix field. 
The explicit
factor of $N_c/2$ that multiplies the action 
invites a semiclassical approximation in the large-$N_c$ limit which we do in Section~\ref{sec:Nc_infty}.
This of course neglects the $N_c$-dependence of the interaction term, but we recall that for $N_c\gg 1$, the couplings of the quantum Hamiltonian in Chapter~\ref{chap:Heff} are $J_j\sim 1/\left(g^2 N_c^2\right)$. This means that the couplings of the classical Hamiltonian in \Eq{eq:H_sigma}, multiplied by $2/N_c$ in \Eq{eq:action_sigma} scale like $1/g^2N_c$, which we then fix when taking $N_c\rightarrow \infty$.

We conclude this section by noting that the form of the Berry phase of \Eq{eq:action_sigma} is exactly the reason why the representation of $\sigma$ and $U$ in term of the variable $\phi$ given in \Eq{eq:Uphi} becomes useful. In terms of this variable the kinetic term becomes
\begin{equation}
S_B=\frac{N_c}2 \int d\tau \sum_\bn \Tr \left[ \phi^\dag_\bn \cdot \partial_\tau \phi_\bn - \partial_\tau  \phi_\bn^\dag \cdot \phi_\bn \right]. \label{eq:Skin_phi}
\end{equation}
This simple {\it quadratic} form is not obtained by expanding $U$ in terms of $a$ or $Z$. 

\section{The classical counterpart of the sigma model}
\label{sec:classical}

In this section we make a pause from the discussion on the real quantum sigma model and investigate its classical counterpart by ignoring  
the Berry phase $S_B$. The classical sigma model provides us insight about the way the system behaves at high temperatures, where $S_B$ should not be very important. Also, thermal fluctuations may partly replace the quantum fluctuations of the real system which, as discussed in Chapter~\ref{chap:Qfluctuation}, are very important.

The classical model is much more accessible than the real quantum one because its action is real, and relying on convexity arguments, we can look for a mean field theory that mimics its dynamics. This classical action is also amenable to Monte Carlo simulations, but we defer any progress in that direction to future research. 

The partition function of this classical model is given by
\begin{equation}
Z_{\text{classical}}=\int D\sigma \, e^{-\beta \, S_{\text{classical}}}, \label{eq:Z_classical} 
\end{equation}
in $d=3$, with $S_{\text{classical}}=H(\sigma)$ defined by \Eq{eq:H_sigma}. In particular we concentrate on the nearest-neighbor (NN) theory only, and have 
\begin{equation}
S_{\text{classical}}=\frac {J_1}2 \left( \frac{N_c}2 \right)^2 \sum_{\bn,\mu} \Tr \sigma_\bn \sigma_{\bn+\muhat}.
\end{equation}
Using mean field (MF) theory, we analyze a toy sigma model that has $N=3$. We restrict to a set of configurations with different values of baryon number density $n_B$, and derive the classical MF phase diagram in the temperature-density plane. We will show that the MF ground states at zero temperature break the $U(3)$ symmetry, and that as we increase the temperature, thermal fluctuations melt the condensate and restore the symmetry above a certain temperature $T_c$. We will show that $T_c$ decreases with increasing $n_B$.

\subsection{Mean field theory} 

Our starting point to mean field theory is the partition function \Eq{eq:Z_classical} that we rewrite
\begin{equation}
Z=\int D\sigma \, \, e^{-S_0(\sigma)}\times e^{-\left[S_{\text classical}(\sigma)-S_0(\sigma)\right]}\equiv Z_0 \cdot \< \, e^{-\left[S_{\text classical}(\sigma)-S_0(\sigma)\right]} \, \>_0.
\end{equation}
Here $S_0$ is a trial MF action, and $Z_0$ is its partition function. Also the brackets $\< ,\>_{0}$ denote the MF average of any functional $A$ that depends on the sigma fields.
\begin{equation}
\< \, A\left( \{ \sigma_\bn \}\right) \, \>_0 \equiv Z^{-1}_0 \int \, D\sigma \,\, e^{-S_0(\sigma)} A\left( \{ \sigma_\bn \} \right).
\end{equation}

Using standard convexity arguments one has 
\begin{equation}
\< \, e^{-\left[ S_{\text classical}-S_0 \right]} \, \> \le e^{-\< \, \left[ S_{\text classical}-S_0 \right] \, \> }.
\end{equation}
This means that the free energy $F=-\log Z$ has a upper bound, and obeys the Bogoliubov inequality,
\begin{equation}
F \le F_0 + \< \, S_{\text classical}-S_0  \, \>_0 \equiv \Phi.
\end{equation}
This bound is useful since we can now guess a MF action $S_0$, minimize it with respect to its parameters and obtain an evaluation for the real free energy of the system. The value of the upper bound $\Phi$ determines how close the MF guess is to capture the dynamics of the system.

In our case we choose the simplest possible MF action that corresponds to decoupling all the nearest-neighbor interactions in $S_{\text classical}$. We first choose to proceed in the basis of $U(N)$ generators $M^\eta$ that obey
\begin{equation}
\Tr M^\eta M^{\eta'}=\frac12 \delta_{\eta,\eta'}, \qquad
\sum_\eta \left( M^\eta \right)_{ij} \left( M^\eta \right)_{kl} = \frac12 \delta_{il} \delta_{jk}. 
\end{equation}
In this basis the sigma field is given by its $N^2$ components $\sigma^\eta$,
\begin{equation}
\sigma= 2\sum_\eta \sigma^\eta M^{\eta T},
\end{equation}
and we write the MF action as 
\begin{equation}
S_{\text{MF}} = \beta J_1 \left( \frac{N_c}{2} \right)^2 \sum_{\bf n} \vec{\sigma}_{\bf n} \cdot \vec{h}_{\bf n},
\end{equation}
parametrized by the set $\vec{h}_{\bf n}$. The function $\Phi$ is then
\begin{equation}
\Phi = -\sum_{\bf n} {\rm log} \left \{ \int d\sigma_{\bf n} \, e^{-K \vec{\sigma}_{\bf n} \cdot \vec{h}_{\bf n}} \right \} + K\sum_{\langle {\bf n m} \rangle } \vec{\mu}_{\bf n} \cdot \vec{\mu}_{\bf n} - 
K\sum_{\bf n} \vec{\mu}_{\bf n} \cdot \vec{h}_{\bf n},
\end{equation}
where the first contribution is $F_0$, and the next two terms correspond to $\< \, S_{\text classical} \, \>_0$ and $\< \, S_0 \, \>_0$. $K=\beta J_1 \left( \frac{N_c}{2} \right)^2 $ and $\vec{\mu}$ is the magnetization
\begin{equation}
\vec{\mu}_{\bf n} \equiv \frac{\int d\sigma_{\bf n} e^{-K\vec{\sigma}_{\bf n} \cdot \vec{h}_{\bf n}} \vec{\sigma}_{\bf n}}{\int d\sigma_{\bf n} e^{-K\vec{\sigma}_{\bf n} \cdot \vec{h}_{\bf n}}}.
\label{eq:magnetization}
\end{equation}
Note that the integration measure $d\sigma_{\bf n}$ depends on the $U(N)$ representation chosen on site $\bn$.
We minimize $\Phi$ with respect to $\{ \vec{h}_{\bf n} \}$ to obtain the MF equations,
\begin{equation}
\vec{\mu}_{\bf n} = \frac{\int d\sigma_{\bf n} e^{-K\vec{\sigma}_{\bf n} \cdot \sum_{{\bf m}({\bf n})} \vec{\mu}_{\bf m}} \vec{\sigma}_{\bf n}}{\int d\sigma_{\bf n} e^{-K\vec{\sigma}_{bf n} \cdot \sum_{{\bf m}({\bf n})} \vec{\mu}_{\bf m}}}, \label{eq:MF_mag}
\end{equation}
which are extrema (in $\vec \mu$ space) of $\Phi$. This allows us to write $\Phi$ as a function of $\vec \mu$ only,
\begin{equation}
\Phi = -\sum_{\bf n} {\rm log} \left \{ \int d\sigma \, e^{-K \vec{\sigma}_{\bf n} \cdot \sum_{{\bf m}({\bf n})} \vec{\mu}_{\bf m}} \right \} - 
K\sum_{\langle {\bf n m} \rangle} \vec{\mu}_{\bf n} \cdot \vec{\mu}_{\bf m}. \label{eq:MF_Phi}
\end{equation}

The MF equations \Eq{eq:MF_mag} and the function $\Phi$ in \Eq{eq:MF_Phi} constitute our classical MF theory. For each value of $K\sim1/T$, $\Phi$ has a set of extrema points that obey the MF equations. A subset of these are minima that correspond to classical ground states of the MF theory. The magnetizations, $m_{\bf n} =2\vec{\mu}_{\bf n} \cdot \vec{M}^T$, of each minimum determine the symmetry properties of the corresponding ground state.  

\subsection{The toy model of $U(3)$}

\begin{figure}[htb]
\begin{center}
\epsfig{width=13cm,file=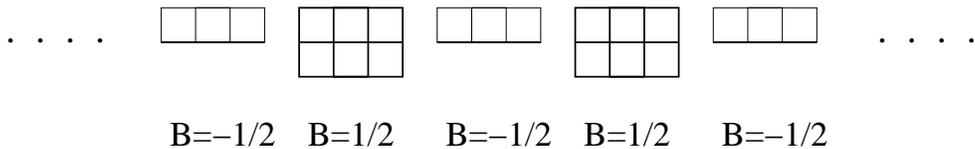}
\caption{The most symmetric $B=0$ state in the $U(3)$ toy model}
\label{fig:zeroB_U3}
\end{center}
\end{figure}

\begin{figure}[htb]
\begin{center}
\epsfig{width=13cm,file=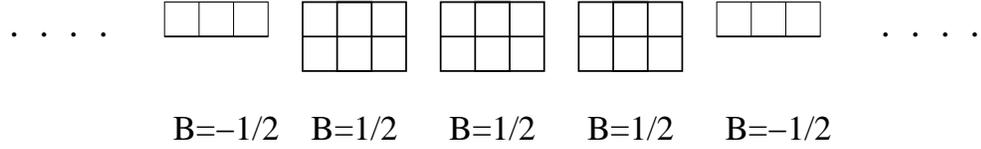}
\caption{Adding a baryon in the $U(3)$ toy model.}
\end{center}
\label{fig:nonzeroB_U3}
\end{figure}

The $N_f=1$ theory is a $U(4)$ antiferromagnet and contains many
degrees of freedom in which to do mean field theory.
We treat the {\em simplest\/} non-trivial model which is a toy model with $U(3)$ symmetry. It does not correspond to an actual value of $N_f$, but allows for analytical evaluation of the MF quantities. 
The most symmetric $B=0$ state in this model cannot have $B=0$ on every site,
but must alternate between the representations corresponding to $B=\pm1/2$ (see Fig.~\ref{fig:zeroB_U3}).
Increasing $B$ is done by replacing some of the $B=-1/2$ representations with $B=+1/2$ representations as in Fig.~\ref{fig:nonzeroB_U3}. Another option is to replace either $B=-1/2$ representations or $B=+1/2$ representations by $U(3)$ singlets with $B=+3/2$ and three rows.

As mentioned, several MF theory quantities such as $\Phi$ can be calculated analytically for this toy model. The reason is that for $N=3$, and $m=2$, \Eq{Ua} takes the simple form
\begin{equation}
U_\bn=\exp \left( \begin{array}{ccc}
        0       &       0       &       a_1     \\
        0       &       0       &       a_2     \\
        -a^{*}_1 &      -a^{*}_2 &      0       \end{array}     \right).
\end{equation}
We use the definitions
\begin{eqnarray}
a_1 &=& \rho \, {\rm cos} (\theta/2)    e^{i\alpha_1},   \nonumber       \\
a_2 &=& \rho \, {\rm sin} (\theta/2) e^{i\alpha_2},
\end{eqnarray}
with $0<\rho<\pi$, $0<\theta<\pi$ and $0<\alpha_{1,2}<2\pi$, and get 
\begin{equation}
\vec{\sigma} = -2\left( \begin{array}{c}
                -\frac{1}{\sqrt{6}}     \\
                {\rm sin}^2 \rho \, {\rm sin} \theta \, {\rm cos} \phi        \\
                -{\rm sin}^2 \rho \, {\rm sin} \theta \, {\rm sin} \phi      \\                {\rm sin}^2 \rho \, {\rm cos} \theta                       \\
                {\rm sin} 2\rho \, {\rm cos} \theta/2 \, {\rm cos}(\frac{1}{2}(\psi+\phi))     \\
                -{\rm sin} 2\rho \, {\rm cos} \theta/2 \, {\rm sin}(\frac{1}{2}(\psi+\phi))    \\
                {\rm sin} 2\rho \, {\rm sin} \theta/2 \, {\rm cos}(\frac{1}{2}(\psi-\phi))    \\
                -{\rm sin} 2\rho \, {\rm sin} \theta/2 \, {\rm sin}(\frac{1}{2}(\psi-\phi))    \\
                -\sqrt{3}({\rm cos}^2 \rho -\frac{1}{3})
                \end{array}     \right),
\label{Q_form}
\end{equation}
here $\phi \equiv \alpha_1-\alpha_2$ and $\psi = \alpha_1 + \alpha_2$, with $0<\psi<4\pi$ and $0<\phi<2\pi$.

At zero density we assume that the magnetizations on even and odd sites are $\vec{\mu}_1$ and $\vec{\mu}_2$ respectively. This results in the following MF equations 
\begin{eqnarray}
\vec{\mu}_1 &=& \frac{\int d\sigma e^{-6K\vec{\sigma} \cdot \vec{\mu}_2} \vec{\sigma}}{\int d\sigma
 e^{-6K\vec{\sigma} \cdot \vec{\mu}_2}},      \\      
\vec{\mu}_2 &=& -\frac{\int d\sigma e^{+6K\vec{\sigma} \cdot \vec{\mu}_1} \vec{\sigma}}{\int d\sigma e^{+6K\vec{\sigma} \cdot \vec{\mu}_1}}.
\end{eqnarray}

The measure of the manifold $U(3)/U(2)\times U(1)$ is the measure of the
manifold of $3\times 3$ hermitian matrices $d\vec{\sigma}$ modified to account for the dependence~(\ref{Q_form}) of $\vec{\sigma}$ on the manifold coordinates. The induced metric of the space is 
\begin{equation}
g^{{\rm induced}}_{\alpha\beta} = \sum_{\mu,\nu} g_{\mu \nu} \cdot \frac{\partial \sigma_{\mu}}{\partial \theta_{\alpha}} \frac{\partial \sigma_{\nu}}{\partial \theta_{\beta}},
\end{equation}
where $g_{\mu\nu}=\delta_{\mu\nu}$, and the coordinates $\theta_{\alpha}$ are the coset coordinates that appear in $\sigma^{\mu}$. The measure of integration is thus 
\begin{equation}
d\sigma = d\vec{\theta} \sqrt{{\rm det}(g_{\rm induced})}.
\end{equation}
For $U(3)/U(2)\times U(1)$, we find that 
\begin{equation}
d\sigma = d({\rm sin}^4\rho) \,d({\rm cos} \theta) \,\frac{d\phi}{2\pi} \, \frac{d\Psi}{4\pi}.
\end{equation}

In order to solve the MF equations we choose a basis that diagonalizes $m_1\equiv 2\sum_\eta \mu^\eta_1 M^{\eta T}$. Since the measure in the angles $\Psi$ and $\phi$ is flat, $m_2\equiv 2\sum_\eta \mu^\eta_2 M^{\eta T}$ is diagonal as well, and only $\mu^0_2$, $\mu^3_2$, and $\mu^8_2$ are non-zero. Since $\mu^0$ is fixed [see \Eq{Q_form}], one is left with the following MF equations for $\vec \mu_i=\left(\mu^3_i,\mu^8_i\right)$ with $i=1,2$,
\begin{eqnarray}
\mu^3_1&=&f_3(6K\mu^3_2,6K\mu^8_2), \hskip 1.82cm \mu^8_1=f_8(6K\mu^3_2,6K\mu^8_2), \label{eq:MF_mu1} \\
\mu^3_2&=&-f_3(-6K\mu^3_1,-6K\mu^8_1), \qquad \mu^8_2=-f_8(-6K\mu^3_1,-6K\mu^8_1).  \label{eq:MF_mu2}
\end{eqnarray}
Here
\begin{eqnarray}
f_3(x,y)&=&\left[ \frac{e^{x_+}}{x_+} + \frac{e^{x_-}}{x_-} - \frac{e^{x_-}-1}{x_-}\left(\frac1{x^3}-\frac1{x_-}\right) - \frac{e^{x_+}-1)}{x_+}\left(\frac1{x^3}+\frac1{x_+} \right) \right]\times \nonumber \\ 
&&\qquad \left[ \frac{e^{x_-}-1}{x_-}-\frac{e^{x_+}-1}{x_+} \right]^{-1}, \label{eq:f3} \\
f_8(x,y)&=&\frac{2}{\sqrt{3}} + \sqrt{3}\left[ \frac{e^{x_+}}{x_+} - \frac{e^{x_-}}{x_-} - \frac{e^{x_+}-1}{x^2_+} + \frac{e^{x_-}-1)}{x^2_-} \right]\times \nonumber \\&&\qquad \left[ \frac{e^{x_-}-1}{x_-}-\frac{e^{x_+}-1}{x_+} \right]^{-1}, \label{eq:f8}
\end{eqnarray}
with $x_\pm=\sqrt{3}y\pm x$.

\begin{figure}[htb]
\begin{center}
\epsfig{width=13cm,file=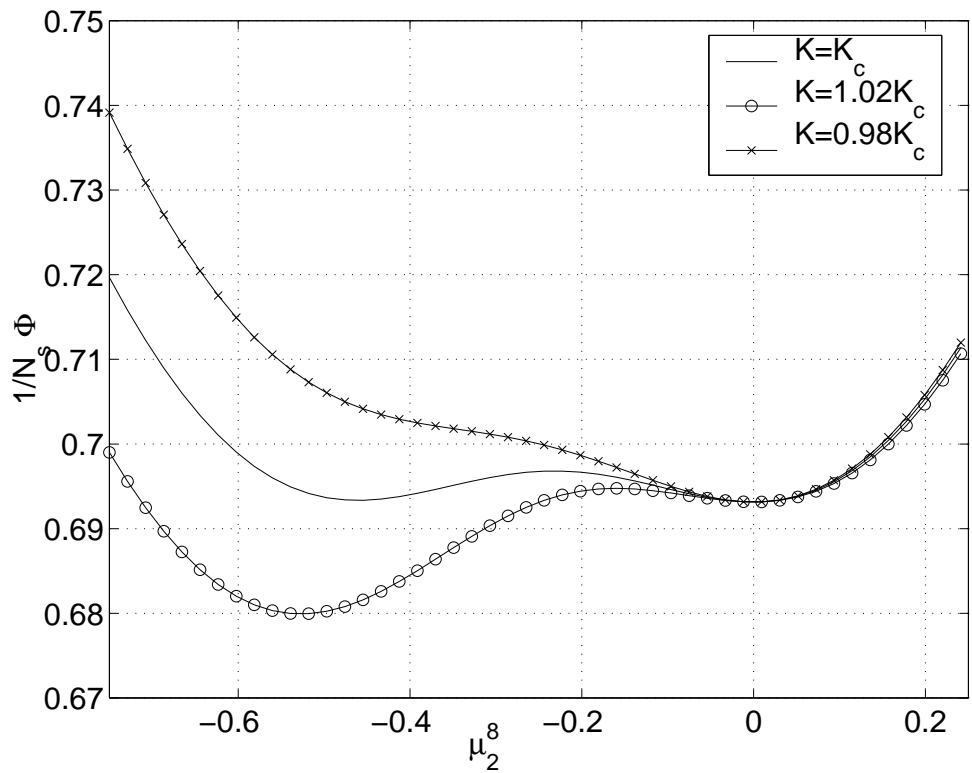}
\caption[$\Phi$ at zero density for $N=3$.]{$\Phi$ for zero density near the first order phase transition.}
\label{fig:F_MFB0}
\end{center}
\end{figure}

We use \Eq{eq:MF_mu1} to evaluate $\Phi(\vec\mu_{i})$ as a function of $\vec \mu_2$,
\begin{eqnarray}
\Phi(\vec \mu_i)/N_s&=& -3K \, \, \vec \mu_1\cdot \vec \mu_2   -2K\sqrt{3}(\mu^8_1-\mu^8_2) - \varphi(6K\, \vec \mu_i), \\
\varphi(\vec x_i) &=& \frac12 \left[ \log \left( \frac{e^{-x_{+,1}}-1}{x_{+,1}x^3_{1}}-\frac{e^{-x_{-,1}}-1}{x_{-,1}x^3_{1}}\right) 
- \log \left( \frac{e^{x_{+,2}}-1}{x_{+,2}x^3_{2}} -\frac{e^{x_{+,2}}-1}{x_{+,2}x^3_{2}} \right) \right]. \nonumber \\ \label{eq:phi_MF}
\end{eqnarray}
Here $N_s$ is the number of sites and $x_{\pm,i}=\sqrt{3} \mu^8_{i}\pm \mu^3_{i}$. It is easy to check that the MF equations have a symmetric solution with $\vec \mu_{2}=0$ (and hence $\vec \mu_1=0$) for all values of $K$. The magnetizations of this solution are 
\begin{equation}
m_1=-m_2 \sim {\bm 1}, \label{eq:m12}
\end{equation}
and are invariant under $U(3)$. As we increase $K$, three additional nontrivial minima appear. In the $\mu^3$--$\mu^8$ plane, the solutions are found on the following three lines.
\begin{eqnarray}
\mu^3_i&=&0, \label{eq:sol1} \\
\mu^3_i&=&+ \sqrt{3}\mu^8_i, \label{eq:sol2} \\
\mu^3_i&=&- \sqrt{3}\mu^8_i, \label{eq:sol3}
\end{eqnarray}
They have the same $\Phi$. The magnetizations of each solutions are invariant under a certain $U(2)\times U(1)$ subgroup of $U(3)$. For example the solution with $\mu^3=0$ has
\begin{equation}
m_1\sim -m_2 \sim \left( \begin{array}{ccc} 1 & 0 & 0 \\ 0 & 1 & 0 \\ 0 & 0 & -2 \end{array} \right).
\end{equation}
For $K>K_c$ these solutions have lower free energy than the symmetric solution~(\ref{eq:m12}) and $U(3)$ is spontaneously broken to $U(2)\times U(1)$. In Fig.~\ref{fig:F_MFB0} we present $\Phi$ on the line $\mu^3_2=0$ for $K/K_c=0.98,1,1.02$. 
We find that there exists a first order phase transition at $K_c=0.9$.  
\clearpage
Next we performed the MF analysis for a set of configurations that have non-zero baryon number. We construct these by putting net baryon number on the sites of $2\times 2\times 2$ unit cells. This set is listed in Table~\ref{table:config}, where we give a pictorial description of the unit cells of each configuration.

\begin{longtable}{@{\extracolsep{\fill}}c||ccc}
\caption[Baryon number configurations in the $U(3)$ toy model.]{The set of configurations with different baryon number density for the $U(3)$ toy model. Here $\rho_{\text{max}}=3/2$. A $\pm$ represents $B=\pm 1/2$, and a $\bullet$ represents $B=+3/2$ with $\sigma=\bm{1}$.}  \label{table:config} \\ 
Configuration & $\rho/\rho_{\text{max}}$ &   Unit cell & $K_c$ \\ \\
\hline \hline \\
\endfirsthead 
\caption[]{(continued)}\\
Configuration & $\rho/\rho_{\text{max}}$ &   Unit cell & $K_c$ \\ \\
\hline \hline \\
\endhead
\hline
\endfoot
1 & 0 & \qquad  \parbox{1.4in}{\epsfig{width=2cm,file=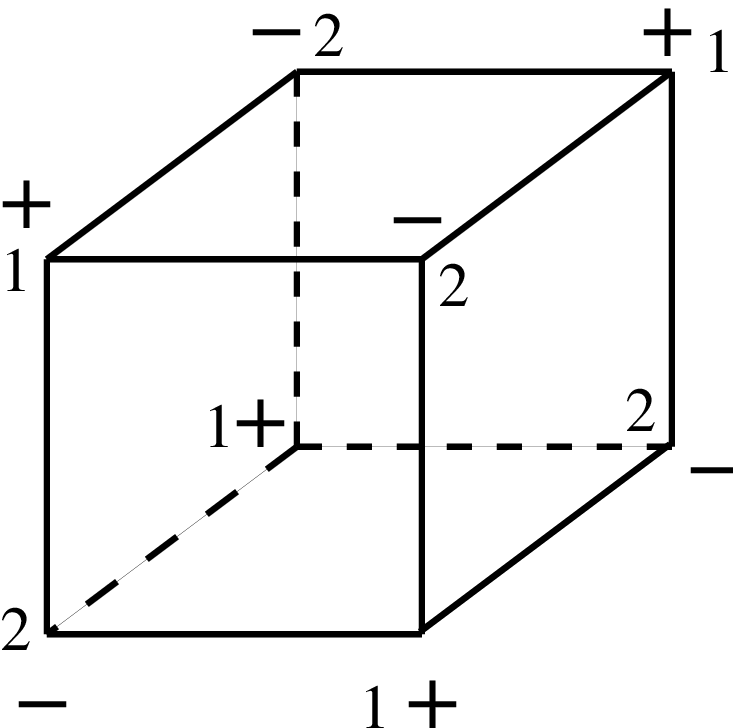}} & 0.9 \\ \\ 
2 & $\frac1{12}$ & \qquad \parbox{1.4in}{\epsfig{width=2cm,file=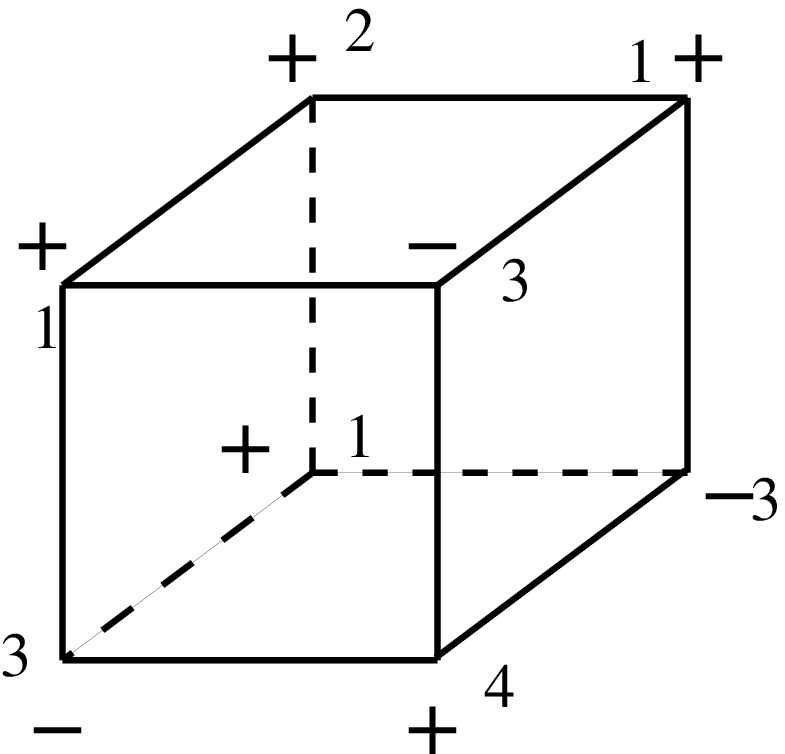}} & 0.94 \\ \\
3 & $\frac16$ & \qquad \parbox{1.4in}{\epsfig{width=2cm,file=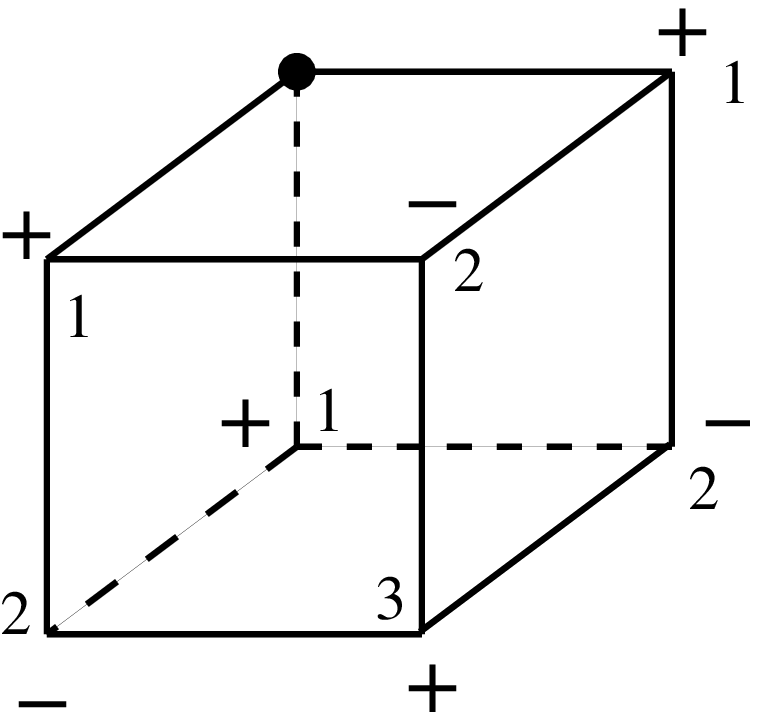}} & 1.03 \\ \\
4 & $\frac13$ & \qquad  \parbox{1.4in}{\epsfig{width=2cm,file=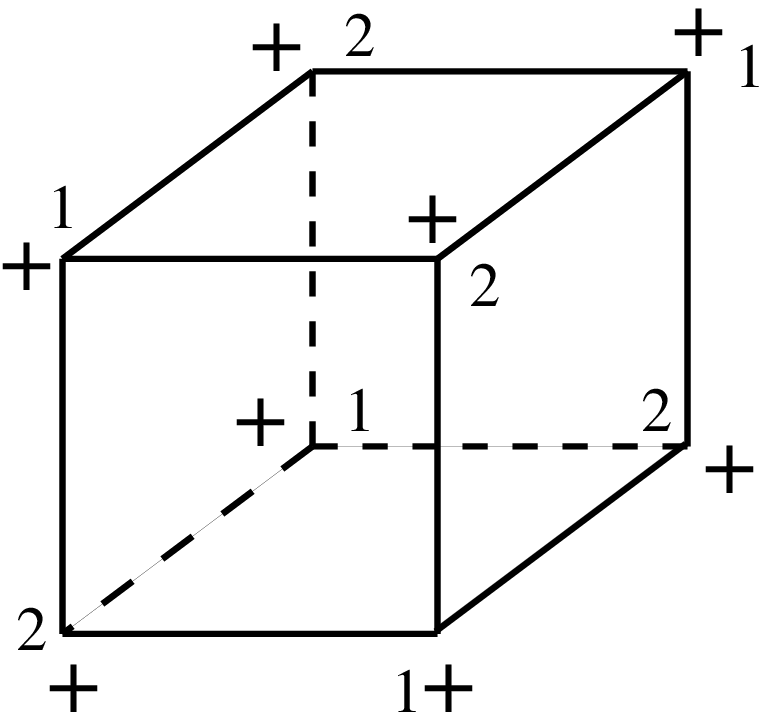}} & 1 \\ \\
5 & $\frac5{12}$ & \qquad  \parbox{1.4in}{\epsfig{width=2cm,file=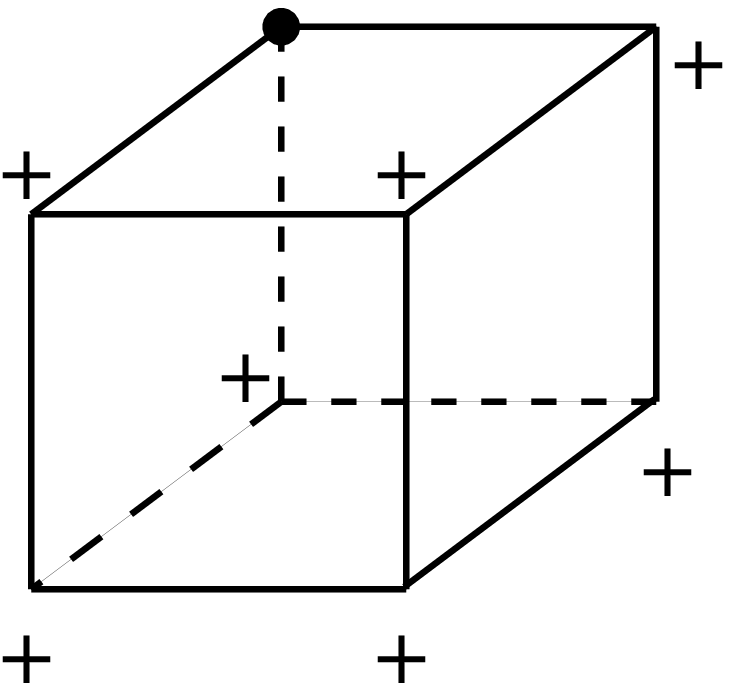}} & 1.14 \\ \\
6 & $\frac7{12}$ & \qquad  \parbox{1.4in}{\epsfig{width=2cm,file=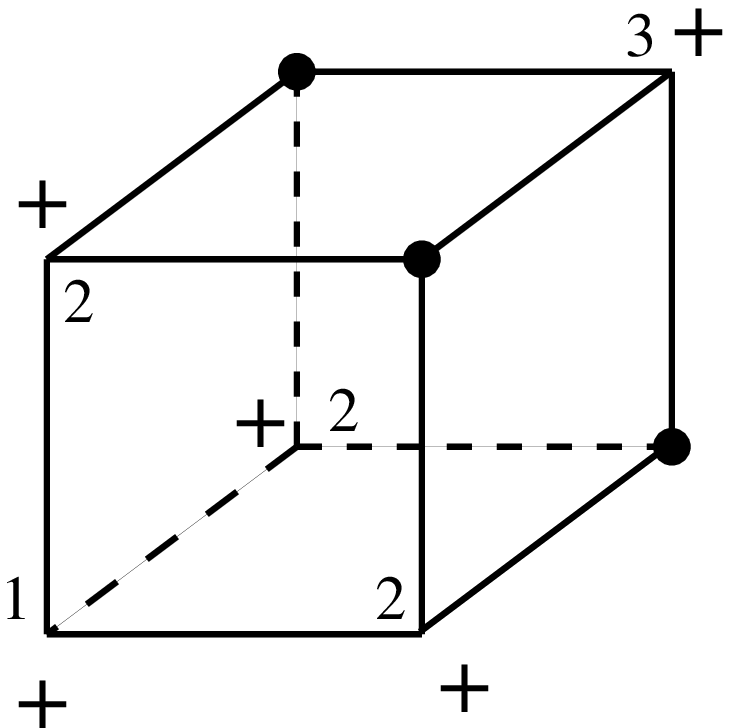}} & 1.72 \\ \\
7 & $\frac23$ & \qquad  \parbox{1.4in}{\epsfig{width=2cm,file=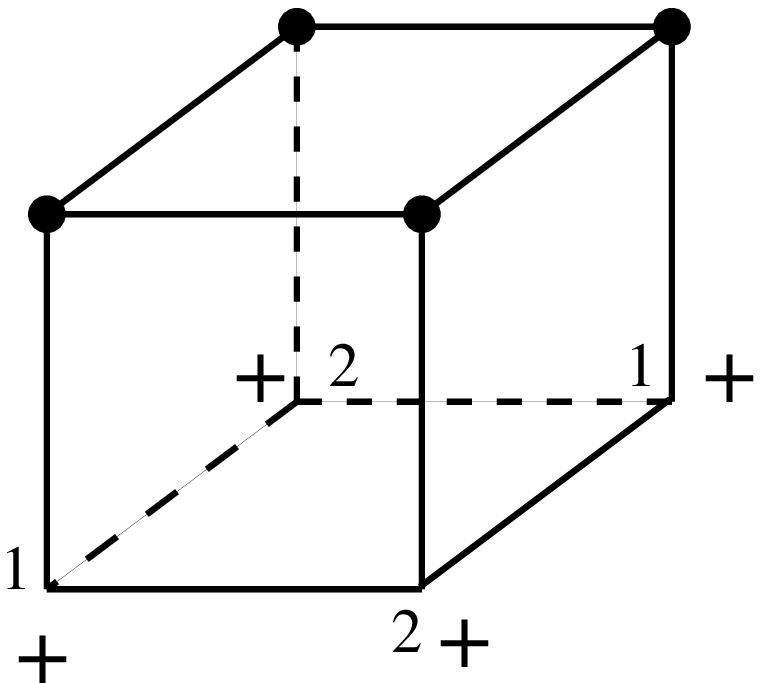}} & 1.5 \\ \\
8 & $\frac56$ & \qquad  \parbox{1.4in}{\epsfig{width=2cm,file=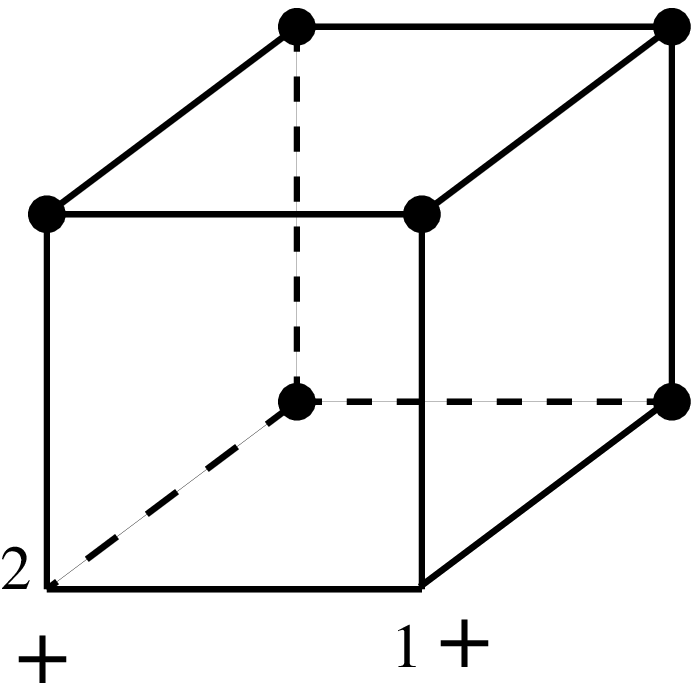}} & 3 \\ \\ \hline
\end{longtable}
The numbering of the sites in the unit cells denotes the magnetization distribution of the ansatz used for each configuration. For example, in configuration no.~1 the ansatz consists of having two magnetizations denoted by $m_1$ and $m_2$ on the even and odd sites denoted by $1$ and $2$. 
We present the MF equations and corresponding $\Phi$'s for all configurations in Appendix~\ref{app:MFeqU3}. 

Except for configuration no.~2, the MF equations are consistent with the assumption that all magnetizations commute and are diagonalized simultaneously. Therefore as for zero baryon number, the MF equations and $\Phi$'s can be written for $\mu^3$, and $\mu^8$ only. Also, in all these configurations $\Phi$ becomes a function of two variables, allowing a simple graphical minimization of $\Phi$. For configuration no.~2 $\Phi$ is a function of four variables, and we make an ansatz that only $\mu^8$ is non-zero in order to treat this case graphically as well.
 
\begin{figure}[htb]
\begin{center}
\epsfig{width=13cm,file=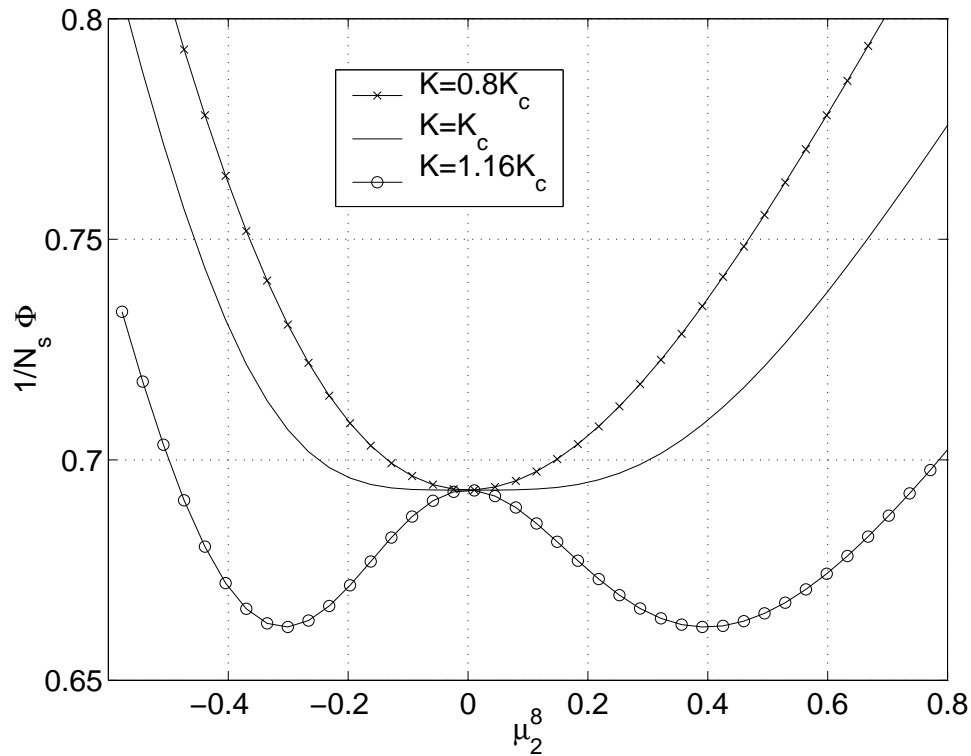}
\caption[$\Phi$ non-zero density for $N=3$]{$\Phi$ for non-zero density near the second order phase transition.}
\label{fig:F_MFB12}
\end{center}
\end{figure}

Except for the symmetric solution that exists for all values of $K$, nontrivial minima of $\Phi$ appear on the lines~(\ref{eq:sol1})--(\ref{eq:sol3}). In configurations~3, 5, 6 one has three equivalent minima that have the same symmetry and $\Phi$. As for zero baryon number $U(3)$ is spontaneously broken to $U(2)\times U(1)$ above a certain value of $K$, and the phase transition is first order. We give the values of $K_c$ in Table~\ref{table:config}. In configurations~4, 7, 8 we find six minima. Four of these correspond to breaking $U(3)$ into $U(2)\times U(1)$, and the remaining two correspond to breaking $U(3)$ to $U(1)\times U(1)\times U(1)$\footnote{This is the symmetry of the ground state of the {\it quantum} sigma model at these configurations, as predicted in the large $N_c$ limit. See Chapter~\ref{chap:Qfluctuation}.}. All six solutions have the same $\Phi$, and become preferable over the symmetric solution above the same value of $K_c$. Here the phase transition is second order, and we illustrate it in Fig.~\ref{fig:F_MFB12} where we plot $\Phi$ on the line~(\ref{eq:restrict1}) for $K/K_c=0.8,1,1.16$ (for configuration~4). 

As mentioned, for configuration no.~2 we take an ansatz that $\mu^3=0$, and find similar behavior to that of the zero density system. In view of what happens in the other cases, we believe that there are at least two more equivalent minima with the same free energy and the same symmetry properties. There is also a possibility that there are other nonequivalent minima, with different symmetries as in configurations~4, 7, 8.

\begin{figure}[htb]
\begin{center}
\epsfig{width=12cm,file=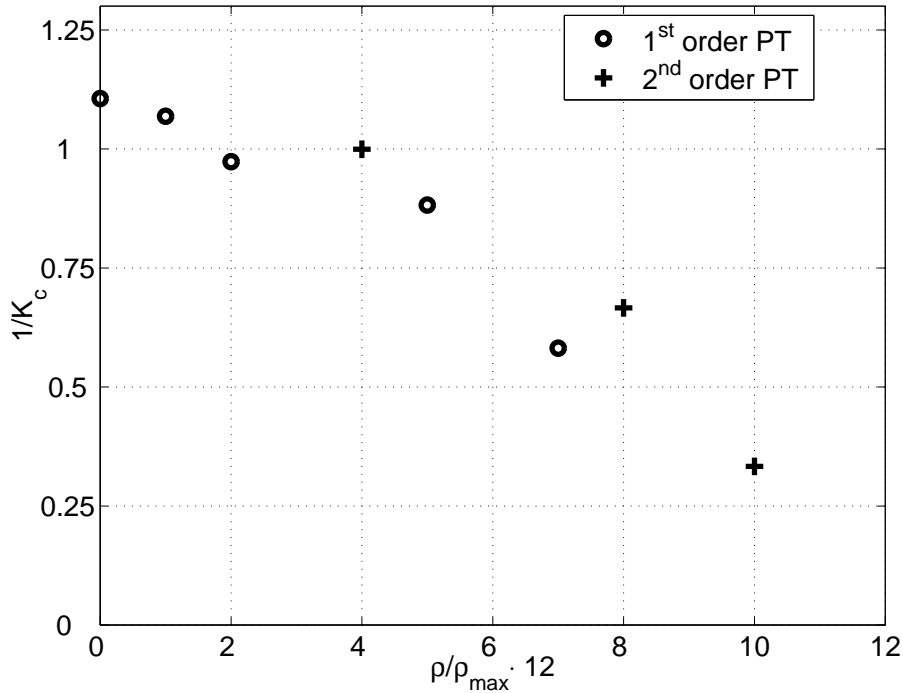}
\caption[Classical phase diagram of the $U(3)$ toy model]{Classical phase diagram of the $U(3)$ antiferromagnet at finite temperature and density. We find that when the phase transition is second order, $U(3)$ is spontaneously broken either to $U(2)\times U(1)$ or $U(1)\times U(1) \times U(1)$. In all other cases $U(3)$ is broken to $U(2)\times U(1)$.}
\label{fig:PT_classical}
\end{center}
\end{figure}

We give the results of these MF calculation in the phase diagram of Fig.~\ref{fig:PT_classical}. Increasing the baryon number density decreases the transition temperature. Note however that we do not find a restoration of the symmetry at zero temperature for any density since this is a MF theory which always breaks the symmetry at $T=0$, even at one dimension.

\section{The semiclassical large $N_c$ limit}
\label{sec:Nc_infty}

We now treat the full quantum sigma model in the large $N_c$ limit by performing a saddle point evaluation of the ground state of the system. We will see that the ground state is degenerate. For $B=0$ this degeneracy is discrete, while for $B>0$ the ground state has a huge continuous degeneracy, exponential in the lattice volume. In general these degeneracies are removed only by $1/N_c$ corrections that we calculate in the next chapter.

\subsection{Zero density \label{sec:zeroB}}

\begin{figure}[htb]
\begin{center}
\epsfig{width=13cm,file=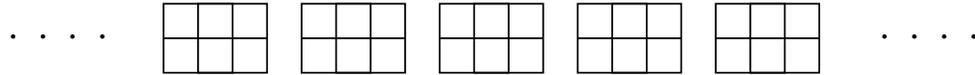}
\caption[A configuration with zero baryon number on all sites.]{$B=0$ configuration, for the case of $N_f=1$.}
\label{fig:zeroB}
\end{center}
\end{figure}

\begin{figure}[htb]
\begin{center}
\epsfig{width=13cm,file=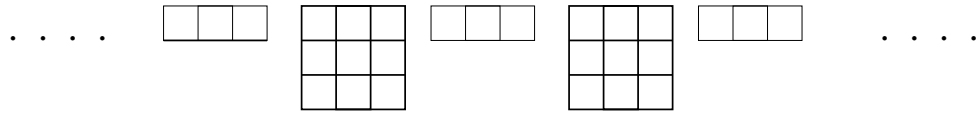}
\caption[A configuration with alternating baryon number on adjacent sites.]{$B=0$ state, for the case of $N_f=1$, and $b=\pm 1$.}
\label{fig:zeroB1}
\end{center}
\end{figure}

The simplest way to set the baryon density to zero is just to
choose $B_\bn=0$ on each site, meaning $m=N/2=2N_f$ (see Fig.~\ref{fig:zeroB}).
It turns out
to be just as easy to consider a slightly generalized case
\cite{Smit}, in which $B_\bn$ is chosen to alternate, $B_\bn=\pm
b$, on even and odd sublattices (see Fig~.\ref{fig:zeroB1}).
 This means to choose a
representation with $m=N/2+b$ rows on even sites and $N-m=N/2-b$
rows on odd sites, which gives a pair of conjugate representations
of $U(N)$. In view of \Eq{SQminus}, we can  substitute
$\sigma\to-\sigma$ on the odd sites and thus have identical
manifolds on all sites. The Hamiltonian is then
\begin{equation}
H(\sigma)=\left(\frac{N_c}2\right)^2\sum_{\bn\mu \atop
j\not=0}J_j(-1)^j
\sigma^\eta_{\bn}\sigma^\eta_{\bn+j\muhat}\left(s^\mu_\eta\right)^{j+1}.
\label{Hsigmaminus}
\end{equation}
$m$ is a new parameter in the theory, and we can ask what value of
$m$ gives the lowest energy for the ground state. 

In the large-$N_c$ limit, we seek the classical saddle
point\footnote{Note that the kinetic term is pure imaginary.} of
$S$. We assume the saddle is at a configuration
$\sigma_{\bn}(\tau)$ that is independent of time, and so we drop
the time derivative. We begin with the nearest-neighbor
Hamiltonian,\footnote{ This classical Hamiltonian is
ferromagnetic; the antiferromagnetic nature of the quantum
Hamiltonian (\ref{Hquantum}) is preserved by the alternating signs
in the time-derivative term.}
\begin{eqnarray}
H&=&-J_1\left( \frac{N_c}2\right)^2\sum \sigma^\eta_{\bn}\sigma^\eta_{\bn+\muhat}\nonumber\\
&=&-\frac {J_1}2 \left( \frac{N_c}2\right)^2\sum \Tr \sigma_\bn \sigma_{\bn+\muhat}. \label{eq:eps_NN}
\end{eqnarray}
Again, the matrices $\sigma_\bn$ are Hermitian and unitary, and
the expansion coefficients $\sigma^\eta_{\bn}$ are real and
satisfy $\sum(\sigma^\eta)^2=N/2$. The minimum of $H$ is clearly
at a constant field, $\sigma_\bn=\sigma_0$, which can be
diagonalized to $\sigma_\bn=\Lambda$ by a $U(N)$ rotation. This is
a N\'eel state in the original variables. The $U(N)$ symmetry is
broken to $U(m)\times U(N-m)$ and there are $2m(N-m)$ Goldstone
bosons.

The classical energy density (per link) is $\epsilon_0=-J_1N^2_cN/8$,
independent of $m$. Thus at leading order in $1/N_c$, the optimal
value of $m$ is undetermined, and any alternating background of
baryon number is equally good.

Since it turns out that the $1/N_c$ corrections select $m=N/2$ [see Section~\ref{sec:Rem_B0}
let us consider the effect of the next-nearest-neighbor (NNN) term in $H$ for this case
only. The perturbation is
\begin{equation}
H'=J_2\left( \frac{N_c}2\right)^2\sum_{\bn\mu}
\sigma^\eta_{\bn}\sigma^\eta_{\bn+2\muhat}s^\mu_\eta.
\end{equation}
It breaks the $U(N)$ symmetry to $SU(N_f)_L\times SU(N_f)_R\times
U(1)_A\times U(1)_B$. Assuming that $J_2\ll J_1$, we again seek the
minimum energy configuration in the form of a constant field; we
minimize
\begin{equation}
\epsilon_2=\sum_{\mu\eta} \sigma^\eta \sigma^\eta s^\mu_\eta
\label{epsprime}
\end{equation}
among the $U(N)$-equivalent $\sigma=\sigma_0$ states that minimize
the nearest-neighbor action. It is not hard to show (see Appendix
\ref{app:nnn}) that $\epsilon_2$ is minimized for
$\sigma=\gamma_0\otimes\bm{1}$. This is a condensate that is
symmetric under the vector generators
$M^\eta=\bm{1}\otimes\lambda^a$ but not under the axial generators
$\gamma_5\otimes\lambda^a$, and thus it breaks the chiral symmetry
to the vector subgroup, $SU(N_f)_V\times U(1)_B$.

It is interesting to note that the discrete degeneracy in the number $m$ can be removed at $O(1/N^0_c)$ as well by the NNN terms themselves. To see this we calculate the NNN energy \Eq{epsprime} for the configurations that minimize the NN enegry \Eq{eq:eps_NN}. It is easiest to do the calculation by writing the NNN energy in terms of the Dirac matrices $\alpha_\mu$ as
\begin{equation}
\epsilon_2 = \frac12 \sum_\mu \Tr \left[ \sigma
\alpha_\mu \sigma \alpha_\mu    \right]. \label{eq:eps_NNN}
\end{equation}
We show in Chapter~\ref{chap:NNN} [\Eq{bound}] that \Eq{eq:eps_NNN} has a lower bound 
\begin{equation}
\epsilon_2 \ge \frac32 (4m-3N), \label{eq:bound_B0}
\end{equation}
which means that the energy from NN and NNN interactions is bounded by
\begin{equation}
E(m) \ge \frac{3N^2_c}8 \left[ -J_1N + \frac{J_2}2(4m-3N)\right]. \label{eq:NN_n_NNN}
\end{equation}
Since $m\ge N/2$ then $E(m)\ge E(N/2)$. In Appendix~\ref{app:nnn} we show that the ground state for $m=N/2$ saturates $E(N/2)=-3N/2$. This means that the ground state of $m=N/2$, is also the overall ground state (out of values of $m$). As a result we find that the discrete degeneracy of the classical NN energy in the number $m$ is removed by NNN as well as by $1/N_c$ corrections.

\subsection{Non-zero density\label{sec:nonzeroB}}

The zero-density theories considered in the preceding section were
defined by selecting representations with $m$ and $N-m$ rows on
alternating sites. For any $m$, this led to a $\sigma$ model with
identical degrees of freedom on all sites---after redefinition of
the spins on the odd sublattice---and ferromagnetic
couplings.\footnote{The $(-1)^\bn$ factor in the kinetic energy
retained information about the antiferromagnetic nature of the
quantum problem; it did not affect the classical analysis.} We
eventually settled on $m=N/2$ as the background that gives the
ground state of lowest energy.

\begin{figure}[htb]
\begin{center}
\epsfig{width=13cm,file=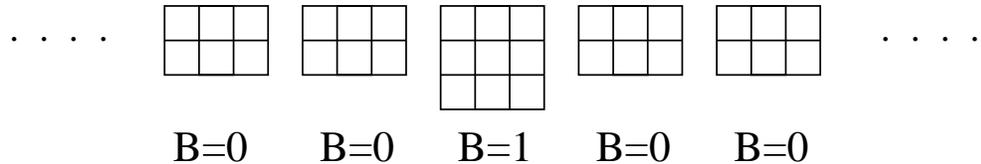}
\caption[Adding one baryon to the lattice.]{Adding one baryon to a site in the $N_f=1$ theory
[a $U(4)$ antiferromagnet]}
\label{fig:single_baryon}
\end{center}
\end{figure}

Introducing non-zero baryon density means changing $m$ on some
sites of the lattice. 
Choosing a different representation adds (or subtracts) baryons on a site-by-site
basis.
For instance, we can add a single baryon by adding a row to the Young
tableau as in Fig.~\ref{fig:single_baryon}.
Since in general there will be
representations on different sites that are not mutually
conjugate, different sites will carry $\sigma$ variables that do
not live in the same submanifold of $U(N)$. We here limit
ourselves to the simpler case of uniform $m$, where adjacent sites
carry identical spins---but the coupling is {\em
antiferromagnetic}.

In order to learn how to work with such a theory, we begin by
studying the two-site problem. The results of this study will lead
directly to an {\em ansatz\/} for the ground state of a lattice
with a fixed density of baryons.

\subsubsection{Two site problem}

\begin{figure}
\begin{center}
\epsfig{width=10cm,file=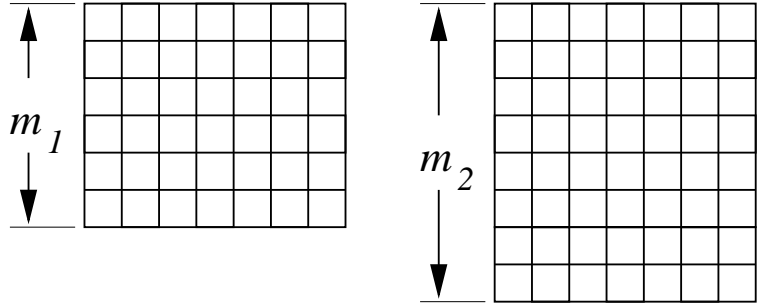} 
\caption{Two sites that carry $U(N)$ representations with $m_1$ and $m_2$ rows.} \label{fig:twospins}
\end{center}
\end{figure}

Consider, therefore, two sites with quantum spins $Q_1$ and~$Q_2$
that carry representations of $U(N)$ with $m_1$ and~$m_2$ rows,
and $N_c$ columns (see Fig.~\ref{fig:twospins}).

The quantum Hamiltonian is
\begin{equation}
H=J_1\sum_\eta Q_1^\eta Q_2^\eta, \label{2site_qH}
\end{equation}
an antiferromagnetic coupling. The corresponding classical
$\sigma$ model has the interaction Hamiltonian
\begin{equation}
H(\sigma)=\frac {J_1}2 \left(\frac{N_c}2\right)^2 \Tr
\sigma_1\sigma_2,
\end{equation}
where
\begin{equation}
\sigma_i=U_i\Lambda_iU_i^\dag.
\end{equation}
The two $\Lambda$ matrices reflect the different values of $m_i$
according to
\begin{equation}
\Lambda_i=\left(
\begin{array}{cc}
\bm1_{m_i}&0\\
0&-\bm1_{N-m_i}
\end{array}\right).
\end{equation}

The $N_c\to\infty$ limit is the classical limit, in which we seek
values of $\sigma_{1,2}$ that minimize $H(\sigma)$. A global
$U(N)$ rotation, {\em viz.},
\begin{equation}
\sigma_i\to V\sigma_i V^\dag,
\end{equation}
can be used to diagonalize $\sigma_1$ so that
$\sigma_1=\Lambda_1$. Now we have to minimize $\Tr
\Lambda_1\sigma_2$. The case of conjugate representations,
$m_2=N-m_1$, is easy: $\sigma_2$ is a unitary rotation of
$\Lambda_2$, which (in this case) can be rotated into
$-\Lambda_1$. This is the unique antiferromagnetic ground state.
$\sigma_1$ and~$\sigma_2$ can be copied to the odd and even
sublattices of an infinite lattice to give the classical N\'eel
state considered in the preceding section.

The case $m_1=m_2=m$ is more complex. We consider $m>N/2$ for
definiteness; the other case is similarly handled. We write
explicitly 
\begin{equation}
\sigma_2=\left( \begin{array}{cc} \cos\left(2\sqrt{a
a^\dag}\right) & -a\frac{\displaystyle\sin\left(2\sqrt{a^\dag
a}\right)}
      {\displaystyle\sqrt{a^\dag a}}       \\[2pt]
-\frac{\displaystyle\sin\left(2\sqrt{a^\dag a}\right)}
            {\displaystyle\sqrt{a a^\dag}}\,a^\dag &
-\cos\left(2\sqrt{a^\dag a}\right)
\end{array}\right).
\label{eq:sigma_2_a}
\end{equation}
Since $\sigma_2$ is
a rotation of $\Lambda$,
\begin{equation}
2m-N=\Tr\sigma_2= \Tr\cos\left(2\sqrt{a
a^\dag}\right)-\Tr\cos\left(2\sqrt{a^\dag a}\right),
\label{eq:B_restrict}
\end{equation}
and hence the energy is
\begin{eqnarray}
E&=&\frac{J_1}2 \left(\frac{N_c}2\right)^2\Tr\Lambda\sigma_2\\
&=&J_1\left(\frac{N_c}2\right)^2\left[\Tr\cos\left(2\sqrt{a^\dag
a}\right) +m-N/2\right].
\end{eqnarray}
$E$ is minimized when all the eigenvalues of $a^\dag a$ are equal
to $\pi^2/4$. This means that the $N-m$ column vectors $\ba_i$
form an orthogonal set in $m$ dimensions, with
\begin{equation}
\ba_i^\dag\cdot\ba_j=\left(\frac\pi2\right)^2\delta_{ij}.
\end{equation}
Since $m>N-m$ by assumption, such a set of vectors can always be
found.

Since $a^\dag a=(\pi^2/4)\bm1_{N-m}$, we have
$\sin\left(2\sqrt{a^\dag a}\right)=0$ and so the off-diagonal
blocks of \Eq{eq:sigma_2_a} vanish. The lower-right block of
$\sigma_2$ is the unit matrix $\bm1_{N-m}$. We know that
$(\sigma_2)^2=1$ since $\Lambda^2=1$ and thus the upper-left block
must have eigenvalues $\pm1$. Equating traces of $\sigma_2$ and
$\Lambda$, we find that the upper-left $m\times m$ block must take
the form
\begin{equation}
\sigma^{(m)}=U^{(m)}\Lambda^{(m)}U^{(m)\dag}, \label{eq:sigma_m}
\end{equation}
where
\begin{equation}
\Lambda^{(m)}=\left( \begin{array}{cc}
\bm1_{2m-N}&0\\
0&-\bm1_{N-m}
\end{array}\right)
\label{eq:Lm}
\end{equation}
and $U^{(m)}\in U(m)$. $\sigma^{(m)}$ represents the coset
$U(m)/[U(2m-N)\times U(N-m)]$.

We conclude that the classical ground state of this $B\not=0$
two-site problem is degenerate, even beyond breaking the overall
$U(N)$ symmetry. The solutions can be written as
\begin{eqnarray}
\sigma_1&=&\Lambda_1=\left( \begin{array}{cc}
\bm1_m&0\\
0&-\bm1_{N-m}\end{array} \right), \nonumber\\[2pt]
\sigma_2&=&\left( \begin{array}{cc}
\sigma^{(m)}     &       0       \\
0       &       \bm1_{N-m} \end{array} \right) \label{eq:Q2}
\end{eqnarray}
(to which a global $U(N)$ rotation can be applied). A particular
instance of $\sigma^{(m)}$ is $\Lambda^{(m)}$, given by
\Eq{eq:Lm}. The symmetry of the vacuum is the set of rotations
that leaves both $\sigma_1$ and $\sigma_2$ invariant, namely,
$U(2m-N)\times U(N-m)\times U(N-m)$.

\subsubsection{Infinite lattice}

At $N_c=\infty$ we seek the saddle point of the action, which we
assume to be a time-independent configuration. The classical
Hamiltonian of the $\sigma$ model is
\begin{equation}
H=\frac {J_1}2 \left( \frac{N_c}2 \right)^2 \sum_{\bn,\mu} \Tr \sigma_\bn \sigma_{\bn+\muhat}.
\end{equation}
Seeking an antiferromagnetic ground state, we set
$\sigma_\bn=\Lambda$ on the sublattice of even sites. The odd
sites are then governed by
\begin{equation}
H^{\text{odd}}=J_1d\left( \frac{N_c}2 \right)^2 \sum_{\bn\ \text{odd}}\Tr\Lambda\sigma_\bn.
\label{Hodd}
\end{equation}
This is just the two-site problem studied above, replicated over
the lattice. As we saw above, the ground state energy is
degenerate with respect to the configuration at each odd site,
\begin{equation}
\sigma_\bn=\left( \begin{array}{cc}
\sigma_\bn^{(m)}     &       0       \\
0       &       \bm1_{N-m} \end{array} \right), \label{submanifold}
\end{equation}
and we find that at $N_c=\infty$ there is a large number of ground state configurations with the same classical energy, that differ in the choice  $\sigma^{(m)}_\bn$ on each of the odd sites. These ground states are continuously connected by the coordinates of $\sigma^{(m)}_\bn$ that furnish $U(m)/[U(2m-N)\times U(N-m)]$. They have different realizations of the global symmetry of $U(N)$;
A uniform choice for the odd sites,
$\sigma_\bn^{(m)}=\Lambda^{(m)}$ for instance, breaks the $U(N)$
symmetry to $U(2m-N)\times U(N-m)\times U(N-m)$; a non-uniform
choice can break the symmetry all the way to $U(N-m)$. Furthermore, one can contemplate making a non-uniform {\em
ansatz\/} for the {\em even\/} spins as well. 
The entropy
of this classical ground state is evidently proportional to the
volume.


The situation is reminiscent of
that in the antiferromagnetic Potts model \cite{AFPotts}. 
The crucial difference is, however, that here the degeneracy is continuous, and the sigma fields are allowed to oscillate around the classical ground state, even at zero temperature. In fact this makes the problem more similar to problems encountered in condensed matter physics in the context of frustrated magnetic systems. These systems are known to experience the phenomenon of ``order from disorder'', \cite{Aharony,Ord_disorder,double_Xchange,Kagome,Sachdev,Henley} where fluctuations remove a classical degeneracy and choose a ground state. We discuss how order from disorder happens in our problem in Chapter~\ref{chap:Qfluctuation}. There we show that when $N_c=\infty$, and one keeps $\sigma_\bn$ on the even sites fixed to be $\Lambda$, it costs no energy to change $\sigma_\bn$ on an odd site by changing its direction within $U(m)/[U(2m-N)\times U(N-m)]$. Fluctuations in these directions are ``zero modes'' and have zero energy for any momentum. In principle they can lead to instabilities of the classical ground state, but as we will show, order from disorder gives the zero modes non-zero energy of $O(1/N_c)$, removes the classical degeneracy, and stabilizes a canted ground state in which all the even sites point to $\Lambda$, and {\em all} the odd sites align uniformly to have the same $\sigma^{(m)}$.

A final note here is that different from the zero baryon case, the NNN interactions do not remove the degeneracy of the NN theory. In particular the zero modes are not removed when one takes a non-zero value for $J_2$. This can be seen in Section~\ref{sec:nnn_excite} that gives the dispersion relations of excitations in the sigma model with NN and NNN interaction. Taking $N_c=\infty$ and keeping $J_2\neq 0$ still results in zero modes.
\clearpage
\chapter[Quantum fluctuations in the sigma model]{Quantum fluctuations in the nonlinear sigma model}
\label{chap:Qfluctuation}

In this chapter we investigate the effects that quantum fluctuations of $O(1/N_c)$ have on the classical ground state of the NLSM introduced in Chapter~\ref{chap:NLSM}. We restrict ourselves to nearest-neighbor interactions only and show that the degeneracies present in the classical $N_c=\infty$ limit are removed at this order. We first show that for zero baryon density fluctuations remove the discrete degeneracy in the parameter $m$ [See Section~\ref{sec:zeroB}] and choose the configuration that has zero baryon number on each site. We proceed to perform a similar calculation for non-zero density and see that in that case fluctuations remove the infinite continuous degeneracy present in the classical ground state. This degeneracy is represented by the fact that in the classical level, the sigma fields on the odd sites can move freely independent of each other in certain flat directions. Fluctuations in these directions have zero energy at $O(1)$ for all momentum and are zero modes. We find that quantum fluctuations give these modes
non-zero energy of $O(1/N_c)$, and stabilized the classical ground state. Furthermore we find that these energies are quadratic in momentum, which makes them type II Goldstone bosons.

\section{Removal of classical degeneracy at zero density}
\label{sec:Rem_B0}

At zero density we consider fluctuations
around the $\sigma_\bn=\Lambda$ minimum of $S$. 
\begin{equation}
S=\frac{N_c}2\int_0^{\beta} d\tau\left[-\sum_{\bn}(-1)^{\bn}
\Tr\Lambda U^{\dag}_{\bn} \partial_\tau U_{\bn} -\frac
{J'_1}2 \sum_{\bn\mu} \Tr \sigma_\bn \sigma_{\bn+\muhat}\right].
\label{eq:Srescaled}
\end{equation}
with $J'_1=2J_1/N_c \sim 1/g^2N_c$. (From here on we drop the prime)
Recalling \Eq{Ua}, we write
\begin{equation}
U_\bn=e^{A_\bn}, \label{eq:expand1}
\end{equation}
where $A_\bn$ is anti-Hermitian and anticommutes with $\Lambda$.
It is more convenient to work with the Hermitian matrix
\begin{equation}
L_\bn=2A_\bn\Lambda,
\end{equation}
in terms of which we expand
\begin{equation}
\sigma_\bn=\Lambda+L_\bn-\frac12L^2\Lambda. \label{eq:sigmaL}
\end{equation}
If we further expand $L_\bn$ in the basis of generators of $U(N)$,
\begin{equation}
L_\bn=\sum_\eta l^\eta M^\eta, \label{eq:Ll_exp}
\end{equation}
we find that the $l^\eta$ corresponding to generators of
$U(m)\times U(N-m)$ vanish; this is the subgroup under which the
vacuum is symmetric. The field $L_\bn$ thus contains $2m(N-m)$
real degrees of freedom, corresponding to the Goldstone bosons.

We expand $U_\bn$ and $\sigma_\bn$ in powers of $L_\bn$; using
\Eq{eq:Ll_exp}, we obtain to second order
\begin{equation}
S=S_0+\frac{N_c}2\int d\tau \sum_\bn\left[
\frac{(-1)^\bn}8C^{\eta\eta'}l_\bn^\eta \partial_\tau
l_\bn^{\eta'} +\frac {J_1}8\sum_\mu\left(l_{\bn+\muhat}-l_\bn
\right)^2\right].
\end{equation}
The coefficient matrix is
\begin{equation}
C^{\eta\eta'}=\Tr(\Lambda[M^\eta,M^{\eta'}]), \label{eq:Cetaeta}
\end{equation}
and the classical energy is $S_0=-\frac
{J_1}2\left(\frac{N_c}2\right)^2NN_sd\beta$. $C$ is antisymmetric and
purely imaginary; we show in Appendix \ref{app:CD} that $C$ has
eigenvalues $\pm1$, each with degeneracy $m(N-m)$. We change basis
so as to diagonalize $C$, and write the index $\eta$ as the
compound $(\alpha,\pm)$ with $\alpha=1,\ldots,m(N-m)$ and the
$\pm$ corresponding to the eigenvalue of $C$. Since the original
$l^\eta$ are real, we have
\begin{equation}
(l^{\alpha+})^*=l^{\alpha-}.
\end{equation}
Thus we eliminate $l^{\alpha-}$ and write
\begin{equation}
S=S_0+\frac{N_c}8 \int d\tau \sum_\bn\left[ (-1)^\bn i\, \Im
l_\bn^{\alpha+*}\partial_\tau l_\bn^{\alpha+} +\frac
{J_1}2\sum_\mu\left|l_{\bn+\muhat}^{\alpha+}-l_\bn^{\alpha+}\right|^2\right].
\label{eq:Sln}
\end{equation}

The alternating sign in \Eq{eq:Sln} is what makes the theory
antiferromagnetic. It forces us to differentiate between even and
odd sites, and we transform to momentum space as follows (dropping
the $+$ superscript):
\begin{equation}
l^\alpha_\bn=\sqrt{\frac2{N_s}}\sum_\bk\left\{
\begin{array}{ll}
l^\alpha_{1,\bk}e^{i\bk\cdot\bn}&\quad \bn \quad \text{even,}\\[2pt]
l^\alpha_{2,\bk}e^{i\bk\cdot(\bn-\hat z)}&\quad \bn \quad \text{odd.}
\end{array}\right.
\end{equation}
The even sites comprise an {\em fcc\/} lattice with lattice
constant 2, and the momenta $\bk$ take values in its Brillouin
zone. We obtain
\begin{equation}
S=S_0+\frac{N_c}8\sum_\bk\int d\tau\, (l_1^*\quad
l_2^*)_{\bk}^\alpha \,\mathcal{M}(\bk) \left(\begin{array}{c}
l_1\\l_2
\end{array}\right)_{\bk}^\alpha.
\end{equation}
Here
\begin{equation}
\mathcal{M}(\bk)=\left({\renewcommand{\arraystretch}{1.5}
\begin{array}{cc}
-\partial_\tau + J_1d &-J_1d\gamma(\bk)\\[2pt]
-J_1d\gamma(\bk)&+\partial_\tau + J_1d
\end{array}}\right)
\end{equation}
and  $\gamma(\bk)=\frac1d\sum_\mu \cos k_\mu$.

The Gaussian path integral over the $l$ field is very similar to the path integrations that appear in Chapter~\ref{chap:MF_schwinger} and gives the free
energy,
\begin{equation}
F = F_0+ m(N-m)\frac{N_c}2\sum_\bk \left[ \frac1{\beta}\log
\left( 2\sinh \frac{\beta\omega(\bk)}2\right) - \frac{J_1d}2
\right],
\end{equation}
where $\omega(\bk) = dJ_1N_c\sqrt{1-\gamma^2(\bk)}$ and $F_0 =
-\frac12 J_1dNN_s(N_c/2)^2$. For the ground state energy, we take
$\beta\to\infty$ and obtain (restoring all constants)
\begin{equation}
E_0 = -J_1N_sNd\left(\frac{N_c}{2} \right)^2 \left[ 1 +
\frac{1}{N_c} \frac{m(N-m) }{N} \int_{\rm BZ} \left(
\frac{dk}{2\pi} \right)^d \left( 1-\sqrt{1-\gamma^2(\bk) } \right)
\right]. \label{eq:E0}
\end{equation}
This is exactly the result of Smit \cite{Smit}. The $O(1/N_c)$
corrections lift the degeneracy of the ground states with
different values of $m$. The integral in \Eq{eq:E0} is positive
and its coefficient contains the number of Goldstone bosons. Thus
the state of lowest energy is that with $m=N/2=2N_f$, and the symmetry
breaking scheme is $U(4N_f)\to U(2N_f)\times U(2N_f)$. Further breaking
by the next-nearest-neighbor terms was discussed in Chapter~\ref{chap:NLSM} where we extended this result to the next-nearest-neighbor theory and found that its
$U(N_f)\times U(N_f)$ chiral symmetry is broken to the vector
$U(N_f)$ flavor subgroup.

\section[Removal of classical degeneracy at non-zero density]{Removal of classical degeneracy at non-zero density}
\label{sec:Rem_Bn0}

Recall the classical solution of the nonzero-density two-site problem, given in Chapter~\ref{chap:NLSM}. We found that the sigma fields are
\begin{eqnarray}
\sigma_1&=&\Lambda_1=\left( \begin{array}{cc}
\bm1_m&0\\
0&-\bm1_{N-m}\end{array} \right), \nonumber\\[2pt]
\sigma_2&=&\left( \begin{array}{cc}
\sigma^{(m)}     &       0       \\
0       &       \bm1_{N-m} \end{array} \right) \label{eq:sigma_2_sig_m},
\end{eqnarray}
and $\sigma^{m}$ is given in \Eq{eq:sigma_m}. The symmetry of the vacuum is the set of rotations
that leaves both $\sigma_1$ and $\sigma_2$ invariant, namely,
$U(2m-N)\times U(N-m)\times U(N-m)$.
 There are two
kinds of degeneracy in this two-site problem: that which results
from breaking the global $U(N)$ to $U(m)\times U(N-m)$, and that
which comes of breaking the $U(m)$ subgroup to $U(2m-N)\times
U(N-m).$ The latter degeneracy is connected with freedom in
choosing the orientation of $\sigma_2$ {\em relative to\/}
$\sigma_1$. It is instructive to see how quantum mechanical
fluctuations lift the classical degeneracies.

\begin{figure}
\begin{center}
\epsfig{width=15cm,file=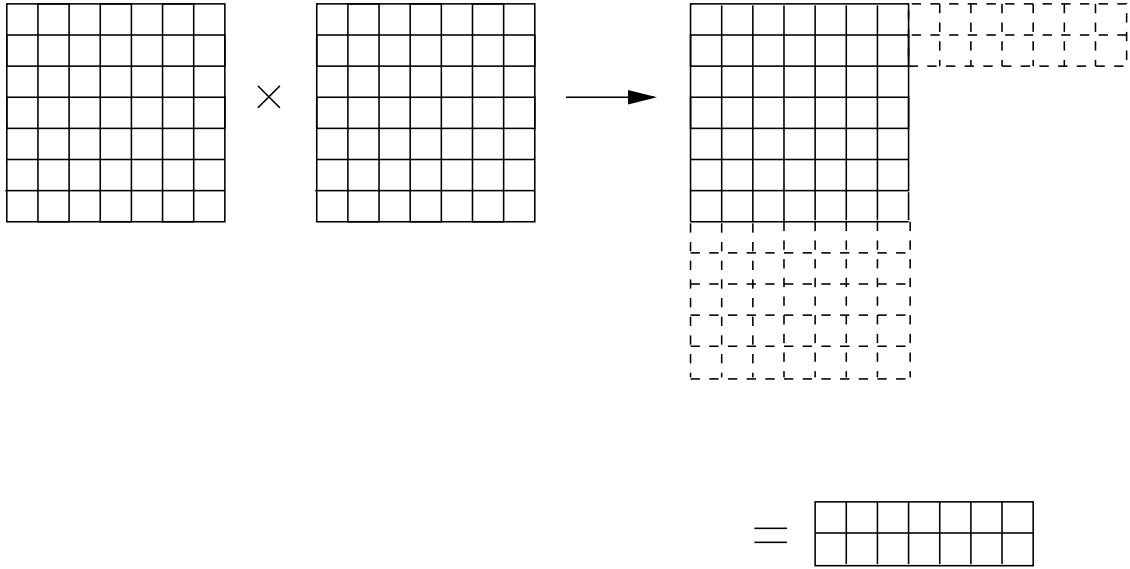} \caption[Coupling two 7-row reps of $U(12)$.]{Coupling two spins
in a 7-row representation of $U(12)$ to the representation with
minimal Casimir operator} \label{fig:twosite}
\end{center}
\end{figure}

The quantum two-site problem is easy to solve. We rewrite the
Hamiltonian (\ref{2site_qH}) as
\begin{equation}
H=\frac {J_1}2\left[(Q_1+Q_2)^2-Q_1^2-Q_2^2\right]. \label{eq:2site_qH2}
\end{equation}
$Q_1^2$ and $Q_2^2$ are constants, the quadratic Casimir operator
in the $m$-row, $N_c$-column representation of $U(N)$. The first
term in \Eq{eq:2site_qH2} is minimized by coupling $Q_1$ and $Q_2$ to
the representation that minimizes the Casimir, which is the
representation with $2m-N$ rows and $N_c$ columns (see
Fig.~\ref{fig:twosite}).
The ground state has discrete degeneracy equal to the dimension of this representation.

The exact quantum solution naturally shows no sign of spontaneous
symmetry breaking and hence it is not of much relevance to the
infinite volume problem. More interesting is the problem where the
state of $Q_1$ is {\em fixed\/} and $Q_2$ is allowed to vary. This
breaks by hand the global $U(N)$ while allowing quantum
fluctuations to lift any remaining degeneracy in the relative
orientation of the two spins, so it can be regarded as
quantization in the presence of spontaneous symmetry breaking. In
effect, this is mean field theory.

\begin{figure}[htb]
\begin{center}
\epsfig{file=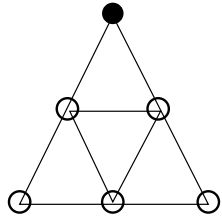} \caption[Weight diagram for $N=3$, $m=1$, $N_c=2$]{Weight diagram for $N=3$,
$m=1$, $N_c=2$ [the sextet of $SU(3)$]. The highest-weight state
lies at the top of the triangle; all states at the base minimize
$H^{MF}$. All values of $N_c$ give the same triangular shape, with
multiplicity of one along the boundary. There are $N_c+1$ states
at the base. For $N=4$ the triangle becomes a tetrahedron, and
there are $(N_c+1)(N_c+2)/2$ states at its base.}
\label{fig:twositeweights}
\end{center}
\end{figure}

We replace the Hamiltonian (\ref{2site_qH}) by
\begin{equation}
H^{MF}=J_1\sum_{\eta=1}^{N^2}\langle Q_1^\eta\rangle Q_2^\eta.
\end{equation}
To minimize the energy we maximize the mean field by choosing the
state of $Q_1$ to be the highest-weight state. This state
diagonalizes the generators $H_\alpha$ of the Cartan subalgebra while
other generators of $U(N)$ have expectation value zero. Thus
\begin{equation}
H^{MF}=J_1\sum_{\alpha=1}^N\langle H_{1\alpha}\rangle H_{2\alpha}.
\end{equation}
The operators $H_\alpha$ all commute, and their eigenvalues make up the
weight diagram of the representation.\footnote{More precisely, the
weight diagram shows eigenvalues of the $N-1$ traceless diagonal
generators of $SU(N)$.  These can be obtained by isolating the
$U(1)$ member of the set $H_\alpha$ and taking linear combinations of
the rest.} $H^{MF}$ is a dot product of the weight vectors of the
two spins. The energy is minimized by choosing for $Q_2$ a state
that lies opposite the highest-weight state in the weight diagram.
As shown in the example of Fig.~\ref{fig:twositeweights}, this
still leaves a degeneracy, albeit a discrete one. We stress that
this degeneracy comes from freedom in the relative orientation of
$Q_1$ and~$Q_2$.

In the $N_c\to\infty$ limit, the discrete degeneracy becomes
infinite and presumably it is well described by the continuous
degeneracy of the classical problem. 
As mentioned this is similar to problems of frustrated magnetic systems, where a local group of spins will have a degenerate ground state, with a continuous degeneracy. 

This degeneracy, however, is an artifact of
the mean-field approach that, like the classical {\em ansatz\/},
assumes a fixed state for the spins on the even sites. In fact the sigma fields will
fluctuate as soon as they are allowed to do so, i.e., when $N_c<\infty$. A given odd spin
will {\em not\/} be surrounded by a uniform fixed background; the
neighboring even spins will be influenced by {\em all\/} their odd
neighbors, and will induce an interaction among the odd spins. This should reduce the entropy of the ground state to zero. 


The systematic way to see this effect is to
carry out an expansion in $1/N_c$ and verify the stability of the classical {\em ansatz\/}. We do that for the canted ground state, which is the solution to the two-site problem,
replicated {\em uniformly\/} over the lattice:
\begin{equation}
\begin{array}{rcll}
\sigma_\bn&=&\Lambda,\hfill&\ \bn \quad \text{even}, \label{eq:gs} \\[2pt]
\sigma_\bn&=&\left(\begin{array}{cc}
\sigma^{(m)}&0\\
0&\bm1_{N-m}
\end{array}\right),&\ \bn \quad \text{odd}, 
\end{array}
\end{equation}
where $\sigma^{(m)}=U^{(m)}\Lambda^{(m)}U^{(m)\dag}$ is now a {\em
global\/} degree of freedom. 

The meaning of the classical local degeneracy is that at $O(1)$ there is a set of directions in which one can move from one classical vacuum to another without cost of energy. Fluctuations in these directions are zero modes, that have zero energy for all momentum, and we discuss them in Section~\ref{sec:O1}. 
The sign for stability must therefore be the removal of these zero modes at $O(1/N_c)$, which we demonstrate in Section~\ref{sec:Self_E} along the lines of the phenomenon of ``order from disorder'' \cite{Aharony,Ord_disorder,double_Xchange,Kagome,Sachdev,Henley}
in frustrated magnetic systems.

The essence of order from disorder is that fluctuations of
quantum, thermal or even quenched nature can lift a classical
degeneracy. A famous example is the Kagom{\' e}
antiferromagnet \cite{Kagome} whose classical ground state energy is invariant
under correlated rotations of local groups of spins, leading to a
degeneracy exponential in the volume. This system also has modes with zero
energy for all momentum. These zero modes obtain non-zero energy due to quantum
fluctuations.
Most discussions of order from disorder work in a Hamiltonian framework (but see \cite{Sachdev,Henley}).
We employ instead the Euclidean path integral. In Appendix~\ref{app:Hamiltonian_odo} we also present a 
Hamiltonian calculation using
a generalized Holstein-Primakoff transformation. The latter can only be constructed easily for a subset of the
cases we consider \cite{Smit,Salam}. The Euclidean calculations are less restrictive in their applicability.

Finally, as an exercise, we present a derivation of the effective
ferromagnetic interaction among the $\sigma_\bn$ on the odd sites in Appendix~\ref{app:corrections}, which in fact motivated us to investigate the canted ground state in particular.

\subsection{Zero modes at {\bm $O(1)$} \label{sec:O1}}


We begin to investigate the stability of the canted ansatz, where the vacuum expectation values (vevs) of the $\sigma$
fields on the odd sites point in the same direction (a ferromagnetic alignment).
Without loss of generality the ground state can be chosen to have
\begin{eqnarray}
\< \sigma_{\text{even}} \>&\propto&\Lambda,     \nonumber       \\
\< \sigma_{\text{odd}} \>&\propto&\Lambda_2 \equiv \left( \begin{array}{ccc}
        1_{2m-N} & 0 & 0 \\
        0 & -1_{N-m} & 0 \\
        0 & 0 & 1_{N-m}  \end{array} \right)=V\Lambda V^{\dag}, \label{eq:gs_vev}
\end{eqnarray}
with 
\begin{equation}
\label{eq:V}
V=\left( \begin{array}{ccc}
        1 & 0 & 0       \\
        0 & 0 & 1       \\
        0 & -1 & 0 \end{array}  \right).
\end{equation}
These vevs break $U(N)$ to $U(2m-N)\times U(N-m) \times U(N-m)$, with \linebreak $n_G=2(3m-N)(N-m)$ broken generators. Here we write an effective action for the fluctuations around the up to $O(1/N_c)$. Working at $O(1)$
first, we identify the zero modes that correspond to the classical
degeneracy. We then show how quantum fluctuations give them non-zero energy.

In this and next sections we choose to work with the parametrization of the coset space $U(N)/[U(m)\times U(N-m)]$ \cite{Salam} of Eqs.~(\ref{eq:Uphi})--(\ref{eq:sig_phi}), in terms of the $m \times (N-m)$ complex matrix $\phi$.  As mentioned, the main
  advantage of this $\phi$ parameterization is that the
kinetic part of the action is quadratic in $\phi$
[see~(\ref{eq:Skin_phi})].

The fields $\phi$, suitably shifted, can be identified with Goldstone bosons around the ground state~(\ref{eq:gs}). For
the even sites, $\phi=0$ indeed gives $\sigma=\Lambda$. For the odd
sites, the vacuum is at 
$$
\phi=\left( \begin{array}{c}
            0 \\ 1_{N-m} \end{array} \right),
$$
where the upper part of $\phi$ is a $(2m-N)\times (N-m)$ matrix.
We therefore shift $\phi$ on the odd sites according to
\begin{eqnarray}
\label{eq:diff_odd}
U'&\equiv &VU, \nonumber \\
\sigma'(\phi)&\equiv &V\sigma(\phi) V^{\dag}.
\end{eqnarray}
This gives $\sigma'(0)=\Lambda_2$. We drop the primes henceforth.
For later convenience we write $\phi$ as
\begin{equation}
\label{eq:phi}
\phi=\left( \begin{array}{c}    \chi \\ \pi   \end{array} \right).
\end{equation}
Here $\pi$ is an $(N-m)-$~dimensional square matrix while $\chi$ has $(2m-N)$ rows
and $(N-m)$ columns. Both are complex. Thus
\begin{equation}
\label{eq:even}
\sigma_{\text{even}}=\left( \begin{array}{ccc}
        1-2\chi \chi^{\dag} & -2\chi \pi^{\dag} & -2\chi \sqrt{1-\phi^{\dag}\phi}        \vspace{0.3cm} \\
        -2\pi \chi^{\dag} & 1-2\pi \pi^{\dag} & -2\pi \sqrt{1-\phi^{\dag}\phi} \vspace{0.3cm} \\
        -2\sqrt{1-\phi^{\dag}\phi} \chi^{\dag} & -2\sqrt{1-\phi^{\dag}\phi} \pi^{\dag} & -1+2\phi^{\dag} \phi \\      \end{array}     \right),
\end{equation}
and
\begin{equation}
\label{eq:odd}
\sigma_{\text{odd}}=\left( \begin{array}{ccc}
       1-2\chi \chi^{\dag} & -2\chi \sqrt{1-\phi^{\dag}\phi} & 2\chi \pi^{\dag} \vspace{0.3cm} \\
        -2\sqrt{1-\phi^{\dag}\phi} \chi^{\dag} & -1+2\phi^{\dag} \phi & 2\sqrt{1-\phi^{\dag}\phi} \pi^{\dag}  \vspace{0.3cm} \\
        2\pi \chi^{\dag} & 2\pi \sqrt{1-\phi^{\dag}\phi} & 1-2\pi \pi^{\dag} \\ \end{array} \right).
\end{equation}
Each submatrix in Eqs.~(\ref{eq:even})--(\ref{eq:odd}) has dimensions as
indicated in~(\ref{eq:gs}).

After scaling $\phi \rightarrow \phi/\sqrt{N_c}$ the
action takes the form
\begin{eqnarray}
S&=&\Tr \int d\tau\sum_\bn \left[ \chi^{\dag}_\bn \partial_{\tau} \chi_\bn +\pi^{\dag}_\bn \partial_{\tau} \pi_\bn \right] \nonumber \\
&&\qquad +J_1\sum_{\bn \muhat} \left[ \pi_\bn \pi^{\dag}_\bn +\pi_{\bn+\muhat} \pi^{\dag}_{\bn+\muhat} -\pi_\bn \pi_{\bn+\muhat} -\pi^{\dag}_\bn \pi^{\dag}_{\bn+\muhat} \right] \nonumber \\
&&\qquad +\frac{J_1}{\sqrt{N_c}} \sum_{\bn \muhat} \left[ \chi^{\dag}_\bn \chi_{\bn+\muhat} \pi_\bn  - \chi^{\dag}_{\bn+\muhat} \chi_\bn \pi_{\bn+\muhat}  +h.c. \right] \label{eq:SeffX} \\
&&\qquad +\frac{J_1}{N_c} \sum_{\bn \muhat} \left[\chi_\bn \chi^{\dag}_\bn \chi_{\bn+\muhat} \chi^{\dag}_{\bn+\muhat} -\pi_\bn \pi_{\bn+\muhat}^{\dag} \pi_{\bn+\muhat}^{\dag} \pi_{\bn+\muhat} -\pi_{\bn+\muhat} \pi_{\bn+\muhat}^{\dag} \pi_\bn^{\dag} \pi_\bn \right.  \nonumber \\ 
&& \qquad \qquad \qquad -\pi_\bn \pi_\bn^{\dag} \chi^{\dag}_{\bn+\muhat} \chi_{\bn+\muhat} -\pi_{\bn+\muhat} \pi_{\bn+\muhat}^{\dag} \chi^{\dag}_\bn \chi_\bn \nonumber \\
&& \qquad \qquad \qquad +\frac12 \left( \pi_{\bn+\muhat} \pi_{\bn+\muhat}^{\dag} \pi_{\bn+\muhat} \pi_\bn +\pi_{\bn+\muhat} \pi_\bn \pi_\bn^{\dag} \pi_\bn \nonumber \right. \\
&& \qquad \qquad \qquad \qquad \left. \left. +\pi_{\bn+\muhat} \chi_{\bn+\muhat}^{\dag}
    \chi_{\bn+\muhat} \pi_\bn + \pi_{\bn+\muhat} \pi_\bn
    \chi^{\dag}_\bn \chi_\bn + h.c. \right) \right]  \nonumber \\ 
&& \qquad +O(1/N_c^{3/2}). \nonumber 
\end{eqnarray}
The AF structure of the action demands the introduction of an fcc
lattice. We write the fields $\phi_\bn \equiv \phi^A_{\bN}$ with
$A=(\text{even},\text{odd})$ and $\bN$ belonging to an fcc
lattice. We Fourier transform according to 
\begin{equation}
\label{eq:FT_phi}
\phi^A_{\bN} (\tau)=\sqrt{\frac{2}{N_s}} \sqrt{\frac1{\beta}}
\sum_{\bk,\omega} \phi^A_k e^{i(\bk \cdot \bN-\omega \tau)} \times \left\{ \begin{array}{ll} 1 & A=\text{even} \\ e^{-ik_z/2} & A=\text{odd} \end{array} \right. .
\end{equation}
Here $k\equiv(\bk,\omega)$. In momentum space the action is given by
$S=S_2+S_3+S_4$ with (discarding {\it Umklapp} terms)
\begin{eqnarray}
S_2&=&\Tr \sum_k \left( \begin{array}{cc} \pi_k^{\e \dag} & \pi_{-k}^\o \end{array} \right)
\left( \begin{array}{cc} -i\omega + 2dJ_1 & -2J_1d \gamma_\bk \\
        -2J_1d \gamma_\bk & i\omega + 2J_1d \end{array} \right)
\left( \begin{array}{c} \pi_k^\e \\ \pi_{-k}^{\o \dag} \end{array} \right) \nonumber \\
&&\qquad \quad +\left( \begin{array}{cc} \chi_k^{\e \dag} & \left(\chi_{-k}^\o \right)^T \end{array} \right)
\left( \begin{array}{cc} -i\omega & 0  \label{eq:S2} \\
        0 & i\omega \end{array} \right)
\left( \begin{array}{c} \chi_k^\e \\ \left( \chi_{-k}^{\o *} \right) \end{array} \right), \\
S_3&=&2J_1d \sqrt{\frac2{N_s N_c \beta}} \Tr \sum_{kp} \left(
  \chi_p^{\o \dag} \chi_k^\e \pi_{p-k}^\o \gamma_\bk - \chi_k^{\e \dag}
  \chi_{k-p}^\o \pi_p^\e \gamma_{\bk-\bp} + h.c. \right), \label{eq:S3} \\
S_4&=&2J_1d \frac2{N_s N_c \beta}  \Tr \sum_{kpq}  \left(
    \chi_k^\e \chi_p^{\e \dag} \chi_q^\o \chi_{q+k-p}^{\o \dag}
    -\pi_k^\e \pi_p^{\e \dag} \pi_q^{\o \dag} \pi^\o_{q-k+p}
    -\pi_k^{\e \dag} \pi_p^\e \pi_q^\o \pi^{\o \dag}_{q-k+p} \right. \nonumber \\
&&\qquad \qquad \qquad \qquad \left. -\chi_k^{\e \dag} \chi_p^\e \pi_q^\o \pi_{q+k-p}^{\o \dag} -
  \pi_k^\e \pi_p^{\e \dag} \chi_q^{\o \dag} \chi_{q-k+p}^\o \right) \gamma_{\bk-\bp} \\
&&\qquad \qquad \qquad \qquad +\frac12 \left( \left( \pi_k^{\e \dag} \pi_p^\e + \chi_k^{\e \dag}
    \chi_p^\e \right) \left( \pi^\o_{k-p-q} \pi_q^\e + \pi_{-q}^{\e
      \dag} \pi_{-(k-p-q)}^{\o \dag} \right) \right. \nonumber \\
&&\qquad \qquad \qquad \qquad \qquad  \left. \left( \pi_k^{\o \dag} \pi_p^\o + \chi_k^{\o \dag} \chi_p^\o
    \right) \left( \pi^\e_{k-p-q} \pi_q^\o + \pi_{-q}^{\o \dag}
      \pi_{-(k-p-q)}^{\e \dag} \right) \right) \gamma_{\bk-\bp-\bq} \nonumber
 \label{eq:S4} .
\end{eqnarray}
$\gamma_\bk$ is given by
\begin{equation}
\label{eq:gamma}
\gamma_{\bk}=\frac1{d} \sum_{\mu=1}^d \cos{k_{\mu}/2}.
\end{equation}

At large $N_c$, $S_3$ and $S_4$ are small perturbations. The bare propagators can be read from $S_2$,
\begin{eqnarray}
\left( G^{\pi}_k \right)_{A B} &=& \frac1{\omega^2+4J^2_1 d^2 E^2_\bk} \left( \begin{array}{cc}
                i\omega + 2J_1d & 2J_1d \gamma_\bk  \\
                2J_1d \gamma_\bk    & -i\omega + 2J_1d \end{array}
            \right)_{A B} ,  \label{eq:Gopi} \\
\left( G^{\chi}_k \right)_{A B} &=& \left( \begin{array}{cc}
                -1/i\omega & 0 \\
                0       & 1/i\omega \end{array} \right)_{A B}. \label{eq:Gochi}  
\end{eqnarray}
(The propagators are diagonal in
the internal group indices.) Here $E_\bk=\sqrt{1-\gamma_\bk^2}$. The poles of the propagators give the
dispersion relations of the various bosons at $O(1)$. The result is
\begin{equation}
\omega^2+(2J_1d E_\bk)^2=0 \qquad \qquad \textrm{for} \,\, \pi,     \label{eq:Eopi}
\end{equation}
and
\begin{equation}
\omega=0 \qquad \qquad \qquad \qquad \qquad \textrm{for} \,\, \chi. \label{eq:Eochi}
\end{equation}

There are $4(N-m)^2$ real fields per momentum of the $\pi$ kind on the even and odd sites, and the propagator~(\ref{eq:Gopi}) has $4(N-m)^2$ poles that obey~(\ref{eq:Eopi}). We pair these into $2(N-m)^2$ physical massless excitations with 
\begin{equation}
i\omega = \pm 2J_1dE_\bk, \label{eq:pair_poles}
\end{equation}
which are $2(N-m)^2$ spin waves. It is easy to verify that the Euclidean dispersion relation is $\omega^2 \sim -|\bk|^2$ at low momentum. We note that at $B=0$ they are the only excitations.

Let us pause to see how the above discussion would change if the interaction between the $\pi$ fields were ferromagnetic. First the propagator~(\ref{eq:Gopi}) would be replaced by 
\begin{equation}
G^{-1}_{\text{FM}}=i\omega - 2J_1d(1-\gamma_\bk),
\label{eq:GoFM}
\end{equation}
where $\bk$ is now a momentum in the first BZ of a simple cubic lattice (the fcc structure is redundant for a ferromagnet). Here one has $2(N-m)^2$ fields per momentum, all of them are massless and obey $i\omega\sim J_1 |\bk|^2$. Pairing negative and positive frequencies together, one has $(N-m)^2$ massless excitations only. To see the connection to the antiferromagnetic case write the ferromagnetic action and propagator in terms of fcc degrees of freedom as
\begin{eqnarray}
S_{\text{FM}}&\sim& \sum_\bk \left( \begin{array}{cc} \pi^{\dag e}_k & \pi^{\dag o}_k \end{array} \right) G^{-1}_{\text{FM}} \left( \begin{array}{c} \pi^e_k \\ \pi^o_k \end{array} \right), \\
G^{-1}_{\text{FM}}&=&\left( \begin{array}{cc} i\omega - 2J_1d & 2J_1d\gamma_\bk \\
2J_1d\gamma_\bk & i\omega - 2J_1d \end{array} \right).
\label{eq:GoFMfcc}
\end{eqnarray}
Here the zeros obey $(i\omega-2J_1d)^2=(2J_1d\gamma_\bk)^2$, or
\begin{equation}
i\omega=2J_1d(1\pm \gamma_\bk).
\end{equation}
So one has again $4(N-m)^2$ fields per momentum, but half of them have a mass equal to $4J_1d$. Next we pair positive and negative frequencies together and find that half of the remaining $2(N-m)^2$ massless real fields describe $(N-m)^2$ massless particles. The difference between the two cases is that while the number of broken real fields is the same $n_G=2(N-m)^2$, the number of physical excitations is different. It is $n_G$ for the antiferromagnet, and $n_G/2$ for the ferromagnet. 

The $\chi$ fields are $4(N-m)(2m-N)$ zero modes (``soft'' modes). Their energy is zero for
all momentum, a sign of the local degeneracy of the classical ground
state discussed above. 

We conclude this subsection by classifying the different fields according to their $U(2m-N)\times U(N-m) \times U(N-m)$ representations. We denote a representation by
\begin{equation}
(r_1,r_2,r_3)^{(q_1,q_2,q_3)}.
\end{equation}
Here $r_1$ denotes the representation of $SU(2m-N)$, and $r_{2,3}$ denote the representations of the two $SU(N-m)$ subgroups. $q_i$ are the charges of the excitations under the remaining $U(1)$ factors of the unbroken subgroup. These are generated by the following diagonal matrices.
\begin{equation}
1_N, \qquad \left( \begin{array}{ccc} 1_{2m-N} & 0 & 0 \\
                                                0 & 0 & 0 \\ 
                                                0 & 0 & 0 \end{array} \right), \quad \text{and} \quad \left( \begin{array}{ccc} 0 & 0 & 0 \\
               0 & 1_{N-m} & 0 \\
               0 & 0 & -1_{N-m} \end{array} \right).
\end{equation}
We give the different representations in Table~\ref{reps}. The crucial point is that the AF spin waves and the zero modes reside in completely different representations. This means that the separation between these excitations in the $O(1)$ spectrum will survive at higher orders and that mixing can not occur (unless there is further spontaneous symmetry breakdown in higher orders).

\begin{table}[ht]
\begin{center}
\begin{tabular}{|c|c|c|} \hline
Field  & Representation & Dimension \\ \hline \hline
$\pi$ & $\left (1,N-m,\overline{N-m} \right)^{(0,0,+2)}$ & $(N-m)^2$\\ \hline
$\chi_{\text{even}}$ & $\left( 2m-N,1,\overline{N-m} \right)^{(0,+1,+1)}$ & $(2m-N)(N-m)$\\ \hline
$\chi_{\text{odd}}$ & $\left( 2m-N,\overline{N-m},1 \right)^{(0,+1,-1)}$ & $(2m-N)(N-m)$ \\ \hline
\end{tabular}
\caption[Field classification under the unbroken group of the nearest-neighbor theory.]{$U(2m-N)\times U(N-m) \times U(N-m)$ representations of the excitations. The conjugate fields belong to the conjugate representations. $n(\overline{n})$ stands for the fundamental (its conjugate) representation of $SU(n)$ and $1$ stands for a singlet. } \label{reps}
\end{center}
\end{table}

To be precise we give the transformation laws of each of the fields with respect to a $U(2m-N)\times U(N-m)\times U(N-m)$ element given by
\begin{equation}
U=\left( \begin{array}{ccc}
u_1 & 0 & 0 \\ 0 & u_2 & 0 \\ 0 & 0 & u_3 \end{array}\right).
\end{equation}
Here $u_{2,3}\in U(N-m)$, and $u_1\in U(2m-N)$. The fields $\pi^e$ and $\pi^{\dag o}$ transform as
\begin{equation}
\pi^e \rightarrow u_2 \pi^e u^\dag_3, \qquad \qquad \pi^{\dag o} \rightarrow u_2 \pi^{\dag o} u^\dag_3, \label{eq:pi_trns}
\end{equation}
while the $\chi$ fields transform as
\begin{equation}
\chi^e \rightarrow u_1 \chi^e u^\dag_3, \qquad \qquad \chi^o \rightarrow u_1 \chi^o u^\dag_2. \label{eq:chi_trns}
\end{equation}
From \Eq{eq:SeffX} we see that the action is invariant under these transformations.

\subsection{Removal of zero modes at $O(1/N_c)$ -- the self energy calculation\label{sec:Self_E}}
\begin{figure}[ht]
\begin{center}
\includegraphics[width=15cm]{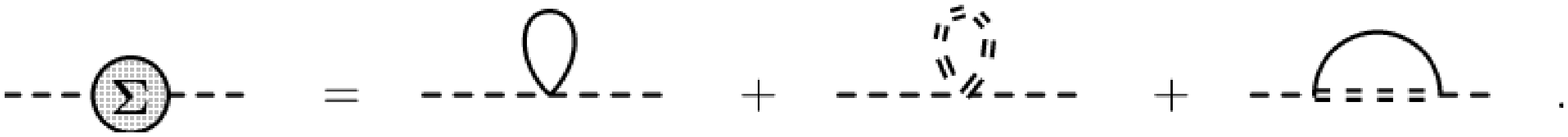}
\caption[Schwinger-Dyson equation for the $\chi$ fields self energy.]{Schwinger-Dyson equation for the self energy of the $\chi$ fields in one loop. Solid lines correspond to the $\pi$ propagators, and double dashed lines represent the {\it full} $\chi$ propagator. Single dashed lines are amputated external legs.} \label{fig:dyson}
\end{center}
\end{figure}

In this section we calculate the self energy of the $\chi$ fields to
first order in $1/N_c$. We shall see that the poles in the soft modes propagators
move away from zero energy in this order. The Feynman rules
for this problem are presented in Appendix~\ref{app:Feynman}. We treat $S_3$ and $S_4$ as
perturbations to $S_2$ and present the contributions to the self energy in one loop in Fig.~\ref{fig:dyson}.
  
Since the bare $\chi$ propagator is divergent at $\omega=0$ for all $\bk$, we insert its self energy self-consistently, making the replacement
\begin{equation}
\left( G^{\chi}_k \right)^{-1} = \left( G^{\chi}_{0\,k}
\right)^{-1} - \Sigma_k. \nonumber
\end{equation}
$G_0$ is the bare propagator given in \Eq{eq:Gochi} and $\Sigma$ is a matrix coupling odd and even degrees of freedom
according\footnote{We assume that the $\chi$ propagator has a similar
  structure as \Eq{eq:Gopi}.} to
\begin{equation}
\Sigma=\left( \begin{array}{cc}
    \Sigma_1 & \Sigma_2 \\
    \Sigma_2^* & \Sigma_1^* \end{array} \right). \label{eq:Sigma_ansatz}
\end{equation}
Thus the full $\chi$ propagator is 
\begin{equation}
\label{eq:Gchi}
G^{\chi}_k = \frac1{(-i\omega - \Sigma_{1,k})(i\omega - \Sigma_{1,k}^*)-|\Sigma_{2,k}|^2} \left( \begin{array}{cc}
        i\omega - \Sigma_{1,k}^* & \Sigma_{2,k} \\
        \Sigma_{2,k}^*   &       -i\omega - \Sigma_{1,k}
      \end{array} \right) .
\end{equation}
Here we assume that both matrices $\Sigma_{1,k}$ and $\Sigma_{2,k}$
are diagonal in group space and that
\begin{equation}
\label{eq:stability}
\Re  \Sigma_{1,k}  < 0
\end{equation}
for the stability of the path integral. 

At this point we note that the ansatz~(\ref{eq:Sigma_ansatz}) is invariant under~(\ref{eq:chi_trns}) only for $\Sigma_2=0$. A non-zero $\Sigma_2$ leads to the following non-zero vev 
\begin{equation}
\< \chi^e(k)_{\alpha i} \chi^o(-k)_{\beta j} \> \sim \Delta_2(k) \delta_{\alpha \beta} \delta_{ij}, \label{eq:vev2}
\end{equation}
which is of $O(1/N_c)$. This vev is invariant under \Eq{eq:chi_trns} only for $u_1=u^T_1$ and $u_2=u^*_3$. This in turn means that when the Schwinger-Dyson equation has a solution with $\Sigma_2\neq 0$, then the stationary subgroup $U(2m-N)\times U(N-m) \times U(N-m)$ of the vevs in~(\ref{eq:gs}) is broken by the $O(1/N_c)$ correlation function~(\ref{eq:vev2}) to 
\begin{equation}
O(2m-N)\times U(N-m). \label{eq:gs_symm}
\end{equation}
A different ansatz can lead to a different symmetry. For example, a self-energy structure that couples $\chi^{\dag e}$ to $\chi^o$ leaves the $U(2m-N)$ subgroup intact. In principle one should compare the free energy of these two ground states. We do not investigate in this direction any further since the only difference will be in the symmetry structure of condensates of $O(1/N_c)$. The symmetry structure at $O(1)$ will be the same.

In fact because the additional condensates here are of $O(1/N_c)$, the physics related to it (Goldstone modes corresponding to the generators it breaks) is weakly coupled to the $O(1)$ physics related to the vevs in~(\ref{eq:gs}), and we ignore it in this work. 


In Appendix~\ref{app:Intequation} we derive self-consistent
equations for $\Sigma_{1,k}$ and $\Sigma_{2,k}$. Defining 
$\tanh{\theta_\bk}=-\Re \Sigma_{2,(0,\bk)} / \Re \Sigma_{1,(0,\bk)}$,
we write these
equations in the form of a single integral equation,
\begin{equation}
\label{eq:Inteq}
(N-m)\tanh{\theta_\bk}=\frac{ {\displaystyle \int_{\text{BZ}} \left( \frac{dq}{4\pi}
  \right)^d I_2(\bq,\bk) \sinh{\theta_\bq} } }{ {\displaystyle \int_{\text{BZ}} \left(
    \frac{dq}{4\pi} \right)^d 
  I_1(\bq,\bk) \cosh{\theta_\bq} -\eta(\bk) } } \equiv \frac{\beta_\bk}{\alpha_\bk}.
\end{equation}
$\eta(\bk)$ and $I_{1,2}(\bq,\bk)$ are defined in appendix~\ref{app:Intequation}. The
poles of the propagators may then be obtained from \Eq{eq:Gchi} [see
\Eq{eq:Eq4poles}]. The dispersion relations turn out to be
\begin{eqnarray}
\pm i\omega &=&\frac{2J_1d}{N_c}\sqrt{(N-m)^2\alpha^2_\bk-\beta^2_\bk}
\nonumber \\
&=&\frac{2J_1d(N-m)}{N_c}\alpha_\bk \sqrt{1-\tanh^2{\theta_\bk}}
 \\
&\equiv& \frac{2J_1d(N-m)}{N_c} \epsilon_\bk. \nonumber
\end{eqnarray} 

The solution of \Eq{eq:Inteq} for $d=3$ can be obtained numerically by assuming
that the function $\theta_\bk$ depends only on $|\bk|$. We plot the solution in
 Fig.~\ref{fig:tanh} for $N-m=1,2,3,4,5$ (for details see Appendix~\ref{app:Intequation}). Recall that the baryon
 density is given by $B=m-N/2$. Thus for $N_f \le 3$ we cover all
 baryon density short of saturation. 

\begin{figure}[htb]
\begin{center}
\resizebox{130mm}{!}{\includegraphics{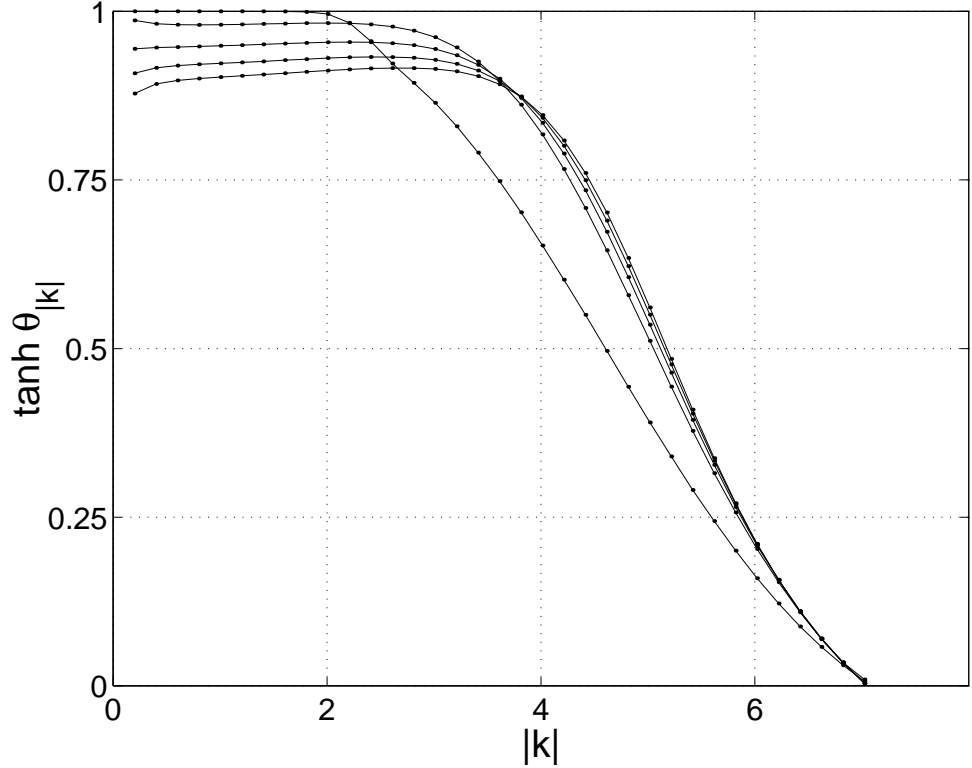}}
\caption[Solution of integral equation]{Solution of \Eq{eq:Inteq} for $N-m=1,2,3,4,5$. The value of $\tanh \theta$ for $|\bk| \rightarrow 0$ drops with increasing $(N-m)$.} \label{fig:tanh}
\end{center}
\end{figure}

\begin{figure}[htb]
\begin{center}
\resizebox{120mm}{!}{\includegraphics{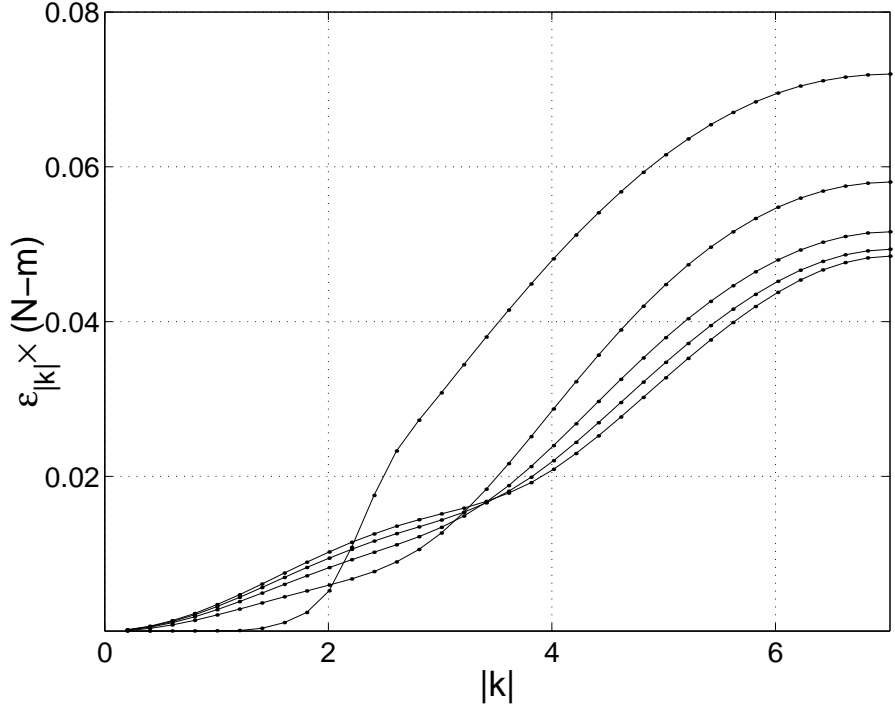}}
\caption[Dispersion relation of the $\chi$ fields.]{Rescaled energy of the $\chi$ bosons for 
  for $N-m=1,2,3,4,5$. The value of $(N-m)\times \epsilon_{\bk}$ at the boundary $|\bk|=\sqrt{5}\pi$ of the Brillouin zone increases with $N-m$.} \label{fig:Ek}
\end{center}
\end{figure}

The form of $\epsilon_\bk$ is shown in Fig.~\ref{fig:Ek}.
It is easy to check that the bosons are massless, since
$\eta(0)=I_{1,2}(\bq,0)=0$ implies that $\alpha_{\bk}$ and $\beta_{\bk}$ 
vanish  at $\bk=0$. This is a direct result of the Ward identities
concerning the global $U(N)$ symmetry. Moreover, near $\bk=0$ both $\alpha$ and
$\beta$ have a quadratic dependence on $|\bk|$. This means that at low
momenta $\epsilon_\bk \sim |\bk|^{2}$, characteristic of
ferromagnetic magnons. A possible exception is the case $N-m=1$. There we see that at low momenta, $\tanh{\theta_\bk}\rightarrow
1$ and thus it is possible that $\epsilon_\bk \sim |\bk|^{2+p}$ with
$p$ a positive integer\footnote{$\epsilon_\bk$ cannot be nonanalytic
  at $\bk=0$ since the integrals in \Eq{eq:Inteq} are regular there.}.

Finally we recall that the relation between the number of $\chi$
fields and
the number of physical excitations depends on the dispersion
relation. Since these soft modes obey $\epsilon_\bk \sim |\bk|^2$, the
$4(N-m)(2m-N)$ fields describe only half that many ferromagnetic magnons. The case of $N-m=1$ might be different
depending on whether $p$ is odd or even. In order to determine $p$ it is necessary to improve the resolution of the momentum mesh. (This may explain the different behavior of $\tanh \theta$ in Fig.~(\ref{fig:tanh}) and of the resulting $\epsilon_\bk$ for this case)

The $\phi$ fields become two kinds of Goldstone boson. The $\pi$ fields are antiferromagnetic spin waves with a linear dispersion
relation. The $\chi$ fields are ferromagnetic
magnons with a quadratic dispersion relation. This
is consistent with the loss of Lorentz invariance due to finite
density. 
This set of excitations falls into the pattern described by Chadha and Nielsen \cite{Nielsen:hm} and by Leutwyler 
\cite{Leutwyler:1993gf} in their
studies of nonrelativistic field theories. 

In general, a nonrelativistic system that undergoes spontaneous symmetry breakdown
can posses two types of Goldstone boson. The energy of type I bosons is an odd power of their momentum.
 Their effective field theory has a second time derivative which means that
each excitation is described by one real field as in the case of a relativistic scalar field. 
Type II bosons have a dispersion relation 
that contains an even power of the momentum. They can appear only in nonrelativistic theories having properties similar to a Schr\"{o}dinger field theory, namely theories whose action has only a first time 
derivative. As in the Schr\"{o}dinger case, each excitation is
described by a complex field, or two real fields. 
The number of the $\phi$ Goldstone fields is of course equal to the number $n_G$ of generators broken by \Eq{eq:gs_symm}. The counting of these massless excitations is summarized by the Chadha--Nielsen counting rule, 
\begin{equation}
\label{eq:count}
n_{I}+2n_{II} \geq n_G,
\end{equation}
where $n_I(n_{II})$ are the number of type I and II Goldstone bosons. A well-known example 
is the $SU(2)$ spin system with a collinear ground state, where $n_G=2$. The antiferromagnet has two spin
waves with a linear dispersion relation ($n_I=2,n_{II}=0$) while the
ferromagnet has one magnon with a quadratic dispersion relation
($n_I=0,n_{II}=1$). As mentioned above in our case both $n_I$ and $n_{II}$ are non-zero.

In addition to these massless fields there exist another set of massless fields. These are the Goldstone bosons that correspond to the breakdown of $U(2m-N)$ to $O(2m-N)$, and of $U(N-m)\times U(N-m)$ to $U(N-m)$, by the $O(1/N_c)$ condensate~(\ref{eq:vev2}). In this case the Goldstone bosons are not represented by any of the fundamental $\chi$ or $\pi$ fields. Similar to the BCS case, they are collective excitations that correspond to poles in two-particle propagators of the generic form $\< \chi^{*e} \chi^{*o} \chi^e \chi^o \>$. These poles describe zero mass particles which are bound states of two $\chi$ particles. We do not pursue their properties (symmetry structure, dispersion relations, etc.) in this work since as mentioned they are weakly coupled to the excitations we treat.

The energy scale of the type II Goldstone bosons $\chi$ is smaller by a factor of $1/N_c$ compared to that of the type I bosons $\pi$.
 This points to a possible hierarchy of phase transitions at
finite temperature which can be described by a classical model that has the Hamiltonian
\begin{equation}
\label{eq:Hmodel}
H=J_1 \hspace{0.1cm} \Tr \sum_{\bn \muhat} \left[ \sigma_\bn \sigma_{\bn +\muhat}
\right] - J_2 \hspace{0.1cm} \Tr \sum_{\bn \muhat} \left[ \sigma_\bn \sigma_{\bn+2\muhat} \right].
\end{equation}

For $J_2=0$, the ground state
will be highly degenerate as above. A small positive value of $J_2$ removes this
degeneracy and picks out a ferromagnetic alignment of the next-nearest-neighbor $\sigma$ fields. Thus for $T \ll J_1,J_2$ the ground state is 
invariant only under\footnote{We note here that order from disorder is not necessary here, and therefore there is no condensate analogous to~(\ref{eq:vev2}). This means that the physics described by this classical model is {\it different} than the real model.}
\begin{equation}
U(N-m)\times U(N-m) \times U(2m-N).
\end{equation}
This symmetry pattern will persists until some finite temperature $T_c^{\text{FM}} \sim J_2$.
For $T > T_c^{\text{FM}}$ the ferromagnetic magnons melt the
magnetization and restore some of the symmetry.
This situation will persists until a second temperature $T_c^{\text{AFM}} \sim J_1$ where the symmetry will be restored completely. As long
as $J_1 \gg J_2$ (which corresponds to $N_c \gg 1$ in our system), the hierarchy of phase transitions is well defined. 

Finally we mention that recent work \cite{Continuum_type2} on effective field theories for dense QCD also predicts the
existence of type II Goldstone bosons. It is tempting to identify these with our ferromagnetic magnons. First, however, 
one must reduce the artificial $U(4N_f)$ 
symmetry of the action to the physical chiral symmetry. This important issue is discussed in the next chapter.

\clearpage
\chapter[Next-nearest-neighbor interactions in the sigma model]{Next-nearest-neighbor interactions in the nonlinear sigma model}
\label{chap:NNN}
The previous chapters are all mainly restricted to nearest-neighbor (NN) interactions only, which were symmetric under the global group of $U(4N_f)$. In this chapter we add the sigma model next-nearest-neighbor (NNN) interactions that reduce this symmetry to the chiral symmetry $U(N_f)\times U(N_f)$, thus completing our picture of the lattice theory that has the symmetry
of continuum QCD. Taking the NNN coupling to be much smaller than the NN one, we perturb the NN ground state and its excitations, found in Chapters~\ref{chap:NLSM} and~\ref{chap:Qfluctuation} with these NNN terms. These interactions reduce some of the global degeneracy of the NN vacuum, and leave only a part of chiral symmetry intact. We find that the global symmetry of the ground state varies with $N_f$ and with the background baryon density. In all cases the condensate breaks the discrete rotational symmetry of the lattice as well as part of the chiral symmetry group. We also find two types of massless excitations: type I Goldstone bosons with linear dispersion relations and type II Goldstone bosons with quadratic dispersion relations. Some of the type I bosons exist also at zero density. The others were originally type II bosons that received energy contributions, linear in momentum, from the NNN terms, and became type I bosons as well. 
Both kinds of bosons can have anisotropic dispersion relations.

\section{Action and symmetries of next-nearest-neighbor interactions}
\label{sec:NNN_action}

The NN effective Hamiltonian and sigma model discussed up to now in Chapters~\ref{chap:MF_schwinger}--\ref{chap:Qfluctuation} correspond to an underlying kernel of naive fermions. Although these fermions suffer from species doubling in weak coupling, this doubling is not seen directly in strong coupling. The reason is that their we do not have fermion excitations, but rather only gauge invariant degrees of freedom, i.e., the mesonic $Q^\eta_\bn$ operators and the baryonic $b^I_\bn$ operators (see Chapter~\ref{chap:Heff}). Doubling of species can therefore reflect itself only in the number of excitations that correspond to these color singlet field, and is therefore determined by the global symmetry of the effective Hamiltonian. 

The global symmetry group of the $\sigma$ model depends on the underlying fermion kernel of the Hamiltonian. For $N_f$ flavors of naive fermions we get an interaction between nearest-neighbor (NN) sites that is invariant under $U(N)$ with
\begin{equation}
N=4N_f.
\end{equation}
This symmetry is too large and is indeed indicative of species doubling. This symmetry is also present in the original Hamiltonian, as can be seen by spin-diagonalizing the interaction. In order to extract physical consequences from our model, it is imperative that it has the same global symmetry as QCD. We achieve this goal by taking a different underlying fermion kernel in \Eq{eq:H_F} with the reduced, correct symmetry. We add next-nearest-neighbor (NNN) interactions to the kernel and reduce the symmetry to
\begin{equation}
U(N_f)_L \times U(N_f)_R,
\end{equation}
which is almost the symmetry of the continuum theory. The additional $U(1)_A$ 
is inevitable if one starts with a local, chirally symmetric, underlying theory \cite{NN}.
It can be easily removed by hand in the $\sigma$ model by adding $U(1)_A$-breaking terms 
\cite{U1A}, and we ignore it.



The action of the $\sigma$ model with nearest-neighbor next-nearest neighbor interactions is [See Chapter~\ref{chap:NLSM}]
\begin{equation}
\label{S} S=\frac{N_c}2\int d\tau\left[-\sum_{\bn} \Tr\Lambda
U^{\dag}_{\bn} \partial_\tau U_{\bn} +\frac {J'_1}2\sum_{\bn i} \Tr
\left( \sigma_\bn \sigma_{\bn+\muhat} \right) + \frac{J'_2}2
\sum_{\bn \mu} \Tr \left( \sigma_\bn \alpha_\mu \sigma_{\bn+2\muhat}
\alpha_\mu \right)  \right].
\end{equation}
Here $J'_1$ and $J'_2$ are the couplings of the classical Hamiltonian \Eq{eq:H_sigma}, multiplied by $2/N_c$. They scale as $1/g^2N_c$ [recall the discussion in at the end of Section~\ref{sec:action_NLSM}]. For brevity we drop the primes henceforth.
If one derives the fermion Hamiltonian by
truncating the SLAC Hamiltonian \cite{DWY}, then the couplings $J_1$ and
$J_2$ obey $\hbox{$J_2=J_1/8$}$. If we argue,
however, that the strong-coupling Hamiltonian is derived by
block-spin transformations applied to a short-distance
Hamiltonian, then we cannot say much about the couplings that
appear in it.  We will assume that couplings in the effective
Hamiltonian fall off strongly with distance, that is $\hbox{$0<J_2 \ll J_1$}$.

The NN term is invariant under the global $U(N)$ transformation
$U\to VU$, or $\sigma\to V\sigma V^\dag$.  The NNN term is only
invariant if $V^\dag \alpha_\mu V=\alpha_\mu$ for all $\mu$.  This
restricts $V$ to the form
\begin{equation}
V=\exp
\left[i\left(\theta_V^a+\gamma_5\theta_A^a\right)\lambda^a\right],
\end{equation}
where $\lambda^a$ are flavor generators.  This is
a chiral transformation in $U(N_f)\times U(N_f)$.
The NNN term couples (discrete) rotational symmetry to the internal symmetry, {\it viz.}
\begin{equation}
\sigma_\bn \to R^{\dag} \sigma_{\bn'} R, \qquad \bn'={\cal R} \bn.
\label{rotattions}
\end{equation}
Here ${\cal R}$ is a $90^o$ lattice rotation and $R$ represents it according to
\begin{equation}
R=\exp \left[ i \frac{\pi}4 \left( \begin{array}{cc}
         \sigma_\mu & 0 \\
        0 & \sigma_\mu \end{array} \right) \right] \otimes {\bf 1}_{N_f}.
\label{R}
\end{equation}


\section{Ground state of the next-nearest-neighbor theory}
\label{sec:gs}

The overall $N_c/2$ factor in \Eq{S} allows a systematic treatment
in orders of $1/N_c$.  In leading order, the ground state is found
by minimizing the action, which gives field configurations that
are $\tau$ independent and that minimize the interaction terms. For the NN
theory, minimizing the single link interaction
\begin{equation}
\label{site2} E=\frac{J_1}2 \Tr \sigma_1 \sigma_2
\end{equation}
allows us to construct the vacuum by placing $\sigma_1$ and
$\sigma_2$ on the even and odd sites.

We impose a uniform baryon density, $B_\bn=B>0$, on the effective
Hamiltonian by setting a fixed value of $m>2N_f$ on every site. To
minimize the single-link energy~(\ref{site2}) we first choose a
basis where
\begin{equation}
\sigma_1=\Lambda=\left( \begin{array}{cc}
\bm1_m&0\\
0&-\bm1_{N-m}\end{array} \right). \label{twospin1}
\end{equation}
The analysis in chapter~\ref{chap:Qfluctuation} shows that at the classical level the ground state is locally degenerate, but taking into consideration $O(1/N_c)$ fluctuations lifts this degeneracy. As a result we can choose the the value of $\sigma_2$ with out loss of generality to be
\begin{equation}
\sigma_2=\left( \begin{array}{ccc}
\bm1_{2m-N} & 0     &       0       \\
0 & -\bm1_{N-m} & 0 \\
0       &  0 &      \bm1_{N-m} \end{array} \right). \label{twospin2}
\end{equation}
This is a N\'eel structure, with two sublattices. The vacuum expectation values (vevs) of the even sigma fields break $U(N)$ to $\hbox{$U(m)\times U(N-m)$}$ and then the vevs of the odd sigma fields break the symmetry further to $\hbox{$U(2m-N)\times
U(N-m) \times U(N-m)$}$.


Now we add the NNN interactions to the effective
action.  At the classical level, they do not by themselves remove
the classical degeneracy of the NN theory (for nonzero density). We have to
introduce the $O(1/N_c)$ fluctuations \textit{first} in order to
stabilize the N\'eel ground state of the NN Hamiltonian. More precisely we assume that 
\begin{equation}
\label{strengths}
0<J_2 \ll J_1/N_c.
\end{equation}
This means that we take as our starting point the (globally degenerate)
vacuum determined in Chapter~\ref{chap:Qfluctuation} for the NN theory with $O(1/N_c)$
corrections.
The NNN interaction is a perturbation on this vacuum and its
excitations.

We begin, then, by assuming a N\'eel ansatz that minimizes the NN
term in the action (\ref{S}). The NNN term acts within each of the
two sublattices.  Writing $\sigma_{e,o}$ for the sublattice
fields, the NNN contribution to the energy per $2\times 2 \times 2$ lattice cell is
\begin{equation}
E_{\nnn} = \frac{J_2}2 \sum_{a=e,o} \sum_\mu \Tr \left[ \sigma_a
\alpha_\mu \sigma_a \alpha_\mu    \right]. \label{eq:E}
\end{equation}
Here $\sigma_a$ is a global unitary rotation of the solution to the NN theory given by Eqs.~(\ref{twospin1})--(\ref{twospin2}).

We can find a lower bound for $E_{\nnn}$. Writing in each term $\Sigma_1^a=\sigma_a$
and $\Sigma_2^{a\mu}=\alpha_{a \mu} \sigma_a \alpha_{a\mu}^\dag$, we have
\begin{equation}
E_{\nnn}=\frac{J_2}2 \sum_{a \mu} \Tr
\Sigma_1^a\Sigma_2^{a\mu}.\label{minim}
\end{equation}
Note that $\Sigma_{1,2}$ are unitary rotations of the reference matrix $\Lambda$.
Each term in \Eq{minim} may be bounded from below by allowing these unitary
rotations to vary independently over the
entire $U(N)$ group. This is just
the minimization problem posed in \Eq{site2} above. The solution is
given by Eqs.~(\ref{twospin1})--(\ref{twospin2}), whence the bound
\begin{equation}
E_{\nnn}\ge 3J_2(4m-3N). \label{bound}
\end{equation}

We minimize $E_{\nnn}$ by writing an ansatz for $\sigma_a$ that
saturates the lower bound. At this point we choose to work in a
basis where $\gamma_5$ is diagonal,
\begin{equation}
\gamma_5=\left( \begin{array}{cc}
    \bm1_{N/2} & 0   \\
    0 & -\bm1_{N/2}  \end{array} \right).
\end{equation}
Our ansatz is
\begin{eqnarray}
\sigma_e&=&U\Lambda_e U^\dag= U\left( \begin{array}{cc}
    \bm1_m & 0   \\
    0 & -\bm1_{N-m}  \end{array} \right)U^\dag,\nonumber\\[2pt]
\sigma_o&=&U\Lambda_o U^\dag= U\left( \begin{array}{ccc}
    \bm1_{2m-N} & 0 &0  \\
    0 & -\bm1_{N-m}&0\\
    0&0&\bm1_{N-m}  \end{array} \right)U^\dag. \label{ansatz}
\end{eqnarray}
(Note that $\Lambda_o$ is a rotation of $\Lambda_e$.) This is a global rotation (via $U$) of a
particular configuration that minimizes the NN energy, as seen
above.  We further suppose that $U$ takes the form
\begin{equation}
 U=\frac1{\sqrt{2}} \left( \begin{array}{cc}
    u & u   \\
    -u & u  \end{array} \right),
\label{eq:Uo}
\end{equation}
with $u\in U(N/2)$.

The minimization of the energy via the ansatz is a problem of
vacuum alignment \cite{Peskin}. We begin with the unperturbed NN
problem, where the symmetry group $G\equiv U(N)$ is broken to
$H\equiv \hbox{$U(2m-N)\times U(N-m)\times U(N-m)$}$. Our
reference vacuum is given by $\sigma_{e,o}=\Lambda_{e,o}$, giving
a specific alignment of $H$ as the invariance group of this
vacuum. We determine the rotation matrix $U$ [within the ansatz
(\ref{eq:Uo})] that minimizes the energy of the perturbation
$E_{\nnn}$, which we write as
\begin{equation}
\label{E_bar} E_{\nnn}=\frac{J_2}2 \sum_{a \mu} \Tr
\Lambda_a \bar{\alpha}_{\mu} \Lambda_a \bar{\alpha}_{\mu}^\dag,
\end{equation}
where $\bar{\alpha}_{\mu}=U^{\dag} \alpha_\mu U=\bar{\alpha}_{\mu}^\dag$.

We denote the generators of $H$ collectively as $T$ and the
remaining generators of $G$ as $X$.  The $T$ matrices commute with
both $\Lambda_e$ and $\Lambda_o$, while the $X$ matrices do not.
The symmetry-breaking term in the energy is $E_{\nnn}$, given by
\Eq{E_bar} in terms of the rotated Hermitian matrices
$\bar{\alpha}_{\mu}$. We project each
$\bar{\alpha}_{\mu}$ onto the $T$ and $X$ subspaces, giving the
decomposition
\begin{equation}
\bar{\alpha}_\mu = \bar{\alpha}_\mu^{T} + \bar{\alpha}_\mu^X.
\end{equation}
Using the invariance of $\Lambda_{e,o}$ under $T$,
\begin{equation}
\left[ \Lambda_a , \bar{\alpha}_\mu^T \right]=0,
\end{equation}
and the orthogonality of the $T$ and $X$ subspaces,
\begin{equation}
\Tr \left[ \bar{\alpha}_\mu^T \bar{\alpha}_\mu^X \right]=0,
\label{eq:relations}
\end{equation}
we have
\begin{equation}
E_{\nnn}=\frac{J_2}2 \sum_{i,a} \Tr \left[ \left( \bar{\alpha}_\mu^T
\right)^2 + \Lambda_a \bar{\alpha}_\mu^X \Lambda_a \bar{\alpha}_\mu^X
\right] \label{eq:Etx}
\end{equation}
Following the block form of $\Lambda_{e,o}$, we divide the broken
generators $X$ into 3 sets, denoting them as $X_a$ with $a=1,2,3$.
Their structures are respectively
\begin{equation}
\left( \begin{array}{ccc}
    0 & \tilde{X}_1 & 0 \\
    \tilde{X}_1^{\dag} & 0 & 0  \\
    0 & 0 & 0   \\  \end{array} \right) ,\quad
\left( \begin{array}{ccc}
    0 & 0 & \tilde{X}_2 \\
    0 & 0 & 0   \\
    \tilde{X}_2^{\dag} & 0 & 0  \\  \end{array} \right)
    ,\quad\textrm{and}\
\left( \begin{array}{ccc}
    0 & 0 & 0   \\
    0 & 0 & \tilde{X}_3 \\
    0 & \tilde{X}_3^{\dag} & 0  \\  \end{array} \right).
\end{equation}
Writing $\bar{\alpha}_\mu^X=\sum_a \bar{\alpha}_\mu^{X_a}$, the
following relations can be proved easily:
\begin{equation}\begin{array}{c @{\qquad} c}
\left[ \Lambda_1 , \bar{\alpha}_\mu^{X_1} \right]=0 & \left\{\Lambda_2 , \bar{\alpha}_\mu^{X_1} \right\}=0
\\[4pt]
\left\{ \Lambda_1 , \bar{\alpha}_\mu^{X_2} \right\}=0 & \left[\Lambda_2 , \bar{\alpha}_\mu^{X_2} \right]=0
\\[4pt]
\left\{ \Lambda_1 , \bar{\alpha}_\mu^{X_3} \right\}=0 &
\left\{\Lambda_2 , \bar{\alpha}_\mu^{X_3} \right\}=0.
\end{array}\label{eq:com}
\end{equation}
Using these together with the further orthogonality conditions,
\begin{equation}
\Tr \left[ \bar{\alpha}_\mu^{X_a} \bar{\alpha}_\mu^{X_b} \right]
=0,\qquad a \neq b, \label{eq:indep}
\end{equation}
we bring \Eq{eq:Etx} to the form
\begin{equation}
E_{\nnn}=J_2 \sum_\mu \Tr \left[ \left( \bar{\alpha}_\mu^{T} \right)^2
-\left( \bar{\alpha}_\mu^{X_3} \right)^2 \right]. \label{eq:E_TX3}
\end{equation}

The rotation $U$
given in \Eq{eq:Uo} saturates the lower
bound for the energy. We will proceed to prove this for the case
$m \ge 3N/4$. In view of  \Eq{eq:Uo} we can write $\bar{\alpha}_\mu$
in the form
\begin{equation}
\label{al_bar} \bar{\alpha}_\mu = \left( \begin{array}{cc}
                        0 & \bar{\sigma}_\mu \\
                        \bar{\sigma}_\mu & 0 \end{array} \right),
\end{equation}
where $\bar{\sigma}_\mu\equiv u^{\dag} \sigma_\mu u$. It is
straightforward to check that for $m \ge 3N/4$ we have
\begin{eqnarray}
\bar{\alpha}_\mu^{X_3} &=& 0, \label{al_X3_above} \\
\bar{\alpha}^T_\mu &=& \left( \begin{array}{cc}
                0 & \bar{\sigma}'_\mu \\
                \bar{\sigma}'^{\dag}_\mu &  0  \end{array} \right), \label{al_T_above}
\end{eqnarray}
with $\bar{\sigma}'_\mu$ composed of the first $(2m-3N/2)$ columns
of $\bar{\sigma}_\mu$,
\begin{equation}
\left( \bar{\sigma}'_\mu \right)_{pq} = \left\{ \begin{array}{ll}
\left( \bar{\sigma}_\mu \right)_{pq}
& \quad \textrm{for} \quad q=1,\dots,2m-3N/2, \\[2pt]
0 & \quad \textrm{else}. \end{array} \right.
\end{equation}
The energy is
\begin{equation}
\label{E_m_above} E_{\nnn}=2J_2 \sum_{i,pq} |(\bar{\sigma}'_\mu)_{pq}|^2
= 2J_2 \sum_{i} \sum_{q=1}^{2m-3N/2} \left( \bar{\sigma}^{\dag}_\mu
\bar{\sigma}_\mu \right)_{qq} = 3J_2 (4m-3N),
\end{equation}
which is exactly the lower bound. Note that according to
\Eq{E_m_above}, any value $u$ within $U(N/2)$ saturates the lower bound. Nevertheless, some values of $u$ give the same vacuum as does $u={\bf 1}$. More precisely, all the elements $u\in U(2m-3N/2) \times U(N-m) \times U(N-m)$ act trivially
in \Eq{ansatz}.
Vacua that are associated with different nontrivial choices of $u$ are
in general inequivalent, and give different realizations of the
$U(N_f)\times U(N_f)$ symmetry of the theory.
Since these vacua are not related by symmetry transformations, there is nothing
to prevent lifting of the degeneracy in higher orders in $1/N_c$.
In the sequel, we set $u={\bf 1}_{N/2}$.
This gives the vacuum with the largest symmetry accessible via the ansatz
(\ref{eq:Uo}).\footnote{We have found, however, one exception where another ansatz, with different symmetries, gives larger symmetry than that of \Eq{ansatz} for the case of $B=2$ and $N_f=3$.}

The simplicity of this calculation depends on the assumption
$m\ge3N/4$. For $m < 3N/4$, both $\bar{\alpha}^{X_3}_\mu$ and
$\bar{\alpha}^T_\mu$ are non-zero. Moreover the index structure of
the projections is more complex.\footnote{An exception is the
$m=N/2$ case ($B=0$), which was solved in Chapter~\ref{chap:Qfluctuation}.} Therefore
in these cases we resort to numerical minimization of \Eq{E_bar}
over $u\in U(N/2)$.  In each case we find that the ansatz
(\ref{ansatz})--(\ref{eq:Uo}) yields a minimum that saturates the
lower bound (\ref{bound}).
In view of what happens for $m \ge 3N/4$ this may be only one point in a 
degenerate manifold of ground states.
We emphasize that the degeneracy of these vacua is not related to the global
$U(N_f)\times U(N_f)$ chiral symmetry. It is an accidental global degeneracy of 
the ground state.

Upon calculating the $\sigma$ fields using
Eqs.~(\ref{ansatz})--(\ref{eq:Uo}), it is straightforward to
ascertain the symmetry of the vacuum. The $U(N_f)\times U(N_f)$
generators that commute with both $\sigma_e$ and~$\sigma_o$ form
the unbroken subgroup of the NNN theory. The rest are broken
generators that correspond to Goldstone bosons.

The symmetry of these ground states is summarized in Table \ref{table1}.
In general, both chiral symmetry and discrete lattice rotations are broken;
in some cases a symmetry under rotations around the $z$ axis survives. Note that if we remove the unphysical axial $U(1)$ symmetry from the $\sigma$ model, all its realizations will also drop from Table \ref{table1},  namely, there will be no axial $U(1)$ unbroken symmetries (third column) and no Goldstone bosons corresponding to a broken axial $U(1)$ (fourth column). Anticipating these simple modifications we see no reason to proceed in this direction.

\begin{table}[htb]
\caption[Realization of chiral symmetry for different densities]{Breaking of $SU(N_f)_L\times SU(N_f)_R\times U(1)_A$ for
all baryon densities (per site) accessible for $N_f\le3$. Results
for $B=0$ are from Chapter~\ref{chap:Qfluctuation}. \label{table1}}
\begin{center}
\begin{tabular}{cccc}

$N_f$   &   $|B|$   &   Unbroken symmetry & Broken charges\\
\hline
    &   0   & $-$  & 1  \\
1   &   1   & $-$  & 1  \\
    &   2   & $U(1)_A$ & 0  \\  \hline
    &   0   & $SU(2)_V$ & 4  \\
    &   1   & $U(1)_{I_3}$ & 6 \\
2   &   2   & $SU(2)_V$ & 4  \\
    &   3   & $U(1)_{I_3}$ & 6  \\
    &   4   & $SU(2)_L\times SU(2)_R\times U(1)_A$ & 0  \\  \hline
    &   0   & $SU(3)_V$ & 9  \\
    &   1   & $U(1)_Y\times SU(2)_V$ & 13\\
    &   2   & $U(1)_Y$ & 16 \\
3   &   3   & $SU(3)_V$ & 9  \\
    &   4   & $U(1)_{I_3}\times U(1)_Y$ & 15 \\
    &   5   & $U(1)_{I_3}\times U(1)_Y\times U(1)_{A'}$ & 14  \\
    &   6   & $SU(3)_L\times SU(3)_R\times U(1)_A$ & 0 \\
\end{tabular}
\end{center}
\end{table}

In addition we note the following:
\begin{itemize}
\item   Since the baryon background is fixed, we cannot tell whether
the $U(1)$ corresponding to baryon number is broken. [The $U(1)_B$
group acts trivially on $Q^\eta_\bn$ and on $\sigma_\bn$.] 
\item   For each value of $N_f$, the case $B=2N_f$ is a completely
saturated lattice.  Each site is in a singlet under the chiral
group, and there is no spontaneous symmetry breaking.
\item   The axial $U(1)$ is not a symmetry of the continuum, and must be
broken by hand. Wherever it appears in Table \ref{table1} it should
be neglected, whether as an unbroken symmetry or as a broken
charge. 
\item   The $U(1)_{A'}$ appearing in the table for $N_f=3$
is not the original $U(1)_A$ group but rather is generated by
$\gamma_5 \otimes \lambda'$ with
\begin{equation}
\lambda'= \left( \begin{array}{ccc}
    1 & 0 & 0 \\
    0 & 0 & 0 \\
    0 & 0 & 0 \end{array} \right).
\end{equation}
This is the only case where an axial symmetry survives spontaneous
symmetry breaking. If $U(1)_A$ is broken by hand, so is
$U(1)_{A'}$.
\end{itemize}

For all non-zero densities short of saturation the vacuum breaks
rotational  invariance. This is easily checked for $m \ge 3N/4$ by
noting that the ansatz for $\sigma_{e,o}$ fails to commute with
the rotation operator~(\ref{R}). For the other cases this is
easily checked numerically. (In some cases a discrete symmetry
around the $z$ axis remains unbroken.) Since this is not a
continuous symmetry it will not give rise to additional Goldstone
bosons. The broken rotational invariance will of course affect the
excitation spectrum. In particular, whereas the NN theory
possesses excitations with linear and quadratic dispersion
relations \cite{odo}, the NNN theory will produce anisotropic dispersion relations, like those seen
in \cite{Sannino} and Section~\ref{sec:nnn_excite} below.


In comparing our results to those of continuum CSC calculations,
one must keep in mind that we study systems with large, fixed, and
discrete values of $B$, rather than with large, continuous $\mu$.
Moreover, we use large-$N_c$ approximations which necessarily ignore the
discrete properties of the $SU(N_c)$ group that are essential to
baryons.  Quantum effects at finite $N_c$, treated correctly, should
yield effects that are not accessible through the $1/N_c$
expansion.

The values of $(N_c,N_f)$ that are of interest for CSC are $N_c=3$
and~$N_f=2$ or~3.  In the two-flavor case the favored $qq$
condensate is a flavor singlet and a color triplet, so that while color
is partially broken, chiral symmetry is unbroken.  We do not
see this for any density at $N_f=2$.  Plainly our results are due
to a $\bar qq$ condensate; whether there is a $qq$ condensate as
well cannot be ascertained.

For $N_f>2$ the situation is similar.  Sch\"afer \cite{Schafer}
has considered the color--flavor structure of the condensates that
arise for $N_c=3$ and $N_f\ge 3$, and he has found that both color
and flavor are partially broken, with a condensate that locks one
or more subgroups of the flavor group to the color group.  Since
we work at large $N_c$, we should stand the argument on its head.
A plausible $qq$ condensate would lock successive subgroups of the
{\em color\/} group to the flavor group and hence to each other,
leaving unbroken the diagonal $SU(N_f)_{L+R+C}$ and some leftover
color symmetry.  Judging by the global symmetry of the vacuum,
perhaps we see this for $(N_f=3,B=3)$.  The other cases could
conceivably arise from a combination of $\bar qq$ and~$qq$
condensates, but whether the latter actually occur is an open
question.


Finally we note that according to diagrammatic power-counting arguments
\cite{DGR,ShusterSon}, CSC is suppressed in the 't~Hooft limit, where $N_c\to\infty$ and $g^2N_c$ is kept fixed. Deryagin, Grigoriev, and Rubakov showed that in the perturbative regime of small $g^2N_c$, a Fermi sea is unstable towards formation of a chiral density wave with  
momentum $|\bp|=2p_F$, where $p_F$ is the fermi momentum \cite{DGR}. 
Our ansatz does not permit this possibility.

\section{Excitations in the next-nearest-neighbor theory}
\label{sec:nnn_excite}

The Goldstone bosons of the NN theory were discussed in above.
There are $2(N-m)^2$ bosons of type I with $\omega \sim J_1 |\bk|$ at low
momenta;
these are generalized antiferromagnetic spin waves (and are the only
excitations at zero density).
There are also $2(2m-N)(N-m)$ bosons of type II, that derive their energy
from quantum fluctuations in $O(1/N_c)$.
These are generalized {\em ferromagnetic} magnons
with $\omega \sim (J_1/N_c) |\bk|^2$.


Now we calculate the effects of the NNN interactions on the
spectrum. In view of \Eq{strengths}, the NN contributions to the
propagators, found in above, remain unchanged. In particular
we can take over the self-consistent determination of the
self-energy of the type II bosons. We need consider the NNN
contributions to the propagators in tree level only. We proceed to
calculate these for $m \ge 3N/4$. In these cases, the calculations
simplify and we perform them
analytically.\footnote{An exception is $(N=12,m=10)$, where we
have no analytic solution.  See below.} We believe that the
spectra of the other cases have similar features.

In the NN theory the $\sigma$ fields represent fluctuations around the
vacuum (\ref{ansatz}) with $U=1$.
We parametrize them as in Eqs.~(\ref{eq:even})--(\ref{eq:odd}).
If $\phi=0$, we have
$\sigma_{\text{even,odd}}=\Lambda_{\text{even,odd}}$, which is the
ground state of the NN theory. We adapt
Eqs.~(\ref{eq:even})--(\ref{eq:odd}) to the NNN theory by rotating them,
\begin{equation}
\label{sig_nnn}
\sigma \to U \sigma U^{\dag},
\end{equation}
with $U$ as given in \Eq{eq:Uo}.
Now $\phi=0$ corresponds to the ground state of the NNN theory.

We substitute Eqs.~(\ref{eq:even})--(\ref{eq:odd}) into the action (\ref{S}).
The rotation $U$ disappears from the kinetic term
and from the NN interaction---they are both $U(N)$ invariant.
This means that the bare
spectra found in Chapter~\ref{chap:Qfluctuation}, when $U$ was absent, remain intact.

We write the NNN energy as 
\begin{equation}
\label{Ennn_fluc}
E_{\text{nnn}}=\frac{N_cJ_2}4 \sum_{a \bN \mu} \Tr \alpha_\mu
\sigma_{a,\bN} \alpha_\mu \sigma_{a,\bN + 2\muhat},
\end{equation}
where $a=(\text{even, odd})$ and $\bN$ denotes a site on the corresponding fcc sublattice. We rescale $\phi
\rightarrow \phi/\sqrt{N_c}$ and expand \Eq{Ennn_fluc} to
second order,
\begin{eqnarray}
E_{\text{nnn}}&=&N_c E_0 +
\frac{J_2}4 \sqrt{N_c} \sum_{a \bN \mu} \Tr \bar{\alpha}_\mu \Lambda_a
\bar{\alpha}_\mu
 \left( \Delta^{(1)}_{a\bN} + \Delta^{(1)}_{a, \bN+2 \muhat} \right) \nonumber \\
               &&+\frac{J_2}4 \sum_{a \bN \mu} \Tr \bar{\alpha}_\mu
\Delta^{(1)}_{a\bN} \bar{\alpha}_\mu \Delta^{(1)}_{a,\bN+2\muhat}
               +\frac{J_2}4 \sum_{a \bN \mu} \Tr \bar{\alpha}_\mu
\Lambda_a \bar{\alpha}_\mu \left( \Delta^{(2)}_{a\bN} +
\Delta^{(2)}_{a,\bN+2\muhat} \right)
\nonumber \\
               &&\quad \qquad +O\left(\frac1{\sqrt{N_c}}\right).
\label{Ennn_1overNc}
\end{eqnarray}
We have defined $\bar{\alpha}_\mu=U^{\dag} \alpha_\mu U$, and
$\Delta^{(1,2)}$ correspond to the linear and quadratic deviations
of the $\sigma$ fields from their ground state values. The latter
are given by
\begin{equation} \Delta^{(1)}_{e}=\left(
\begin{array}{ccc}
    0 & 0 & -2\chi \\
    0 & 0 & -2\pi  \\
    -2\chi^{\dag} & -2\pi^{\dag} & 0 \end{array} \right) ,
\qquad  \Delta^{(2)}_{e}=\left( \begin{array}{ccc}
    -2\chi \chi^{\dag} & -2\chi \pi^{\dag} & 0 \\
    -2\pi \chi^{\dag} & -2\pi \pi^{\dag} & 0  \\
    0 & 0 & +2\phi^{\dag} \phi \end{array} \right)
\label{Delta_even}
\end{equation}
on the even sites, and
\begin{equation} \Delta^{(1)}_{o}=V\left(
\begin{array}{ccc}
    0 & 0 & -2\chi \\
    0 & 0 & -2\pi  \\
    -2\chi^{\dag} & -2\pi^{\dag} & 0 \end{array} \right)V^\dag ,
\qquad  \Delta^{(2)}_{o}=V\left( \begin{array}{ccc}
    -2\chi \chi^{\dag} & -2\chi \pi^{\dag} & 0 \\
    -2\pi \chi^{\dag} & -2\pi \pi^{\dag} & 0  \\
    0 & 0 & +2\phi^{\dag} \phi \end{array} \right)V^\dag
\label{Delta_odd}
\end{equation}
on the odd sites.
Here
\begin{equation}
V=\left( \begin{array}{ccc}
    1 & 0 & 0 \\
    0 & 0 & 1 \\
    0 & -1 & 0 \end{array} \right)
\end{equation}
is the matrix that rotates $\Lambda_{\text{even}}$ to
$\Lambda_{\text{odd}}$. It is easy to show that the terms linear
in $\Delta^{(1)}$ vanish.

In view of the block structure of $\Lambda_{\text{even,odd}}$ and
of $U$, as given in Eqs.~(\ref{ansatz})--(\ref{eq:Uo}), and of
$\alpha_\mu$, it is convenient to decompose $\chi$ for $m\ge3N/4$ as
\begin{equation}
\label{chi}
\chi=\left( \begin{array}{c} \chi_1 \\ \chi_2 \end{array} \right)
\end{equation}
Here $\chi_{1}$ has $N/2$ rows, and $\chi_2$ has $2m-3N/2$ rows.
Both have $N-m$ columns. Substituting into \Eq{Ennn_1overNc} and
omitting the ground state energy we find that the $O(1)$
contribution of the NNN energy depends only on $\chi_1$. The $\pi$
and $\chi_2$ fields do not enter the NNN energy at this order and
remain of type I and of type II, respectively.

We now define a new $N\times N$ matrix $\hat{\chi}$ that contains
only $\chi_1$,
\begin{equation}
\hat{\chi}=\left( \begin{array}{ccc} 0 & 0 & \chi_1 \\
0 & 0 &  0 \\
\chi_1^{\dag} & 0 & 0 \end{array} \right), \label{chi_1}
\end{equation}
and use it to write
\begin{eqnarray}
E_{\text{nnn}}&=& \frac{J_2}4 \sum_{\bN \in \text{fcc}\atop \mu}
\bigg\lbrace4\Tr\left[ \bar{\alpha}_\mu \hat{\chi}^e_{\bN}
\bar{\alpha}_\mu \hat{\chi}^e_{\bN+2\hat{\mu}}+\bar{\alpha}_{\mu}
\left(V\hat{\chi}^o_{\bN} V^{\dag}\right)\bar{\alpha}_{\mu}
\left(V\hat{\chi}^o_{\bN+2\hat{\mu}} V^{\dag} \right) \right]
\nonumber \\
&& \hskip 2cm
-4\Tr\left[\bar{\alpha}_{\mu}\Lambda_e\bar{\alpha}_{\mu}\Lambda_e
\left( \hat{\chi}^e_{\bN} \right)^2 +
\bar{\alpha}_{\mu}\Lambda_o\bar{\alpha}_{\mu}\Lambda_o \left(
V\hat{\chi}^o_{\bN} V^{\dag}\right)^2 \right]\bigg\rbrace.
\end{eqnarray}
Next we expand $\hat{\chi}=\chi^\eta\Gamma^\eta$, where
$\chi^{\eta}$ are real and $\Gamma^{\eta}$ are the hermitian
generators of $U(N)$, normalized to
\begin{equation}
\tr \left[ \Gamma^{\eta} \Gamma^{\eta'} \right] = \delta^{\eta \eta'}.
\end{equation}
The form~(\ref{chi_1}) of $\hat\chi$ implies that $\chi^{\eta}\neq
0$ for those generators whose elements
$\left(\Gamma^{\eta}\right)_{\alpha \beta}$ are non-zero for
$\alpha\in [1,N/2]$ and $\beta\in [m+1,N]$ or vice versa. Thus
\begin{equation}
E_{\text{nnn}}= \sum_{a\bN}\sum_{\eta,\eta'} \left[
\chi^{a\eta}_{\bN} \left(N_a\right)^{\eta\eta'}\chi^{a\eta'}_{\bN}
+ \sum_\mu \chi^{a\eta}_{\bN}
\left(M_{a\mu}\right)^{\eta\eta'}\chi^{a\eta'}_{\bN+2\hat{\mu}}\right],
\end{equation}
where
\begin{eqnarray}
\left(N_e\right)^{\eta \eta'}&=&-J_2\sum_\mu \Tr \left[
\Gamma^{\eta} \Gamma^{\eta'} \bar{\alpha}_\mu\Lambda_a
\bar{\alpha}_\mu \Lambda_a \right], \\
\left(N_o\right)^{\eta \eta'}&=&-J_2\sum_\mu \Tr \left[
V \Gamma^{\eta} \Gamma^{\eta'} V^\dag \bar{\alpha}_\mu\Lambda_a
\bar{\alpha}_\mu \Lambda_a \right], \\
\left(M_{e\mu}\right)^{\eta \eta'}&=&J_2 \Tr \left[ \Gamma^{\eta}
\bar{\alpha}_\mu \Gamma^{\eta'} \bar{\alpha}_\mu \right], \\
\left(M_{o\mu}\right)^{\eta \eta'}&=& J_2 \Tr \left[ V \Gamma^{\eta} V^\dag
\bar{\alpha}_\mu V \Gamma^{\eta'} V^\dag \bar{\alpha}_\mu \right].
\end{eqnarray}
With a Fourier transform,
\begin{equation}
\chi^{a\eta}_{\bN}=\sqrt{\frac2{N_s\beta}}\sum_{\bk\in
BZ\atop\omega} \chi^{a\eta}_k e^{i\bk \cdot \bN+i\omega \tau},
\label{FT}
\end{equation}
we write the energy in momentum space as
\begin{eqnarray}
E_{\text{nnn}}&=&\sum_{\bk \in BZ \atop \omega >0 }
\chi^{e\dag}_{k}  \left[N_e+N^T_e +
M_e(\bk)+M^\dag_e(\bk)\right]\chi^{e}_{k} \nonumber \\ &&\hskip
1cm+\chi^{o\dag}_{-k} \left[N_o+N^T_o +
M_o(-\bk)+M^\dag_o(-\bk)\right] \chi^{o}_{-k}. \label{SNNNmom}
\end{eqnarray}
Here $M_a(\bk)=\sum_\mu M_{a\mu} e^{ik_\mu}$.

The NN action, including the time derivative and $O(1/N_c)$
self-energy, was written down in Chapter~\ref{chap:Qfluctuation} in terms of the
Fourier transform $\tilde\chi$ of the $(2m-N)\times (N-m)$ matrix
field $\chi$ [\Eq{chi}]:
\begin{eqnarray}
S_{\text{nn}}&=&\sum_{\omega,\bk} \Tr \left[
\left(i\omega-\Sigma_{1,\bk}\right) \tilde\chi^{e\dag}_{k}
\tilde\chi^e_{k}+\left(-i\omega-\Sigma_{1,\bk}\right)
\tilde\chi^{o\dag}_{-k}
\tilde\chi^o_{-k} \right. \nonumber \\
&& \left. \hskip 1.4cm - \Sigma_{2,\bk} \left( \tilde\chi^e_{k}
\tilde\chi^{oT}_{-k}+c.c.\right) \right] .
\end{eqnarray}
The self-energies $\Sigma_a$ are of order $J_1/N_c$ and depend on
$N$ and~$m$.
We set $\chi_2=0$ and repeat the steps leading to \Eq{SNNNmom} to
write $S_{\text{nn}}$ in terms of $\chi_k^{\eta}$,
\begin{equation}
S_{\text{nn}}=\sum_k \chi^{e\dag}_k K_{ee}\chi^{e}_k +
\chi^{o\dag}_{-k} K_{oo} \chi^{o}_{-k} + \chi^{eT}_k
K_{eo}\chi^{o}_{-k}, \label{SNN}
\end{equation}
with the matrices $K_{ee}$, $K_{oo}$, and $K_{eo}$ given by
\begin{eqnarray}
\left( K_{ee} \right)^{\eta\eta'}&=&\frac{i\omega}2 \Tr \left[
\Lambda_e \Gamma^{\eta}\Gamma^{\eta'} \right] -\frac12
\Sigma_{1,\bk} \delta^{\eta\eta'}, \\
\left( K_{oo} \right)^{\eta\eta'}&=&-\frac{i\omega}2 \Tr \left[
\Lambda_e \Gamma^{\eta}\Gamma^{\eta'} \right] -\frac12
\Sigma_{1,\bk} \delta^{\eta\eta'}, \\
\left( K_{eo} \right)^{\eta\eta'}&=&-\Sigma_{2,\bk} \Tr \left[
\Gamma^{\eta}\Gamma^{\eta' T} \right].
\end{eqnarray}
Equation (\ref{SNN}) is to be added to \Eq{SNNNmom} to give the
quadratic action of the type II Goldstone bosons $\chi_1$.

Diagonalizing the quadratic form is straightforward but tedious.
As noted above, only a subset of the generators $\Gamma^\eta$ of $U(N)$
appear in the expansion of \Eq{chi_1}.
Since $\chi_1$ has dimensions $N/2\times(N-m)$, there are
$N(N-m)/2$ pairs of generators in the sum, which we write (similar to
the Pauli matrices $\sigma_x,\sigma_y$) as
$\tilde{\Gamma}_x^{\eta},\tilde{\Gamma}_y^{\eta}$, with $\eta=1,\dots,N(N-m)/2$.
Their coefficients are similarly written as
$\chi_x^{\eta},\chi_y^{\eta}$.
Thus for each $k,\eta$ we have
\begin{equation}
\chi^{\eta}_k=\left( \begin{array}{c}
\chi^{e\eta}_{xk} \\
\chi^{e\eta}_{yk} \\
\chi^{o\eta *}_{x,-k} \\
\chi^{o\eta *}_{y,-k}
\end{array} \right),
\end{equation}
and the action is
\begin{equation}
S=\sum_{\omega>0}\sum_{\bk} \chi^\dag_k \, G^{-1}_k \, \chi_k,
\end{equation}
The inverse propagator is 
\begin{equation}
G^{-1}_k = \left( \begin{array}{cccc}
-\Sigma_{1,\bk} + n_e +m_e  & -\omega   & -\Sigma_{2,\bk}   & 0 \\
\omega   & -\Sigma_{1,\bk} +n_e-m_e & 0 & \Sigma_{2,\bk}  \\
-\Sigma_{2,\bk}   & 0 & -\Sigma_{1,\bk} +n_o+m_o & -\omega   \\
0 & \Sigma_{2,\bk}   & \omega   & -\Sigma_{1,\bk} +n_o-m_o
\end{array} \right) \label{invG},
\end{equation}
where
\begin{eqnarray}
\left( n_a \right)^{\eta\eta'}&=&-2J_2\sum_{\mu} \tr \left[ (\Gamma_x^{\eta})^2 \bar{\alpha}_\mu\Lambda_a \bar{\alpha}_\mu \Lambda_a \right] \delta^{\eta\eta'}, \\
\left( m_a \right)^{\eta\eta'}&=&2J_2 \sum_\mu \tr \left[ \Gamma_x^{\eta} \bar{\alpha}_\mu \Gamma_x^{\eta'} \bar{\alpha}_\mu \right] \cos k_\mu.
\end{eqnarray}
$\Sigma_{1,2}$ and $\omega$ contain a factor of $\delta_{\eta,\eta'}$.
If we write the $4\times4$ matrix $G^{-1}_k$ in terms of $2\times2$ blocks,
\begin{equation}
G^{-1}_k=\left(
\begin{array}{cc}
A&B\\C&D
\end{array}\right),
\end{equation}
then its determinant is easily calculated via 
\begin{equation}
| G^{-1}_k| = |C|\,|B-AC^{-1}D|\nonumber\\[3pt]
\end{equation}
that gives the following determinant.
\begin{equation}
\left|\begin{array}{cc}
\omega^2+(n_e+m_e-\Sigma_1)(n_o+m_o-\Sigma_1)-\Sigma_2^2 & \omega(n_e+m_e-n_o+m_o) \\
\omega(n_o+m_o-n_e+m_e) & \omega^2+(n_e-m_e-\Sigma_1)(n_o-m_o-\Sigma_1)-\Sigma_2^2 \end{array} \right|.
\nonumber \\ \nonumber \\ \label{detG}
\end{equation}

For $N=4N_f\le12$ (and $m\ge3N/4$), the matrices
$n_e,n_o,m_e,m_o$ all commute, except for the case ($N=12$, $m=10$).  
Dropping this last from consideration, we are left with the values
of $(N_f,B)$ listed in Table \ref{table2}.
For each case, the simultaneous diagonalization of $n_{e,o}$ and $m_{e,o}$
gives the eigenvalues shown.
\begin{table}[htb]
\begin{center}
\caption[Eigenvalues of $n_{e,o}$ and $m_{e,o}$]{Simultaneous eigenvalues of $n_{e,o}$ and $m_{e,o}$ (in units of
$2J_2$) for all values of $N_f$ and $B$ considered.
The resulting spectra fall into four classes.
Here $v_z=\cos k_z$ and
$v_{\perp}=2\sin \left(\frac{k_x+k_y}2\right)\sin\left(\frac{k_x-k_y}2\right)$
\label{table2}}
\begin{tabular}{cccccccc}
$N_f$ & $B$ & $n_e$ & $n_o$ & $m_e$ & $m_o$& degeneracy & class \\
\hline
1&1&1&2&$ v_z    $&$ v_\perp$&         &4\\
 & &2&1&$ v_\perp$&$ v_z$    &         &4\\
2&2&1&2&$ v_z    $&$ v_\perp$&$\times3$&4\\
 & &2&1&$ v_\perp$&$ v_z$    &$\times3$&4\\
 & &1&2&$-v_z    $&$-v_\perp$&         &4\\
 & &2&1&$-v_\perp$&$-v_z$    &         &4\\
2&3&0&1&0        &$v_z$      &         &2\\
 & &1&0&$v_z$    &0          &         &2\\
 & &0&2&0        &$v_\perp$  &         &3\\
 & &2&0&$v_\perp$&0          &         &3\\
3&3&1&2&$ v_z    $&$ v_\perp$&$\times6$&4\\
 & &2&1&$ v_\perp$&$ v_z$    &$\times6$&4\\
 & &1&2&$-v_z    $&$-v_\perp$&$\times3$&4\\
 & &2&1&$-v_\perp$&$-v_z$    &$\times3$&4\\
3&5&0&0&0        &0          &$\times2$&1\\
 & &0&1&0        &$v_z$      &         &2\\
 & &1&0&$v_z$    &0          &         &2\\
 & &0&2&0        &$v_\perp$  &         &3\\
 & &2&0&$v_\perp$&0          &         &3\\
\end{tabular}
\end{center}
\end{table}

The zeros
of the determinant~(\ref{detG}) determine the spectrum $\omega(\bk)$, giving
(after $\omega\to i\omega$)
\begin{eqnarray}
\omega^2&=&{\Sigma_1}^2-{\Sigma_2}^2+\frac12\left(n^2_e-m^2_e+n^2_o-m^2_o\right)-\Sigma_1(n_e+n_o) \nonumber \\
&&\pm \Biggl\{\left[ \Sigma_1(n_o-n_e)+\frac12 (n^2_e-m^2_e-n^2_o+m^2_o)\right]^2
\nonumber\\
&&\qquad\qquad+{\Sigma_2}^2\left[(m_e+m_o)^2-(n_e-n_o)^2 \right]\Biggr\}^{1/2}.
\label{poles}
\end{eqnarray}
Because of the symmetries of \Eq{poles}, the spectra of the various cases
shown in Table \ref{table2} fall into four classes.
We examine each class in turn.

\paragraph*{Class 1:}
Here there is no contribution at all from the NNN interaction.  As shown
in \cite{odo}, 
$\Sigma_{1,2}$ are proportional to $J_1/N_c$ and for small momenta they
are quadratic in $|\bk|$; the same holds for the NN
energy $\sqrt{{\Sigma_1}^2-{\Sigma_2}^2}$.
These fields remain isotropic type II Goldstone bosons as in the NN theory.

\paragraph*{Class 2:}
Fields that correspond to the minus sign in \Eq{poles} remain type II, but with
anisotropic dispersion laws of the form
\begin{equation}
\omega^2=c^2\bk^4 \left(1+a\delta \frac{k_z^2}{k^2}\right). \label{case2a}
\end{equation}
The plus sign in \Eq{poles} gives a linear dispersion law, again anisotropic,
\begin{equation}
\label{poles_expand1}
\omega^2=4c_1J_2\left[k_x^2+k_y^2+(1+b\delta)k_z^2\right] \label{case2b}
\end{equation}
The anisotropy in both cases is proportional
to the ratio $\delta \equiv J_2/(12J_1/N_c)$ of NNN to NN couplings.
The coefficients $c$ and $c_1$
are defined as
\begin{eqnarray}
c&=&\left(\frac{d}{d\bk^2}\sqrt{{\Sigma_1}^2-{\Sigma_2}^2}\right)_{\bk=0}, \\
c_1&=&-\left(\frac{d\Sigma_{1}}{d\bk^2}\right)_{\bk=0}>0.
\end{eqnarray}
They are of order $J_1/N_c$.
The coefficients
\begin{equation}
a=\frac{N_c}{6J_1}\frac{c_1^2-c^2}{2c_1c^2}\quad{\text and}\quad
b=\frac{N_c}{6J_1}\frac2{c_1}
\end{equation}
are of order $10^2$ for most cases at hand. These coefficients as well as $c$ and $c_1$ are given in Appendix~\ref{app:constants}.

\paragraph*{Classes 3 and 4:}
Taking $\bk=0$ in \Eq{poles} we find that the fields that correspond to the
plus sign get a mass equal to $2J_2$. This is a result of the explicit
breaking of the $U(4N_f)$ symmetry by the NNN interaction terms; these
particles are no longer Goldstone bosons.
The massless bosons in Class 3, corresponding to the minus sign, are
type II bosons described by
\begin{equation}
\label{case3}
\omega^2 = c_1^2\bk^4.
\end{equation}
(This is different from Class 1 where $\omega^2=c^2\bk^4$.)
The massless bosons in Class 4 are again anisotropic, obeying
\Eq{case2b}, and are of type I.

These dispersion relations are correct to $O(\delta)$ for momenta
of $O(\delta)$ or smaller.
In all cases the dispersion relation to $O(1)$ for $\bk^2 \gg \delta$ is
quadratic and
isotropic, unchanged from the NN result presented in Chapter~\ref{chap:Qfluctuation}.
Since $\delta$ is a small parameter, this means that in most of the Brillouin
zone the propagator maintains its NN form.
This is the reason why the self consistent calculations in Chapter~\ref{chap:Qfluctuation} that
yield $\Sigma_{1,2}$ do not change when we add the NNN
interactions.

\clearpage

%
\chapter{Summary and discussion}
\label{chap:summary}

In this work we have studied the strong coupling limit of lattice QCD with non-zero baryon density. We used the Hamiltonian formalism with nearest-neighbor (NN) fermions and next-nearest-neighbor (NNN) fermions. Also we do not take the continuum limit, but rather we regard this Hamiltonian as an effective Hamiltonian that describes QCD at large distances.

Expanding in powers of the inverse coupling one obtains an effective Hamiltonian for color singlet objects. In this work we restrict ourselves to second order in the expansion, where the effective Hamiltonian is a generalized antiferromagnet for $U(4N_f)$ spins. It has NN and NNN interactions, which are invariant under $U(4N_f)$ and $U(N_f)\times U(N_f)$ respectively. Also at this order the {\em local} baryon number $B_\bn$ is conserved and one diagonalizes the antiferromagnet for a fixed background of baryon number, which fixes the representations of the $U(4N_f)$ spins.  

When we write the partition function of the effective Hamiltonian with boson coherent states, it becomes tractable in the limit of large $N_c$ and $N_f$, with the ratio $r=N_c/N_f$ kept fixed. In this limit we extend Arovas and Auerbach's result for zero density and dimension $d=2$ to three dimensions. We find that for $r>r_c$, the antiferromagnet is in a N\'eel ordered phase, which corresponds to a phase with spontaneous broken chiral symmetry. The value of $r_c$ in three dimensions is found to be $0.31$.
When we analyze this limit for a periodic configuration of baryon number, where a concentration $c=\frac18$ and $c=\frac14$ of the lattice sites is saturated with baryons, we find that $r_c$ is $0.35$ and $0.47$ respectively. This means that the N\'eel phase is shrinking with increasing density. 

For a random distribution of sites saturated with baryons, we average over the sites' positions in a quenched approximation. This is valid only if the concentration $c$ of these sites obeys $c\ll 1$, and we calculate the free energy to first order in the concentration. Comparing this to the unquenched calculations at $c=\frac18,\,\frac14$, we find that the two calculations deviate near the phase transition point $r=r_c$. In fact, in the quenched calculation, $r_c$ decreases with increasing density.

Despite this deviation, we note that the two approaches are in good agreement far from $r_c$. This opens the way to the systematic investigation of the system's properties with the quenched approach. For example, relying on the condensed matter analog of magnetic impurities, low lying excitations should develop non-zero decay widths and decreased velocities for non-zero density. Note that in Section~\ref{sec:quenched} we were concerned with the disordered phase, where the excitations are massive. In certain cases, however, there exists a Holstein-Primakoff transformation for the generalized spins in terms of bosons (see Appendix~\ref{app:Hamiltonian_odo}). Using this transformation and taking the limit of large $N_c$ and fixed $N_f$, one can perform the quenched calculation in the ordered phase, on lines similar to Section~\ref{sec:quenched}. It will be interesting therefore to see how the Goldstone bosons decay and slow down in the presence of baryons.

Next we wrote the partition function using generalized spin coherent states to obtain a nonlinear sigma model. In Section~\ref{sec:classical} we analyzed the classical counterpart of the sigma model by dropping the Berry phase term. This analysis presumably describes the high temperature region of the system, where quantum fluctuations are not very important. It also serves as a tool to see how the symmetry realization changes as we add baryons to the lattice. Although the classical sigma model can be treated numerically (its action is real), we postponed this to future research and used mean field theory instead for a toy model that has $N=3$. In that case we extracted the {\em classical} phase diagram in temperature-density plane. We found that $U(3)$ can be broken either to $U(2)\times U(1)$ or to $U(1)\times U(1)\times U(1)$, depending on the density. The order of the phase transition also depends on the density, as does the critical temperature, which decreases with increasing density.

We studied the full quantum sigma model in its semiclassical limit of large $N_c$ and finite $N_f$.
In the vacuum sector, we rederived Smit's result for the
lowest-energy configuration of alternating $B_\bn=\pm(m-N/2)$ sites. The classical degeneracy in the number $m$ is removed by quantum fluctuations of $O(1/N_c)$, and the ground state indeed has $B_\bn=0$; the $U(4N_f)$ symmetry of the nearest-neighbor
theory is spontaneously broken to $U(2N_f)\times U(2N_f)$.
We extended this result to the \nnn\ theory and found that its
$U(N_f)\times U(N_f)$ chiral symmetry is broken to the vector
$U(N_f)$ flavor subgroup. We also found that the classical degeneracy of the nearest-neighbor interactions can be removed by the \nnn\ terms themselves, with no quantum corrections. These two procedures for lifting the degeneracy led to the same ground state.

Adding net baryon number to the system, we examined the case of
uniform baryon density, $B_\bn=B=m-2N_f>0$. 
When we limit ourselves to the nearest-neighbor theory, we
find a N\'eel-like ground state that breaks
$U(N)$ to $U(2m-N)\times U(N-m) \times U(N-m)$.
The number of broken generators $n_{G}$ thus depends on the baryon number $B$ as 
\begin{equation}
n_{G}=2(2N_f-B)(2N_f+3B).
\end{equation}

When we calculate the low energy excitations of the NN theory, we find two types of Goldstone bosons \cite{Nielsen:hm,Leutwyler:1993gf}. The type I
bosons are $n_I=2(2N_f-B)^2$ antiferromagnetic spin
waves that have a linear dispersion relation. We also have $n_{II}=4B(2N_f-B)$ type II bosons\footnote{Recall that the local baryon number $B$ ranges from $0$ to $2N_f$ with unit increments.}. The numbers $n_I$, $n_{II}$, and $n_G$ obey the Nielsen-Chadha counting rule
\begin{equation}
n_I+2n_{II}\ge n_G.
\end{equation}

We call the type II bosons ferromagnetic magnons since they emerge due to a ferromagnetic alignment of \nnn spins. As typical for magnons, their energy is quadratic in momentum\footnote{A possible exception is for $N-m=1$, where the energy may depend on a different power of the momentum. Correspondingly, the number of excitations will be
  different.} and of $O(J_1/N_c)$, where $J_1$ is the coupling of the nearest-neighbor interactions.

The mechanism that removed the classical degeneracy at zero and non-zero density, and gave rise to type II Goldstone bosons in the latter case, is known in the
 condensed
matter literature as order from disorder.
 It takes place in systems that possess a classically degenerate ground state. For example we mention the double exchange model \cite{double_Xchange} and the Kagom{\' e} antiferromagnet \cite{Kagome}. There, the classical ground state energy is invariant under a rotation of local groups of spins, making it exponentially degenerate as in our problem. 

Finally when we perturb the nearest-neighbor ground state with the NNN interactions, we find that some of the global degeneracy of the nearest-neighbor ground state is removed, and that the vacuum expectation values of the sigma fields break chiral symmetry in a way that depends on $N_f$ and the baryon number. In all cases with non-zero $B$, discrete lattice rotations are spontaneously broken as well. We summarize the patterns of spontaneous symmetry breaking in Table~\ref{table1}. 

In comparing Table~\ref{table1} to results of continuum CSC calculations,
one must keep in mind that we study systems with large, fixed, and
discrete values of $B$, rather than with large, continuous $\mu$.
Moreover, we use large-$N_c$ approximations which necessarily ignore the
discrete properties of the $SU(N_c)$ group that are essential to
baryons. 

The values of $(N_c,N_f)$ that are of interest for CSC are $N_c=3$
and~$N_f=2$ or~3.  In the two-flavor the ground state does not break chiral symmetry. We do not
see this for any density at $N_f=2$.  Our results are due to a $\bar qq$ condensate; whether there is a $qq$ condensate as well cannot be ascertained.

For $N_f>2$ the situation is similar.  The CSC ground state for $N_c=3$ and $N_f\ge 3$ partially breaks color and flavor, with a condensate that locks one or more subgroups of the flavor group to the color group \cite{Schafer}. Since we work at large $N_c$, we should stand the argument on its head. A plausible $qq$ condensate would lock successive subgroups of the
{\em color\/} group to the flavor group and hence to each other,
leaving unbroken the diagonal $SU(N_f)_{L+R+C}$ and some leftover color symmetry.  Judging by the global symmetry of the vacuum, perhaps we see this for $(N_f=3,B=3)$.  The other cases could conceivably arise from a combination of $\bar qq$ and~$qq$ condensates, but whether the latter actually occur is an open question.


Finally we note that according to diagrammatic power-counting arguments
\cite{DGR,ShusterSon}, CSC is suppressed in the 't~Hooft limit, where $N_c\to\infty$ and $g^2N_c$ is kept fixed. In the perturbative regime of small $g^2N_c$, a Fermi sea prefers pairing of particle and holes that gives a chiral density wave with momentum 
$|\bp|\simeq 2\mu$, where $\mu$ is the chemical potential. Our ansatz for the ground state does not permit this possibility.



When we determined how the NNN terms change the dispersion relations for the Goldstone bosons we found that the physics of the NN theory is mostly undisturbed by the NNN interaction. 
At leading order, the properties of the type I bosons ($\pi$) and of some of the type II bosons ($\chi_2$) do not change.
The type II bosons grouped in the $\chi_1$ field suffer a variety of fates,
falling into four classes that appear for different values of
$N_f$ and $m$.
Class 1 bosons are unaffected by the NNN perturbation.
Class 2 bosons split into type I and type II, and all 
gain anisotropic contributions of $O(J_2/\left(J_1/N_c\right))$ to
their energies. Here $J_2$ is the coupling of the NNN interactions in the antiferromagnet.
Some of the Class 3 bosons become massive while other remain unaffected.
In Class 4, some become massive while the others become anisotropic
type I bosons.

The symmetry of the theory, in all cases, is severely broken by the NNN terms---from
$SU(4N_f)$ to $SU(N_f)\times SU(N_f)\times U(1)_A$.
Not surprisingly,
a simple count shows that the total number of massless real fields is far
greater than the number of spontaneously
broken generators of $SU(N_f)\times SU(N_f)\times U(1)_A$,
as shown in Table~\ref{table1}.
The particular NNN interaction we use is simply unable to generate masses
{\em in lowest order\/} for many of the
particles unprotected by Goldstone's Theorem.
This is partly reflected in the accidental
degeneracy of the ground state, which we mentioned below
\Eq{eq:Uo}.
Just as this degeneracy should be lifted in higher orders in $1/N_c$
[beginning with $O(J_2/N_c)$], the corresponding massless excitations
should develop masses.
The only particles protected from mass generation are the minimal number
needed to satisfy Goldstone's theorem (or the Nielsen-Chadha variant).

Another effect that is missing is the mixing of type I and type II Goldstone
bosons, which is certainly permitted when the NNN interaction is turned on.
In the NN case we proved that such a mixing is forbidden, since
the two types of boson belong to different representations of the unbroken
subgroup.
 We present the classification in the NNN theory in Appendix~\ref{app:clasification}, where it is easy to see that it is less restrictive, and permits mixing
of the bosons. Whether mixing occurs is a dynamical issue that can be
settled only by
calculating to higher order in $1/N_c$. 

To conclude we note that other recent work \cite{Continuum_type2,Sannino} on the high
density regime of QCD---in the continuum---also predicts type II Goldstone bosons and anisotropic dispersion. There, the starting point is 
an effective field theory that describes the low energy dynamics of QCD with non-zero chemical potential $\mu$. For $\mu\neq0$, Lorentz invariance is broken, and the field equations become nonrelativistic. This leads to the emergence of type II Goldstone bosons. In addition, the ground state in \cite{Sannino} can support a non-zero expectation value of vector fields. This breaks rotational
symmetry and makes some of the dispersion relations anisotropic.

Possible directions of future research begin with gaining more freedom in fixing the baryon configurations. In this work we restricted ourselves to either having sites saturated with baryons (in the boson coherent states approach) or with having the same baryon number on all sites (in the nonlinear sigma model approach). It is interesting to see what is the ground state of the antiferromagnet for other baryon configurations, that have $2N_f>B_\bn>0$ only on a sparse sublattice, along the lines of Section~\ref{sec:classical}. This can be easily investigated for the classical counterpart of the sigma model, whose action is real, by Monte Carlo simulations. A general baryon distribution has a classical degeneracy which may be removed by order from disorder. This can be seen by calculating $1/N_c$ corrections that correspond to quantum order from disorder, as in Chapter~\ref{chap:Qfluctuation}. Another way to lift the degeneracy is with thermal fluctuations. These can be studied in the classical counterpart of the sigma model as well.

Another interesting direction is to fix the chemical potential $\mu$ instead of the baryon number. This requires calculating the grand canonical partition function $Z_{GC}(\mu)$, which is related to the canonical partition function $Z(\{B_\bn\})$, treated in this work, by 
\begin{equation}
Z_{GC}=\sum_{\{B_\bn\}} e^{\beta \mu \sum_\bn B_\bn}Z(\{B_\bn\}).
\end{equation}

This will make the comparison to results of continuum and other strong coupling calculations more straightforward. Again a first step can be to calculate $Z_{GC}$ for the classical sigma model. Also it will be interesting to see how order from disorder manifests itself in the grand canonical ensemble. 

An ultimate goal is the incorporation of the third-order term in the effective Hamiltonian in order to have a theory with dynamical baryons. This Hamiltonian looks like the $t$--$J$ model of condensed matter physics, but is in fact more complicated since the baryon operators do not obey canonical commutation relations. An instructive half-measure would be to study the antiferromagnet in the presence of a disordered baryon background. 

One can also replace the baryon operators with simpler operators that mimic only some of the baryons' features. Reintroducing all the properties of the real operators can be done using methods of Hilbert space mapping (see for example \cite{Meyer:mu}), which were introduced in the context of atomic and nuclear physics to treat quantum fields that do not obey canonical commutation relations. 
\clearpage
%
%
\renewcommand{\chaptermark}[1]{\markboth{\appendixname\ \thechapter.\ #1}{}} 
\renewcommand{\sectionmark}[1]{\markright{\thesection.\ #1}}   
\fancyhead[RE]{\bfseries\leftmark}      
\fancyhead[LO]{\bfseries\rightmark}     
\appendix
\chapter{Anticommutation relations of baryon operators\label{app:anticomm}}

In this appendix we present the calculation of the anticommutation relations of the baryon composite operators introduced in Chapter~\ref{chap:Heff}. For simplicity we slightly deviate from the notations of Chapter~\ref{chap:Heff} and denote color indices by $a,b,c,\dots=1,2,3$ and Dirac-flavor indices by $i,j,k,\dots=1,\dots,4N_f$. The baryon operators are given by
\begin{equation}
b_{ijk}=\epsilon_{abc} \psi^a_i \psi^b_j \psi^c_k,
\end{equation}
where $\psi^a_i$ are fermion operators obeying 
\begin{equation}
\left\{ \psi^a_i , \psi^{\dag b}_j \right\} = \delta_{ab} \delta_{ij}. \label{anti0}
\end{equation}
We wish to calculate the following anticommutation relations of the baryon operators. Note that we are only interested in operators residing on the same site, and only in the anticommutators between $b$ and $b^\dag$. All other relations are trivial.

We wish to calculate 
\begin{eqnarray}
\left\{ b^\dag_{ijk} , b_{mnl} \right\} &=& \epsilon_{abc} \epsilon_{a'b'c'} \left( \psi^{\dag c}_k \psi^{\dag b}_j \psi^{\dag a}_i \psi^{a'}_{m} \psi^{b'}_{n} \psi^{c'}_{l}+\psi^{a'}_{m} \psi^{b'}_{n} \psi^{c'}_{l} \psi^{\dag c}_k \psi^{\dag b}_j \psi^{\dag a}_i \right). \label{anti1}
\end{eqnarray}
We now concentrate on the second term, which we want to bring to a form similar to the first term, with a minus sign. The two will then cancel, and leave the expression we are after. We do that by moving the $\psi^\dag$'s in the second term of \Eq{anti1} to the left, and have 
\begin{eqnarray}
\psi^{a'}_{m'} \psi^{b'}_{n'} \psi^{c'}_{l'} \psi^{\dag c}_k \psi^{\dag b}_j \psi^{\dag a}_i &=& \left( \psi^{a'}_m \psi^{\dag c}_k \psi^{\dag b}_j \psi^{b'}_{n} \psi^{c'}_{l}\psi^{\dag a}_{i}+\psi^{a'}_{m} \left[ \psi^{b'}_{n} \psi^{c'}_{l} , \psi^{\dag c}_k \psi^{\dag b}_j \right] \psi^{\dag a}_i \right) \nonumber \\
&=&\left(  -\psi^{\dag c}_k \psi^{a'}_m + \delta_{km} \delta_{a'c}  \right)\psi^{\dag b}_j \psi^{b'}_n \left( -\psi^{\dag a}_i \psi^{c'}_l + \delta_{il} \delta_{ac'} \right) \nonumber \\ 
&&+ \psi^{a'}_{m} \left( \psi^{b'}_{n} \left[ \psi^{c'}_{l} , \psi^{\dag c}_k \psi^{\dag b}_j \right] + \left[ \psi^{b'}_{n} , \psi^{\dag c}_k \psi^{\dag b}_j \right] \psi^{c'}_{l}  \right) \psi^{\dag a}_i \nonumber \\
&=&\psi^{\dag c}_k \left( -\psi^{\dag b}_j\psi^{a'}_m+\delta_{mj}\delta_{a'b} \right)\left(-\psi^{\dag a}_i \psi^{b'}_n + \delta_{ni}\delta_{b'a} \right)\psi^{c'}_l \nonumber \\
&&-\delta_{km}\delta{a'c}\psi^{\dag b}_j \psi^{b'}_n \psi^{\dag a}_i \psi^{c'}_l - \psi^{\dag c}_k \psi^{a'}_m \psi^{\dag b}_j \psi^{b'}_n \delta_{il}\delta_{ac'} \nonumber \\
&&+\delta_{a'c}\delta_{km}\delta_{il}\delta_{ac'}\psi^{\dag b}_j \psi^{b'}_n + \psi^{a'}_m\left[ \psi^{b'}_n\left( \delta_{kl}\delta_{cc'}\psi^{\dag b}_j-\delta_{lj}\delta_{c'b}\psi^{\dag c}_k \right) \right. \nonumber \\
&& \left. + \left( \delta_{nk}\delta_{b'c} \psi^{\dag b}_j -\delta_{nj} \delta_{b'b} \psi^{\dag c}_k \right)\psi^{c'}_l \right] \psi^{\dag a}_i.
\label{anti2}
\end{eqnarray} 
Moving $\psi^{\dag a}_i$ in the first term of \Eq{anti2} leftward, we cancel the first term in \Eq{anti1} and obtain
\begin{eqnarray}
\left\{ b^\dag_{ijk} , b_{mnl} \right\} &=& \epsilon_{abc} \epsilon_{a'b'c'} \left[ \psi^{\dag c}_k \psi^{\dag b}_j \psi^{b'}_{n} \psi^{c'}_{l} \delta_{im} \delta_{aa'}- \psi^{\dag c}_k \psi^{\dag b}_j  \psi^{a'}_{m} \psi^{c'}_{l} \delta_{in}\delta_{b'a}
\right. \nonumber \\
&& - \psi^{\dag c}_k \psi^{\dag a}_i \psi^{b'}_{n} \psi^{c'}_{l} \delta_{jm} \delta_{ba'} + \psi^{\dag c}_k \psi^{c'}_l \delta_{in} \delta_{jm} \delta_{ab'} \delta_{a'b} \nonumber \\ 
&& - \psi^{\dag b}_j \psi^{b'}_n \psi^{\dag a}_{i} \psi^{c'}_{l} \delta_{km} \delta_{ca'} - \psi^{\dag c}_k \psi^{a'}_m \psi^{\dag b}_{j} \psi^{b'}_{n} \delta_{il} \delta_{ac'} \nonumber \\
&& + \psi^{\dag b}_j \psi^{b'}_n \delta_{a'c} \delta_{km} \delta_{il} \delta_{ac'} + \psi^{a'}_m \psi^{b'}_n \psi^{\dag b}_{j} \psi^{dag a}_{i} \delta_{kl} \delta_{cc'} \nonumber \\
&& - \psi^{a'}_m \psi^{b'}_n \psi^{\dag c}_{k} \psi^{dag a}_{i} \delta_{jl} \delta_{bc'} + \psi^{a'}_m \psi^{b}_j \psi^{c'}_{l} \psi^{\dag a}_{i} \delta_{nk} \delta_{b'c} \nonumber \\
&& \left. - \psi^{a'}_m \psi^{\dag c}_k \psi^{c'}_{l} \psi^{\dag a}_{i} \delta_{nj} \delta_{bb'} \right].
\label{anti3}
\end{eqnarray}
Using the $\delta$ functions and the two $\epsilon$'s in front of the brackets in \Eq{anti3}, we have the following result for the anti-commutation relations,
\begin{eqnarray}
\left\{ b^\dag_{ijk} , b_{mnl} \right\} &=& 6 \delta_{(ijk,lnm)} -2\left( \delta_{kl} \delta_{im} Q_{jn}+\delta_{kl}\delta_{jm}Q_{in} + \delta_{jl}\delta_{km}Q_{in}\right) \nonumber \\
&&-3\left( \delta_{km} \delta_{il} Q_{jn}+\delta_{km}\delta_{in}Q_{jl} + \delta_{km} \delta_{in} Q_{jl}+\delta_{kl}\delta_{jn}Q_{im} \right. \nonumber \\
&&\left. +\delta_{kl} \delta_{in} Q_{jm}+\delta_{jl}\delta_{kn}Q_{im} + \delta_{nk} \delta_{il} Q_{jm}+\delta_{nk}\delta_{im}Q_{jl} + \delta_{jl}\delta_{im}Q_{kn} \right) \nonumber \\
&&-4\left( \delta_{im} \delta_{jn} Q_{kl}+\delta_{in}\delta_{jl}Q_{km} + \delta_{in} \delta_{jm} Q_{kl}+\delta_{jm}\delta_{il}Q_{kn} \right. \nonumber \\
&&\left. +\delta_{il} \delta_{jn} Q_{km}+\delta_{nk}\delta_{jm}Q_{il} + \delta_{nj} \delta_{km} Q_{il} \right) \nonumber \\
&& + \delta_{im}\left( Q_{kl}Q_{jn}+Q_{kn}Q_{jl}\right) + \delta_{in}\left( Q_{kl}Q_{jm}+Q_{km}Q_{jl}\right) \nonumber \\
&& + \delta_{il}\left( Q_{km}Q_{jn}+Q_{kn}Q_{jm}\right) + \delta_{jm}\left( Q_{kl}Q_{in}+Q_{kn}Q_{il}\right) \nonumber \\
&& + \delta_{jn}\left( Q_{km}Q_{il}+Q_{kl}Q_{im}\right) + \delta_{jl}\left( Q_{kn}Q_{im}+Q_{in}Q_{km}\right) \nonumber \\
&& + \delta_{kn}\left( Q_{jm}Q_{il}+Q_{jl}Q_{im}\right) + \delta_{km}\left( Q_{jn}Q_{il}+Q_{jl}Q_{in}\right) \nonumber \\
&& + \delta_{kl}\left( Q_{jn}Q_{im}+Q_{in}Q_{jm}\right). \label{anti4}
\end{eqnarray}
Here we define 
\begin{eqnarray}
\delta_{(ijk,mnl)}&=&\delta_{kl}\delta_{jm}\delta_{in} + \delta_{kl}\delta_{jn}\delta_{im} + \delta_{jl}\delta_{km}\delta_{in}+ \delta_{jl}\delta_{kn}\delta_{im}+\delta_{kn}\delta_{jm}\delta_{il}+\delta_{nj}\delta_{km}\delta_{il}, \nonumber  \\
Q_{ij}&=&\sum_{a=1}^3\psi^{\dag a}_i \psi^a_j.
\end{eqnarray}
We note in passing that in order to get \Eq{anti4} we used the following relations
\begin{eqnarray}
\psi^a_i \psi^{\dag a}_j &=& - \psi^{\dag a}_j \psi^a_i + 3\delta_{ij}, \\
\left[ Q_{ij} , \psi^a_k \right] &=& - \delta_{ik} \psi^a_j, \\
\psi^{\dag c}_k Q_{ij} \psi^c_l &=& Q_{kl}Q_{ij}-\delta_{il} Q_{kj}, \\
\psi^{c}_k Q_{ij} \psi^{\dag c}_l &=& -Q_{ij}Q_{lk}+3\delta_{lk} Q_{ij} + 3\delta_{ik}\delta_{lj}-\delta_{ik} Q_{lj}
\label{anti5}
\end{eqnarray}
As mentioned in Chapter~\ref{chap:Heff}, the baryon operators are {\it not} canonical anticommuting operators, and their anti-commutation relations obey $\left\{ b_I , b^\dag_{I'} \right\} = \Delta_{II'} \neq \delta_{II'}$.
\clearpage
\chapter{Discretization of Euclidean time\label{app:discretization}}

In this section we stress the importance of the slicing prescription one uses for Euclidean time. In particular we work with a nearest-neighbor action defined on a bipartite lattice. We show the difference between using the exact prescription that results from the coherent state formalism of Section~\ref{sec:schwinger}, and a naive prescription.

The coherent state formalism results in an action of the form
\begin{eqnarray}
S &=& \sum_{\tau=1}^{N_\tau} \left\{ \sum_{\bn}  z^{e*}_{\tau,\bn} \left( z^e_{\tau,\bn} - z^e_{\tau+1,\bn} \right) + \epsilon \lambda z^{e*}_{\tau,\bn} z^e_{\tau+1,\bn}  \right. \nonumber \\
&&\hskip 1cm + \sum_{\bmm}  z^{o*}_{\tau,\bmm} \left( z^o_{\tau,\bmm} - z^o_{\tau+1,\bmm} \right) + \epsilon \lambda z^{o*}_{\tau,\bmm} z^o_{\tau+1,\bmm} \nonumber \\
&&\left. \hskip 1cm- \epsilon \sum_{\bn,\bmm} z^e_{\tau,\bn} Q_{\bn\bmm} z^o_{\tau,\bmm} + z^{o*}_{\tau,\bmm} Q^*_{\bmm\bn} z^{e*}_{\tau,\bn} \right\} \nonumber \\
&\equiv& f^* G^{-1} f.
\end{eqnarray}
Here the index $\bn(\bmm)$ runs from $1$ to $N_e(N_o)$ and denotes the sites on the even(odd) sublattice. $f$ has two indices; $\tau=1,\dots,N_\tau-1$, and a site index $\bn$ that takes values from $1$ to $N_e$ or $N_o$,
\begin{equation}
f_{\tau,\bn} = \left( \begin{array}{c} z^e_{\tau,1} \\ \vdots \\ z^e_{\tau,N_e} \\ \hline z^{o*}_{\tau,1} \\ \vdots \\ z^{o*}_{\tau,N_o} \end{array} \right).
\end{equation}
In this basis the inverse propagator $G^{-1}$ is 
\begin{equation}
G^{-1}=\left( \begin{array}{c|c} -\hat{\partial} + \lambda \hat{D} & Q \\ \hline Q^{\dag} & -\hat{\partial}^\dag + \lambda \hat{D}^\dag \end{array} \right), \label{eq:invG_dis}
\end{equation}
where the diagonal blocks have dimensions $N_e N_\tau$, and $N_o N_\tau$. The matrices $\hat{\partial}$, and $\hat{D}$ are 
\begin{equation}
\hat{\partial}_{\tau,\tau '} = \delta_{\tau ,\tau '-1} - \delta_{\tau ,\tau '}, \hskip 3cm \hat{D}_{\tau,\tau '} = \delta_{\tau ,\tau'-1}.
\end{equation}
times the unity in the lattice indices. $\hat{Q}$ is a unit matrix in Euclidean time indices multiplied by an $N_e\times N_o$ matrix that describes the interactions between the two sublattices. Note that we do not take $N_e=N_o$, to account for systems with vacancies. In order to calculate the determinant of the propagator, it is convenient to work in Matsubara space. There we have 
\begin{equation}
\hat{\partial}_{n,n'} = \delta_{n,n'} (-1+e^{-i\omega_n\epsilon}), \hskip 3cm \hat{D}_{n,n'} = e^{-i\omega_n\epsilon},
\end{equation}
and the determinant of \Eq{eq:invG_dis} becomes
\begin{eqnarray}
\det G^{-1} &=& \prod_{n=-\infty}^{\infty} \det \left( \begin{array}{cc} c_n {\bm 1}_{N_e} & \epsilon Q \\ \epsilon Q^\dag & c^*_n {\bm 1} \end{array} \right) \nonumber \\
&=& \prod_n \det \left( c^*_n {\bm 1}_{N_o} \right) \det \left( c_n {\bm 1}_{N_e} - \epsilon^2 Q (c^*_n)^{-1} {\bm 1} Q^\dag \right) \nonumber \\
&=& \prod_n \left( c^*_n \right)^{N_o-N_e} \det \left( |c_n|^2 {\bm 1}_{N_e} - \epsilon^2 Q Q^\dag \right).
\end{eqnarray}
Here $c_n=(1-e^{-i\omega_n\epsilon}) + e^{-i\omega_n\epsilon} \epsilon \lambda$. Next we diagonalize the $N_e\times N_e$ matrix $QQ^\dag$ and write the determinant as
\begin{equation}
\det G^{-1} = \prod_n |c_n|^{N_o-N_e} \prod_{\alpha=1}^{N_e} \left( |c_n|^2-\epsilon^2 q^2_\alpha \right). \label{eq:invdet}
\end{equation}
Here $q^2_\alpha$ are the $N_e$ eigenvalues of $QQ^\dag$. 

We now calculate the determinant for the continuum naive prescription, where the time derivative $\hat{\partial} \rightarrow -i \omega_n \epsilon \delta_{n,n'}$, and $\hat{D}\rightarrow \delta_{n,n'}$. In this case we have $c_n = \epsilon (i\omega_n + \lambda)$, and the products in \Eq{eq:invdet} become
\begin{eqnarray}
\prod_n |c_n| &\propto& \exp \left[ \frac12 \sum_n \log \left( \omega_n^2+\lambda^2 \right) \right], \\
\prod_n \left( |c_n|^2 - \epsilon^2 q_\alpha^2 \right) &\propto& \exp \left[ \sum_n \log \left( \omega_n^2+\omega_\alpha^2 \right)\right].
\end{eqnarray}
Using the following relation \cite{Kapusta}
\begin{equation}
\sum_{n=-\infty}^{\infty} \log \left( \omega_n^2+E^2 \right) = - 2 \log 2\sinh \beta E/2 + \text{constant}, \label{eq:kapusta}
\end{equation}
where the constant depends only on $\beta$, we find
\begin{equation}
\det G^{-1}_{\text{continuum}} = \left( \frac{1}{2\sinh \left(\beta\lambda/2 \right) } \right)^{N_o-N_e} \prod_{\alpha} \left( \frac{1}{2\sinh \left( \beta\omega_\alpha /2 \right)} \right)^2.
\end{equation}

Returning to the exact prescription we see that the products in \Eq{eq:invdet} become
\begin{eqnarray}
\prod_n |c_n| &\propto& \exp \left[ \frac12 \sum_n \log \left( 4(1-\epsilon \lambda) \sin^2 (\omega_n\epsilon/2) +\epsilon^2\lambda^2 \right) \right], \\
\prod_n \left( |c_n|^2 - \epsilon^2 q_\alpha^2 \right) &\propto& \exp \left[ \sum_n \log \left( 4(1-\epsilon \lambda) \sin^2 (\omega_n\epsilon/2)+\epsilon^2 \omega_\alpha^2 \right)\right].
\end{eqnarray}
We separate the sums of the form given above to
\begin{equation}
\sum_n \log \left( 4(1-\epsilon \lambda) \sin^2 (\omega_n\epsilon/2)+\epsilon^2 X^2 \right) = -\beta \lambda + \sum_n \log \left( 4\sin^2 (\omega_n\epsilon/2)+\epsilon^2 Y^2 \right). \label{eq:step1}
\end{equation}
Here $Y^2=X^2/(1-\epsilon \lambda)$. We have verified numerically that the sum on the right hand side of \Eq{eq:step1} is given by the sum in \Eq{eq:kapusta}, by replacing $E$ with $X$. One can understand this point as follows. Divide the sum into two,
\begin{equation}
\sum_n \log \left( 4\sin^2 (\omega_n\epsilon/2)+\epsilon^2 Y^2 \right) = \sum_{|n|\le m} + \sum_{|n|>m},
\end{equation}
where $m\ll N_\tau$. In the first sum we replace $4\sin^2 (\omega_n \epsilon/2)$ with $\omega^2_n$, and $Y$ with $X$. For $N_\tau \rightarrow \infty$, $m$ becomes very large, and since this sum is convergent, one can replace $m$ with $\infty$. 

The second sum the $\sin^2(\omega_n \epsilon /2)$ term is no longer of $O(\epsilon^2)$, and $Y^2$ can be dropped. Collecting only the $X$ dependent term we are left with a sum of the form of \Eq{eq:kapusta}. This gives
\begin{equation}
\det G^{-1} = \left( \frac{1}{1-e^{-\beta \lambda}} \right)^{N_o-N_e} \prod_{\alpha} \left( \frac{e^{\beta\lambda/2}}{2\sinh \left( \beta\omega_\alpha /2 \right)} \right)^2, \label{eq:invdet1}
\end{equation}
with $\omega_\alpha = \sqrt{\lambda^2-q^2_\alpha}$.

As a result we see that the effective action received after integrating the bosons $S_{\text{eff}}\sim \beta^{-1} \log \det G^{-1}$ obeys
\begin{equation}
S^{\text{exact}}_{\text{eff}}=S^{\text{continuum}}_{\text{eff}} - (N_e+N_o) \lambda /2. \label{eq:thumb}
\end{equation}

For practical reasons it is more convenient to perform calculations with the naive prescription of $\hat{\partial} \rightarrow -i\omega$, and in order to restore the correct result we use the rule of thumb given in \Eq{eq:thumb}.
\clearpage
\chapter{Minimizing the next-nearest-neighbor term \label{app:nnn}}
\label{sec:App1}

The $s^\mu_\eta$ signs are defined only when the $\sigma^\eta$ are
written in the basis
$M^\eta=\Gamma^A\otimes\lambda^a=\rho^\alpha\otimes\sigma^\beta
\otimes\lambda^a$. We choose a chiral basis for the gamma
matrices, so that $\gamma_5=\rho^3$, $\alpha_i=\rho^3\sigma^i$,
and $\beta=\rho^1$. The energy (\ref{epsprime}) is a sum of
squares,
\begin{equation}
\epsilon'=\sum_{\eta} A_\eta \left(\sigma^\eta\right)^2,
\end{equation}
with the constraint $\sum_\eta (\sigma^\eta)^2=N/2$. The
coefficients $A_\eta=\sum_\mu s^\mu_\eta$ take on the values
$\{-3,-1,1,3\}$. The minimum of $\epsilon'$ occurs when all
$\sigma^\eta$ are zero except those corresponding to $A_\eta=-3$,
namely, those for
$M^\eta=\beta\otimes\lambda^a=\rho^1\otimes\lambda^a$ and
$M^\eta=\beta\gamma_5\otimes\lambda^a=\rho^2\otimes\lambda^a$; the
energy is independent of these $\sigma^\eta$, and will be equal to 
\begin{equation}
\epsilon'=-3\sum_\eta \left(\sigma^\eta\right)^2 = -3N/2.
\end{equation}
Thus the set of
solutions can be written in the form
\begin{equation}
\sigma_0=\left(
\begin{array}{cccc}
0&0&U&0\\
0&0&0&U\\
U^{\dag}&0&0&0\\
0&U^{\dag}&0&0
\end{array}
\right) =\frac{\rho^1+i\rho^2}2\otimes U
+\frac{\rho^1-i\rho^2}2\otimes U^{\dag}.
\end{equation}
Recalling that $\sigma_0^2=1$, we have $UU^\dag=\bm1$, so $U\in
U(N_f)$. A chiral rotation $\sigma_0\to V^{\dag}\sigma_0V$, with
\begin{equation}
V= \left( \begin{array}{cccc}
U&0&0&0\\
0&U&0&0\\
0&0&{\bf 1}&0\\
0&0&0&{\bf 1}
\end{array}\right)=
\frac12(1+\rho^3)\otimes U+\frac12(1+\rho^3)\otimes{\bf 1},
\end{equation}
turns $\sigma_0$ into $\rho^1=\gamma_0$, which is invariant only
under vector transformations generated by $\bm1\otimes\lambda^a$. 
\clearpage
\chapter[The function $\Phi$ for the $U(3)$ toy model at non-zero density]{The function $\Phi$ for mean field theory of the $U(3)$ toy model at non-zero density\label{app:MFeqU3}}

In this section we give $\Phi$ for all configurations that appear in Table~\ref{table:config} with non-zero baryon number. The lower indices of the magnetizations correspond to the numbering in Table~\ref{table:config}.

As for zero density, we reduce the number of independent variables in $\Phi$ by substituting for some of them with their corresponding MF equations. Also we assume that all magnetizations commute, and are therefore diagonal in some basis. This restriction is justified for all configurations except no.~2 where we have to assume this as an ansatz.

\subsubsection{\underline{Configuration no.~2}}

In this case the form of the MF equations does not allow us to take the magnetizations as diagonal, and we make the ansatz that in this case
\begin{equation}
m_i \sim \left( \begin{array}{ccc} 1 & 0 & 0 \\ 0 & 1 & 0 \\ 0 & 0 & -2 \end{array} \right), \quad i=1,2,3,4,
\end{equation}
which means that we allow $\mu^8_i\neq 0 $ only. This gives the following form for $\Phi$ as a function of $\mu^8_i$.
\begin{eqnarray}
\Phi/N_s &=& \sqrt{3}/4 K (\mu_1-3\mu_3-\mu_2+\mu_4) - K/4 (3\mu_1\mu_2+3\mu_3\mu_4+6\mu_1\mu_3) \nonumber 
-\varphi(6K\sqrt{3}\mu_i), \\
\varphi(x_i)&=&\frac18 \left[ \log \left( \frac{e^{-x_1}-1}{x^2_1}+\frac1{x_1} \right) +\log \left( \frac{e^{-x_3}-1}{x^2_3}+\frac1{x_3} \right) \right.  \\
&&\qquad +3\log \left( \frac{e^{-(\frac23 x_3+\frac13 x_2)}-1}{(\frac23 x_3+\frac13 x_2)^2}+\frac1{\frac23 x_3+\frac13 x_2} \right) \nonumber \\
&&\left. \qquad +3\log \left( \frac{e^{(\frac23 x_1+\frac13 x_4)}-1}{(\frac23 x_1+\frac13 x_4)^2}+\frac1{\frac23 x_1+\frac13 x_4} \right) \right], \nonumber \\
\mu_2&=&f(6K\sqrt{3}\mu_1), \\
\mu_3&=&-f(-2K\sqrt{3}(2\mu_1+\mu_4)), 
\end{eqnarray}
where we have dropped the ``8'' index from all $\mu$'s, and 
\begin{equation}
f(x)=\frac2{\sqrt{3}} - \sqrt{3}\frac{e^x(1-2/x+2/x^2)-2/x^2}{e^x (1-1/x) + 1/x}.
\end{equation}

We find that there is a first order phase transition, similar to the zero density case, at $K_c=0.94$.

\subsubsection{\underline{Configurations no.~3, 5, 6}}
Here taking only diagonal magnetizations into account is justified and we obtain the following $\Phi$'s. (From here on we denote $(\mu^3,\mu^8)$ by $\vec \mu$, for all magnetizations)
\begin{enumerate}
\item   \underline{Configuration no.~3}

There are three independent magnetizations $\vec\mu_i;
\,i=1,2,3$ and we have
\begin{eqnarray}
\Phi/N_s &=& \sqrt{3}/2 K (3\mu^8_2-2\mu^8_1-\mu^8_3) - 3K/4 \left( 2 \vec \mu_1 \cdot \vec \mu_2 
+ \vec\mu_2\cdot \vec \mu_3\right) -\varphi(6K\vec \mu_i), \\
\varphi(\vec x_{i})&=&\frac18 \left[ \log \left( \frac{e^{x_{+,2}}-1}{x_{+,2}x^3_2} - \frac{e^{x_{-,2}}-1}{x_{-,2}x^3_2} \right) +3\log \left( \frac{e^{\frac23 x_{+,2}}-1}{\frac23 x_{+,2}x^3_2}-\frac{e^{\frac23 x_{-,2}}-1}{\frac23 x_{-,2}x^3_2} \right) \right. \nonumber \\
&&\left. \qquad +3\log \left( \frac{e^{-\frac13(x_{+,3}+2x_{+,1})}-1}{\frac19 (x_{+,3}+2x_{+,1})(x^2_3+2x^3_4)}-\frac{e^{-\frac13(x_{-,3}+2x_{-,1})}-1}{\frac19 (x_{-,3}+2x_{-,1})(x^2_3+2x^3_4)}\right) \right]. \nonumber \\ \label{eq:varphi3} \\
\mu^3_1 &=& f_3(4K\mu^3_2,4K\mu^8_2), \qquad \mu^8_1 = f_8(4K\mu^3_2,4K\mu^8_2), \label{eq:mu1_38} \\
\mu^3_3 &=& f_3(6K\mu^3_2,6K\mu^8_2), \qquad \mu^8_3 = f_8(6K\mu^3_2,6K\mu^8_2), \label{eq:mu3_38}
\end{eqnarray}
where $f_{3,8}$ are given in Eqs.~(\ref{eq:f3})--(\ref{eq:f8}). Also $x_{\pm,i}=\sqrt{3}x^8_i\pm x^3_i$, and we use this definition in further cases as well.

\item   \underline{Configuration no.~5}

There are three independent magnetizations $\vec \mu_i;\, i=1,2,3$ and we have
\begin{equation}
\Phi/N_s = \sqrt{3}/2 K (3\mu^8_2+2\mu^8_1+\mu^8_3) - 3K/4 \left( 2\vec \mu_1\cdot \vec \mu_2 + \vec \mu_2\cdot \vec \mu_3 \right)-\varphi(6K\vec \mu_i).
\end{equation}
Here $\varphi$, and $\vec{\mu}_{1,3}$ are given by Eqs.~(\ref{eq:mu1_38})--(\ref{eq:mu3_38}).

\item   \underline{Configuration no.~6}

Here we have two magnetizations $\vec \mu_i;\, i=1,2$ and we have
\begin{eqnarray}
\Phi/N_s &=& \sqrt{3}/2 K (\mu^8_2+\mu^8_1) - 3K/4 \,\, \vec \mu_1\cdot \vec \mu_2 -\varphi(6K\vec \mu_i), \\
\varphi(\vec x_i)&=&\frac18 \left[ \log \left( \frac{e^{x_{+,2}}-1}{x_{+,2}x^3_2} - \frac{e^{x_{-,2}}-1}{x_{-,2}x^3_2} \right) +3\log \left( \frac{e^{\frac13 x_{+,1}}-1}{\frac23 x_{+,1}x^3_1}-\frac{e^{\frac13 x_{-,1}}-1}{\frac23 x_{-,1}x^3_1} \right) \right], \nonumber \\ \\
\mu^3_1 &=& f_3(6K\mu^3_2,6K\mu^8_2), \qquad \mu^8_1 = f_8(6K\mu^3_2,6K\mu^8_2).
\end{eqnarray}
\end{enumerate}

In these three configurations we find that $\Phi$ has three equivalent nontrivial minima on the lines Eqs.~(\ref{eq:restrict1})--(\ref{eq:restrict3}) in the $\mu^3$--$\mu^8$ plane. These become lower in energy than the symmetric solution for $K>K_c$. The phase transition is first order and $K_c$ is 1.03, 1.14, and 1.72 for configurations~3, 5, and 6.

\subsubsection{\underline{Configuration no.~4, 7, 8}}
Here we also find that one can take the two  magnetizations to be diagonal. In all these configurations we have two magnetizations $\vec \mu_i;\, i=1,2$
\begin{enumerate}
\item   \underline{Configuration no.~4}

$\Phi$ is given by
\begin{eqnarray}
\Phi(\vec \mu_i)/N_s&=& -3K\,\, \vec \mu_1 \cdot \vec \mu_2 -2K\sqrt{3}(\mu^8_1+\mu^8_2) - \varphi(6K\vec \mu_i), \\
\varphi(\vec x_i) &=& \frac12 \left[ \log \left( \frac{e^{x_{+,1}}-1}{x_{+,1}x^3_{1}}-\frac{e^{-x_{-,1}}-1}{x_{-,1}x^3_{1}}\right) 
- \log \left( \frac{e^{x_{+,2}}-1}{x_{+,2}x^3_{2}} -\frac{e^{x_{+,2}}-1}{x_{+,2}x^3_{2}} \right) \right], \nonumber \\ 
\mu^3_1&=&f_3(6K\mu^3_2,6K\mu^8_2), \hskip 1.82cm \mu^8_1=f_8(6K\mu^3_2,6K\mu^8_2).
\end{eqnarray}

In this case we find that the phase transition is of second order. Also, there are six nontrivial minima. All of them have the same $\Phi$, and become preferable over the symmetric solution for $K>1$. Four of these minima are invariant under $U(2)\times U(1)$, and two are invariant only under $U(1)\times U(1)\times U(1)$.

\item   \underline{Configurations no.~7,8}

Since these configurations are the two and one dimensional counterparts of configuration~4 [see Table~\ref{table:config}], the MF equations and $\Phi$ are given by replacing $6K$ in the MF equations and $\Phi$ of configuration no.~4 by $4K$ and $2K$, respectively. This replacement simply increases the value of $K_c$ by factors $3/2$ and $3$.
\end{enumerate}

\clearpage
\chapter{The matrices $C$ and $D$ \label{app:CD} in the $O(1/N_c)$ calculations}

To analyze the matrices $C$ and $D_\bn$, given by
Eqs.~(\ref{eq:Cetaeta}) and~(\ref{eq:Dmatrix}), we begin with the $U(N)$
generators $M^\eta$ that lie outside the subalgebra $H=U(m)\times
U(N-m)$ that commutes with
\begin{equation}
\Lambda=\left(\begin{array}{cc}
\bm1_m&0\\
0&-\bm1_{N-m}
\end{array}\right).
\end{equation}
We choose for them a basis $M^{pq1}$ and $M^{pq2}$ with
$p=1,\dots,m$ and $q=m+1,\dots,N$, given by
\begin{eqnarray}
(M^{pq1})_{\alpha \beta} &=& \frac 12\left(\delta_{p\alpha}\delta_{q\beta}
+\delta_{p \beta}\delta_{q\alpha}\right) \label{eq:M1} \\
(M^{pq2})_{\alpha \beta} &=& \frac i2\left(\delta_{p\alpha}\delta_{q\beta} -\delta_{p\beta}\delta_{q\alpha}\right). \label{eq:M2}
\end{eqnarray}
Since the coset space $U(N)/H$ is a symmetric space, the
commutator $[M^{pqa},M^{p'q'a'}]$ lies in $H$; in order for
$\Tr(\Lambda[M^{pqa},M^{p'q'a'}])$ to be nonzero, the commutator
must have a nonzero component in the Cartan subalgebra of $H$.
This is only possible if $a\not=a'$ and $(p,q)=(p',q')$. Thus in
this basis $C$ takes the form
\begin{equation}
C=i\left(\begin{array}{cc}
0&\bm1_{m(N-m)}\\
-\bm1_{m(N-m)}&0
\end{array}\right).
\end{equation}
Diagonalizing $C$ gives eigenvalues $\pm1$. The generators
corresponding to the basis that diagonalizes $C$ are
\begin{eqnarray}
(M^{pq+})_{\alpha \beta} &=& (M^{pq1} + i M^{pq2})_{\alpha \beta} = \delta_{p\beta}\delta_{q\alpha}     \\
(M^{pq-})_{\alpha  \beta} &=& (M^{pq1} - i M^{pq2})_{\alpha \beta} =
\delta_{p\alpha}\delta_{q\beta}.
\end{eqnarray}

As for $D$: The only anticommutators among the $M^{pq\pm}$ that do
not vanish [note the bounds on $(p,q)$] are between $M^{pq+}$ and
$M^{p'q'-}$, {\em viz.,}
\begin{equation}
\{M^{pq+},M^{p'q'-}\}_{\alpha \beta} =\delta_{qq'}\delta_{p\alpha}\delta_{p'\beta} +
\delta_{pp'}\delta_{q'\alpha}\delta_{q\beta}. \label{eq:antcomm}
\end{equation}
Noting that $M^{pq-}=(M^{pq+})^\dag$, we find that $D_\bn$ takes
the block-diagonal form
\begin{eqnarray}
D_\bn^{pq\pm,p'q'\pm}&=& -\sum_{\bmm(\bn)}\Tr \left[\{
(M^{pq\pm})^\dag,M^{p'q'\pm}\} \left( \begin{array}{cc}
        \sigma^{(m)}_\bmm      &       0       \\
        0       &       - \bm1_{N-m}  \end{array} \right) \right]     \\
&=&\delta_{qq'} \sum_{\bmm(\bn)}
(\bm1_{m}-\sigma^{(m)}_\bmm)_{pp'}.
\end{eqnarray}
We summarize this by writing
\begin{equation}
D =  \left( \begin{array}{cc}
        E_\bn &  0       \\
        0       &       E_\bn \end{array} \right) \otimes 1_{N-m},
\end{equation}
where $E_\bn$ is an $m\times m$ matrix given by
\begin{equation}
E_\bn = \sum_{\bmm(\bn)} \left( \bm1_m - \sigma^{(m)}_\bmm
\right). \label{eq:E_mat}
\end{equation}
It is easy to prove that the eigenvalues of $E$ range from $0$ to
$4d$. In particular, $D$ is positive.
\clearpage
\chapter[Feynman rules for the effective action of the sigma model]{Feynman rules to $O(1/N_c)$ for the effective action of the nonlinear sigma model\label{app:Feynman}}

\begin{figure}[htb]
\begin{center}
\epsfig{width=8cm,file=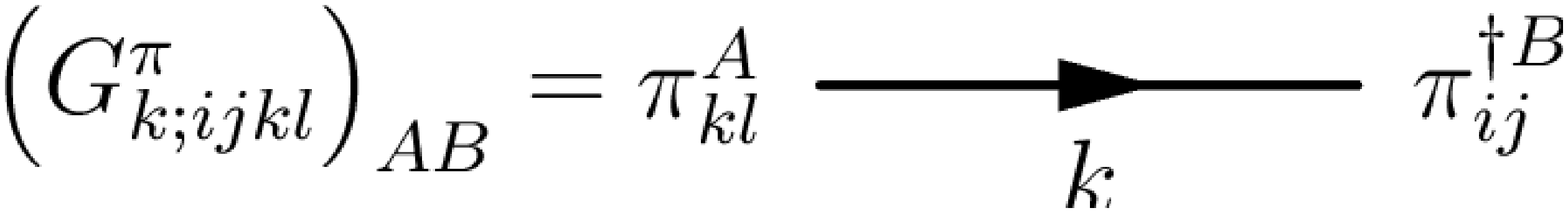}
\caption{Propagator of the $\pi$ fields.}
\label{fig:Gpi}
\end{center}
\end{figure}

\begin{figure}[htb]
\begin{center}
\epsfig{width=8cm,file=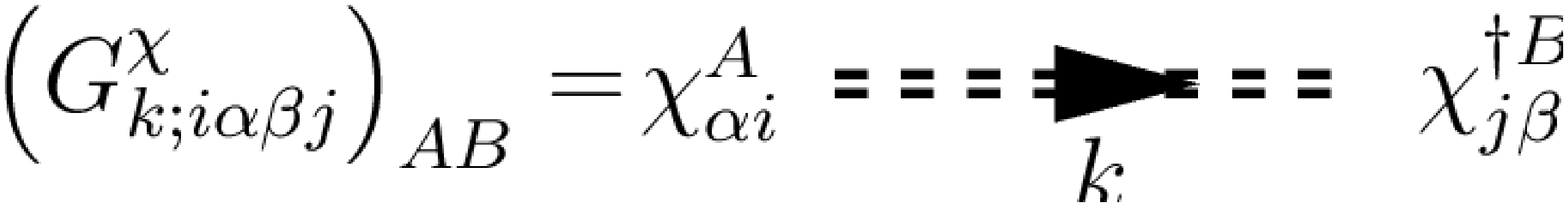}
\caption{Propagator of the $\chi$ fields.}
\label{fig:Gchi}
\end{center}
\end{figure}

\begin{figure}[htb]
\begin{center}
\epsfig{width=14cm,height=22cm,file=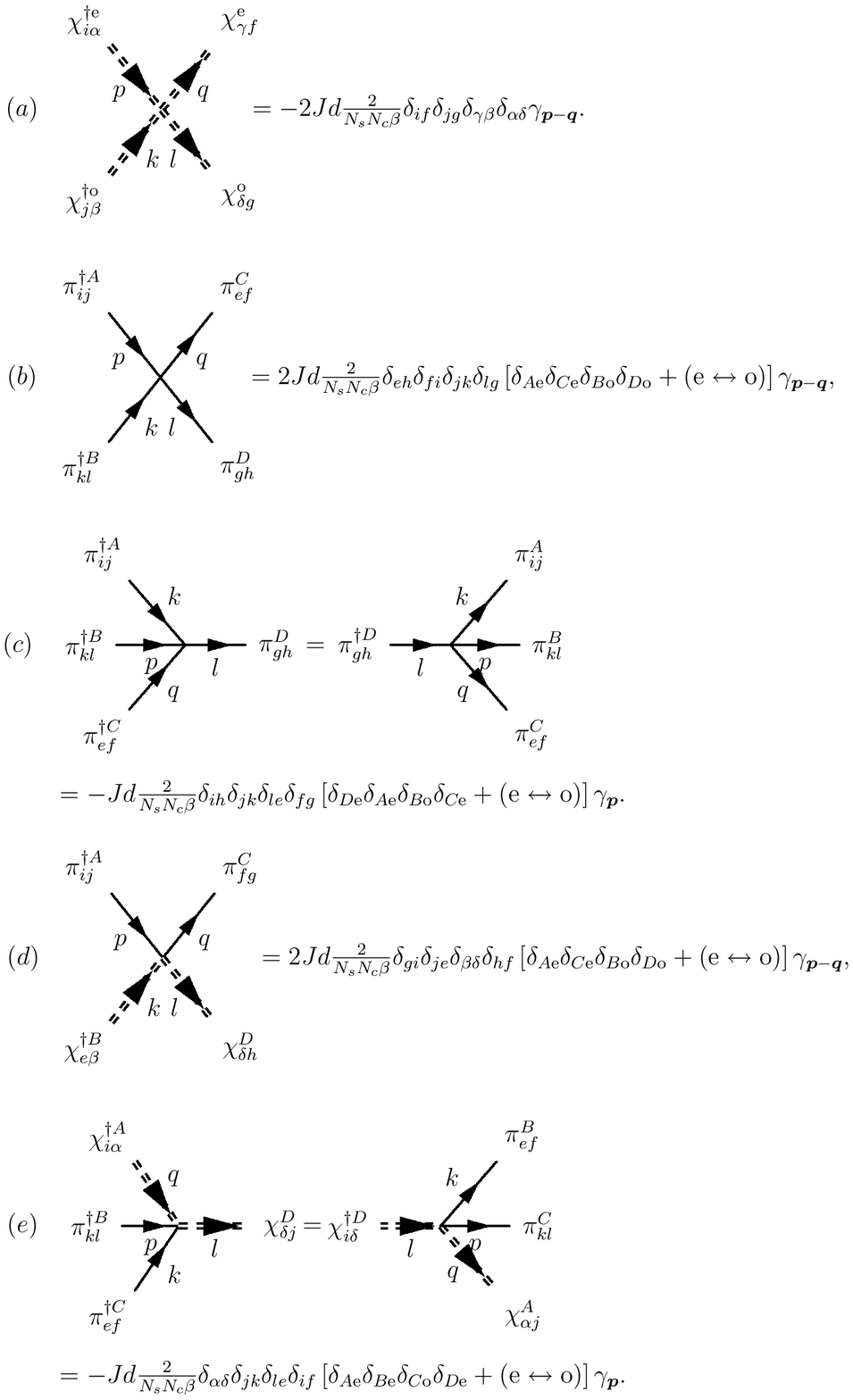}
\label{fig:V4}
\caption{Vertices corresponding to the quartic interaction.}
\end{center}
\end{figure}

\begin{figure}[htb]
\begin{center}
\epsfig{width=10cm,file=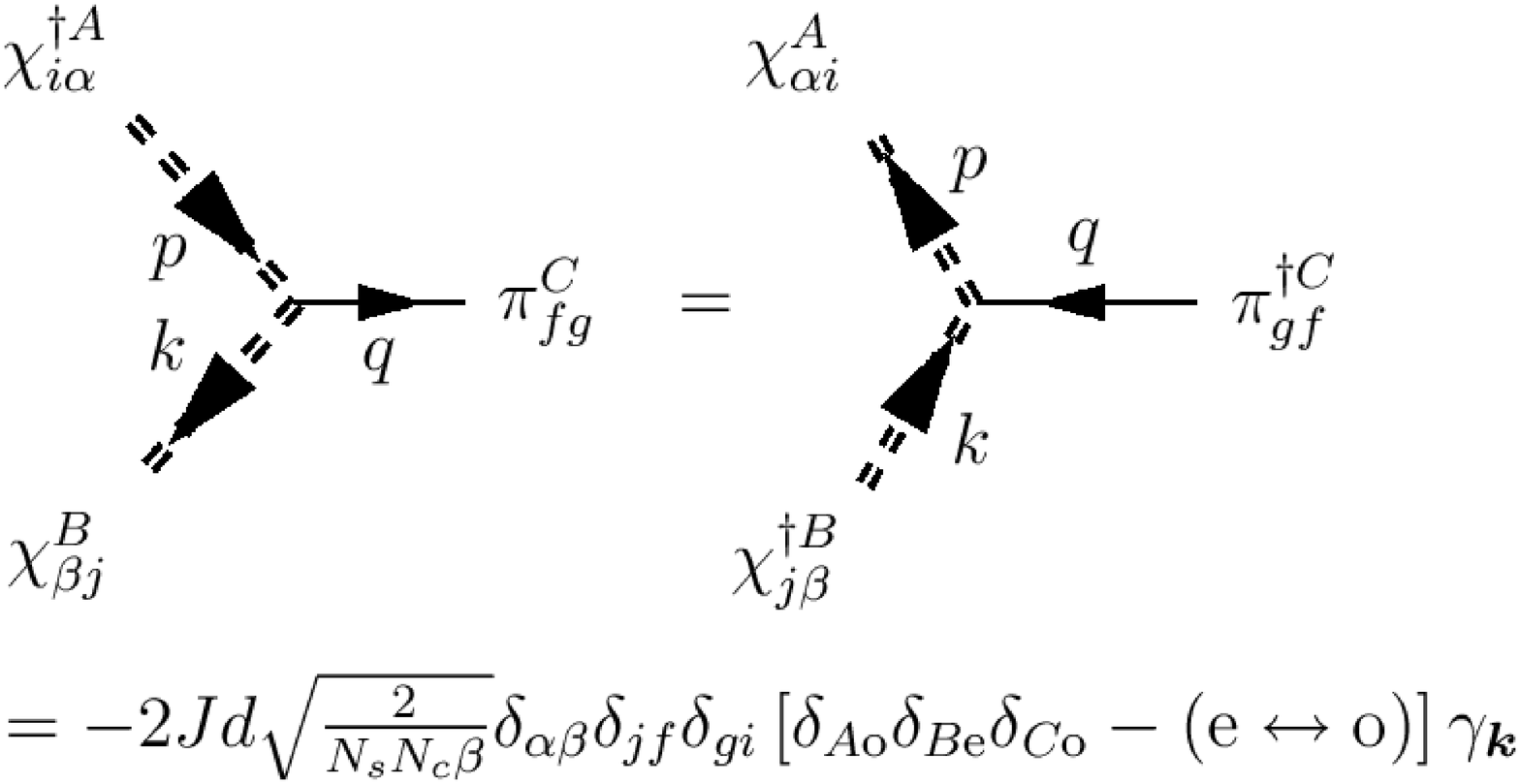}
\caption{Vertices corresponding to the cubic interaction.}
\label{fig:V3}
\end{center}
\end{figure}

The following are the Feynman rules extracted
from Eqs.~(\ref{eq:S2})--(\ref{eq:S4}). $A=(\e, \o)$ stands for even and odd
respectively. Latin indices take values in the range $[1,N-m]$ while Greek indices take values in the
range $[1,2m-N]$. The propagators are given in Figs.~\ref{fig:Gpi} and~\ref{fig:Gchi}. The vertices extracted from the cubic
 and quartic interactions [Eqs.~(\ref{eq:S3})--(\ref{eq:S4})] are given in Figs.~\ref{fig:V4}.3--\ref{fig:V3}.
\clearpage
\chapter[Integral equations for the self energy of the zero modes]{Integral equations for the self energy of the zero modes\label{app:Intequation}}

In this appendix we derive the integral equations represented by Fig.~\ref{fig:dyson} and solve them approximately in order to obtain the dispersion relations of the $\chi$ bosons to
$O(1/N_c)$.
In analogy with the propagators of the antiferromagnetic spin waves we
assume the following structure for $\Sigma_{1,k}$ and $\Sigma_{2,k}$ (we denote the frequency by $\Omega$), 
\begin{eqnarray}
\Sigma_{1,k}&=&A_1(\Omega^2;\bk)+i\Omega B_1(\Omega^2;\bk),  \label{eq:Sig1} \\
\Sigma_{2,k}&=&A_2(\Omega^2;\bk)+i\Omega B_2(\Omega^2;\bk) \label{eq:Sig2} ,
\end{eqnarray}
where $A_{1,2},B_{1,2}$ are real and $A^2_1 \geq A^2_2$ for
stability. In order to perform the frequency integrals we make the following rational ansatz. 
\begin{eqnarray}
A_1&=&-\frac{2J_1d}{N_c}
\frac{P_1(\Omega^2;\bk)}{Q_1(\Omega^2;\bk)} < 0,
\label{eq:A1} \\
A_2&=&\frac{2J_1d}{N_c}
\frac{P_2(\Omega^2;\bk)}{Q_2(\Omega^2;\bk)} \label{eq:A2},
\end{eqnarray}
Where $P_{1,2}(\Omega^2;\bk)$ and $Q_{1,2}(\Omega^2;\bk)$
are polynomials of the same order in $\omega^2$. Thus we have 
\begin{equation}
\label{eq:Gchi_AB}
(G^{\chi}_k)_{A B} = \frac1{\Omega^2[(1+B_1)^2-B^2_2]+A_1^2-A_2^2} \left( \begin{array}{cc}
        i\Omega(1+B_1) - A_1 & A_2 +i\Omega B_2 \\
        A_2-i\Omega B_2   &       -i\Omega(1+B_1) - A_1 \end{array}
    \right)_{A B}. \qquad
\end{equation}
Since $B_{1,2}\sim O(1/N_c)$ we drop them in the denominator. The
self-consistent equations for $A$ and $B$ then decouple, while only $A_{1,2}$ are needed to find the poles of the $\chi$ propagators to leading order. 

\section{Evaluation of diagrams}
\label{sec:diag_cal}

\begin{figure}[htb]
\begin{center}
\includegraphics[width=15cm]{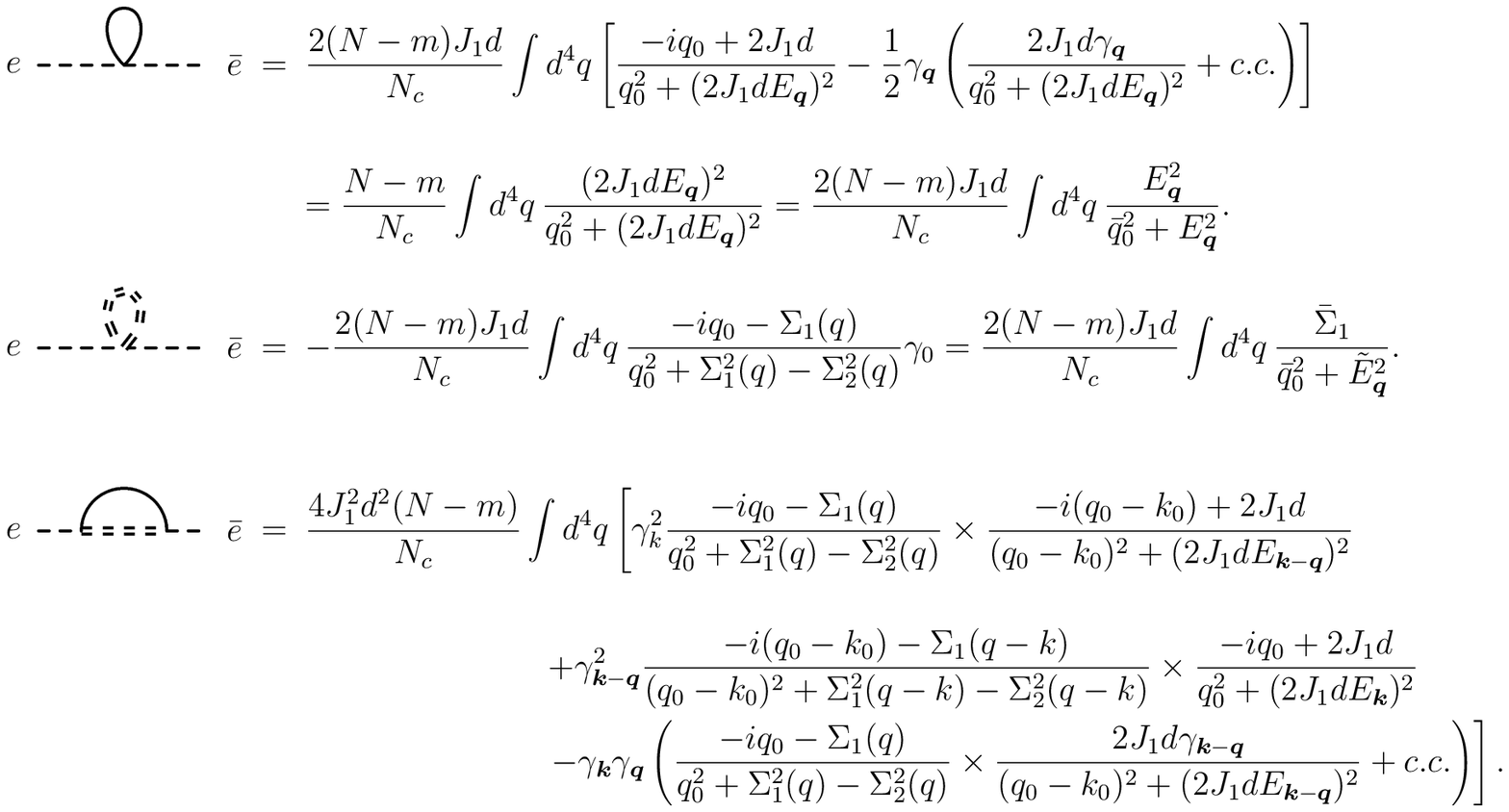}
\caption[Evaluation of $\Sigma_1$.]{Schwinger-Dyson equation for $\Sigma_1$.} \label{fig:sigma1}
\end{center}
\end{figure}

We begin with the diagrams that contribute to $\Sigma_1$. In this case we have contributions from all three diagrams in Fig.~\ref{fig:dyson}. Using the Feynman rules we evaluate each diagram and present the results for the self-energy matrix's upper left block with $\chi^e$ external legs  in Fig.~\ref{fig:sigma1}. (The result for external legs of $\chi^o$ is the complex conjugate of that) \footnote{Here we define $\int d^4q \equiv \int_{\text{BZ}} \left( \frac{dq}{4\pi}\right)^d \int_{-\infty}^\infty \frac{dq_0}{2\pi}$.} 

\begin{figure}[htb]
\begin{center}
\includegraphics[width=15cm]{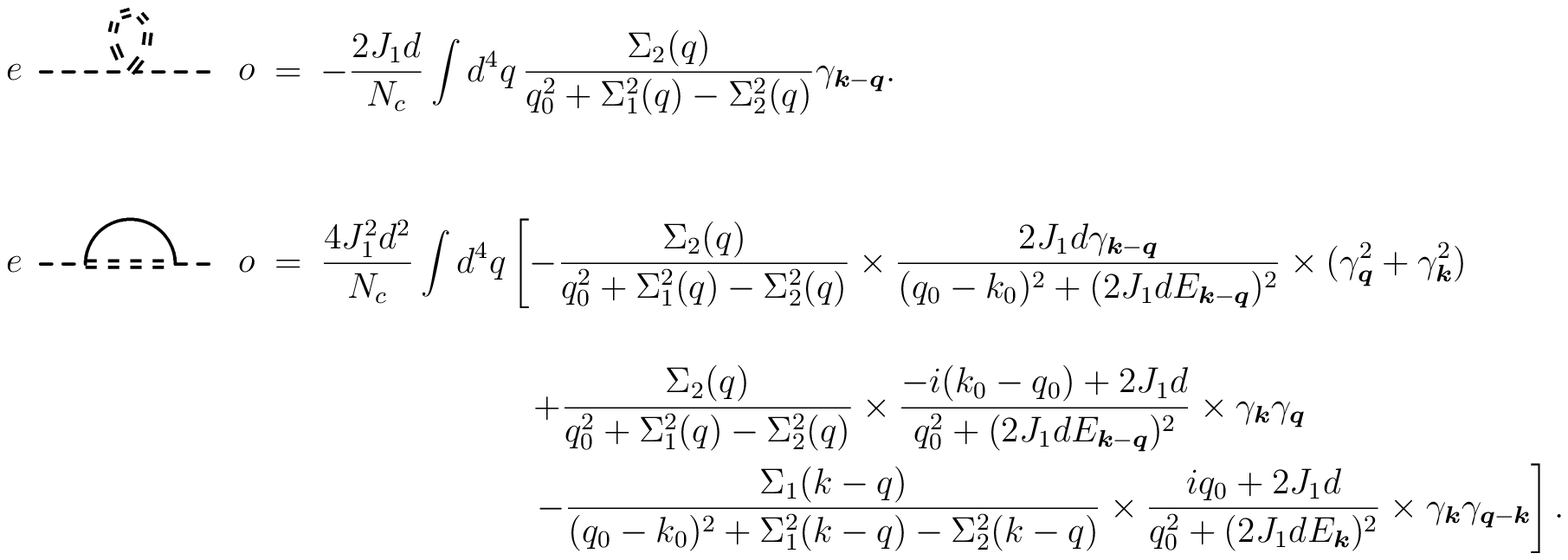}
\caption[Evaluation of $\Sigma_2$.]{Schwinger-Dyson equation for $\Sigma_2$.} \label{fig:sigma2}
\end{center}
\end{figure}

Here
\begin{equation}
E_\bq=\sqrt{1-\gamma^2_\bq}. \label{eq:E_pi}
\end{equation}
The factor of $(N-m)$ is the multiplicity of the diagram, and represents the number of particles that flow in the loop and have the same contribution. The diagrams that contribute to $\Sigma_2$ are only the second and third in Fig.~\ref{fig:dyson}. The result is given in Fig.~\ref{fig:sigma2}.
Note that in this case there is no multiplicity to the diagrams. This is a result of the structure of the vertices. For example the first term in \Eq{eq:S4} that contributes to the second and first diagram in Fig.~\ref{fig:sigma1}, and Fig.~\ref{fig:sigma2} has the following form.
\begin{equation}
\Tr \chi^e \chi^{\dag e} \chi^o \chi^{\dag o}= \sum_{\alpha \beta \in [1,2m-N] \atop ij \in [1,N-m]} (\chi^e)_{\alpha i} (\chi^{* e})_{\beta i} (\chi^o)_{\beta j} (\chi^{* o})_{\alpha j}.
\end{equation}
Its contribution to $\Sigma_1$ (of the even sites) comes from the contracting $\chi^o$ and $\chi^{*o}$. Since this contraction is proportional to $\delta_{\alpha \beta} \delta_{jj}$, a sum over $j$ gives a factor of $N-m$. The contribution to $\Sigma_2$ comes from contracting $\chi^{*o}$ to $\chi^{*e}$. This gives $\delta_{\alpha \beta} \delta_{ij}$, and has multiplicity 1. The same reasoning applies to all the diagrams in the figures we present. 

Using the parametrization~(\ref{eq:A1})--(\ref{eq:A2}) we get the following integral equations\footnote{We have defined $q_0 =2J_1d \omega$ and
  $k_O=2J_1d \Omega$.}.
\begin{eqnarray}
\frac{P_1(\Omega^2;\bk)}{Q_1(\Omega^2;\bk)}&=&2(N-m) \int_{\text{BZ}} \left( \frac{dq}{4\pi} \right)^d \int_{-\infty}^{\infty} \frac{d\omega}{2\pi} \left[ -\frac{E^2_\bq}{(\omega-\Omega)^2+E^2_\bq} + \frac{\frac1{N_c} P_1(\omega^2;\bq)/Q_1(\omega^2;\bq)}{\omega^2+\tilde{E}^2(\omega^2;\bq)} \right. \nonumber \\
&& \left. -\frac{\omega(\omega-\Omega) \left(
      \gamma^2_\bq-\gamma^2_\bk \right)
    +\frac1{N_c}P_1(\omega^2;\bq)/Q_1(\omega^2;\bq) \left(
      2\gamma_\bk \gamma_\bq \gamma_{\bq-\bk} -
      \gamma^2_\bq-\gamma^2_\bk \right) }{\left[
      \omega^2+\tilde{E}^2(\omega^2;\bq) \right]  \left[
      (\omega-\Omega)^2+E^2_{\bq-\bk} \right] } \right],
\qquad \nonumber \\ \label{eq:Ineq1_wq}
\end{eqnarray}
\newpage
where the three terms come from the three diagrams in Fig.~\ref{fig:sigma1} and
\begin{eqnarray}
\frac{P_2(\Omega^2;\bk)}{Q_2(\Omega^2;\bk)}&=&2 \int_{\text{BZ}} \left( \frac{dq}{4\pi} \right)^d \int_{-\infty}^{\infty} \frac{d\omega}{2\pi} \frac{\frac1{N_c} P_2(\omega^2;\bq)/Q_2(\omega^2;\bq)}{\omega^2+\tilde{E}^2(\omega^2;\bq)}
 \nonumber \\
 && \qquad \qquad \qquad \times \left[ \frac{2\gamma_\bk \gamma_\bq - \gamma_{\bq-\bk} \left(
      \gamma^2_\bq+\gamma^2_\bk \right)
    }{(\omega-\Omega)^2+E_{\bq-\bk}^2 } - \gamma_{\bq-\bk}
\right], \label{eq:Ineq2_wq} \quad \qquad
\end{eqnarray}
which comes from the second and third diagrams in Fig.~\ref{fig:sigma2}. 
Here
\begin{equation}
\tilde{E}(\omega^2;\bq)=\frac1{N_c} \sqrt{
  \frac{P^2_1}{Q^2_1} - \frac{P^2_2}{Q^2_2} } \equiv \frac1{N_c}
\frac{P}{Q}. \label{eq:E_tilde}
\end{equation}

\section{Evaluation of frequency integrals}
\label{sec:matzubara_calc}

It remains to evaluate the $O(1/N_c)$ contribution of the following integrals 
\begin{eqnarray}
I_1^i&=&\frac1{N_c}\int_{-\infty}^{\infty} \frac{d\omega}{2\pi}
\frac{P_i/Q_i}{\omega^2+\tilde{E}^2}, \label{eq:I1} \\
I_2^i&=&\frac1{N_c}\int_{-\infty}^{\infty} \frac{d\omega}{2\pi}
\frac{P_i/Q_i}{\left( \omega^2+\tilde{E}^2 \right) \left[
    (\omega-\Omega)^2 + E^2 \right] }, \label{eq:I2} \\
I_3&=&\int_{-\infty}^{\infty} \frac{d\omega}{2\pi} \frac{\omega ( \omega - \Omega) }{\left( \omega^2+\tilde{E}^2 \right) \left[ (\omega-\Omega)^2 + E^2
\right] } \label{eq:I3} .
\end{eqnarray}
All these integrals are convergent at large $\omega$ and can be evaluated with residue calculus.
Beginning with $I_1^i$,
\begin{equation}
\label{eq:I1cal0}
I_1^i=\frac1{N_c}\int_{-\infty}^{\infty} \frac{d\omega}{2\pi}
\frac{P_i Q_i Q^2_{j\neq i}}{\omega^2 Q^2 + \frac1{N_c^2}P^2}.
\end{equation}
The poles of the integrands are given by
\begin{equation}
\label{eq:Eq4poles}
\omega^2 Q^2(\omega^2) + \frac1{N_c^2} P^2(\omega^2) = 0.
\end{equation}
To leading order in $1/N_c$ the roots are determined by either
\begin{equation}
\label{eq:pole1}
\omega^2=-\frac1{N^2_c}P^2(0)/Q^2(0)\equiv \omega^2_0
\end{equation}
or by
\begin{equation}
\label{eq:more_poles}
Q^2(\omega^2)=0.
\end{equation}
Since the polynomials $P_i$ and $Q_i$ do not depend on $N_c$ and
$Q_i(0)$ can be chosen to be 1, the solutions of \Eq{eq:more_poles} are
all $O(1)$. Moreover since $Q^2 > 0$, all these roots appear in complex
conjugate pairs $(\omega,\omega^*)$. For definiteness we choose
$\Im \omega >0$ and  close the contours from above.

 The result for $I_1^i$ is \footnote{Note that the denominator $Q^2 \omega^2 + 1/N^2_c P^2>0$ is a polynomial in $\omega^2$ and can be written as $\prod_m (\omega^2-\omega^2_m)(\omega^2-\omega^{2*}_m)$, where $\pm \omega_m$ are the (complex) zeros of \Eq{eq:Eq4poles}. We also have $Q^2=\prod'_m(\omega^2-\omega^2_m)(\omega^2-\omega^{2*}_m)$, where the product includes only the zeros of \Eq{eq:more_poles}.}
\begin{eqnarray}
I^i_1&=&\frac1{N_c} \frac{2\pi i}{2\pi} \sum_n \frac{P_i(\omega_n)Q_i(\omega_n)Q^2_{j\neq i}(\omega_n)}{2i\Im \omega_n \prod_{m\neq n} (\omega^2_n -\omega^2_m)(\omega^2_n -\omega^{*2}_m)} \nonumber \\
&&=\frac1{2N_c} \left[ \frac{P_i(\omega_0)Q_i(\omega_0)Q^2_{j\neq i}(\omega_0)}{\Im \omega_0 \prod_{\omega_m\neq \omega_0} (\omega^2_0 -\omega^2_m)(\omega^2_0-\omega^{*2}_m)} + \sum_{\omega_n\neq \omega_0} \frac{P_i(\omega_n)Q_i(\omega_n)Q^2_{j\neq i}(\omega_m)}{\Im \omega_n \prod_{m\neq n} (\omega^2_n -\omega^2_m)(\omega^2_n-\omega^{*2}_m)} \right]\nonumber \\
&&\simeq \frac1{2N_c} \left[ \frac{P_i(0)Q_i(0)Q^2_{j\neq i}(0)}{1/N_c P(0)/Q(0)\prod_{\omega_m\neq \omega_0} |\omega^{*2}_m|^2} + \sum_{\omega_n\neq \omega_0} \frac{P_i(\omega_n)Q_i(\omega_n)Q^2_{j\neq i}(\omega_0)}{\Im \omega_n \prod_{m\neq n} (\omega^2_n -\omega^2_m)(\omega^2_n-\omega^{*2}_m)} \right]\nonumber \\
&&=\frac1{2N_c} \left[ N_c \frac{P_i(0)Q_i(0)Q^2_{j\neq i}(0)}{P(0)/Q(0)\cdot Q^2(0)} + O(1) \right]=\frac12 \frac{P_i(0)}{P(0)} \frac{Q_i(0)Q_{i\neq j}(0)}{Q(0)}+ O(1/N_c) \nonumber \\
&&=\frac12 \frac{P_i(0)}{Q_i(0)}\frac1{\sqrt{\left( \frac{P_1(0)}{Q_1(0)}\right)^2 - \left( \frac{P_2(0)}{Q_2(0)}\right)^2 }}+ O(1/N_c).
\end{eqnarray}
The leading
contribution to \Eq{eq:I1} comes from the poles given by
\Eq{eq:Eq4poles}. For the fourth row we used the fact that all other poles are of $O(1)$, and in the fact that $Q^2(0)$ can be written in terms of a product of its zeros. For the fifth row we used the definitions of $Q(0)$ and $P(0)$. 

The calculation of $I_2$ and $I_3$ is similar, but one
has two more poles to consider, at $\omega'=\Omega \pm iE$. In the case of $I_2$, the leading contribution is again due to the pole at $\omega_0$ only. 
\begin{eqnarray}
I^i_2&=&\frac1{N_c}\int\frac{d\omega}{2\pi}
\frac{P_i Q_i Q^2_{j\neq i}}{\left( \omega^2 Q^2 + \frac1{N_c^2}P^2 \right)\left[ (\omega-\Omega)^2+E^2 \right]}\nonumber \\
&&= \frac1{N_c} \frac{2\pi i}{2\pi} \left\{ \sum_n \frac{P_i(\omega_n)Q_i(\omega_n)Q^2_{j\neq i}(\omega_n)}{2i\Im \omega_n \prod_{m\neq n} (\omega^2_n -\omega^2_m)(\omega^2_n -\omega^{*2}_m)} \times \frac1{(\omega_n-\Omega)^2+E^2} \right. \nonumber \\
&&\qquad \qquad \qquad \left. + \frac{P_i(\omega')Q_{j\neq i}(\omega')}{2iE \left[ \omega'^2Q^2(\omega') + 1/N^2_c P^2(\omega') \right] } \right\} \nonumber \\
&&=\frac12 \frac{P_i(0)}{Q_i(0)}\frac1{\sqrt{\left( \frac{P_1(0)}{Q_1(0)}\right)^2 - \left( \frac{P_2(0)}{Q_2(0)}\right)^2 }}\times \frac1{\Omega^2+E^2}+ O(1/N_c).
\end{eqnarray}

The integral $I_3$ is written as
\begin{eqnarray}
I_3&=&\int\frac{d\omega}{2\pi}
\frac{\omega(\omega-\Omega)Q^2}{\left( \omega^2 Q^2 + \frac1{N_c^2}P^2 \right)\left[ (\omega-\Omega)^2+E^2 \right]}\nonumber \\
&&=\frac{2\pi i}{2\pi} \left\{ \sum_n \frac{\omega_n(\omega_n-\Omega)Q^2(\omega_n)}{2i\Im \omega_n \prod_{m\neq n} (\omega^2_n -\omega^2_m)(\omega^2_n -\omega^{*2}_m)}\times \frac1{(\omega_n-\Omega)^2+E^2} \right. \nonumber \\
&&\qquad \qquad \left. + \frac{iE \omega' Q^2(\omega')}{2iE \left[ \omega'^2Q^2(\omega') + 1/N^2_c P^2(\omega') \right] } \right\} \nonumber \\
&&\simeq\frac12\left\{ \frac{\omega_0(\omega_0-\Omega)Q^2(\omega_0)}{\Im \omega_0 \prod_{\omega_m\neq \omega_0} (\omega^2_0 -\omega^2_m)(\omega^2_0 -\omega^{*2}_m)}\times \frac1{(\omega_0-\Omega)^2+E^2} \right.\nonumber \\
&&\left. \qquad \quad + \sum_{\omega_n\neq \omega_0} \frac{\omega_n(\omega_n-\Omega)Q^2(\omega_n)}{2i\Im \omega_n \prod_{m\neq n} (\omega^2_n -\omega^2_m)(\omega^2_n -\omega^{*2}_m)}\times \frac1{(\omega_n-\Omega)^2+E^2} + \frac{i}{\Omega+iE} \right\} \nonumber \\
&&\simeq\frac12\left[ \frac{-i\Omega}{\Omega^2+E^2} + \frac{i}{\Omega+iE} \right] = \frac12 \frac{E}{\Omega^2+E^2} +O(1/N_c).
\end{eqnarray}
Here we used the the fact that $\omega_{n\neq0}$ are the zeros of $Q^2$, which means that the residue of these poles is zero. 

Using our freedom to normalize the polynomials $Q_{1,2}(0)>0$, we choose $Q_{1,2}(0)=1$ and write the integrals to leading order as
\begin{eqnarray}
I_1^i&=&\frac12 \frac{P_i(0)}{\sqrt{P_1^2(0)-P_2^2(0)}}
\label{eq:I1_result}, \\
I_2^i&=&\frac12 \frac{P_i(0)}{\sqrt{P_1^2(0)-P_2^2(0)}} \frac1{\Omega^2+E^2}, \label{eq:I2_result} \\
I_3&=&\frac12 \frac{E}{\Omega^2+E^2} \label{eq:I3_result}.
\end{eqnarray}
Thus we find that to $O(1/N_c)$, all integrals depend
only on the values of the polynomials at
$\omega=0$. Finally, Eqs.~(\ref{eq:Ineq1_wq})--(\ref{eq:Ineq2_wq}) simplify to
\begin{eqnarray}
\frac{P_1(\Omega^2,\bk)}{Q_1(\Omega^2,\bk)}&=&(N-m) \int_{\text{BZ}} \left( \frac{dq}{4\pi} \right)^d \left[ \cosh{\theta_\bq} \left( 1+\frac{2\gamma_\bk \gamma_\bq \gamma_{\bq-\bk} - \gamma^2_\bq - \gamma^2_\bk}{E^2_{\bq-\bk} +\Omega^2} \right)
\right. \nonumber \\
&& \left. -\frac{E_{\bq-\bk} \left( \gamma^2_\bq-\gamma^2_\bk \right)}{E^2_{\bq-\bk}+\Omega^2} - E_\bq \right],
\label{eq:Inteq1_q} \\
\frac{P_2(\Omega^2,\bk)}{Q_2(\Omega^2,\bk)}&=& \int_{\text{BZ}}
\left( \frac{dq}{4\pi} \right)^d \left[ \sinh{\theta_\bq} \left(
    \frac{2\gamma_\bq \gamma_\bk-\gamma_{\bq-\bk} \left( \gamma^2_\bq
        + \gamma^2_\bk \right)}{E^2_{\bq-\bk}+\Omega^2} -
    \gamma_{\bq-\bk} \right) \right]. \label{eq:Inteq2_q} \qquad
\end{eqnarray}
Where we have defined 
\begin{equation}
\label{eq:tanh}
\tanh{\theta_\bq}=\frac{P_2(0,\bq)}{P_1(0,\bq)}.
\end{equation}
Here we also point out that $\Sigma_{1,2}$ have zeros at $k=0$. For example the function multiplying the the $\cosh$ in \Eq{eq:Inteq1_q} is exactly zero at $\Omega=0$, and $\bq=0$. This is also the case for the rest of \Eq{eq:Inteq1_q}, and for the function multiplying the $\sinh$ in \Eq{eq:Inteq2_q}. This is a direct result of the Ward identities regarding the $U(N)$ generators that correspond to fluctuations in the $\chi$ directions.

Taking $\Omega=0$, and dividing \Eq{eq:Inteq2_q}
by \Eq{eq:Inteq1_q}, we have 
\begin{equation}
(N-m)\tanh{\theta_\bk}=\frac{{\displaystyle \int_{\text{BZ}} \left( \frac{dq}{4\pi}
  \right)^d I_2(\bq,\bk) \sinh{\theta_\bq}  }}{{\displaystyle \int_{\text{BZ}} \left(
    \frac{dq}{4\pi} \right)^d I_1(\bq,\bk)\cosh{\theta_\bq} -\eta(\bk) }} \equiv \frac{\beta_\bk}{\alpha_\bk}.
\end{equation}
Here we have further defined 
\begin{eqnarray}
\eta(\bk)&=& \int_{\text{BZ}} \left( \frac{dq}{4\pi} \right)^d
\left[ \frac{\gamma^2_{\bq-\bk}-\gamma^2_\bk}{E_{\bq}} + E_\bq \right] \label{eq:al0},       \\
I_1(\bq,\bk)&=& \left( 1+\frac{2\gamma_\bq \gamma_\bk \gamma_{\bq-\bk}-\gamma^2_\bq-\gamma^2_\bk}{E^2_{\bq-\bk}} \right), \label{eq:a1}  \\
I_2(\bq,\bk)&=& \left(\frac{2\gamma_\bq
    \gamma_\bk-\gamma_{\bq-\bk}\left( \gamma^2_\bq+\gamma^2_\bk
    \right) }{E^2_{\bq-\bk}}-\gamma_{\bq-\bk} \right) \label{eq:a2}.
\end{eqnarray}
A solution of this equation must obey
\begin{equation}
\label{eq:stability1}
\int_{\text{BZ}} \left( \frac{dq}{4\pi} \right)^d I_1(\bq,\bk)
\cosh{\theta_\bq} -\eta(\bk) \geq 0,
\end{equation}
so that the stability condition~(\ref{eq:stability}) will be satisfied. 

After solving
 the integral equation~(\ref{eq:Inteq}) one can extract the
spectrum of the $\chi$ bosons by calculating the poles of the
propagator~(\ref{eq:Gchi}). These poles are given by
equation~(\ref{eq:Eq4poles}). As seen there are two kinds of
solutions. The $O(1/N_c)$ poles of the form~(\ref{eq:pole1}) give rise to
a low energy band [see \Eq{eq:E_tilde}]
\begin{equation}
\label{eq:E1k}
\pm i
\omega=\frac{2J_1d}{N_c}\sqrt{P_1^2(0,\bk)-P_2^2(0,\bk)}
\end{equation}
Using \Eq{eq:tanh} and \Eq{eq:Inteq} we have 
\begin{equation}
\pm
i\omega=\frac{2J_1d(N-m)}{N_c}\alpha_{\bk}\sqrt{1-\tanh^2{\theta_\bk}}
\equiv \frac{2J_1d(N-m)}{N_c} \epsilon_\bk.
\end{equation}
The $O(1)$ poles that solve \Eq{eq:more_poles} represent uninteresting
massive excitations.

\section{Solution of integral equations}
\label{sec:Inteq_sol}

\begin{figure}[ht]
\begin{center}
\includegraphics[width=10cm]{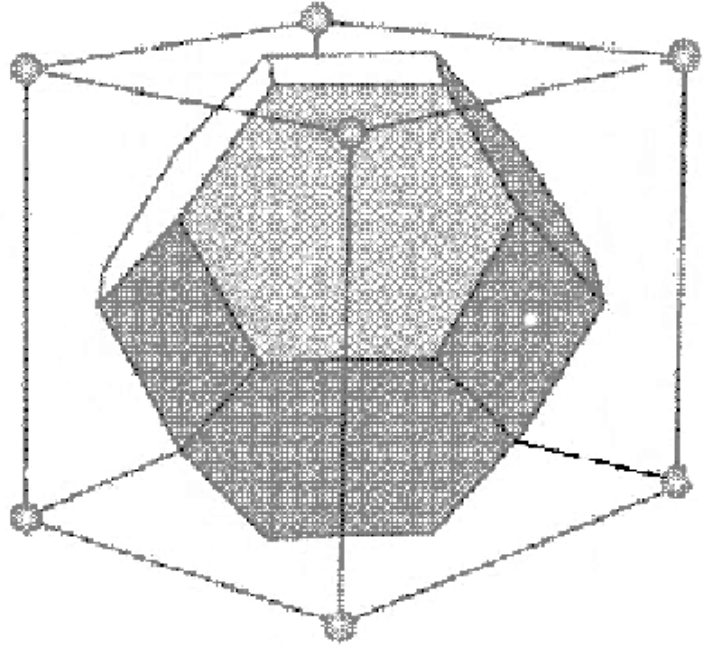}
\caption[The first Brillouin zone (BZ) of the fcc lattice.]{The first Brillouin zone of the fcc lattice. The side of the outer cube is $4\pi$. The BZ surface is composed out of six small and eight larger faces. The distance between the origin and the of the faces are $\sqrt{3}\pi$ and $2\pi$. (Reproduced from http://140.114.18.41/micro/chap2/fig/fig0206.jpg)}
\label{fig:fcc_BZ}
\end{center}
\end{figure}

In order to solve the integral equations we assumed that the function $\theta_\bq$ is a function of the absolute value $q=|\bq|$. 
Taking $\Omega=0$, and dividing \Eq{eq:Inteq2_q}
by \Eq{eq:Inteq1_q}, we have (for $d=3$) 
\begin{equation}
(N-m)\tanh{\theta_\bk}=\frac{{\displaystyle \int_{\text{BZ}} dq \, q^2 d\cos(\phi) \, d\varphi \, I_2(\bq,\bk) \sinh{\theta_q}  }}{{\displaystyle \int_{\text{BZ}} dq \, q^2 \, d\cos (\phi) \, d\varphi I_1(\bq,\bk)\cosh{\theta_q} -\eta(\bk) }}.
\end{equation}
and $\bq=q\left(\sin(\phi) \cos(\varphi),\sin(\phi) \sin(\varphi),\cos(\phi) \right)$, where $0<\varphi<2\pi$, $0<\phi<\pi$. The integration is over the first Brillouin zone of the fcc lattice. The BZ is presented in Fig.~\ref{fig:fcc_BZ}. Since it is not spherically symmetric, the maximum value that $|\bq|$ and $|\bk|$ can take depend on the direction of the vectors. In direction $(1,0,0)$ it is $2\pi$, whereas in direction $(1,1,1)$ it is $\sqrt{3}\pi$. The maximum allowed value is $\sqrt{5}\pi$, and is attained for example in direction $(0,1,2)$. In order to perform the momentum integrations we have generated a spherical mesh of the $6$ coordinates $\bk$ and $\bq$ that has a radius of $\sqrt{5}\pi$ while restricting it to the BZ by multiplying all integrands by a step function that is $1$ if a point in this six dimensional space obeys the following restrictions, and zero otherwise. We first restrict $|k_{x,y,z}|<2\pi$ and $|q_{x,y,z}|<2\pi$, and further  
\begin{eqnarray}
|q_x+q_y+q_z|<3\pi, &\qquad& |-q_x+q_y+q_z|<3\pi, \nonumber \\
|q_x-q_y+q_z|<3\pi, &\qquad& |q_x+q_y-q_z|<3\pi, 
\end{eqnarray}
and the same for $\bk$. 
This forms the shape shown in Fig.~\ref{fig:fcc_BZ}.
Using the spherical assumption we reduce the integral equation to be one dimensional (with regard to the integration). We define
\begin{equation}
a_1(q,\bk) = q^2\, \int_{{\text BZ}(q,\bk)} d\cos(\phi) \, d\varphi I_1(\bq,\bk), \qquad a_2(q,\bk) = q^2\, \int_{{\text BZ}(q,\bk)} d\cos(\phi) \, d\varphi I_2(\bq,\bk),
\end{equation}
where the boundaries of the integration depend on the value of $q$ and $\bk$, hence the notation ${\text BZ}(q,\bk)$. Finally we have
\begin{equation}
(N-m)\tanh{\theta_\bk}=\frac{{\displaystyle \int_0^{\sqrt{5}\pi} dq \, a_2(q,\bk) \sinh{\theta_q}  }}{{\displaystyle \int_0^{\sqrt{5}\pi} dq \, a_1(q,\bk)\cosh{\theta_q} -\eta(\bk) }}. \label{eq:Inteq_apprx}
\end{equation}

\begin{figure}[htb]
\begin{center}
\resizebox{130mm}{!}{\includegraphics{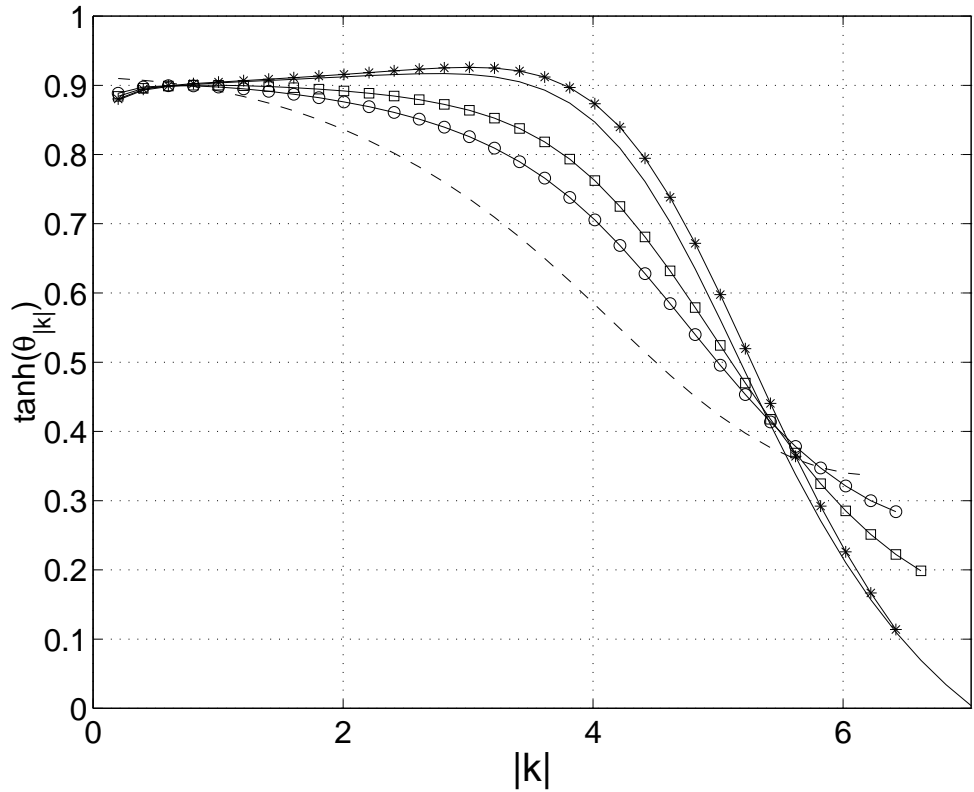}}
\caption[Behavior of the approximated integral equation.]{Inspection of the integral equation with the spherical approximation in different directions. Here we show the case of $c=5$.} \label{fig:tanh_directions}
\end{center}
\end{figure}

\Eq{eq:Inteq_apprx} is only an approximation to \Eq{eq:Inteq}. On the right hand side, there is an explicit dependence on the direction of $\bk$ (via $a_{1,2}$ and $\eta$), while on the left hand side the dependence should vanish by assumption. We first examine the solution to \Eq{eq:Inteq_apprx} for $\bk$ in the direction $(0,1,2)$, so that $|\bk|$ can take all values up to $\sqrt{5}\pi$ like $|\bq|$ on the left hand side. This solution is presented in Fig.~\ref{fig:tanh}. The next step is to see how this solution behaves in other directions. We do not expect the equation to be satisfied for all directions. More exactly we expect that the behavior away from the BZ boundary, near $\bk=0$, will indeed is spherical, so that the equation is approximately correct for this region. We give the results of this check in Fig.~\ref{fig:tanh_directions}.
Here the solid line is the original solution of the integral equation, in direction $(0,1,2)$, with $(\theta,\phi)=(\tan^{-1}(0.5),0)$. There $|\bk|$ has a maximum value of $\sqrt{5}\pi$. The stars, boxes, and circles are for directions of $(\theta,\phi)=(\tan^{-1}(1/2),\pi/10),(\tan^{-1}(2/5),0),(\pi/10,0)$. The dashed line is for $(\theta,\phi)=0$. In all these cases, the solution deviates from the original one for high enough momenta. As we get further away from the original direction, the region in $|\bk|$ space in which the solution is correct becomes smaller. Common to all directions is that near $\bk=0$ they agree, which was expected, and points to a spherical behavior there.
\clearpage
\chapter[Hamiltonian approach to order from disorder]{Hamiltonian approach to order from disorder for $\bm{N-m=1}$  \label{app:Hamiltonian_odo}}
In this appendix we derive the self-consistent equations in a
Hamiltonian formulation using a generalized Holstein-Primakoff
transformation for $N-m=1$.
The Hamiltonian is 
\begin{equation}
\label{eq:Hqmapp}
H=J_1 \sum_{\bn \muhat \alpha \beta} Q_{\alpha \beta}(\bn)
Q_{\beta \alpha}(\bn+\muhat).
\end{equation}
where the $U(N)$ generators $Q_{\alpha \beta}(\bn)$ obey
\begin{equation}
  \label{eq:algebra}
  \left[ Q_{\alpha \beta}(\bn) , Q_{\gamma \delta}(\bmm) \right] =
  \left( Q_{\alpha
    \delta}(\bn) \delta_{\beta \gamma} - Q_{\gamma \beta}(\bn) \delta_{\alpha
    \delta} \right)\delta_{\bn \bmm} . \label{eq:comm_Q}
\end{equation}
For $N-m=1$ there is a simple representation of these operators in terms of $N-1$ boson operators boson operators
\cite{Salam},
\begin{equation}
  \label{eq:realize}
  Q(\bn)_{\alpha \beta} = \frac{N_c}2 \left( \begin{array}{ccc}
      {\bf 1}_{N-1}-2\phi \phi^{\dag} & & 2\phi\sqrt{1-\phi^{\dag} \phi} \vspace{0.3cm} \\
      2\sqrt{1-\phi^{\dag} \phi} \phi^{\dag} & & -1+2\phi^{\dag}\phi
    \end{array} \right)_{\alpha \beta} = Q^\dag(\bn)_{\beta \alpha}. \label{eq:Q_HP}
\end{equation}
Here $\phi$ is a $(N-1)$--component boson field defined on site $\bn$ that obeys 
\begin{equation}
  \label{eq:commutations}
  \left[ \phi(\bn)_i , \phi^\dag(\bmm)_j \right] = \frac1{N_c} \delta_{ij} \delta_{\bn\bmm},\qquad \forall i,j \in [1,N-1]. \label{eq:phi_comm}
\end{equation}
The relation~(\ref{eq:phi_comm}) has an important consequence. It means that at the large $N_c$ limit the field operators $\phi_i$ are classical c-numbers, and the problem is not quantum. In the path integral formalism, \Eq{eq:phi_comm} is equivalent to the fact that the Berry phase of the action is proportional to $S_B\sim N_c \int d\tau \, \phi^\dag_i \partial_\tau \phi_i$. Indeed at large $N_c$, stationary phase arguments hold and the kinetic term is zero, making the system completely classical. 

The Hilbert space is created by repeated applications of $\phi^\dag$ on the no-quantum state $|0\>$, $\phi|0\>=0$. It is restricted to be a $U(N)$ representation with $N_c$ columns and $N-1$ rows, by restricting the eigenvalues of the number operator $\phi^\dag \phi$ to be smaller than or equal to one. This representation is the generalized Holstein-Primakoff transformation. For $N-m>1$ the generalization is more complicated (see the Appendix in \cite{Smit}). 

We first verify that the relations \Eq{eq:comm_Q} are satisfied by \Eq{eq:Q_HP}. We have four kinds of operator given by
\begin{eqnarray}
Q_{ij}&=&\frac{N_c}2 \left( \delta_{ij} -2 \phi_i \phi^\dag_j\right), \\
Q_{NN}&=&\frac{N_c}2 \left( -1 + 2\hat n \right), \\
Q_{iN}&=&N_c \phi_i \sqrt{1-\hat n}, \\
Q_{Ni}&=&N_c \sqrt{1-\hat n} \, \phi^\dag_i = Q^\dag_{iN}.
\end{eqnarray}
Here we define the number operator $\hat n=\sum_i \phi^\dag_i \phi_i$. We need to verify the following commutation relations. (All operators are on the same site.)
\begin{eqnarray}
\left[Q_{ij},Q_{kl}\right]&=&Q_{il}\delta_{jk}-Q_{kj}\delta_{il}, \label{eq:comm1} \\
\left[Q_{ij},Q_{NN}\right]&=&Q_{iN}\delta_{jN}-Q_{Nj}\delta_{iN}=0, \label{eq:comm2} \\ 
\left[Q_{ij},Q_{kN}\right]&=&Q_{iN}\delta_{jk}-Q_{kj}\delta_{iN}=Q_{iN}\delta_{jk}, \label{eq:comm3} \\ 
\left[Q_{NN},Q_{iN}\right]&=&Q_{NN}\delta_{iN}-Q_{iN}\delta_{NN}=-Q_{iN}, \label{eq:comm4} \\
\left[Q_{iN}, Q_{jN}\right]&=&Q_{iN}\delta_{jN}-Q_{jN}\delta_{iN}=0, \label{eq:comm5} \\
\left[Q_{iN},Q_{Nj}\right]&=&Q_{ij}\delta_{NN}-Q_{NN}\delta_{ij}=N_c(1-\hat n) -N_c \phi_i \phi^\dag_j. \label{eq:comm6} 
\end{eqnarray}
All other commutators are automatically satisfied once the above list is satisfied. We first note that $\hat n$ (and therefore $Q_{NN}$) commutes with all operators that do not change the number of bosons. As a result the second commutator is zero, as expected from \Eq{eq:comm2} ($i$ cannot be equal to $N$). We now evaluate the rest of the commutators using \Eq{eq:phi_comm}.
\begin{eqnarray}
\left[Q_{ij},Q_{kl}\right]&=&(-2\frac{N_c}2)^2[\phi_i \phi^\dag_j,\phi_k \phi^\dag_l]=N^2_c \phi_i[\phi^\dag_j,\phi_k]\phi^\dag_l + N^2_c \phi_k[\phi_i,\phi^\dag_l]\phi^\dag_j \nonumber \\
&&=Q_{il}\delta_{jk} - Q_{kj}\delta_{il}, \\
\left[Q_{ij},Q_{kN}\right]&=&-N^2_c[\phi_i \phi^\dag_j,\phi_k]\sqrt{1-\hat n}=N_c \phi_i\delta_{jk} \sqrt{1-\hat n} = Q_{iN} \delta_{jk} - Q_{jk} \delta_{iN}, \\
\left[Q_{NN},Q_{iN}\right]&=&N^2_c[\hat n ,\phi_i] \sqrt{1-\hat n} = -N_c \phi_i \sqrt{1-\hat n} = Q_{NN}\delta_{iN} - Q_{iN} \delta_{NN}, \\
\left[Q_{iN},Q_{jN}\right]&=&N^2_c(1-\hat n) [\phi_i,\phi_j]=0, \\
\left[Q_{iN},Q_{Nj}\right]&=&N^2_c[\phi_i\sqrt{1-\hat n},\sqrt{1-\hat n} \phi^\dag_j]=N^2_c\left( \phi_i(1-\hat n) \phi^\dag_j - \sqrt{1-\hat n} \phi^\dag_j \phi_i \sqrt{1-\hat n} \right) \nonumber \\
&=&N^2_c \left( \phi_i (1-\hat n) \phi^\dag_j - \phi^\dag_j \phi_i \sqrt{1-\hat n} \right) = N^2_c[\phi_i(1-\hat n),\phi^\dag_j] \nonumber \\
&=&N_c(1-\hat n) -N_c \phi_i \phi^\dag_j. \\ \nonumber \\
&& Q.E.D. \nonumber
\end{eqnarray}

Next we write the operator $\phi$ as
\begin{equation}
  \label{eq:phi_H}
  \phi=\left( \begin{array}{c} \chi \\ \pi \end{array} \right).
\end{equation}
The $(N-2)$--component field $\chi$ will represent the zero modes and the fields $\pi$ will constitute the antiferromagnetic spin waves. 
We write the generators on the even sites by inserting \Eq{eq:phi_H} into \Eq{eq:realize}. As in Section~\ref{sec:O1} we replace 
$Q$ by
\begin{equation}
Q'(\phi) \equiv V Q(\phi) V^{\dag}
\end{equation}
 on the odd sites. $V$ is given in \Eq{eq:V}.
It is easy to see that the
  algebra~(\ref{eq:algebra}) on the odd sites is obeyed by $Q'$ as well. 
  
Now, we scale $\phi \rightarrow
\phi/\sqrt{N_c}$ , substitute $Q$ and $Q'$ into \Eq{eq:Hqmapp}, and 
  expand up to $O(1/N_c)$. We are left with the following normal ordered form of the Hamiltonian,
\begin{equation}
  \label{eq:Heff}
H_{\text{eff}}=N_c^2 E_0 + N_c H_2 + \sqrt{N_c} H_3 + H_4, 
  \end{equation}
where
\begin{eqnarray}
  H_2&=&J_1\sum_{\bn \muhat} \left( \pi^\dag_\bn \pi_\bn + \pi^\dag_{\bn+\muhat} \pi_{\bn+\muhat}
    -\pi_\bn \pi_{\bn+\muhat} -\pi^{\dag}_\bn \pi^{\dag}_{\bn+\muhat} \right),
    \label{eq:H2} \\
  H_3&=& J_1\sum_{\bn \muhat,i} \left( \chi^{\dag }_{\bn+\muhat,i} \chi_{\bn i} \pi_{\bn+\muhat} -
  \pi^{\dag }_\bn \chi^{\dag }_{\bn+\muhat,i} \chi_{\bn i} + h.c.
  \right), \label{eq:H3} \\
  H_4&=& J_1\sum_{\bn \muhat} \left\{ \sum_{ij} \chi^\dag_{\bn j} \chi_{\bn i} \chi^\dag_{\bn+\muhat,i} \chi_{\bn+\muhat,j}
  \right. \nonumber \\
  && \qquad \quad \sum_i \left[ \frac12 \left( \chi^{\dag}_{\bn i} \chi_{\bn i} \pi_{\bn+\muhat} \pi_\bn + \chi^{\dag}_{\bn+\muhat,i} \chi_{\bn+\muhat,i}  \pi_\bn \pi_{\bn+\muhat} + h.c.  \right) \right. \nonumber \\
    &&\qquad \qquad \quad \left. - \left(  \chi^{\dag}_{\bn i} \chi_{\bn i} \pi^{\dag}_{\bn+\muhat} \pi_{\bn+\muhat} + (\bn \leftrightarrow \bn+\muhat) \right) \right] \\
&& \qquad \quad \left. + \frac12 \left[ \left( \pi^{\dag}_\bn \pi_\bn \pi_\bn \pi_{\bn+\muhat} - \pi^{\dag}_\bn \pi^{\dag}_{\bn+\muhat} \pi_\bn \pi_{\bn+\muhat}
+ h.c. \right) + (\bn \leftrightarrow \bn+\muhat) \right] \right\} \nonumber
\label{eq:H4}
\end{eqnarray} 

\section{The ground state at $O(1)$}

At lowest order in $1/N_c$, the Hamiltonian is given by $H_2$, which
does not depend on $\chi_\bn$. Next we introduce the spatial Fourier transform
\begin{equation}
\label{eq:FT_spatial}
\phi^A_{\bN}=\sqrt{\frac{2}{N_s}} \sum_{\bk} \phi^A_\bk e^{i\bk \cdot \bN} \times \left\{ \begin{array}{ll} 1 & A=\text{even} \\ e^{-ik_z/2} & A=\text{odd} \end{array} \right. , 
\end{equation}
where $\bN$ is a site in the fcc lattice, and $\bk$ belongs to its first Brillouin zone. In momentum space one diagonalizes $H_2$ with a Bogoliubov transformation to find
\begin{equation}
  \label{eq:H2q}
  N_c H_2=2J_1d N_c \sum_\bq \sqrt{1-\gamma^2_\bq} \left(
  a^{\dag}_\bq a_\bq + b^{\dag}_\bq b_\bq \right). 
\end{equation}
$\gamma_\bq$ is given by \Eq{eq:gamma} and 
\begin{equation}
  \label{eq:Bogoliubov}
  \left( \begin{array}{c} \pi^\e_\bq \\ \pi^{\dag \o}_{-\bq} \end{array}
  \right) = \left( \begin{array}{ccc} \cosh{\varphi_\bq} & &
      \sinh{\varphi_\bq} \\ \sinh{\varphi_\bq} & & \cosh{\varphi_\bq}
    \end{array} \right) \left( \begin{array}{c} a_\bq \\
      b^{\dag}_{-\bq} \end{array} \right),
\end{equation}
where the even function $\phi_\bq=\phi_{-\bq}$ is given by
\begin{equation}
\tanh{2\varphi_\bq}=\gamma_\bq, \label{eq:Bog2}
\end{equation}
and the fields $a_\bq$ and $b_\bq$ obey
the usual commutation relations of ladder operators.
The $O(1)$ ground state $|\tilde 0 \rangle$ is thus given by
\begin{equation}
\label{eq:Hgs}
a_\bq |\tilde 0 \rangle = b_\bq |\tilde 0 \rangle = 0.
\end{equation}
To $O(1)$ the excitations of $|\tilde 0 \rangle$ are antiferromagnetic spin waves with a linear dispersion
relation. The ground state $|\tilde 0 \rangle$ has a local degeneracy, corresponding to creating an arbitrary number of $\chi$ bosons on any site (creation of a $\chi$ boson costs zero energy at this order). It is therefore more appropriate to write 
\begin{equation}
|\tilde 0 \> = |\tilde 0_\pi \> \otimes |\chi\>,
\end{equation}
which emphasizes that the ground state projection to the $\chi$ Fock space is arbitrary. This is exactly the result of Section~\ref{sec:O1}.

\section{Self energy and $O(1/N_c)$ effective Hamiltonian}

We now calculate an effective Hamiltonian that splits the spectrum of the degenerate subspace discussed above and removes the local degeneracy. We use Rayleigh-Schr\"{o}dinger perturbation
 theory \cite{Kato} where $H_4$ and $H_3$ are perturbations to $H_2$. In fact $H_4$ contributes at first order and $H_3$
at second order, and both give a contribution of
$O(1/N_c)$. The effective Hamiltonian is therefore
\begin{equation}
  \label{eq:Hdeg}
  \tilde{H}_{\text{eff}}={\bm P} H_4 {\bm P}+{\bm P} H_3 \frac{{\bm 1}-{\bm P}}{D} H_3 {\bm P} \equiv \tilde H^{(4)}_{\text{eff}} + \tilde H^{(3)}_{\text{eff}}.
\end{equation}
Here ${\bm P}$ is a projection operator to the degenerate sector 
The projection operator is
\begin{equation}
{\bm P}=|\tilde 0_\pi \> \< \tilde 0_\pi | \otimes {\bm 1}_\chi,
\end{equation}
and $D$ is the energy denominator. $H_{\text{eff}}$ is an effective Hamiltonian for the $\chi$ fields alone. Any bilinear of the $\pi$ fields in $\tilde H^{(3,4)}_{\text{eff}}$ is replaced by the bilinear's vacuum expectations value (vev), which we calculate using~(\ref{eq:Bogoliubov}) and~(\ref{eq:Hgs}),
\begin{eqnarray}
\langle \tilde 0 | \pi^{\dag e} \pi^e | \tilde 0 \rangle_\bN &=&\frac2{N_s} \sum_{\bk,\bq} e^{i\bN(\bq-\bk)} \<\tilde 0 |\pi^{\dag e}_\bk \pi^e_{\bq}|\tilde 0\>\nonumber \\
&&=\frac2{N_s}\sum_{\bk,\bq} e^{i\bN(\bq-\bk)} \sinh \varphi_\bk \sinh \varphi_\bq \<\tilde 0 |b_{-\bk} b^\dag_{-\bq}|\tilde 0\> \nonumber \\
&&=\frac2{N_s} \sum_\bq \sinh^2 \varphi_\bq, \label{eq:vev1pi} \\
\langle \tilde 0 | \pi^{\dag o} \pi^o | \tilde 0 \rangle_\bN &=&\frac2{N_s} \sum_{\bk,\bq} e^{i(\bN-\hat z/2)(\bq-\bk)} \<\tilde 0 |\pi^{\dag o}_\bk \pi^o_{\bq}|\tilde 0\> \nonumber \\
&&=\frac2{N_s}\sum_{\bk,\bq} e^{i(\bN-\hat z/2)(\bq-\bk)} \sinh \varphi_{-\bk} \sinh \varphi_{-\bq} \<\tilde 0 |a_{-\bk} a^\dag_{-\bq} |\tilde 0\> \nonumber \\
&&=\frac2{N_s} \sum_\bq \sinh^2 \varphi_{\bq}, \nonumber \\ \label{eq:vev2pi} \\
\langle \tilde 0 | \pi^e_\bN \pi^o_{\bN+\hat \rho} | \tilde 0 \rangle &=&\frac2{N_s} \sum_{\bk,\bq} e^{i\bN(\bq+\bk)+i\bq(\hat \rho-\hat z/2)} \<\tilde 0 |\pi^{e}_\bk \pi^o_{\bq}|\tilde 0\> \nonumber \\
&&=\frac2{N_s}\sum_{\bk,\bq} e^{i\bN(\bq+\bk)+i\bq(\hat \rho-\hat z/2)} \cosh \varphi_\bk \sinh \varphi_{-\bq} \<\tilde 0 |a_\bk a^\dag_{-\bq}|\tilde 0\>\nonumber \\
&&=\frac2{N_s} \sum_\bq \sinh \varphi_\bq \cosh \varphi_{\bq} e^{i\bq(\hat \rho -\hat z/2)}, \label{eq:vev3pi} \\
\langle \tilde 0 | \pi^{\dag e}_\bN \pi^{\dag o}_{\bN+\hat \rho} | \tilde 0 \rangle_\bN &=&\frac2{N_s} \sum_{\bk,\bq} e^{-i\bN(\bq+\bk)-i\bq(\hat \rho-\hat z/2)} \<\tilde 0 |\pi^{e}_\bk \pi^o_{\bq}|\tilde 0\> \nonumber \\
&&=\frac2{N_s}\sum_{\bk,\bq} e^{-i\bN(\bq+\bk)-i\bq(\hat \rho-\hat z/2)} \sinh \varphi_\bk \cosh \varphi_{-\bq} \<\tilde 0 |b_{-\bk} b^\dag_{\bq}|\tilde 0\>\nonumber \\
&&=\frac2{N_s} \sum_\bq \sinh \varphi_\bq \cosh \varphi_{\bq} e^{-i\bq(\hat \rho -\hat z/2)}. \label{eq:vev4pi}
\end{eqnarray}
Here we have used the following notation. $\bN$ is a site on the fcc lattice of even sites. On each fcc site one has two degrees of freedom, $\pi^e$, and $\pi^o$. The neighboring sites of $\bN$ are $\bN+\hat \rho$, with
\begin{equation}
\hat \rho = 0, \hat z, \hat z \pm \hat x/2, \hat z \pm \hat y/2.
\end{equation}

Inserting Eqs.~(\ref{eq:vev1pi})--(\ref{eq:vev4pi}) into ${\bm P}H_4{\bm P}$ we have 
\begin{eqnarray}
{\bm P}H_4{\bm P} &=& |\tilde 0_\pi \> \< \tilde 0_\pi | H_4 | \tilde 0_\pi \> \< \tilde 0_\pi|\\
\< \tilde 0_\pi | H_4 | \tilde 0_\pi \>&=& J_1\sum_{\bN \hat \rho} \left\{ \sum_{ij} \chi^{\dag e}_{\bN j} \chi^e_{\bN i} \chi^{\dag o}_{\bN+\hat \rho,i} \chi^o_{\bN+\hat \rho,j} \right. \nonumber \\
&& - \left. \sum_i \left( \chi^{\dag e}_{\bN i} \chi^e_{\bN i}+ \chi^{\dag o}_{\bN +\hat \rho, i} \chi^o_{\bN+\hat \rho, i}\right) \frac2{N_s} \sum_\bq \sinh^2 \varphi_\bq \right.  \nonumber \\
    && \left. + \sum_i \frac12 \left[ \left( \chi^{\dag e}_{\bN i}  \chi^e_{\bN i} + \chi^{\dag o}_{\bN+\hat \rho,i} \chi^o_{\bN+\hat \rho,i} \right) \frac2{N_s} \sum_\bq \sinh \varphi_\bq \cosh \varphi_\bq e^{i \bq (\hat \rho -\hat z/2)} + h.c. \right] \right\}. \nonumber \\
&=& J_1\frac2{N_s} \sum_{\bk \bq i}  \left\{  2\sum_{\bp j} \chi_{\bp j}^{\e \dag} \chi_{\bk i}^\e \chi_{\bq+\bk-\bp,i}^{\o \dag} \chi_{\bq j}^\o \cdot \gamma_{\bk-\bp} \right.\nonumber \\
&& \left. -2d \, \left( \chi^{\dag e}_{\bk i} \chi^e_{\bk i}+\chi^{\dag o}_{\bk i}  \chi^o_{\bk i}\right) \times \left( \sinh^2 \varphi_\bq - \gamma_\bq \sinh \varphi_\bq \cosh \varphi_\bq \right)\right\}.
\end{eqnarray}
Here we have omitted an irrelevant constant that represents terms quartic in $\pi$. Finally we use \Eq{eq:Bog2} to write $\tilde H^{(4)}_{\text{eff}}$ as
\begin{eqnarray}
\tilde H^{(4)}_{\text{eff}} &=& J_1d \frac{2}{N_s} \left\{ \sum_{\bq \bk ,i} \left( \begin{array}{cc}
    \chi^{\dag \e}_{\bk i} & \chi^{\o}_{-\bk i} \end{array} \right)
\left( \begin{array}{cc} 1-\sqrt{1-\gamma^2_\bq} & 0
    \\
     0  &  1-\sqrt{1-\gamma^2_\bq} \end{array} \right) \left(
  \begin{array}{c}
    \chi^{\e}_{\bk i} \\ \chi^{\dag \o}_{-\bk i} \end{array} \right). \right. \nonumber \\
&&\left. \qquad \qquad + 2\sum_{\bk \bp \bq \atop ij}  \chi_{\bp j}^{\e \dag} \chi_{\bk i}^\e \chi_{\bq+\bk-\bp,i}^{\o \dag} \chi_{\bq j}^\o \cdot \gamma_{\bk-\bp} \right\}
\label{eq:H4dec}
\end{eqnarray}
We see that $\tilde H^{(4)}_{\text{eff}}$ has harmonic and anharmonic terms. In order to evaluate $\tilde H^{(3)}_{\text{eff}}$ we write $({\bm 1}-{\bm P})$ as
\begin{equation}
\label{eq:1_P}
{\bm 1}-{\bm P}={\bm 1}_\chi \otimes \sum_\bk |\bk_a \rangle \langle \bk_a| + |\bk_b \rangle
  \langle \bk_b |,
\end{equation}
where, for example $|\bk_a \rangle \equiv a^{\dag}_\bk |\tilde 0_\pi
\rangle \otimes |\chi \rangle $. Here $ |\chi \rangle$ is $|\bk_a
\rangle$'s component in the $\chi$ Fock space. In general~(\ref{eq:1_P})  is incorrect since it takes into account
excitations of one spin wave only. 
Here it suffices since $H_3$ connects $|\tilde 0_\pi \rangle$ with excitations of that sort only. 
Next we replace the energy denominator with
\begin{equation}
\label{eq:D}
D=-2J_1dN_c\sqrt{1-\gamma^2_\bk},
\end{equation}
and obtain the following expression for $\tilde H^{(3)}_{\text{eff}}$,
\begin{eqnarray}
\tilde H^{(3)}_{\text{eff}}&=&-\frac{(J_1\sqrt{N_c})^2}{2dJ_1N_c}\sum_\bk \frac1{\sqrt{1-\gamma^2_\bk}} {\bm P}\sum_{\bN \hat \rho,i} \left( \chi^{\dag o}_{\bN+\hat \rho,i} \chi^e_{\bN i} \pi^o_{\bN+\hat \rho} -
  \pi^{\dag e}_\bN \chi^{\dag o}_{\bN+\hat \rho,i} \chi^e_{\bN i} + h.c.\right) \times \nonumber \\
&& |\bk_a \rangle \langle \bk_a| + |\bk_b \rangle
  \langle \bk_b | \times \sum_{\bN' \hat \rho' i'} \left( \chi^{\dag o}_{\bN'+\hat \rho',i'} \chi^e_{\bN' i'} \pi^o_{\bN'+\hat \rho'} -
  \pi^{\dag e}_{\bN'} \chi^{\dag o}_{\bN'+\hat \rho',i'} \chi^e_{\bN' i'} + h.c.\right) {\bm P}. \nonumber \\ \label{eq:Heff3a}
\end{eqnarray}

In order to calculate \Eq{eq:Heff3a} we use \Eq{eq:Bogoliubov} and note that
\begin{equation}
\< \tilde 0_\pi | \pi^{\dag e}_\bN | \bk_a \> \sim \< \tilde 0_\pi | \left( a^{\dag} \quad  \text{or} \quad  b \right) a^{\dag}_\bk |\tilde 0_\pi\>=0.
\end{equation}
On similar lines one can easily show that $\< \tilde 0_\pi | \pi^{o}_\bN | \bk_a \>=\< \tilde 0_\pi | \pi^{e}_\bN | \bk_b \>=\< \tilde 0_\pi | \pi^{\dag o}_\bN | \bk_b \>=0$. The only non-zero elements are
\begin{eqnarray}
\< \tilde 0_\pi | \pi^{e}_\bN | \bk_a \> &=&\sqrt{\frac2{N_s}} \cosh \varphi_{\bk} e^{i\bk \bN}, \\
\< \tilde 0_\pi | \pi^{\dag o}_\bN | \bk_a \> &=&\sqrt{\frac2{N_s}} \sinh \varphi_{\bk} e^{-i\bk (\bN-\hat z/2)}, \\
\< \tilde 0_\pi | \pi^{\dag e}_\bN | \bk_b \> &=&\sqrt{\frac2{N_s}} \sinh \varphi_{\bk} e^{i\bk \bN}, \\
\< \tilde 0_\pi | \pi^{o}_\bN | \bk_b \> &=&\sqrt{\frac2{N_s}} \cosh \varphi_{\bk} e^{i\bk (\bN-\hat z/2)}.
\end{eqnarray}
Using these relations in \Eq{eq:Heff3a} we see that $\tilde H^{(3)}_{\text{eff}}$ is a quartic interaction in the $\chi$ fields. For example the contribution of  $|\bk_b \> \< \bk_b |$ is
\begin{eqnarray}
&&-J_1 \frac1{N_s}\sum_\bk \frac1{\sqrt{1-\gamma^2_\bk}} \sum_{\bN \bN' \atop \hat \rho \hat \rho' ii'} \left( \cosh \varphi_{\bk} e^{i\bk(\bN+\hat \rho -\hat z/2)} \chi^{\dag o}_{\bN+\hat \rho,i}  \chi^e_{\bN i} -
  \sinh \varphi_{\bk} e^{i\bk\bN} \chi^{\dag o}_{\bN+\hat \rho,i} \chi^e_{\bN i} \right) \times \nonumber \\
&&\hskip 2cm \left( \cosh \varphi_{\bk} e^{-i\bk(\bN'-\hat z/2+\hat \rho')}\chi^{\dag e}_{\bN' i'} \chi^{o}_{\bN'+\hat \rho',i'} - \sinh \varphi_{\bk} e^{-i\bk\bN'}\chi^{\dag e}_{\bN' i'} \chi^{o}_{\bN'+\hat \rho',i'} \right)\nonumber \\
&&=-J_1\frac1{N_s}\sum_\bk \cosh 2\varphi_\bk \sum_{\bN \bN' \atop \hat \rho \hat \rho' ii'} \chi^{\dag o}_{\bN+\hat \rho,i} \chi^e_{\bN i} \chi^{\dag e}_{\bN' i'} \chi^{o}_{\bN'+\hat \rho',i'} \times e^{i\bk (\bN-\bN')} \times \nonumber \\
&&\hskip 3cm \left( \cosh \varphi_\bk e^{i\bk(\hat \rho-\hat z/2)}-\sinh \varphi_\bk \right) \left( \cosh \varphi_\bk e^{-i\bk(\hat \rho'-\hat z/2)}-\sinh \varphi_\bk \right) \nonumber \\
&&=-J_1\frac1{N_s}\sum_\bk \cosh 2\varphi_\bk \sum_{\bN \bN' \atop \hat \rho \hat \rho' ii'} \chi^{\dag o}_{\bN+\hat \rho,i} \chi^{o}_{\bN'+\hat \rho',i'} \chi^{\dag e}_{\bN' i'} \chi^e_{\bN i} \times e^{i\bk (\bN-\bN')} \times \nonumber \\
&&\hskip 3cm \left( \cosh \varphi_\bk e^{i\bk(\hat \rho-\hat z/2)}-\sinh \varphi_\bk \right) \left( \cosh \varphi_\bk e^{-i\bk(\hat \rho'-\hat z/2)}-\sinh \varphi_\bk \right) \nonumber \\
&& -J_1\frac1{N_s}\sum_\bk \cosh \varphi_\bk \sum_{\bN \hat \rho \hat \rho' i} \chi^{\dag o}_{\bN+\hat \rho,i} \chi^{o}_{\bN+\hat \rho',i} \times \left( \cosh \varphi_\bk e^{i\bk(\hat \rho-\hat z/2)}-\sinh \varphi_\bk \right) \nonumber \\
&&\hskip 3cm \left( \cosh \varphi_\bk e^{-i\bk(\hat \rho'-\hat z/2)}-\sinh \varphi_\bk \right). \label{eq:Heff3b}
\end{eqnarray}
We write the last (quadratic) term as
\begin{eqnarray}
&&-J_1\frac1{N_s}\sum_{\bq \bk i} \cosh 2\varphi_\bk \sum_{\hat \rho \hat \rho'} \chi^{\dag o}_{\bq i} \chi^{o}_{\bq i} \times \left( \cosh^2 \varphi_\bk e^{i(\bk+\bq)(\hat \rho-\hat \rho')}+\sinh^2 \varphi_\bk e^{(\bk+\bq)(\hat \rho-\hat \rho')} \nonumber \right. \\
&&\hskip 1cm \left. -\cosh \varphi_\bk \sinh \varphi_\bk (e^{i(\bk-\bq)(\hat \rho-\hat z/2)+i\bq(\hat \rho'-\hat z/2)}+e^{i(\bq-\bk)(\hat \rho'-\hat z/2)-i\bq(\hat \rho-\hat z/2)})  \right) \nonumber \\
&&=-4d^2J_1\frac1{N_s}\sum_{\bq \bk i} \chi^{\dag o}_{
\bq i} \chi^{o}_{\bq i}  \, \cosh 2\varphi_\bk \left( \gamma^2_{\bq+\bk}(\cosh^2 \varphi_\bk +\sinh^2 \varphi_\bk )  -2\gamma_\bq \gamma_{\bq+\bk}\cosh \varphi_\bk \sinh \varphi_\bk  \right) \nonumber \\
&&=-4d^2J_1\frac1{N_s}\sum_{\bq \bk i} \chi^{\dag o}_{
\bq i} \chi^{o}_{\bq i}  \, \cosh 2\varphi_\bk  \left( \gamma_{\bq+\bk}\cosh2 \varphi_\bk -\gamma_\bq \sinh 2\varphi_\bk \right) \gamma_{\bq+\bk}
\label{eq:Heff3c}
\end{eqnarray}

In order to treat the quartic interactions in $\tilde H_{\text{eff}}$ we make an ansatz for the ground state of the $\chi$ Fock space, and assume the following vacuum expectation values of the $\chi$ fields.
\begin{eqnarray}
\< \tilde 0 |\chi^{\dag e}_{\bk j} \chi^e_{\bk' i} |\tilde 0 \>
&=&\< \tilde 0 |\chi^{\dag o}_{\bk j} \chi^o_{\bk' i} |\tilde 0 \>
\equiv  \frac12 \Delta_{1\bk} \delta_{\bk \bk'} \delta_{ij},
\label{eq:del1} \\
\< \tilde 0 | \chi^e_{\bk i} \chi^o_{-\bk' ,j} | \tilde 0 \> &=& \< \tilde 0 | \chi^{\dag e}_{\bk i} \chi^{\dag o}_{-\bk',j} | \tilde 0 \> ^* \equiv 
-\frac12 \Delta_{2\bk} \delta_{\bk \bk'} \delta_{ij}, \label{eq:del2}
\end{eqnarray}
or in real space
\begin{eqnarray}
\< \tilde 0 |\chi^{\dag e}_{\bN j} \chi^e_{\bN' i} |\tilde 0 \>
&=&\< \tilde 0 |\chi^{\dag o}_{\bk j} \chi^o_{\bk' i} |\tilde 0 \>
\equiv  \frac1{N_s} \sum_{\bk} \Delta_{1\bk} e^{i\bk(\bN'-\bN)} \cdot \delta_{ij},
\label{eq:del1real} \\
\< \tilde 0 | \chi^e_{\bN i} \chi^o_{\bN' ,j} | \tilde 0 \> &=& \< \tilde 0 | \chi^{\dag e}_{\bN i} \chi^{\dag o}_{\bN',j} | \tilde 0 \> ^* \equiv 
-\frac1{N_s} \sum_\bq \Delta_{2\bk} e^{i\bk(\bN-\bN'+\hat z/2)}\delta_{\bk \bk'} \delta_{ij}. \nonumber \\ \label{eq:del2real}
\end{eqnarray}
Here $\Delta_{1,2}$ are assumed to be real and even in the momentum $\bk$. Note that a non-zero value of $\Delta_2$ breaks the stationary group of the condensates of $Q_e$ and $Q_o$, $U(2m-N)\times U(N-m) \times U(N-m)$, to $O(2m-N)\times U(N-m)$. 

As in the treatment of order from disorder in condensed matter (see for example \cite{Aharony}) we decouple all the quartic terms in $\tilde H_{\text{eff}}$ to be quadratic terms times a condensate in all possible ways. The quartic term in $\tilde H^{(4)}_{\text{eff}}$ becomes the following four quadratic terms. (We go back to real space.)
\begin{eqnarray}
&&J_1 \sum_{\bN \hat \rho \atop ij} \left( \chi^\dag_j \chi_i \right)^e_\bN \left( \chi^\dag_i \chi_j \right)^o_{\bN+\hat \rho} \simeq J_1 \sum_{\bN \hat \rho \atop ij} \left[ \< \chi^\dag_j \chi_i \>^e_\bN \left( \chi^\dag_i \chi_j \right)^o_{\bN+\hat \rho} + \left( \chi^\dag_j \chi_i \right)^e_\bN \< \chi^\dag_i \chi_j \>^o_{\bN+\hat \rho} \right. \nonumber \\
&&\hskip 4cm \left. + \chi^e_{\bN i} \chi^o_{\bN + \hat \rho,j} \< \chi^{\dag e}_{\bN j} \chi^{\dag o}_{\bN + \hat \rho,i}\> + \< \chi^e_{\bN i} \chi^o_{\bN + \hat \rho,j} \> \chi^{\dag e}_{\bN j} \chi^{\dag o}_{\bN + \hat \rho,i} \right] \nonumber \\
&&=J_1 \frac1{N_s} \sum_{\bq \bN \hat \rho i} \left[ \Delta_{1\bq} \left( \chi^{\dag o}_{\bN +\hat \rho,i} \chi^o_{\bN+\hat \rho,i} + \chi^{\dag e}_\bN \chi^e_{\bN i} \right) - \Delta_{2\bq} e^{i\bq(\hat \rho-\hat z/2)} \left( \chi^e_{\bN i} \chi^o_{\bN + \hat \rho,i} + \chi^{\dag e}_{\bN j} \chi^{\dag o}_{\bN + \hat \rho,i} \right) \right] \nonumber \\
&&=2dJ_1\frac1{N_s}\sum_{\bq \bk,i} \left[ \Delta_{1\bq} \left( \chi^{\dag e}_{\bk i} \chi^e_{\bk i} + \chi^{\dag o}_{\bk i} \chi^o_{\bk i} \right) -\Delta_{2\bq} \gamma_{\bq-\bk} \left( \chi^e_{\bk i} \chi^o_{-\bk ,i} + \chi^{\dag e}_{\bk i} \chi^{\dag o}_{-\bk, i} \right) \right]. \label{eq:Heff4a}
\end{eqnarray}
The next step is to use the ansatz~(\ref{eq:del1})--(\ref{eq:del2}) in \Eq{eq:Heff4a}. This gives
\begin{equation}
\label{H4dec}
\tilde H^{(4)}_{\text{eff}} = J_1d \frac{2}{N_s} \sum_{\bq \bk ,i} \left( \begin{array}{cc}
    \chi^{\dag \e}_{\bk i} & \chi^{\o}_{-\bk i} \end{array} \right)
\left( \begin{array}{cc} \Delta_{1 \bq} +
    1-\sqrt{1-\gamma^2_q} & - \Delta_{2 \bq} \gamma_{\bq-\bk}
    \\
     - \Delta_{2 \bq} \gamma_{\bq-\bk} &  \Delta_{1 \bq}  +
    1-\sqrt{1-\gamma^2_q} \end{array} \right) \left(
  \begin{array}{c}
    \chi^{\e}_{\bk i} \\ \chi^{\dag \o}_{-\bk i} \end{array} \right). \qquad
\end{equation}
The same procedure is performed to the quartic terms of $\tilde H^{(3)}_{\text{eff}}$. Here the procedure in much more cumbersome, and the four quartic terms of \Eq{eq:Heff3b} alone give $4\times 4=16$ different quadratic terms. The $|\bk_a\>\<\bk_a|$ terms contribute 16 more. For example, in order to calculate the contribution of \Eq{eq:Heff3b} we first decouple the quartic term,
\begin{eqnarray} 
\chi^{\dag o}_{\bN+\hat \rho,i} \chi^{o}_{\bN'+\hat \rho',i'} \chi^{\dag e}_{\bN' i'} \chi^e_{\bN i} &\rightarrow& \< \chi^{\dag o}_{\bN+\hat \rho,i} \chi^{o}_{\bN'+\hat \rho',i'} \> \chi^{\dag e}_{\bN' i'} \chi^e_{\bN i} + \chi^{\dag o}_{\bN+\hat \rho,i} \chi^{o}_{\bN'+\hat \rho',i'} \< \chi^{\dag e}_{\bN' i'} \chi^e_{\bN i} \> \nonumber \\
&&+\<\chi^{\dag o}_{\bN+\hat \rho,i} \chi^{\dag e}_{\bN' i'} \> \chi^{o}_{\bN'+\hat \rho',i'} \chi^e_{\bN i}+\chi^{\dag o}_{\bN+\hat \rho,i} \chi^{\dag e}_{\bN' i'} \< \chi^{o}_{\bN'+\hat \rho',i'} \chi^e_{\bN i} \>. \nonumber \\ \label{eq:decouple}
\end{eqnarray}
Next we substitute~(\ref{eq:del1real})--(\ref{eq:del2real}) into~(\ref{eq:decouple}) and into its $|\bk_a \> \< \bk_a |$ counterpart, and find (after a lot of algebra) 
\begin{equation}
\label{eq:Hresult}
\tilde{H}_{\text{eff}} = J_1d \sum_{\bk,i} \left( \begin{array}{cc}
    \chi^{\dag \e}_{\bk i} & \chi^{\o}_{-\bk i} \end{array} \right)
\left( \begin{array}{cc} {\cal J}_{1 \bk} & {\cal J}_{2 \bk} \\
    {\cal J}_{2 \bk} &  {\cal J}_{1 \bk} \end{array} \right) \left(
  \begin{array}{c}
    \chi^{\e}_{\bk i} \\ \chi^{\dag \o}_{-\bk i} \end{array} \right), \qquad
\end{equation}
with
\begin{eqnarray}
{\cal J}_{1 \bk} &=& \int_{\text{BZ}} \left( \frac{dq}{4\pi} \right) ^d
\left( \Delta_{1 \bq} + 1 \right) I_1(\bq,\bk)
-\eta(\bk), \label{eq:InteqH1} \\
{\cal J}_{2 \bk} &=&  \int_{\text{BZ}} \left( \frac{dq}{4\pi} \right) ^d \Delta_{2 \bq} I_2(\bq,\bk). \label{eq:InteqH2}
\end{eqnarray}
$I_{1,2}(\bq,\bk)$ and $\eta(\bk)$ are as
given in Eqs.~(\ref{eq:a1})--(\ref{eq:a2}). The Hamiltonian~(\ref{eq:Hresult}) can be diagonalized by a Bogoliubov transformation 
\begin{equation}
  \label{eq:Bogoliubov1}
  \left( \begin{array}{c} \chi^\e_\bq \\ \chi^{\dag \o}_{-\bq} \end{array}
  \right) = \left( \begin{array}{ccc} \cosh{\frac12\theta_\bq} & &
      \sinh{\frac12\theta_\bq} \\ \sinh{\frac12\theta_\bq} & & \cosh{\frac12\theta_\bq}
    \end{array} \right) \left( \begin{array}{c} c_\bq \\
      d^\dag_{-\bq} \end{array} \right),
\end{equation}
with 
\begin{equation}
\label{eq:Bog3}
\tanh(\theta_\bk)=-{\cal J}_{2\bk}/{\cal J}_{1\bk}.
\end{equation}
 The result is
\begin{equation}
\tilde H_{\text{eff}}=J_1d\sum_{\bk i} \sqrt{{\cal J}^2_{1\bk}-{\cal J}^2_{\bk 2}} \left( c^\dag_\bk c_\bk + d^\dag_\bk d_\bk \right),
\end{equation}
which means that the energies of the $\chi$ bosons are smaller by a factor of $N_c$ than the $\pi$ fields' energies.

To obtain self-consistent equations we calculate the
vevs~(\ref{eq:del1})--(\ref{eq:del2}) using~(\ref{eq:Hresult}) and as in the calculation of the $\pi$ vevs~(\ref{eq:vev1pi})--(\ref{eq:vev2pi}) one has
\begin{eqnarray}
\langle \tilde 0 | \chi^{\dag e}_\bq \chi^{e}_\bq | \tilde 0 \>
&=&\langle \tilde 0 | \chi^{\dag o}_\bq \chi^{o}_\bq | \tilde 0 \rangle
= \sinh^2{\frac12\theta_\bq}=\frac12(\cosh \theta_\bq -1), \label{eq:vev1chi} \\
\langle \tilde 0 |  \chi^\e_\bq \chi^\o_{-\bq} | \tilde 0 \rangle
&=&\left( \langle \tilde 0 | \chi^{\dag \e}_\bq \chi^{\dag \o}_{-\bq} | \tilde 0 \rangle \right)= \sinh{\frac12\theta_\bq} \cosh{\frac12\theta_\bq}=\frac12 \sinh \theta_\bq , \label{eq:vev2chi}
\end{eqnarray}
which leads us to identify $\Delta_{1\bk}=\cosh \theta_\bk -1$, and $\Delta_{2\bk}=-\sinh \theta_\bk$. Substituting this into Eqs.~(\ref{eq:InteqH1})--(\ref{eq:InteqH2}) we write \Eq{eq:Bog3} as
\begin{equation}
\tanh{\theta_\bk}=\frac{ {\displaystyle \int_{\text{BZ}} \left( \frac{dq}{4\pi}
  \right)^d I_2(\bq,\bk) \sinh{\theta_\bq} } }{ {\displaystyle \int_{\text{BZ}} \left(
    \frac{dq}{4\pi} \right)^d 
  I_1(\bq,\bk) \cosh{\theta_\bq} -\eta(\bk) } },
\end{equation}
which is exactly the integral equation \Eq{eq:Inteq} for $N-m=1$. 

This Hamiltonian calculation sheds light on the meaning of the
function $\theta_\bk$ that appears in Chapter~\ref{chap:Qfluctuation} in terms of vevs of the $\chi$ operators. A stable solution that obeys~(\ref{eq:stability}) has
non-zero vevs. 
Mathematically, the integral
equation~(\ref{eq:Inteq}) has also a trivial solution with $\theta_\bk=0$ that leads to 
$\Delta_{1,2}=0$.  However, this solution does not
obey~(\ref{eq:stability}) and is thus unphysical, giving ${\cal J}_1<0$ and making $\tilde{H}_{\text{eff}}$ not positive definite.

Finally, we note that the Holstein--Primakoff transformation for $N-m>1$ is much more complicated \cite{Smit}, making the Hamiltonian approach unattractive for this case. The path integral approach is more general.

\clearpage
\chapter[Effective interaction between odd sites in the sigma model]{Effective ferromagnetic interaction at $O(1/N_c)$ between odd sites in the nonlinear sigma model with non-zero density\label{app:corrections}}

In this section we derive an effective interaction in $O(1/N_c)$ between the classical parts of the sigma fields on the odd sites, and show that this is a ferromagnetic interaction which prefers that these fields align to the same direction within the degenerate submanifold discussed in Chapter~\ref{chap:NLSM}. The effective interaction comes from integrating over over fluctuations around the classical $N_c=\infty$ ground state and derive an effective action for the classical part (\ref{submanifold})
of the odd spins. We
let the $\sigma_\bn$ on the even sites fluctuate around $\Lambda$;
we let the $\sigma_\bn$ on odd sites roll freely around the
$U(m)/[U(2m-N)\times U(N-m)]$ manifold covered by
\Eq{submanifold}, and also execute small oscillations off the
manifold into the $U(N)/[U(m)\times U(N-m)]$ coset space. 
The counterpart of the action (\ref{eq:Srescaled}) for our problem
has an antiferromagnetic spin-spin interaction, with no $(-1)^\bn$
factors. We separate it into odd, even, and coupled terms,
\begin{eqnarray}
S &=& \frac{N_c}2\left(S^{\text{odd}} +
S^{\text{even}}+S^{AF}\right),
\nonumber   \\
S^{\text{odd}}&=&\int d\tau \sum_{\bn \text{odd}}
\Tr\Lambda U_\bn^{\dag} \partial_{\tau} U_\bn, \nonumber    \\
S^{\text{even}}&=&\int d\tau \sum_{\bn \text{even}}
\Tr\Lambda U_\bn^{\dag} \partial_{\tau} U_\bn, \nonumber    \\
S^{AF} &=& \int d\tau \sum_{\bn \text{even}} \frac {J_1}2 \Tr
\sigma_\bn \sigma_\bn^o.
\end{eqnarray}
Here $\sigma_\bn^o = \sum_{\bmm(\bn)} \sigma_\bmm$, where
$\bmm(\bn)$ are the nearest neighbors of the even site $\bn$. We
expand the field on the even sites around $\sigma_\bn=\Lambda$ in
the manner of \Eq{eq:sigmaL},
\begin{equation}
\sigma_\bn=\Lambda+L_\bn-\frac12L^2\Lambda\qquad\text{(}\bn \quad
\text{even),} \label{sigmaL2}
\end{equation}
while for the odd sites we write (see Appendix \ref{app:oddspins})
\begin{equation}
\sigma_\bn= U_\bn \left( \begin{array}{cc}
        \bm1_{2m-N}        & 0     \\
        0       & -\Lambda' - L' + \frac12 L'^2 \Lambda'
\end{array} \right) U_\bn^{\dag}\qquad \text{(}\bn \quad \text{odd),}
\label{eq:Sig_1}
\end{equation}
with
\begin{equation}
\Lambda' = \left( \begin{array}{cc}
        \bm1_{N-m} &       0       \\
        0       &       -\bm1_{N-m}        \end{array} \right)
\end{equation}
and
\begin{equation}
U_\bn = \left( \begin{array}{cc}
        U^{(m)}_\bn     & 0     \\
        0       & \bm1_{N-m}       \end{array}     \right).
\end{equation}
$L_\bn$ describes the fluctuations of the even spins around their
classical value $\Lambda$. $U_\bn$ rotates the odd spins within
the manifold of their classical values, while $L'_\bn$ describes
their fluctuations outside that manifold. We further define
\begin{equation}
\sigma^{\text{cl}}_\bn=U_\bn \left( \begin{array}{cc}
        \bm1_{2m-N}        & 0     \\
        0       & -\Lambda'
\end{array} \right) U_\bn^{\dag}
=\left(\begin{array}{cc}
\sigma^{(m)}_\bn&0\\
0&\bm1_{N-m}
\end{array} \right),
\end{equation}
the classical field on the odd sites.

We leave $S^{\text{odd}}$ alone and expand $S^{\text{even}}$
and~$S^{AF}$ around the classical values of the fields,
\begin{eqnarray}
S^{\text{even}}&=& - \frac14 \int d\tau \sum_{\bn \text{even}}
\Tr
\Lambda L_\bn \partial_\tau L_\bn,  \\
S^{AF} &=& S_0 + \frac{J_1}2 \int d\tau \sum_{\bn \text{even}}
\left( \Tr L_\bn \bar\sigma_\bn -\frac12\Tr L_\bn^2 \Lambda
\bar\sigma_\bn
- \Tr L_\bn \bar L_\bn \right) \nonumber \\
&&\ +dJ_1 \int d\tau \sum_{\bn \text{odd}} \left(-\Tr\Lambda\tilde
L_\bn +\frac12\Tr\tilde L^2_\bn \right). \label{eq:SAF}
\end{eqnarray}
Here
\begin{equation}
\tilde L_\bn=U_\bn \left( \begin{array}{cc}
                0       &       0       \\
                0       &       L'_\bn\end{array} \right)
U^\dag_\bn
\label{tildeL}
\end{equation}
is the rotated fluctuation field on the odd sites, and the
Hermitian matrices $\bar\sigma_\bn$ and $\bar L_\bn$ are sums over
the odd neighbors of the even site $\bn$,
\begin{eqnarray}
\bar\sigma_\bn &=& \sum_{\bmm(\bn)} \sigma^{\text{cl}}_\bmm,    \\
\bar L_\bn     &=& \sum_{\bmm(\bn)} \tilde L_\bmm. \label{eq:Lo}
\end{eqnarray}
Since both $\bar\sigma_\bn$ and $\Lambda$ are block diagonal, the
first trace in each integral in \Eq{eq:SAF} is zero.

Now we organize the partition function as follows:
\begin{equation}
Z=\int \left(\prod_{\bn \text{odd}}d\sigma_\bn\right)\, \exp
\left[-\frac{N_c}2 \left( S^{\text{odd}} + S_0 + \frac{dJ_1}2 \int
d\tau \sum_{\bn \text{odd}}\Tr \tilde L^2_\bn \right) \right]
Z_{\text{even}}, \label{eq:Z}
\end{equation}
with
\begin{eqnarray}
Z_{\text{even}}&=&\int \left(\prod_{\bn
\text{even}}dL_\bn\right)\, \exp \left[-\frac{N_c}2 \int d\tau \right.
\nonumber\\
&&\left.\quad\times\sum_{\bn \text{even}}\left( -\frac14 \Tr
\Lambda L_\bn \partial_\tau L_\bn -\frac {J_1}4 \Tr L_\bn^2 \Lambda
\bar\sigma_\bn -\frac {J_1}2 \Tr L_\bn \bar L_\bn \right)\right].
\end{eqnarray}
$Z_{\text{even}}$ is a product of decoupled single-site integrals.
Again we expand in the group algebra,
\begin{equation}
L_\bn = l_\bn^\eta M^\eta,
\end{equation}
where the sum is over the $2m(N-m)$ generators of $U(N)$ that are
not in $U(m)\times U(N-m)$. $\bar L_\bn$ can be expanded similarly
and we obtain the following form for the integral over the even
fields:
\begin{equation}
Z_{\text{even}} = \int Dl_\bn \exp \left[ -\frac{N_c}2 \int d\tau
\sum_\bn \left( l_\bn^\eta \mathcal M_\bn^{\eta\eta'}
l_\bn^{\eta'} +\frac {J_1}4 l^{\eta}_\bn \bar l^\eta_\bn \right)
\right],
\end{equation}
where
\begin{equation}
\mathcal M_\bn^{\eta \eta'}=-\frac18C^{\eta\eta'} \partial_{\tau}
+\frac {J_1}8 D^{\eta \eta'}_\bn.
\end{equation}
The matrix $C$ is the same as in \Eq{eq:Cetaeta},
\begin{equation}
C^{\eta \eta'}=\Tr(\Lambda[M^\eta, M^{\eta'}]),
\end{equation}
while the new matrix $D$ varies with the site $\bn$ according to
the average $\bar\sigma_\bn$ of its neighboring spins,
\begin{equation}
D^{\eta\eta'}_\bn=-\Tr(\{M^{\eta},M^{\eta'}\}\Lambda\bar\sigma_\bn).
\label{eq:Dmatrix}
\end{equation}
We study the two matrices in Appendix \ref{app:CD}. Diagonalizing
$C$ as before, we arrive at
\begin{equation}
Z_{\text{even}} = \prod_{\bn q}\left\{ \int Dl\,\exp -\frac{N_c}2
\int d\tau \left[ l^{q\dag} \hat\mathcal{M}_\bn l^q +\frac{J_1}4
\left( l^{q\dag}\bar{l}_\bn^q + \bar{l}_\bn^{q\dag}l^q \right)
\right] \right\},
\end{equation}
where $\hat\mathcal{M}_\bn$ is the $m\times m$ matrix
\begin{equation}
\hat\mathcal{M}_\bn = \frac14 \left( \partial_{\tau} + J_1 E_\bn
\right). \label{eq:M}
\end{equation}
The quantities $l^q$ (and $\bar l^q$) for each $q=1,\ldots,N-m$
are complex $m$-component vectors; they are rotations of the
$2m(N-m)$ real components $l^\eta$ (and $\bar l^\eta$) into the
basis that diagonalizes $C$. The matrix $E_\bn$ is given by
\Eq{eq:E}; it carries the dependence on $\bar\sigma_\bn$.
Performing the gaussian integration we have
\begin{equation}
Z_{\text{even}} = \prod_{\bn q} \frac1{\Det \hat\mathcal{M}_\bn}
\exp \left[ \frac{N_c}2 \left( \frac {J_1}4 \right)^2 \int d\tau\,\bar
l^{q\dag}_\bn \hat\mathcal{M}_\bn^{-1} \bar l_\bn^q \right].
\label{eq:Zeven}
\end{equation}

Finally we separate the integral (\ref{eq:Z}) over the odd spins
into an integral over the classical field $\sigma_\bn^{\text{cl}}$
and an integral over the fluctuations around it.  We obtain
\begin{eqnarray}
Z&=&\int D\sigma_\bmm^{\text{cl}}\, \exp -\frac{N_c}2\left(S_0
+(N-m)\sum_{\bn}\Tr\log\hat\mathcal{M}_\bn \right) \nonumber \\
&& \times\int D\tilde l_\bmm \,\exp -\frac{N_c}2
\left\{S^{\text{odd}} + \int d\tau \left[\frac{dJ_1}2 \sum_{\bmm q}
|\tilde l^q_\bmm|^2 -\left(\frac {J_1}4 \right)^2\sum_{\bn q} \bar
l_\bn^{q\dag}\hat\mathcal{M}_\bn^{-1} \bar l_\bn^{q} \right]
\right\}. \nonumber \\ \label{eq:Z_eff}
\end{eqnarray}
Here $\bmm$ stands for an odd site, $\bn$ for an even one.

Equation~(\ref{eq:Z_eff}) gives an effective action for the classical
odd spins $\sigma_\bmm^{\text{cl}}$.
These enter the exponents through $\hat\mathcal{M}$ [via
Eqs.~(\ref{eq:E}) and~(\ref{eq:M})] and through
$\bar{l}^{q}_\bn$
[via Eqs.~(\ref{tildeL}) and~(\ref{eq:Lo})].
The action in the first exponent
is minimized when each matrix $E_\bn(\sigma_\bmm^{\text{cl}})$
has the largest number of zero
eigenvalues, each of which makes 
$\Tr\log\hat\mathcal{M}_\bn$ approach $-\infty$. It
is easy to check that $E_\bn$ has $2m-N$ zero eigenvalues (the maximal
number) when the $\sigma_\bmm^{\text{cl}}$ on {\em all} the odd
sites $\bmm(\bn)$ align, {\em i.e.,}
\begin{equation}
\sigma^{\rm cl}_\bmm = \sigma_0 \in U(m)/[U(2m-N)\times U(N-m)].
\label{eq:minimum}
\end{equation}
Moreover, when \Eq{eq:minimum} holds, all the $\tilde{l}_\bmm^q$'s
align parallel to each other and $\bar{l}_\bn^{q}$ is maximized;
also the eigenvalues of $\hat\mathcal{M}_\bn^{-1}$ are
maximized (to $+\infty$).
Thus the action in
the second exponent also has a minimum at
this point in configuration space.
These effects add up to an effective {\em ferromagnetic\/}
interaction among the $2d$ nearest neighbors $\bmm$ of each even
site $\bn$. This effective interaction will
align the classical spins on the odd sublattice to the same direction in
their submanifold, $U(m)/[U(2m-N)\times U(N-m)]$.

The divergences in the effective action have their origin in the fact
that the semiclassical corrections are calculated as gaussian integrals
in the even fluctuation fields $L_\bn$, and the coefficient matrix
$\mathcal M_\bn$ acquires zero eigenvalues.
The correct range of integration over $L_\bn$ is of course not infinite,
but rather the volume of the $U(N)/[U(m)\times U(N-m)]$ manifold.
This will regulate the divergences, but leave the effective action for
the odd spins attractive.
\clearpage
\chapter[Fluctuations on the odd sites at non-zero density]{Fluctuations on the odd sites at non-zero density\label{app:oddspins}}

In this section we show how to separate the fluctuation outside the degenerate submanifold of the odd spin. 

In the classical analysis, the fields on the odd sites take values
in the sub-manifold $U(m)/[U(2m-N)\times U(N-m)]$ of the manifold
$U(N)/[U(m)\times U(N-m)]$. We denote these values
$\sigma^{\text{cl}}$,
\begin{equation}
\sigma^{\text{cl}}= \left( \begin{array}{cc}
    \sigma^{(m)} & 0    \\
    0   &   \bm1_{N-m}   \end{array}    \right). \label{sigmcl}
\end{equation}
Here
\begin{equation}
\sigma^{(m)} = U^{(m)} \Lambda^{(m)} U^{(m)\dag}. \label{eq:Sigm}
\end{equation}
with $U^{(m)} \in U(m)$, and
\begin{equation}
\Lambda^{(m)} = \left( \begin{array}{cc}
    \bm1_{2m-N} & \\
    0   & -\bm1_{N-m}   \end{array} \right).
\end{equation}

$\sigma^{(m)}$ contains $2(N-m)(2m-N)$ independent degrees of
freedom. Any $\sigma\in U(N)/[U(m)\times U(N-m)]$ can be written
as
\begin{equation}
\sigma=\left( \begin{array}{cc} \cos\left(2\sqrt{aa^{\dag}}\right)
& -a\,\frac{\displaystyle\sin\left(2\sqrt{a^{\dag}a}\right)}
    {\displaystyle\sqrt{a^{\dag}a}} \\[2pt]
-\frac{\displaystyle\sin\left(2\sqrt{a^{\dag}a}\right)}
    {\displaystyle\sqrt{a^{\dag}a}}\,a^{\dag}   &
-\cos\left(2\sqrt{a^{\dag}a}\right) \end{array} \right)
\label{eq:Sig_gen}
\end{equation}
[cf.~\Eq{eq:sigma_a}], which coincides with \Eq{sigmcl} if
\begin{equation}
a =  U^{(m)}\left( \begin{array}{c}
    0   \\
    (\pi/2) \bm1_{N-m}  \end{array} \right).
\end{equation}
Recall that $a$ is an $m\times(N-m)$ matrix, so the zero block has
dimensions $(2m-N)\times(N-m)$.

We allow motion out of the sub-manifold by allowing $a$ to vary
further,
\begin{equation}
a =  U^{(m)}\left( \begin{array}{c}
        0       \\
        \bar{a} \end{array}     \right).
\end{equation}
The $2(N-m)^2$ degrees of freedom in $\bar a$ complement the
$2(N-m)(2m-N)$ degrees of freedom inherent in $U^{(m)}$ to give
$2m(N-m)$, the dimensionality of the entire $U(N)/[U(m)\times
U(N-m)]$ coset space.

Writing $\sigma$ with the generalized $a$ we have
\begin{equation}
\sigma= U \left( \begin{array}{ccc}
\bm1_{2m-N} &   0   &   0   \\[2pt]
0   & \cos\left(2\sqrt{\bar{a}\bar{a}^{\dag}}\right) &
-\bar{a}\,\frac{\displaystyle\sin\left(2\sqrt{\bar{a}^{\dag}\bar{a}}\right)}
    {\displaystyle\sqrt{\bar{a}^{\dag}\bar{a}}} \\[2pt]
0 &
-\frac{\displaystyle\sin\left(2\sqrt{\bar{a}^{\dag}\bar{a}}\right)}
{\displaystyle\sqrt{\bar{a}^{\dag}\bar{a}}}\,\bar a^\dag    &
-\cos\left(2\sqrt{\bar{a}^{\dag}\bar{a}}\right)
\end{array} \right)
    U^{\dag},
\end{equation}
with
\begin{equation}
U = \left( \begin{array}{cc}
        U^{(m)}     & 0     \\
        0       & \bm1_{N-m}       \end{array}     \right).
\end{equation}
We can also write this as
\begin{equation}
\sigma = U \left( \begin{array}{cc}
    \bm1_{2m-N} & 0 \\
    0   & \sigma^{[2(N-m)]}(\bar{a},\bar{a}^{\dag})
\end{array} \right) U^{\dag}.
\end{equation}
$\sigma^{[2(N-m)]}$ is a matrix in the manifold
$U(2(N-m))/[U(N-m)\times U(N-m)]$. Indeed for $\bar{a}=(\pi/2)
\bm1_{N-m}$, we have
\begin{equation}
\sigma^{[2(N-m)]} = \left( \begin{array}{cc}
    -\bm1_{N-m} &   0   \\
    0   &   \bm1_{N-m}  \end{array} \right) \equiv -\bar{\Lambda}.
\end{equation}
Since $U(2(N-m))/[U(N-m)\times U(N-m)]$ is a self-conjugate
manifold, its structure near $\sigma^{[2(N-m)]}=-\bar\Lambda$ is
the same as its structure near $\sigma^{[2(N-m)]}=\bar\Lambda$,
which corresponds to $\bar{a}=0$. Expanding $\sigma^{[2(N-m)]}$
around $-\bar\Lambda$ gives \Eq{eq:Sig_1}.
\clearpage
\chapter[Coefficients in the dispersions of Goldstone bosons]{Coefficients in the dispersion relations \label{app:constants} of Goldstone bosons}

Here we present the various coefficients that appear in the dispersion relations given by \Eq{poles}. These are $c$, $c_1$, $a$, and $b$ found in Eqs.~(\ref{case2a})--(\ref{case3}). These coefficients are extracted from the nearest-neighbor theory, and depend on the number $N-m$ only. 

The constants $c$ and $c_1$ are the zero momentum curvatures of the following functions,
\begin{eqnarray}
c&=&\left(\frac{d\sqrt{\Sigma^2_{1}-\Sigma^2_{2}}}{d(\bk^2)}\right)_{\bk=0}, \\
c_1&=&-\left(\frac{d\Sigma_{1}}{d(\bk^2)}\right)_{\bk=0}.
\end{eqnarray}
$a$ and $b$ are defined as
\begin{eqnarray}
a&=&\frac{N_c}{6J_1}\frac{c_1^2-c^2}{2c_1c^2}, \\
b&=&\frac{N_c}{6J_1}\frac{2}{c_1}.
\end{eqnarray}

We present the constants in Table~\ref{table:constants}.
\newpage
\begin{table}[htb]
\caption[The coefficients in the dispersion relations of Goldstone bosons]{The constants $c$, $c_1$, $a$, and $b$, for all values of $N-m$\label{table:constants}. Recall that for $N-m=1$, the solution of the integral equation is different from the other cases, and it is possible that the nearest-neighbor spectrum follows a power of momentum higher than $2$. This is the source of the sizeable differences between this case and the other cases.}
\begin{center}
\begin{tabular}{c|cccc} 
$N-m$  & $c$ & $c_1$ & $a$ & $b$ \\
\hline
1 & $8.3\times 10^{-5}$ & $0.12$ & $8.4\times 10^6$ & $17.1$ \\
2 & $0.002$ & $0.01$ & $1.2\times 10^3$ & $187$  \\
3 & $0.003$ & $0.009$ & $490$ & $220$   \\
4 & $0.003$ & $0.008$ & $350$ & $240$  \\ 
5 & $0.003$ & $0.008$ &  $280$ & $250$  \\ 
\end{tabular}
\end{center}
\end{table}

Here we note that the case of $N-m=1$ has an exceptionally large value for $a$, with a very small curvature $c$. This is a result of the different behavior of the solution to the integral equation for this case [see Fig.~\ref{fig:tanh}]. This behavior should disappear if we take a better momentum mesh, especially near $\bk=0$.
\clearpage
\chapter{Classification of fields according to unbroken subgroups \label{app:clasification}}

In this section we consider the $\sigma$ fields that represent
Goldstone bosons of the effective theory.
The $\sigma$ multiplet contains fields called $\pi$ that are type I Goldstone
bosons in the nearest-neighbor theory, and fields called $\chi$ that are type II bosons there.
Moreover, the antiferromagnetic nature of the theory leads us to consider
separately the fields on the even and odd sites of the lattice.

The next-nearest-neighbor theory breaks its chiral $SU(N_f)\times SU(N_f)$ symmetry
spontaneously as shown in Table \ref{table1}.
The following two tables show
the classification of the various fields according to their representations and
charges under the unbroken groups.
\begin{table}[htb]
\caption[Classification of $\sigma$ fields on the even sites]{Classification of $\sigma$ fields on the {\em even\/} sites under the
unbroken symmetry for $3N/4\le m<N$.
As indicated in \Eq{chi}, not all $\chi$ fields are present for a
given combination of $N_f$ and $B$. We neglect $U(1)_A$.
All fields are complex.
\label{bigtable1}}
{
  \renewcommand{\arraystretch}{1.2}
\begin{tabular}{cccccc}
&&&&&Symmetry rep\\
$N_f$&$B$&Symmetry&Field&Matrix structure&or $U(1)$ charge(s)\\
\hline
1&1&$\emptyset$&$\pi,\chi_1$&$1\times1$&-\\
\hline
2&2&$SU(2)$&$\pi$&$2\times2$&${\bf 1}\oplus{\bf 3}$\\
&&&$\chi_{1}$ &   $4\times2$  &  ${\bf 1}\oplus{\bf 1}\oplus{\bf 3}\oplus{\bf 3}$   \\
\hline
2&3&$U(1)_{I_3}$&$\pi$    &   $1\times1$   &  $+1$    \\
&&&$\chi_{1e}$ &   $4\times1$  &  $\left(\begin{array}{c} +1 \\ 0 \\ +1 \\ 0 \end{array} \right)$\\
&&&$\chi_{2e}$&   $2\times1$  & $\left(\begin{array}{c} +1 \\ 0 \end{array} \right)$   \\
\hline
3&3&$SU(3)$&$\pi$&$3\times3$&${\bf 1}\oplus{\bf 8}$\\
&&&$\chi_{1}$ &   $9\times2$  &  ${\bf 1}\oplus{\bf 1}\oplus{\bf 8}\oplus{\bf 8}$   \\
\hline
3&4&$U(1)_{I_3}\times U(1)_{Y}$&$\pi$&$2\times2$&
$\left(\begin{array}{cc}
(+1/2,-3)&(0,0)\\
(+1,0)&(+1/2,+3)
\end{array}\right)$\\
&&&$\chi_{1e}$&$6\times2$&
$\left(\begin{array}{cc}
(+1,0)&(+1/2,+3) \\
(0,0)&(-1/2,+3) \\
(+1/2,-3)&(0,0) \\
(+1,0)&(+1/2,+3) \\
(0,0)&(-1/2,+3) \\
(+1/2,-3)&(0,0)
\end{array}\right)$\\
&&&$\chi_{2e}$ &$2\times2$&
$\left(\begin{array}{cc}
(+1,0)&(+1/2,+3)\\
(0,0)&(-1/2,+3)
\end{array}\right)$\\
\hline
3&5&$U(1)_{I_3}\times U(1)_{Y}$&$\pi$&$1\times1$&$(-1/2,+3)$\\
&&&$\chi_{1e}$&$6\times1$&
$\left(\begin{array}{c}
(+1/2,+3)\\
(-1/2,+3)\\
(0,0)\\
(+1/2,+3)\\
(-1/2,+3)\\
(0,0)
\end{array}\right)$\\
&&&$\chi_2$&$4\times1$&
$\left(\begin{array}{c}
(+1/2,+3)\\
(-1/2,+3)\\
(0,0)\\
(+1/2,3)
\end{array}\right)$
\end{tabular}
}
\end{table}

\begin{table}[htb]
\caption[Classification of $\sigma$ fields on the odd sites]{Classification of $\chi$ fields on the {\em odd} sites. The classification of the odd $\pi$ fields is the same as of the even $\pi$ fields for all values of $N_f$ and $B$. For $N_f=B=1,2,3$, the classification of the odd $\chi$ fields is the same as of the even $\chi$ fields. Here we present the classification of the odd $\chi$'s for the other cases.
All fields are complex.
\label{bigtable2}}
\vskip 1cm
{
  \renewcommand{\arraystretch}{1.5}
\begin{tabular}{cccccc}
$N_f$&$B$&Symmetry&Field&Matrix structure&$U(1)$ charge(s)\\
\hline
2&3&$U(1)_{I_3}$&$\chi_{1o}$ &   $4\times1$  &  $\left(\begin{array}{c} 0 \\ -1 \\ 0 \\ -1 \end{array} \right)$  \\
&&&$\chi_{2o}$ &   $2\times1$  & $\left(\begin{array}{c} 0 \\ -1 \end{array} \right)$   \\
\hline
3&4&$U(1)_{I_3}\times U(1)_Y$&$\chi_{1o}$&$6\times2$&
$\left(\begin{array}{cc}
(+1/2,+3)&(0,0) \\
(-1/2,+3)&(-1,0) \\
(0,0)&(-1/2,-3) \\
(+1/2,+3)&(0,0) \\
(-1/2,+3)&(-1,0) \\
(0,0)&(-1/2,-3)
\end{array}\right)$\\
&&&$\chi_{2o}$ &$2\times2$&
$\left(\begin{array}{cc}
(+1/2,+3)&(0,0)\\
(-1/2,+3)&(-1,0)
\end{array}\right)$\\
\hline
3&5&$U(1)_{I_3}\times U(1)_Y$&$\chi_{1o}$&$6\times1$&
$\left(\begin{array}{c}
(+1,0)\\
(0,0)\\
(+1/2,-3)\\
(+1,0)\\
(0,0)\\
(+1/2,-3)
\end{array}\right)$\\
&&&$\chi_{2o}$&$4\times1$&
$\left(\begin{array}{c}
(+1,0)\\
(0,0)\\
(+1/2,-3)\\
(+1,0)
\end{array}\right)$
\end{tabular}
}
\end{table}

\clearpage

\pagestyle{plain}
\chapter*{Acknowledgments\markboth{Acknowledgments}{Acknowledgments}}
\addcontentsline{toc}{chapter}{Acknowledgments}


It is my pleasure to thank my advisor Prof.~Benjamin Svetitsky, who suggested the idea for this work, supported and guided me continuously during the research, and proofread this manuscript. Ben, with his wide knowledge of physics and deep understanding of quantum field theory, made this research an inspiring experience. He dedicated a lot of time in countless enlightening discussions, in which he shared his ideas. For that I am grateful. 

I have benefited greatly from fruitful discussions with Dr.~Yigal Shamir who was always willing to teach me more about lattice gauge theory.

At different stages of the work I had useful discussions with  Prof.~A.~Aharony, Prof.~A.~Auerbach, Prof.~O.~Entin-Wohlman, Prof.~L.~Frankfurt, Prof.~S.~Nussinov, Prof.~K.~Rajagopal, Prof.~S.~Sachdev, and Dan Gluck.

Finally I would like to thank all my colleagues and friends at Tel Aviv University for the enjoyable working and social atmosphere.

This work was supported by the Israel Science
Foundation under grant no.~222/02-1 and by the Tel Aviv University Research Fund.

\end{document}